\documentclass{aa}  

\usepackage{bm}
\usepackage{subfig}
\usepackage{rotating}
\usepackage{amssymb,amsmath}
\usepackage{graphicx}
\usepackage{txfonts}
\usepackage{color}

\begin{document} 

        \titlerunning{Auto-RSM: an automated parameter-selection framework for the RSM map}
   \title{Auto-RSM: an automated parameter-selection algorithm for the RSM map exoplanet detection algorithm}

   \author{C.-H. Dahlqvist\inst{1}, F. Cantalloube\inst{2}, \and O. Absil\inst{1}}

   \institute{STAR Institute, Universit\'{e} de Li\`{e}ge, All\'{e}e du Six Ao\^{u}t 19c, 4000 Li\`{e}ge, Belgium\\
              \email{carl-henrik.dahlqvist@uliege.be} \and Aix Marseille Univ, CNRS, CNES, LAM, Marseille, France}
   \date{}

\abstract
   {Most of the high-contrast imaging (HCI) data-processing techniques used over the last 15 years have relied on the angular differential imaging (ADI) observing strategy, along with subtraction of a reference point spread function (PSF)  to generate exoplanet detection maps. Recently, a new algorithm called regime switching model (RSM) map has been proposed to take advantage of these numerous PSF-subtraction techniques; RSM uses several of these techniques to generate a single probability map. Selection of the optimal parameters for these PSF-subtraction techniques as well as for the RSM map is not straightforward, is time consuming, and can be biased by assumptions made as to the underlying data set. }
   {We propose a novel optimisation procedure that can be applied to each of the PSF-subtraction techniques alone, or to the entire RSM framework.}
   {The optimisation procedure consists of three main steps: (i) definition of the optimal set of parameters for the PSF-subtraction techniques using the contrast as performance metric, (ii) optimisation of the RSM algorithm, and (iii) selection of the optimal set of PSF-subtraction techniques and ADI sequences used to generate the final RSM probability map.}
   {The optimisation procedure is applied to the data sets of the exoplanet imaging data challenge (EIDC), which provides tools to compare the performance of HCI data-processing techniques. The data sets consist of ADI sequences obtained with three state-of-the-art HCI instruments: SPHERE, NIRC2, and LMIRCam. The results of our analysis demonstrate the interest of the proposed optimisation procedure, with better performance metrics compared to the earlier version of RSM, as well as to other HCI data-processing techniques.}
   {}


   \maketitle
%
%-------------------------------------------------------------------

\section{Introduction}
\label{sec:intro}

In the study of exoplanets, direct imaging provides   information that is complementary to indirect detection techniques, allowing wider orbits to be explored in greater detail, thus giving access to photometric and spectral features of young massive self-luminous companions. High-contrast imaging (HCI) is however one of the most challenging exoplanet detection techniques, as it requires large telescopes, advanced adaptive optics, coronagraphs, and sophisticated data processing to disentangle the faint planetary signal from the bright host star. Besides the large flux ratio between the companion and the host star, HCI techniques have to deal with residual noise due to quasi-static speckles, which originate from the optical train of the instruments or from uncorrected atmospheric turbulence. 

In the last decade, the field of HCI has been very active and a large number of data processing techniques have been developed to detect and characterise planetary candidates. The most common approach combines the angular differential imaging \citep[ADI,][]{Marois06} observing strategy with subtraction of a reference point spread function (PSF). While the ADI observing strategy relies on angular diversity to better average out quasi-static speckles, subtraction of a reference PSF  increases the sensitivity of HCI by modelling and subtracting quasi-static speckles from the original set of frames. The most popular PSF-subtraction methods include ADI median-subtraction, locally optimised combination of images \citep[LOCI,][]{Lafreniere07}, principal component analysis \citep[PCA/KLIP,][]{Soummer12,Amara12}, non-negative matrix factorisation \citep[NMF,][]{Ren18}, and the local low rank plus sparse plus Gaussian decomposition \citep[LLSG,][]{Gonzalez16}. Other algorithms such as ANDROMEDA \citep{Cantalloube15}, KLIP FMMF \citep{Pueyo16,Ruffio17}, PACO \citep{Flasseur18}, and TRAP \citep{Samland21} exploit the inverse problem approach,  which, for HCI,  consists in tracking a model of the expected planetary signal in the set of frames included in the ADI sequence. All these methods rely on signal-to-noise ratio (S/N) maps to detect planetary candidates. These S/N maps are computed in a separate step for PSF-subtraction methods \citep[using algorithms such as those described in][]{Mawet14,Bottom17,Pairet19}, or are generated as a by-product of the algorithm for inverse problem approaches. Recently, a new PSF-subtraction-based approach was proposed to take better advantage of the numerous existing PSF-subtraction techniques. The regime-switching model (RSM) map \citep{Dahlqvist20} replaces the estimation of the S/N map by the computation of a probability map generated using one or several PSF-subtraction techniques. The most recent version of the RSM map \citep{Dahlqvist21} accommodates up to six different PSF-subtraction techniques, including two techniques relying on the forward modelling of an off-axis point source PSF.

The use of one or several PSF-subtraction techniques to generate a S/N map or a probability map via the RSM algorithm, requires the definition of multiple parameters specific to each method and potentially varying  from one ADI sequence to another. The selected set of parameters can have dramatic effects on the final detection map, both in terms of noise and algorithm throughput. The selection of the optimal set of parameters is usually done manually, which requires time and can lead to bias, as the definition of the optimality of the set is driven by the ability of the user to properly  analyse the generated detection maps. The complexity of the optimal parameter selection can be an obstacle to the use of some HCI data-processing techniques, and can lead to unreliable results. This also makes it difficult to properly
compare the performance of HCI data-processing techniques, as their performance is parameter-driven to a large extent, and therefore depends on subjective choices made by the user. To mitigate these issues in the context of the RSM framework, we propose an optimisation procedure called auto-RSM to automatically select the best set of parameters for the PSF-subtraction techniques, as well as for the RSM algorithm itself. To our knowledge, such an extensive optimisation procedure has not yet been proposed in the HCI literature, although some earlier works have already partly addressed the question of parameter optimisation. Approaches such as S/N-based optimisation of the number of components for PCA \citep{Gonzalez17} or direct optimisation of the non-linear S/N function \citep{Tompson21} focus on a single PSF-subtraction technique, whereas here we proposed a more generic framework applicable to most PSF-subtraction techniques. 

The proposed optimisation framework can be divided into three main steps: (i) selection of the optimal set of parameters for the different PSF-subtraction techniques (and ADI sequences) via Bayesian optimisation, (ii) optimal parametrisation of the RSM algorithm, and (iii) selection of the optimal set of PSF-subtraction techniques (and ADI sequences) to be considered for the computation of the final RSM probability map. This last step is motivated by the fact that, when relying on multiple PSF-subtraction techniques and multiple ADI sequences, some sequences may be noisier while some methods may better cope with the noise independently of their parametrisation. Special attention should therefore be paid to the choice of the cubes of residuals generating the final probability map. The optimisation step for the PSF-subtraction technique parameters is not limited to auto-RSM, and can be performed separately if a S/N map is preferred to the RSM probability map. In addition to the development of auto-RSM, we therefore propose a variant of the algorithm adapted to the use of S/N maps instead of RSM probability maps. Auto-S/N relies on the first step of the auto-RSM algorithm, and on a modified selection framework allowing the optimal combination of multiple S/N maps. 

The performance of both optimisation procedures is assessed using the \textit{exoplanet imaging data challenge} (EIDC), which regroups ADI sequences generated by three state-of-the-art HCI instruments: SPHERE, NIRC2, and LMIRCam. The aim  of the EIDC initiative is to provide tools, i.e., the data sets and performance metrics, to properly compare the various HCI data processing techniques that have been developed recently. The EIDC first phase, which ended in October 2020, regrouped and compared the results from 23 different submissions for the ADI subchallenge \citep{Cantalloube20}, making it a great tool to assess the performance of a new approach.

The remainder of this paper is organised as follows. Section 2 describes the procedure used to optimise the parameters of the PSF-subtraction techniques. In Sect.~3, we introduce the RSM map framework and present the next two steps of the auto-RSM framework: the optimal RSM parameter selection, and the selection of the optimal set of cubes of residuals used to generate the final probability map. Section 4 is devoted to the performance assessment of the optimisation procedure along with the comparison of different versions of the optimisation procedure. Finally, Sect.~5 concludes this work.

\section{PSF-subtraction techniques optimisation}
\label{sec:model}

The proposed optimisation procedure relies on the concept of inverted parallactic angles, which have already been used in the HCI literature \citep[e.g.][]{Gonzalez18,Pairet19}. The sign of the parallactic angles used to de-rotate the ADI sequence is flipped, blurring any planetary signal while preserving the noise temporal correlation and statistical properties. The use of ADI sequences with flipped parallactic angles should allow us to avoid biases due to the contribution from potential planetary signals during the optimisation process. Although the inverted parallactic angles approach allows us to blur planetary signals, it is not immune to potential bright artefacts, which implies that particular attention needs to be paid to the elimination of outliers from the computed optimal parameters.

As mentioned in Sect. 1, the RSM map relies on one or several PSF-subtraction techniques to generate a final probability map. Six different PSF-subtraction techniques are currently used with the RSM map: LOCI, annular PCA \citep[APCA,][]{Gonzalez17}, KLIP, NMF, LLSG, and forward-model versions of KLIP and LOCI \citep[see ][for more details]{Dahlqvist21}. Each PSF-subtraction technique is characterised by its own set of parameters, which strongly affect the quality of the reference PSF modelling. Table~\ref{Parameters} presents the parameters that we have identified as the most relevant for the optimisation of the six considered PSF-subtraction techniques. Other parameters, such as the annulus width, were tested during the auto-RSM development, but were discarded from the optimisation framework as their influence was found to be smaller or because of other practical considerations.

\begin{table*}[t]
                        \caption{Set of parameters selected for the optimisation of the six considered PSF-subtraction techniques.}
                        \label{Parameters}
\centering

                        \begin{tabular}{lcccccc}
                        
                        \hline
Parameters &\textbf{APCA}  & \textbf{NMF}  & \textbf{LLSG}& \textbf{LOCI}& \textbf{KLIP-FM}& \textbf{LOCI-FM} \\                           
 \hline
Number of principal components & X & X & X & & X & \\
FOV minimal rotation & X &  &  & X & X & X\\
Number of azimuthal segments & X & & X &  &  & \\
Tolerance for error minimisation &  &  &  & X &  & X\\
\hline
                        \end{tabular}
                                \end{table*}

\subsection{Definition of the loss function}

Parameter optimisation requires the definition of a loss function $f$, which provides, for a given set of parameters $\bm{p}$, an outcome $f(\bm{p})$ that can be maximised or minimised. In the case of reference PSF modelling, the loss function should quantify the ability of the PSF-subtraction technique to remove the residual noise contained in the ADI sequence and to identify potential planetary companions. The definition of the achievable planet/star flux ratio or contrast, for a given detection significance, is therefore a good candidate to measure the PSF-subtraction technique performance. In the context of HCI, the contrast is defined as follows \citep{Jensen_Clem_2017}:

\begin{eqnarray}
\label{contrast}
\text{contrast}=\left( \frac{\text{factor}\times\text{noise} }{\text{stellar aperture photometry}}\right)\left( \frac{1}{\text{throughput}}\right) .
\end{eqnarray}

\begin{figure}[t]
\begin{center}
\includegraphics[width=200pt]{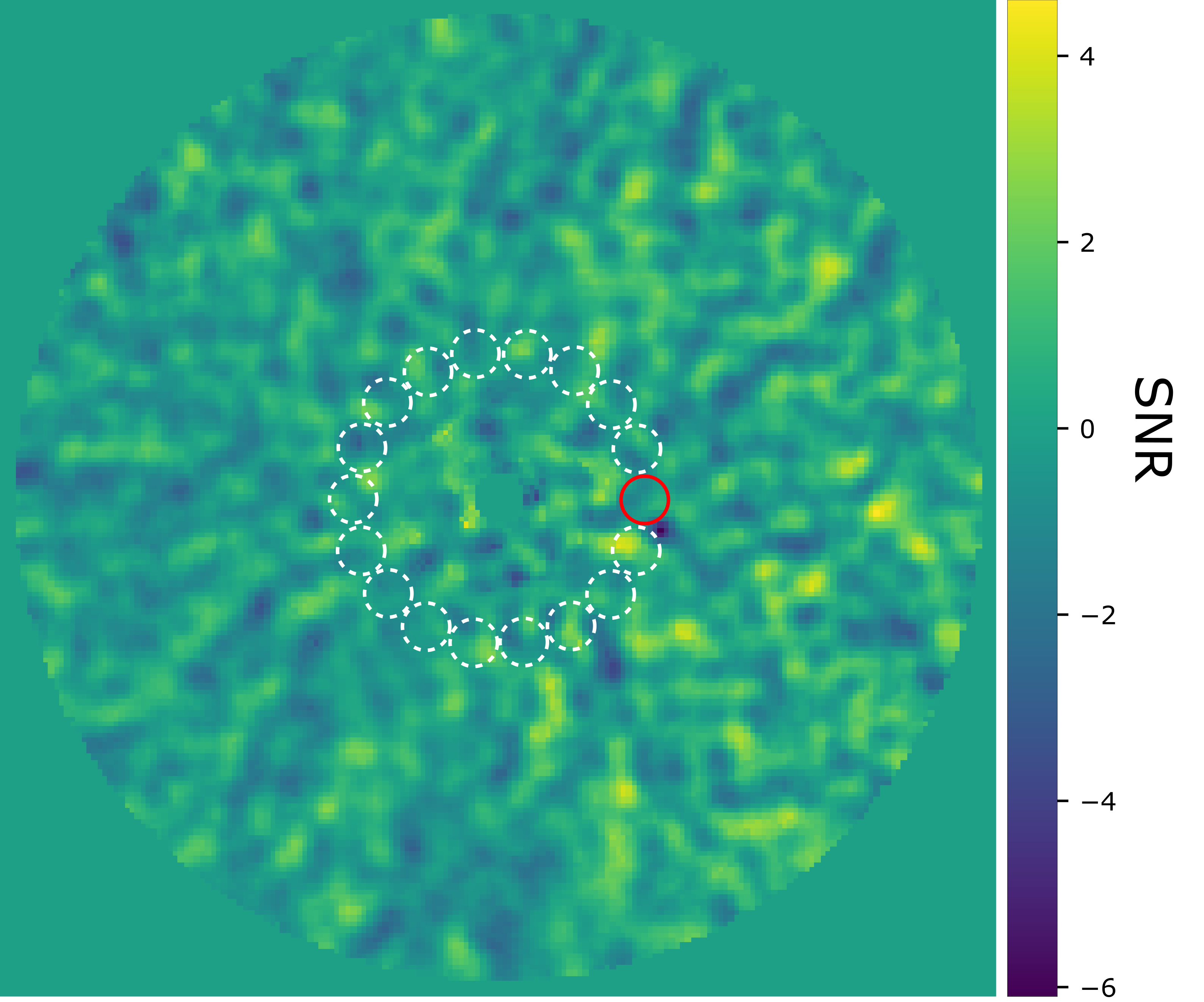} 
                        \caption{\label{noise_esti} Estimation of the noise via the annulus-wise procedure proposed by \citet{Mawet14}. The dotted white circles indicate the apertures whose flux is used for the noise computation, while the red circular region is centred on the pixel for which the noise needs to be estimated.}
                        \end{center}
                        
                          \end{figure}

The contrast is usually defined at a 5 $\sigma$ level which implies $\text{factor}=5$ with $\text{noise}=\sigma$. As the parameter optimisation is done for a single ADI sequence at a time, the stellar aperture photometry does not impact the optimisation process and is therefore irrelevant and set to 1. We rely on the procedure of \citet{Mawet14} illustrated in Fig.\ref{noise_esti} to determine the noise annulus-wise.  For a given annulus and for an aperture centred on the pixel of interest, with a diameter equal to the full width at half maximum (FWHM) of the PSF, the noise is computed by considering the standard deviation of the fluxes in all the non-overlapping apertures (one FWHM in diameter each) included in this annulus. The number of apertures being relatively small for small angular separations, the procedure implements a small statistics correction, relying on a Student t-test to correct the multiplicative factor for the noise.

The throughput quantifies the attenuation of the planetary signal due to reference PSF subtraction. In practice, the throughput is estimated by injecting a fake companion at a predefined position and computing the ratio between the injected aperture flux and the recovered aperture flux after the reference PSF subtraction. Contrast curves can be computed by averaging the sensitivity limit in terms of planet-to-star contrast obtained by injecting several fake companions at different position angles for a series of angular separations. Relying on several azimuthal positions and averaging the associated contrasts reduces the impact of the residual speckles on the estimated contrast. We follow this approach, but instead of injecting individual fake companions separately to compute the average contrast, we inject several fake companions at once, which drastically reduces the computation time. We impose a minimum separation of one FWHM between the apertures containing the fake companions, and a maximum of eight fake companions per annulus, in order to limit potential cross-talk between the injected fake companions. This safety distance, as well as the small intensity of the injected fake companions\footnote{Following the methodology of \cite{Gonzalez17}, the intensity of the injected companions represents only a few percent of the pixel intensity within the ADI sequence for a given annulus, which limits the impact of the multiple injections on the estimation of the reference PSF.}, provides a good approximation of the average contrast (see Appendix \ref{multifc} for a comparison between sequential and multiple injections) while limiting the computation time, which is crucial here as parameter optimisation requires a large number of contrast estimations. The loss function computation may be summarised as follows:

\begin{enumerate}
\item Reference PSF estimation using the selected post-processing technique and set of parameters;
\item reference PSF subtraction from the original ADI sequence, de-rotation of the cube of residuals, and median combination of the resulting frames;
\item computation of the fluxes for the entire set of apertures within the selected annulus in the median-combined frame obtained in step 2, and estimation of the noise relying on a Student t-test;
\item injection of fake companions at the selected set of azimuths with flux value defined as five times the noise computed in step 3;
\item computation of the cube of residuals for the ADI sequence containing the fake companions and median combination;
\item computation of the throughputs by comparing the aperture flux of the injected companions to that of the retrieved companions after PSF subtraction (difference between final frame of step 5 and step 2);
\item estimation of the contrasts via Eq.~\ref{contrast} and computation of the average contrast.
\end{enumerate}

\subsection{Parameter selection via Bayesian optimisation}

The NMF and LLSG PSF-subtraction techniques have integer parameters that are, in practice, restricted to a small range of possible values. One can therefore easily select their optimal parameters by going through their entire parameter space, and simply applying steps 1-7 to compute the contrast for each set of parameters. The optimal set of parameters is the one that minimises the contrast. However, for the other PSF-subtraction techniques, part of the parameter space is continuous, which prevents exploration of the entire parameter space. A more advanced minimisation algorithm is therefore needed. The derivatives and convexity properties of our loss function are unfortunately unknown. However, it is expected that our loss function, i.e. the function describing the evolution of the annulus-wise contrast in terms of the selected parameters, is non-convex and most probably non-linear. This implies that we cannot rely on mainstream minimisation approaches (e.g. Newton-Conjugate-Gradient algorithm or Nelder-Mead Simplex algorithm). In addition, evaluating the  annulus-wise contrast is expensive, because of the numerous steps involved in its estimation. We therefore cannot simply rely on Monte Carlo simulation or random searches to explore the parameter space.

Considering all these constraints, we decided to rely on Bayesian optimisation to select the optimal set of parameters for the remaining PSF-subtraction techniques. Bayesian optimisation is a powerful strategy to limit the number of loss function evaluations needed to reach an extremum \citep[see, e.g. ][]{Mockus78,Jones98}. This strategy belongs to a class of algorithms called sequential model-based optimisation (SMBO) algorithms. This class of algorithms uses previous observations of the loss function to determine the position of the next point inside the parameter space to be evaluated. It is called Bayesian optimisation because it relies on Bayes' theorem to define the posterior probability of the loss function, on which the sampling strategy  is based. Bayes' theorem states that the posterior probability associated with a model, given a set of observations, is proportional to the likelihood of the observations given the model, multiplied by the prior probability of the model:

\begin{eqnarray}
\label{bayesian1}
P(f\mid\mathcal{O}_{1:t}) \propto P(\mathcal{O}_{1:t}\mid f)P(f),
\end{eqnarray}
where $f$ is the loss function to optimise and $\mathcal{O}_{1:t}=\left\lbrace \bm{p}_{1:t}, f(\bm{p}_{1:t})\right\rbrace$ is the set of observations of the loss function, with $\bm{p}_{1:t}$ being the set of tested points (see Appendix \ref{paramdesc} for a summary of all the mathematical notions used throughout the paper). In the case of Bayesian optimisation, we assume a Gaussian likelihood with noise, as follows:

\begin{eqnarray}
\label{bayesian2}
 P(\mathcal{O}_{1:t}\mid f) \sim \mathcal{N}(f(\bm{p}),\sigma^2_{\epsilon}),
\end{eqnarray}
where $\mathcal{O}_{t} = f(\bm{p}_{t}) +\epsilon$ with  $\epsilon \sim \mathcal{N}(0,\sigma^2_{\epsilon})$.\\

Regarding the prior distribution for our loss function, \cite{Mockus94} proposed relying on a Gaussian process (GP) prior as this induces a posterior distribution over the loss function that is analytically tractable\footnote{This implies that it is possible to update the posterior probability with the observations made with a new set of parameters. This will help us to create a continuous function to select the next point to sample in the parameter space.}. A GP is the generalisation of a Gaussian distribution to a function, replacing the distribution over random variables by a distribution over functions. A GP is fully characterised by its mean function $m(\bm{p})$, and its covariance function $\bm{K}$, where

\begin{eqnarray}
\label{bayesian3}
f(\bm{p}_{1:t}) \sim \mathcal{GP}\left( m(\bm{p}_{1:t}),\bm{K}\right) \sim \mathcal{N}( m(\bm{p}_{1:t}),\bm{K}).
\end{eqnarray}

The GP process can be seen as a function that returns the mean and variance of a Gaussian distribution over the possible values of $f$ at $\bm{p}$, instead of returning a scalar $f(\bm{p})$. We make the assumption that the prior mean is the zero function $m(\bm{p})=0$ and we select a commonly used covariance function, the squared exponential function:

\begin{eqnarray}
\label{bayesian4}
\left[ \bm{K}\right]_{i,j}= k(\bm{p}_i,\bm{p}_j)= \exp \left( - \frac{1}{2 l^2} \parallel \bm{p}_i-\bm{p}_j\parallel ^2\right), 
\end{eqnarray}
where $l$ is the length scale of the kernel.\\

Having documented the posterior probability computation for our loss function, we need to define a sampling strategy. Bayesian optimisation relies on an \textit{acquisition function} to define how to sample the parameter space. This function is based on the current knowledge of the loss function, i.e. the posterior probability. The acquisition function is a function of the posterior distribution over the loss function $f$, which provides a performance metric for all new sets of parameters. The set of parameters with the highest performance is then chosen as the next point of the parameter space to be sampled. A popular acquisition function is the expected improvement \citep[EI, ][]{Mockus78,Jones98} which is defined as follows:

\begin{eqnarray}
\label{bayesian5}
\text{EI}(\bm{p}_{t+1})=\mathbb{E} \left[ \max \left\lbrace 0,f(\bm{p}_{t+1})-f(\widehat{\bm{p}})\right\rbrace \right] ,
\end{eqnarray}
where $\mathbb{E}$ is the expected value and $\widehat{\bm{p}}=argmax_{\bm{p}_i \in \bm{p}_{1:t}} f(\bm{p}_i)$ is the current optimal set of parameters. We see that in the case of Bayesian optimisation, we look for the maximum value of the loss function. As we are trying to minimise the contrast for a given set of parameters, we simply define our loss function $f(\bm{p})$ as the inverse of the contrast averaged over the selected set of azimuths (see Section 2.1).

An interesting feature of the EI is that it can be evaluated analytically under the GP model, yielding (see Appendix \ref{gpmodel} for more details about the derivation of these expressions)

\begin{eqnarray}
\label{bayesian6}
&&\text{EI}(\bm{p}_{t+1})= \nonumber \\
  &&\begin{cases}
    (\mu(\bm{p}_{t+1})-f(\widehat{\bm{p}}))\Phi(Z) +\sigma(\bm{p}_{t+1})\phi(Z) & \quad \text{if } \sigma(\bm{p}_{t+1})>0\\
    0      & \quad \text{if } \sigma(\bm{p}_{t+1})=0\\
  \end{cases},
\end{eqnarray}
where $\Phi(Z)$ and $\phi(Z)$ are respectively the cumulative distribution and probability density function of the Gaussian distribution, $\mu(\bm{p}_{t+1})$ and $\sigma(\bm{p}_{t+1})$ are the mean and variance of the Gaussian posterior distribution, and $Z=\left[ \mu(\bm{p}_{t+1})-f(\widehat{\bm{p}})\right] /\sigma(\bm{p}_{t+1})$. We see from this last expression that the EI is high either when the expected value of the loss $\mu(\bm{p})$ is larger than the maximum value of the loss function $f(\widehat{\bm{p}})$ or when the uncertainty $\sigma(\bm{p}_{t+1})$ around the selected set of parameters $\bm{p}_{t+1}$ is high. The EI approach aims to minimise the number of function evaluations by performing a trade-off between exploitation and exploration  at each step. The EI exploits the existing set of observations by favouring the region where the expected value of $f(\bm{p}_{t+1})$ is high, while it also explores unknown regions where the uncertainty associated with the loss function is high. 

Bayesian optimisation starts with the initialisation of the posterior probability by estimating the loss function for several sets of parameters via random search in the parameter space. Once this initial population of observations is computed, the rest of the algorithm can be summarised as follows.

\begin{itemize}
\item Based on the GP model, use random search to find the $\bm{p}_{t+1}$ that maximises the EI, $\bm{p}_{t+1}= argmax \left[ \text{EI}(\bm{p}_{t+1}) \right] $;
\item compute the contrast for the new set of parameters $\bm{p}_{t+1}$;
\item update the posterior expectation of the contrast function using the GP model (see Appendix \ref{gpmodel});
\item repeat the previous steps for a given number of iterations.
\end{itemize}

The number of random searches to compute the initial GP and the number of iterations for the Bayesian optimisation depend on the size of the parameter space associated with the considered PSF-subtraction techniques. A specific number of random searches and iterations are therefore selected for each PSF-subtraction technique. At the end of the Bayesian optimisation, the minimal average contrast for a given annulus $a$ and PSF-subtraction technique $m$ is stored in a matrix element $C_{a,m}$, along with the set of parameters $\bm{p}$ in another matrix $P_{a,m}$.

This first step of the auto-RSM algorithm may be used outside the RSM framework, allowing the production of S/N maps based on the cubes of residuals generated by optimised PSF-subtraction techniques. A S/N-based version of the auto-RSM framework called auto-S/N has been developed and is presented in Appendix \ref{autosn}. Auto-S/N optimally combines S/N maps computed from the cubes of residuals generated by the optimised PSF-subtraction techniques, relying on the same greedy approach as for auto-RSM (see Sect.~\ref{opticombi}). The performance of auto-S/N is assessed in Appendix \ref{autosn}.2 using the same metrics as for auto-RSM (see Sect.~\ref{pa}). The lower performance of auto-S/N implies that auto-RSM should preferred, despite its longer computation time, although the two approaches can be complementary to some extent.

\section{RSM map optimisation}

\subsection{RSM map principles}

The RSM algorithm relies on a two-state Markov chain to model the pixel intensity evolution inside one or multiple de-rotated cubes of residuals generated by PSF subtraction techniques using ADI or SDI observing strategies. The cubes of residuals are treated annulus-wise to account for the radial evolution of the residual speckle noise statistics. For each angular separation $a$, a residual time-series, $\bm{x}_{i_a}$, is built by vectorising the set of patches centred on the annulus of radius $a$, first along the time axis and then the spatial axis. The index $i_a \in \{1,\dots, T \times L_a \}$ gives the position of the considered patches within the cube of residuals, where $L_a$ and $T$ are respectively the number of pixels in the annulus of radius $a$ and the number of frames in the residuals cube. In the case of multiple PSF-subtraction techniques and ADI sequences of the same object, $T$ is defined as the sum of the number of frames per cube multiplied by the number of considered PSF-subtraction techniques. 

The set of patches $\bm{x}_{i_a}$ is described by a probability-weighted sum of the outcomes of two regimes: a noise regime, $S_{i_a}=0$, where $\bm{x}_{i_a}$ is described by the statistics of the quasi-static speckle residuals contained in the annulus; and a planetary regime, $S_{i_a}=1$, where $\bm{x}_{i_a}$ is described by both the residual noise and a planetary signal model\footnote{The off-axis PSF or forward-modelled PSF when using LOCI and KLIP PSF-subtraction techniques.}. The two regimes are characterised by the following equations:

\begin{eqnarray}
\label{maineq}
\bm{x}_{i_a} = \mu+ \beta F_{i_a} \bm{m}+ \bm{\varepsilon_{s,i_a}} =
  \begin{cases}
    \mu+  \bm{\varepsilon_{0,i_a}}  & \quad \text{if } S_{i_a}=0\\
    \mu+ \beta \bm{m}+ \bm{\varepsilon_{1,i_a}}      & \quad \text{if } S_{i_a}=1\\
  \end{cases}
,\end{eqnarray}
where $\mu$ is the mean of the quasi-static speckle residuals, $\bm{\varepsilon_{s,i_a}}$ is their time and space varying part, which is characterised by the quasi-static speckle residuals statistics, and $\beta$ and $\bm{m}$ provide the flux and a model of the planetary signal, respectively. The parameter $F_{i_a}=\left\lbrace 0,1\right\rbrace $ is a realisation of a two-state Markov chain which implies that the probability of being in regime $s$ at index $i_a$ will depend on the probability at the previous step, $i_a-1$. This allows us to better disentangle a planetary signal from bright speckle by providing a short-term memory to the model. The set of patches $\bm{x}_{i_a}$ are described via the probability-weighted sum of the values generated by eq.~\ref{maineq}. 

The probability of being in the regime $s$ at index $i_a$, defined as $\xi_{s,i_a}$, depends on the probability at the previous step, $i_a-1$, on the likelihood of being currently in a given regime, $\eta_{s,i_a}$, and on the transition probability between the regimes given by the matrix, $p_{q,s}$. The probability $\xi_{s,i_a}$  is given by:

\begin{eqnarray}
\label{proba}
\xi_{s,i_a}=\sum^{1}_{q=0} \frac{\eta_{s,i_a} p_{q,s} \; \xi_{1,i_a-1} }{\sum^{1}_{q=0} \sum^1_{s=0} \eta_{s,i_a} p_{q,s} \; \xi_{q,i_a-1}} ,
\end{eqnarray}
where $\sum^{1}_{q=0} \sum^1_{s=0} \eta_{s,i_a} p_{q,s} \; \xi_{q,i_a-1}$ is a normalisation factor and $\eta_{s,i_a}$ is the likelihood associated with the regime $s$, which is given for each patch, $i_a$, in the Gaussian case by
\begin{eqnarray}
\label{like}
\eta_{s,i_a}= \sum^{\theta^2}_n \frac{1}{\theta^2} \frac{1}{\sqrt{2 \pi}\sigma} \exp\left[- \frac{\left[ \bm{x}^n_{i_a} - (F_{i_a} \beta \bm{m}^{n} -\mu)\right] ^2}{2\sigma^2}\right],
\end{eqnarray}
where $\sigma$ is the noise standard deviation (see Section \ref{Paramnoisedist} for more details about its estimation), $\theta$ gives the size in pixels of the planet model, $\bm{m}$, and $n$ is the pixel index within the patch.

As the RSM algorithm relies on a two-state Markov chain, the computation of the probabilities requires the use of an iterative procedure because of the dependence of the probabilities on past observations. Once the probabilities of being in the two regimes have been computed for every pixel of every annulus, the probability of being in the planetary regime, $\xi_{1,i_a}$, is averaged along the time axis to generate the final probability map. A detailed description of the different steps of the RSM algorithm may be found in \cite{Dahlqvist20}. The use of forward-modelled PSFs, as well as the estimation of the probabilities via a forward-backward approach replacing the forward approach presented here, are documented in \cite{Dahlqvist21}.

\subsection{Parameter selection for the RSM map}
\label{RSMparam}

Following the optimisation of the PSF-subtraction techniques to be used in the RSM model (Sect.~\ref{sec:model}), the next step is to consider the parametrisation of the RSM algorithm itself. The use of the RSM algorithm requires the definition of four main parameters. These parameters are (i) the crop-size $\theta$ for the planetary model $\bm{m}$, (ii) the definition of the region of the cube of residuals considered for the computation of the noise properties, whose estimation can be done (iii) empirically or via best fit, and (iv) the method used to compute the intensity of the potential planetary candidate $\beta$. When defining the flux parameter $\beta$ as a multiple of the noise standard deviation, an additional parameter $\delta$ has to be used to determine how far into the noise distribution tail we are looking for potential planetary candidates.

An optimal set of parameters for the RSM algorithm is computed separately for each PSF-subtraction technique, and is based on a performance metric computed using the generated RSM map. We do not rely on multiple simultaneous injections of fake companions at different azimuths, as done previously, as the RSM approach assumes a single planetary signal per annulus. Injecting  the fake companions sequentially would largely increase the computation time. We therefore define, annulus-wise, a single median position in terms of noise intensity, common to all PSF-subtraction techniques. This allows a fair comparison between the PSF-subtraction techniques when selecting the best set of likelihood cubes to generate the final RSM map in the last step of the auto-RSM framework. The determination of this median position starts with de-rotation of the original ADI sequence and the median-combination of the resulting set of frames. We then compute the flux of every aperture contained in the selected annulus, each aperture centre being separated by a single pixel in contrast with the approach of \citet{Mawet14}, where the apertures centre are separated by one FWHM. We define the fake companion injection position as the centre of the aperture for which the flux is the median of all the apertures fluxes. We decided to compute this median-flux position in the original ADI sequence, as the median-flux position inside the PSF-subtracted final frame differs from one PSF subtraction technique to the other, although a single common position is required for the final step of the auto-RSM algorithm.  Regarding the contrast used for the optimisation of the RSM map parameters,  for each PSF-subtraction technique, we select the average contrast $C_{a,m}$ obtained with the optimal set of parameters. Here, we make the assumption that taking the median-flux position and the average contrast should provide a balanced optimised parametrisation that works for brighter as well as fainter planetary signals.

The performance metric used for the RSM algorithm optimisation is then defined as the peak probability in a circular aperture with a diameter of one FWHM centred on the position of the injected fake companion in the final RSM map divided by the maximum probability observed in the remaining part of the annulus of width equal to one FWHM. This allows us to account for potential bright speckles within the probability map as well as for the intensity of the planetary signal. Having defined the loss function used for the RSM parameter selection, we now consider the different parameters that should be optimised. 

\subsubsection{Crop size}

The crop size $\theta$ is one of the parameters affecting the final probability map the most. This is especially true when relying on forward-model versions of the PSF-subtraction techniques, where a larger crop size should be considered to take advantage of the modelling of the negative side lobes appearing on either side of the planetary signal peak , which are due to self-subtraction associated with PSF-subtraction techniques. Self-subtraction depends on the relative position of the planetary candidate compared to the host star, with stronger self-subtraction at small angular separations and almost no self-subtraction at large angular separations. Indeed, the apparent movement of the planetary candidate increases linearly with the distance to the host star as the parallactic angles remain fixed but the radius increases. Larger apparent movement between two frames goes along with reduced self-subtraction. This implies that the optimal crop size for forward-modelled PSF should decrease with angular separation, as the negative side lobes appearing on either side of the planetary signal peak are replaced by noise. The selection of the optimal crop size should account for this effect as well as the range of parallactic angles, which is specific to each data set and also affects self-subtraction patterns. For PSF-subtraction techniques relying on the off-axis PSF to model the planetary signal, we consider a smaller range of crop sizes, as we do not take into account the distortion due to reference PSF subtraction. A maximum size of one FWHM is considered when relying on off-axis PSFs compared to the two FWHMs used for foward-model PSF-subtraction techniques. The definition of a proper crop size is nevertheless still important, because considering the shape of the PSF peak should help in disentangling planetary signals from speckle noise.

\subsubsection{Parametrisation of the noise distribution}
\label{Paramnoisedist}

\begin{figure*}[t]
  \centering
  \subfloat[Spatio-temporal]{\includegraphics[width=105pt]{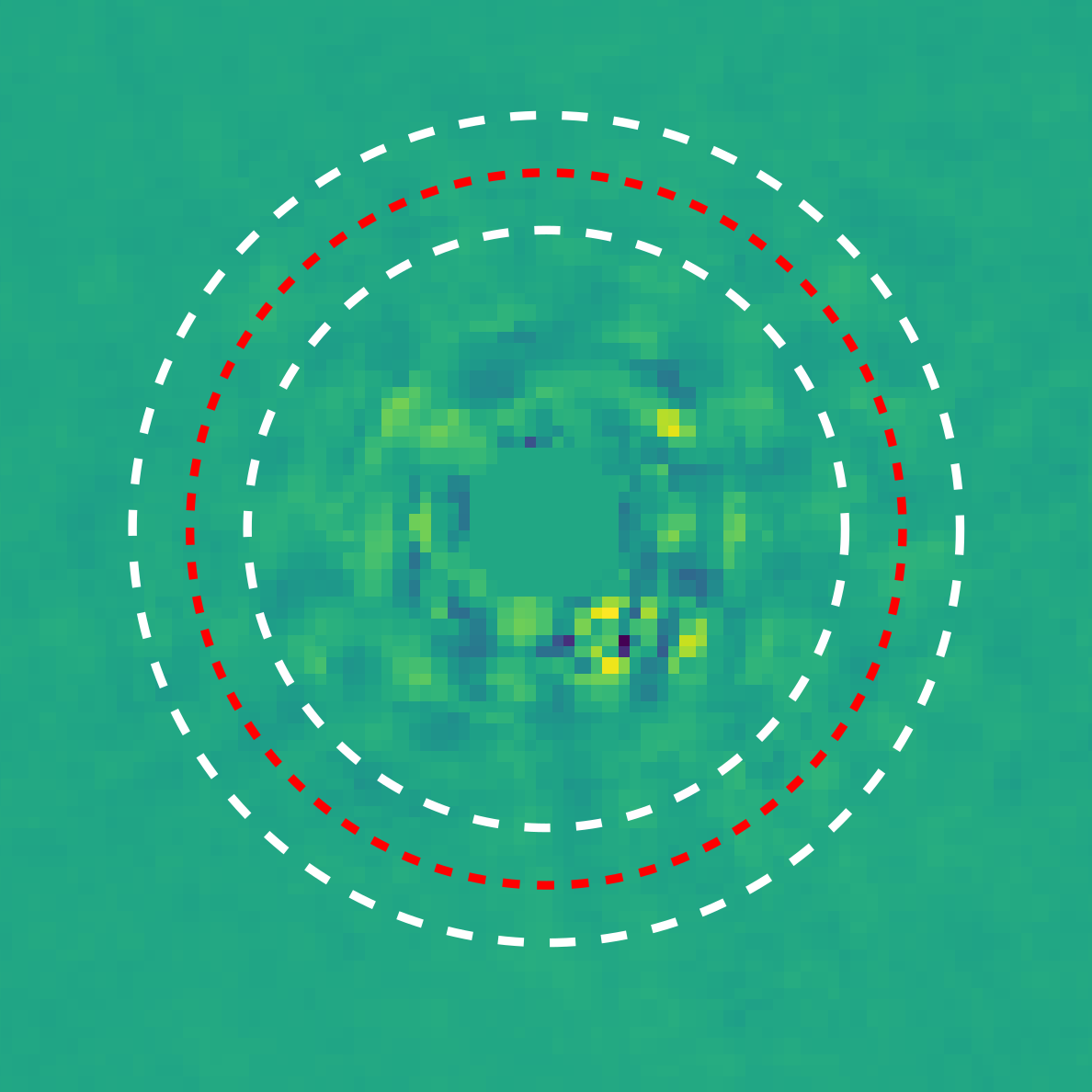}}
  \subfloat[Frame]{\includegraphics[width=105pt]{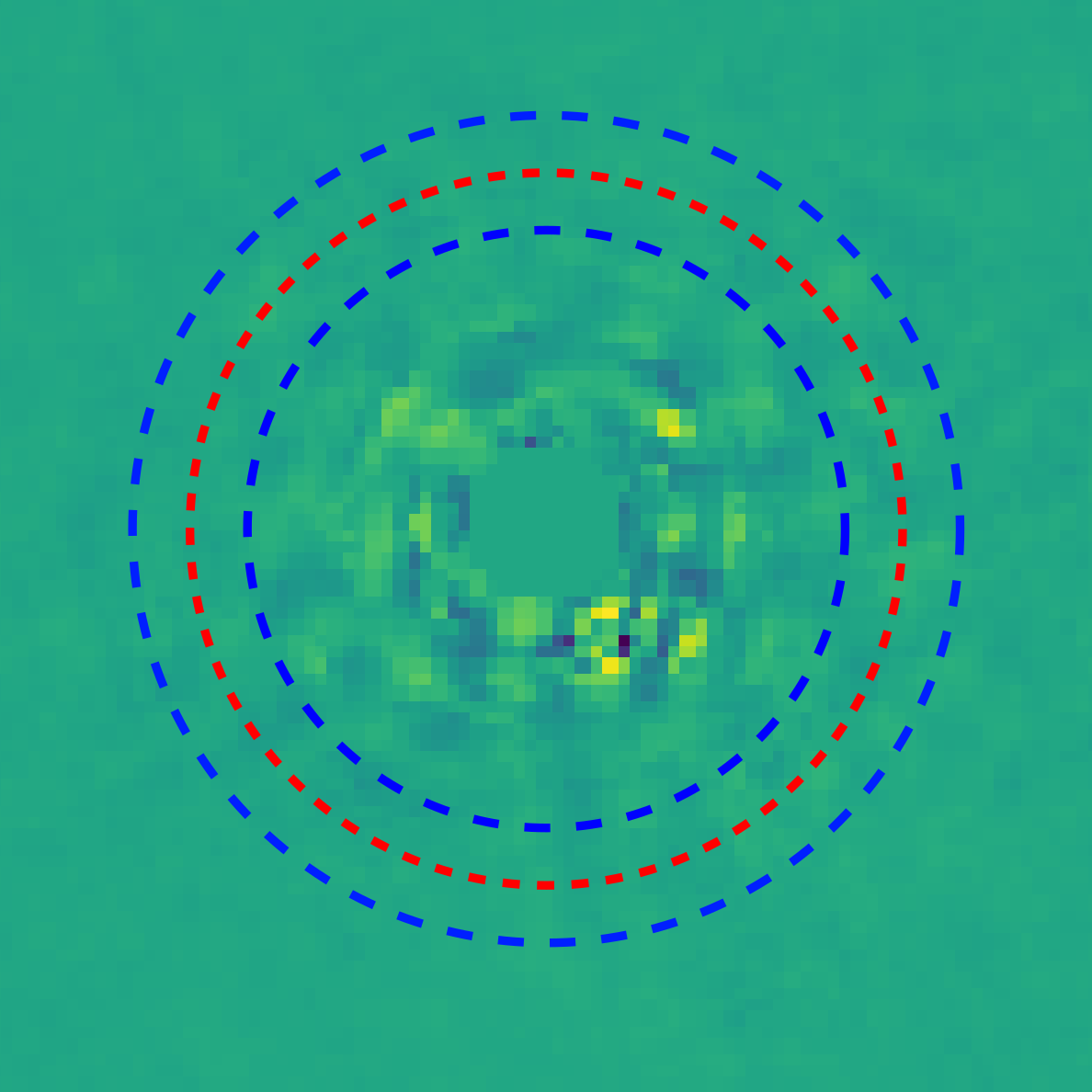}}
  \subfloat[Frame with mask]{\includegraphics[width=105pt]{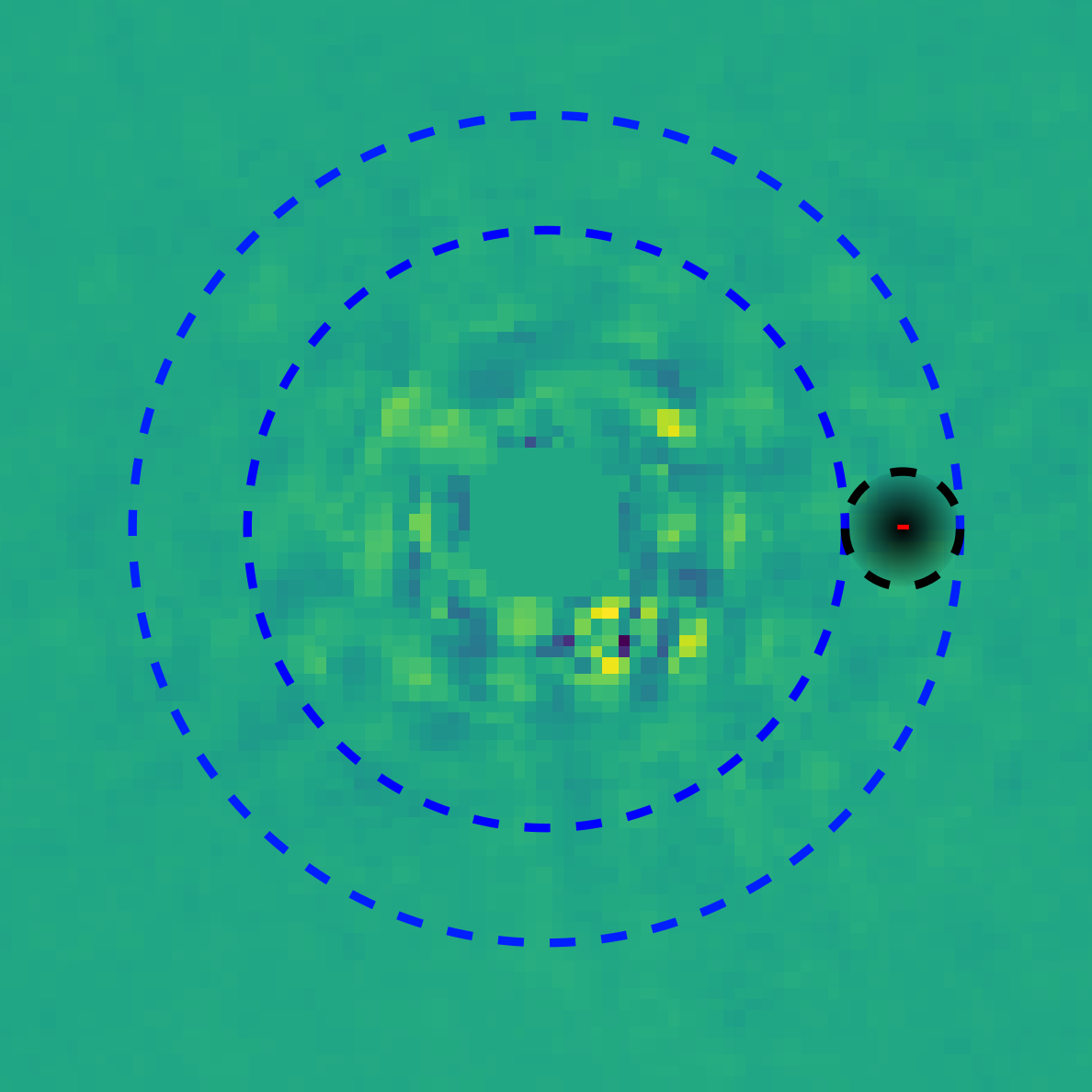}}
  \subfloat[Segment with mask]{\includegraphics[width=105pt]{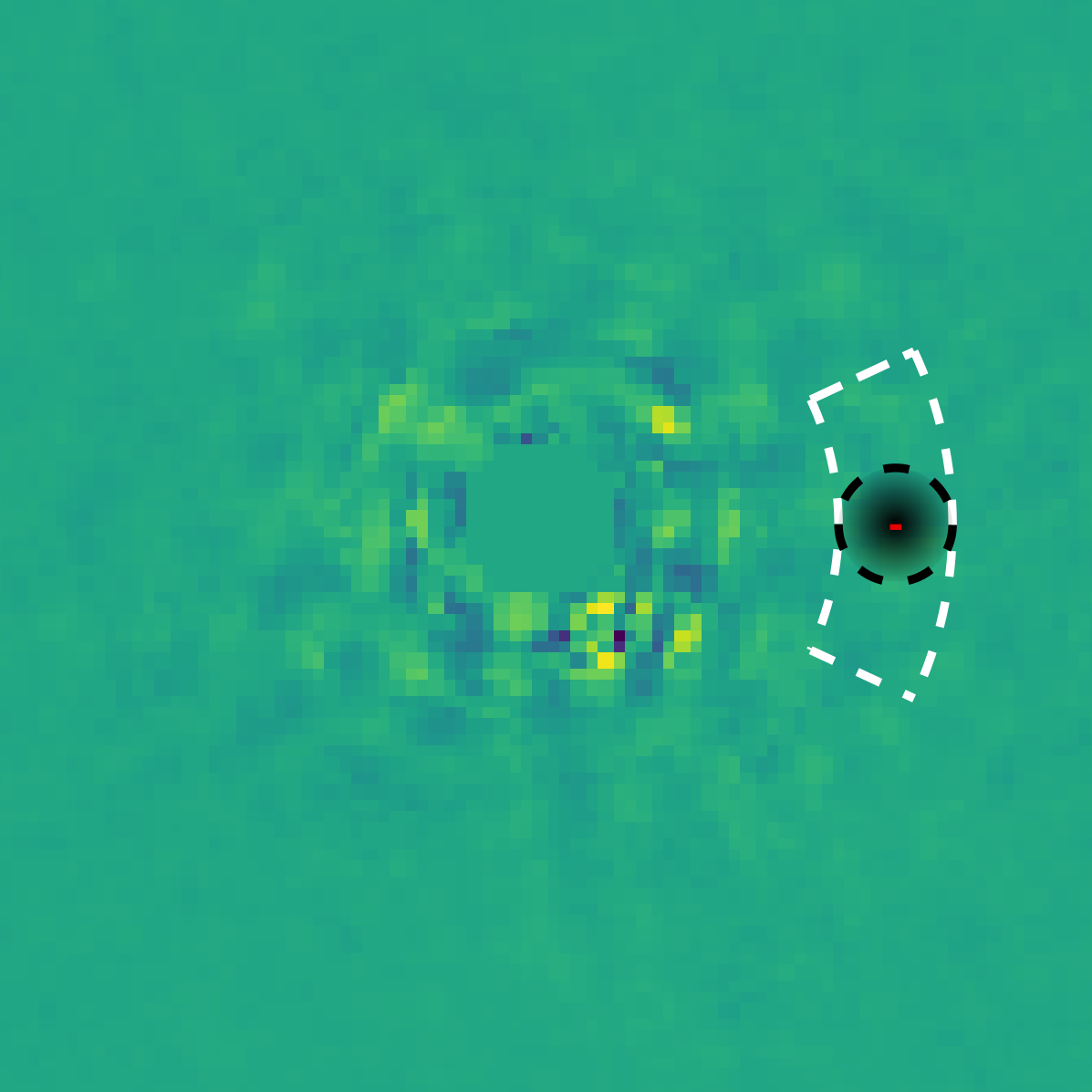}}
  \subfloat[Temporal]{\includegraphics[width=105pt]{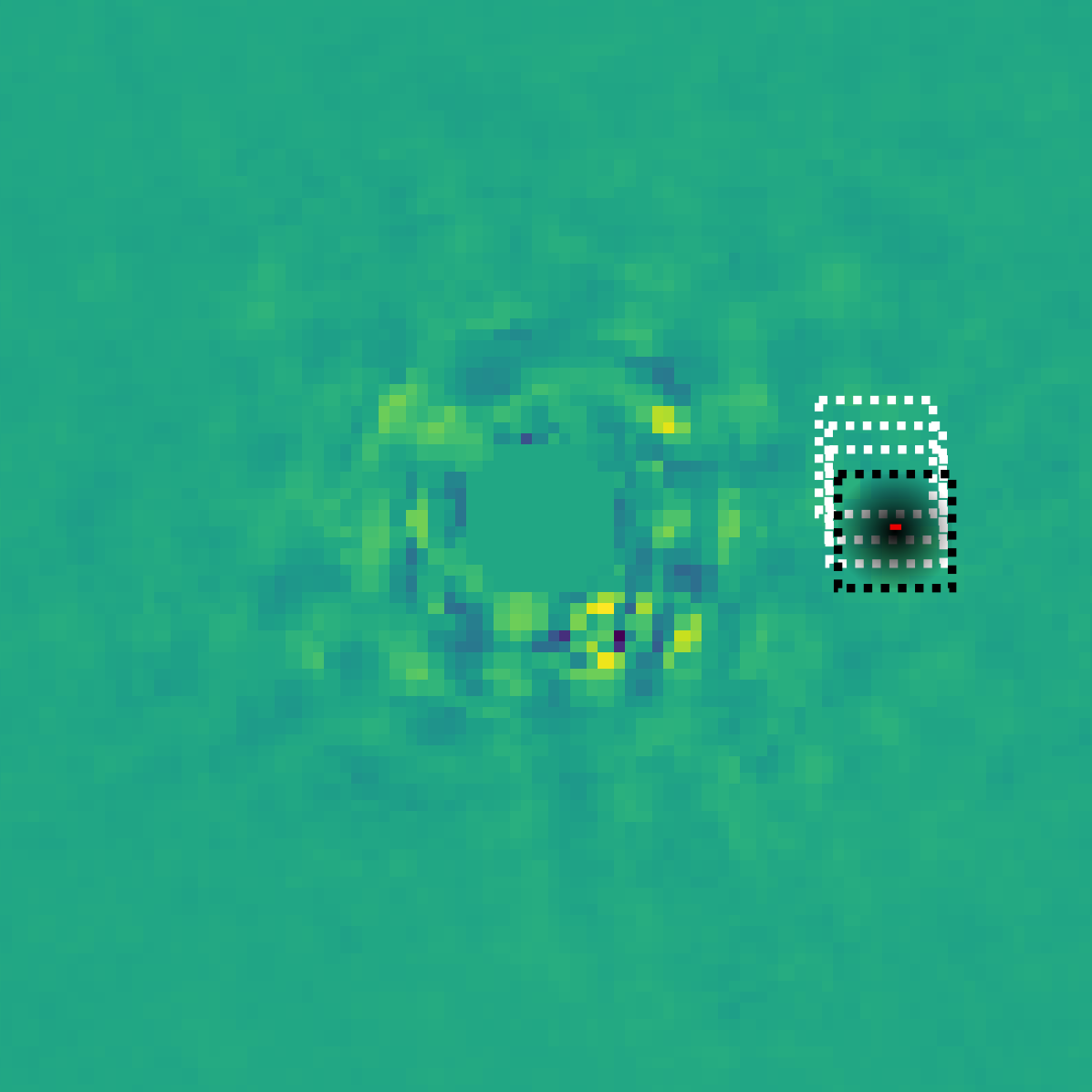}}
  \caption{\label{var_esti} Graphical representation of the estimation of residual noise properties  using the five proposed approaches. The red circle/point indicates the pixel for which the likelihood is estimated. White and blue circles encompass the set of pixels used for computation of the noise properties. White circles indicate that the entire set of frames from the derotated cube are used for the computation, while blue circles indicate that  the estimation is done frame-wise. Black circles define a mask, i.e. pixels that are not considered in the estimation.}
\end{figure*}

One of the corner stones of the RSM algorithm is the proper definition of the likelihood function associated with every patch contained in a given annulus. Four potential noise distribution functions are considered to compute these likelihoods, namely the Gaussian and the Laplacian distribution, the Huber loss \citep{Pairet19}, and a hybrid distribution built as a weighted sum of Gaussian and Laplacian distributions \citep{Dahlqvist20}. The first two noise distribution functions require estimation of the noise mean and variance, whereas the other two require additional parameters. Selection of the optimal distribution is done automatically within the RSM algorithm via a best-fit approach. However, the estimation of the parameters characterising the residual noise distribution function necessitates proper definition of the set of pixels to be considered. Different approaches are tested in auto-RSM to determine the most relevant set of pixels inside the cube of residuals. We have selected five possible ways to evaluate the noise properties:
\begin{itemize}
\item Spatio-temporal estimation: The set of pixels incorporates the pixels inside the selected annulus\footnote{ By selected annulus, we are referring to the annulus of  one FWHM in width centred on the radial distance of interest $a$.} for all the frames contained in the cube of residuals (see 'Spatio-temporal' in Fig.\ref{var_esti}). The distribution function parameters depend solely on the radial distance $a$ ($\mu_a$ and $\sigma^2_a$).
\item Frame-based estimation: The set of pixels incorporates the pixels of a given frame inside the selected annulus (see `Frame' in Fig.\ref{var_esti}). The distribution function parameters depend on both the radial distance $a$ and the time-frame $t$ ($\mu_{a,t}$ and $\sigma^2_{a,t}$).
\item Frame with mask-based estimation: The set of pixels incorporates the pixels of a given frame inside the selected annulus, apart from a region with a diameter of one FWHM centred on the pixels for which the likelihood is estimated (see `Frame with mask' in Fig.\ref{var_esti}). The  distribution function parameters depend on both the radial distance $a$ and the pixels index $i_a$  ($\mu_{a,i_a}$ and $\sigma^2_{a,i_a}$).
\item Segment with mask-based estimation: The set of pixels incorporates the pixels of all frames inside a section (of length equal to three FWHMs) of the selected annulus, apart from a region with a diameter of one FWHM centred on the pixels for which the likelihood is estimated (see `Segment with mask' in Fig.\ref{var_esti}). The  distribution function parameters depend on both the radial distance $a$ and the pixels index $i_a$ ($\mu_{a,i_a}$ and $\sigma^2_{a,i_a}$).
\item Temporal estimation: The last method is inspired by the approach developed in \cite{Flasseur18}. This approach relies on the cube of residuals before de-rotation. For a given patch inside the selected annulus, the pixels selected for computation of the distribution function parameters  are the ones sharing the same position within the cube of residuals before de-rotation but taken at different times (see `Temporal' in Fig.\ref{var_esti}). All the frames except for the frame containing the selected patch are therefore considered. The distribution function parameters depend on both the radial distance $a$ and the pixels index $i_a$  ($\mu_{a,i_a}$ and $\sigma^2_{a,i_a}$).
\end{itemize}
The use of these different methods allows us to investigate which part of the neighbourhood around the patch is relevant in order to correctly estimate the noise profile. This explains the wide variety of proposed methods both in terms of temporal and spatial position. 

Depending on the region selected to compute the noise properties, a specific noise distribution function and parametrisation can be selected for a single patch, a single frame, or the entire set of frames and patches contained in the considered annulus. The estimation of noise distribution parameters can be done empirically or via best fit. The choice between empirical estimation and estimation via best fit represents an additional parameter to be considered during the RSM parameter optimisation.

\subsubsection{Estimation of the planetary intensity}

Two different methods were proposed to compute the planetary intensity parameter $\beta$ \citep{Dahlqvist20,Dahlqvist21}. The first one relies on an additional parameter $\delta$ to define the expected position of the potential planetary signal intensity in the noise distribution. The intensity parameter $\beta$ is defined as $\delta$ multiplied by the estimated noise standard deviation \citep{Dahlqvist20}. A set of $\delta$ is tested and the optimal one is selected via maximisation of the total likelihood associated with a given angular distance (seeEq. \ref{like}). In the case of auto-RSM, this last step is removed and the optimal $\delta$ is selected during the auto-RSM optimisation process. Preliminary tests have shown that the optimisation of $\delta$ using the auto-RSM performance metric can significantly reduce the background noise in the RSM probability map compared to the total likelihood-based optimisation proposed in \cite{Dahlqvist20}, while leaving planetary signals almost unaffected for $\delta \leq 5$. 

The second approach relies on Gaussian maximum likelihood to define a pixel-wise intensity \citep{Dahlqvist21}. The estimation of the intensity parameter $\beta$ via Gaussian maximum likelihood requires the computation of a frame-wise standard deviation. The expression for the pixel-wise intensity is
\begin{eqnarray}
\label{gausslike}
\tilde{\beta}= \frac{ \sum^{T}_j \bm{i}^{\top}_j  \bm{m}_j /\sigma_j}{\sum^{T}_j \bm{m}^{\top}_j  \bm{m}_j /\sigma_j},
\end{eqnarray}
with $\sigma_j$ being the noise standard deviation, $\bm{i}^{\top}_j$ the observed patch, and $\bm{m}_j$ the planet model for frame $j$. Frame-wise computation of the standard deviation implies that the mean and variance computation described in the previous section should be performed via the frame-based estimation, the frame with mask-based estimation, or the temporal estimation, while for the other intensity computation methods, all five approaches can be used. 

\subsubsection{Sequential parameter optimisation}

While optimisation of the PSF-subtraction techniques is done in a single step, optimisation of the RSM parameters is done partly sequentially. The estimation of the RSM map is indeed much slower than the estimation of the contrast. Depending on the region considered for the estimation of the noise properties, the computation time can further increase, especially when relying on the frame with mask-based estimation or the temporal estimation, preventing optimisation of all the parameters in a single step. The selection of the optimal region to compute the noise properties is therefore treated separately. The selection of the optimal set of RSM parameters starts with computation of the RSM map performance metric for the two methods used to determine the intensity parameter $\beta$ using the frame-based estimation of the noise properties\footnote{The frame-based estimation of the noise properties has been selected as initial guess because it is shared by the two approaches used to compute the intensity parameter $\beta$, and is much faster than the frame with mask and the temporal estimations.}. For both methods, a separate performance metric is estimated for the selected range of crop sizes, but also for the selected range of $\delta$ for the method defining the intensity as a multiple of the noise standard deviation. The selection of the optimal value for all three parameters (i.e. $\delta$, the crop size $\theta$ and the method to compute the intensity) is performed by comparing the obtained RSM performance metric. The faster computation of the RSM map when relying on the frame-based estimation of the noise properties allows optimisation of these three parameters in a single step.  The next step involves the selection of the optimal region for estimation of the noise properties. Depending on the method selected to compute the intensity, a reduced set of regions may be considered. Optimisation of the RSM parameters ends by determining whether the noise properties are optimally computed empirically or via a best fit.

\subsection{Optimal combination of the likelihood cubes}
\label{opticombi}

Having optimised the parameters of the PSF-subtraction techniques as well as the ones of the RSM algorithm, we are now left with a series of optimal cubes of likelihoods. One of the most interesting features of the RSM framework is its ability to use several cubes of residuals generated with different PSF-subtraction techniques to maximise the planetary signal, while minimising the residual speckle noise. The RSM algorithm takes advantage of the diversity of noise structures  in the different cubes of residuals. This diversity is reflected in the noise probability distribution but also in the repartition of maxima and minima in the different speckle fields. By taking both aspects  into account, the RSM algorithm is able to better average out the noise and improve the ratio between potential planetary signals and the residual speckle noise. 

Despite optimisation of the parameters, some PSF-subtraction techniques may be less suited to generating a clean cube of residuals for some data sets. Redundancies in the information contained in several cubes of residuals may also degrade the performance of the RSM map by increasing the relative importance of some speckles. When dealing with several ADI sequences of the same object, some sequences can also be much noisier depending on the observing conditions. All these elements necessitate proper selection of the likelihood cubes used to generate the optimal final RSM map. We propose the investigation of two possible approaches to select the set of likelihood cubes used for computation of the final probability map, a bottom-up approach and a top-down approach, making use of a greedy selection framework. 

As the RSM algorithm relies on spatio-temporal series of likelihoods to compute annulus-wise probabilities (see Eq. \ref{proba}), we start by defining the set of available series of likelihoods for a given radius $a$ by $\bm{Y}^a=\left\lbrace Y^a_{c,m}, \forall c \in [0,N_{sequence}], m \in [0,N_{technique}] \right\rbrace $. The $Y^a_{c,m}$ time-series corresponds to the set of likelihoods $\eta_{s,i_a}$ given in the Gaussian case by Eq. \ref{like}, generated for the cube $c$ with the PSF-subtraction techniques $m$ for all pixel indices $i_a$ of the annulus $a$. This last step of auto-RSM is used to define a subset $\bm{Z}^a \subset \bm{Y}^a$ regrouping the series of likelihoods maximising the performance metric of the RSM probability map for annulus $a$. This selection step shares the same performance metric as the RSM parameters optimisation step. To compute the performance metric, a single fake companion injection is used for each annulus for the entire set of PSF-subtraction techniques\footnote{For a given annulus $a^*$, the largest contrast in the set $C_{a^*,m}$ is used for the bottom-up approach and the smallest for the top-down, as they provide the best performance based on tests}. The selected set of time-series of likelihoods $\bm{Z}^a$ are then concatenated to form a single time-series per annulus and used to compute the probabilities via Eq. \ref{proba}. The RSM performance metric, estimated based on these probabilities, allows us to select the optimal set $\bm{Z}^a$.

\subsubsection{Bottom-up approach}

When relying on a bottom-up approach, the iterative selection algorithm starts with an empty set $ \bm{Z}^a$. At each iteration, the series of likelihoods $Y^a_{c,m}$ that leads to the highest performance metric increase is added to the set $ \bm{Z}^a$. The procedure is repeated until no additional series of likelihoods leads to an increase in the performance metric. The bottom-up greedy selection algorithm can be summarised by the following steps.
\begin{itemize}
\item For each series of likelihood contained in $\bm{Y}^a$, compute the corresponding RSM map performance metric using the set of series of likelihoods $ \bm{Z}^a \cup Y^a_{c,m}$ for annulus  $a.$
\item At each iteration, select the series of likelihoods providing the largest incremental performance metric increase and include the considered series of likelihoods $Y^a_{c*,m*}$ in the set of selected series $ \bm{Z}^a$. Remove from $\bm{Y}^a$ the selected series $Y^a_{c*,m*}$, as well as any other series included in $\bm{Y}^a$ that did not lead to an increase in the performance metric.
\item Repeat the previous two steps until $\bm{Y}^a$ is empty. 
\end{itemize}

\subsubsection{Top-down approach}

In contrast with the bottom-up approach, the top-down iterative selection algorithm starts with a set $ \bm{Z}^a=\bm{Y}^a$ and relies on pruning steps to reduce the number of series of likelihoods included in $ \bm{Z}^a$ until an optimum is reached. The steps of the top-down greedy selection algorithm are the following.
\begin{itemize}
\item For each series of likelihood contained in $ \bm{Z}^a$, compute the RSM map performance metric corresponding to the set of series of likelihoods $ \bm{Z}^a  \setminus Y^a_{c,m}$ for annulus $a.$
\item At each iteration, select the series of likelihoods providing the largest incremental performance metric increase and remove the considered series of likelihoods $Y^a_{c*,m*}$ from the set of selected series $ \bm{Z}^a$.
\item Repeat the two previous steps until no more incremental performance metric decrease can be observed.
\end{itemize}

Pseudo codes of both approaches are provided in Tables \ref{bu} and \ref{td}. The potential redundancies in the information contained in different cubes of likelihoods, as well as the iterative procedure used by the RSM algorithm to generate the final probability map, mean that the set of series of likelihoods are not truly independent, which prevents us from finding the global optimum while using a greedy approach. However, these bottom-up and top-down greedy selection algorithms provide a good approximation of the global optimum in a reasonable amount of time.

\subsection{Practical implementation}

\begin{figure}[h!]
  \centering
  \subfloat{\includegraphics[width=260pt]{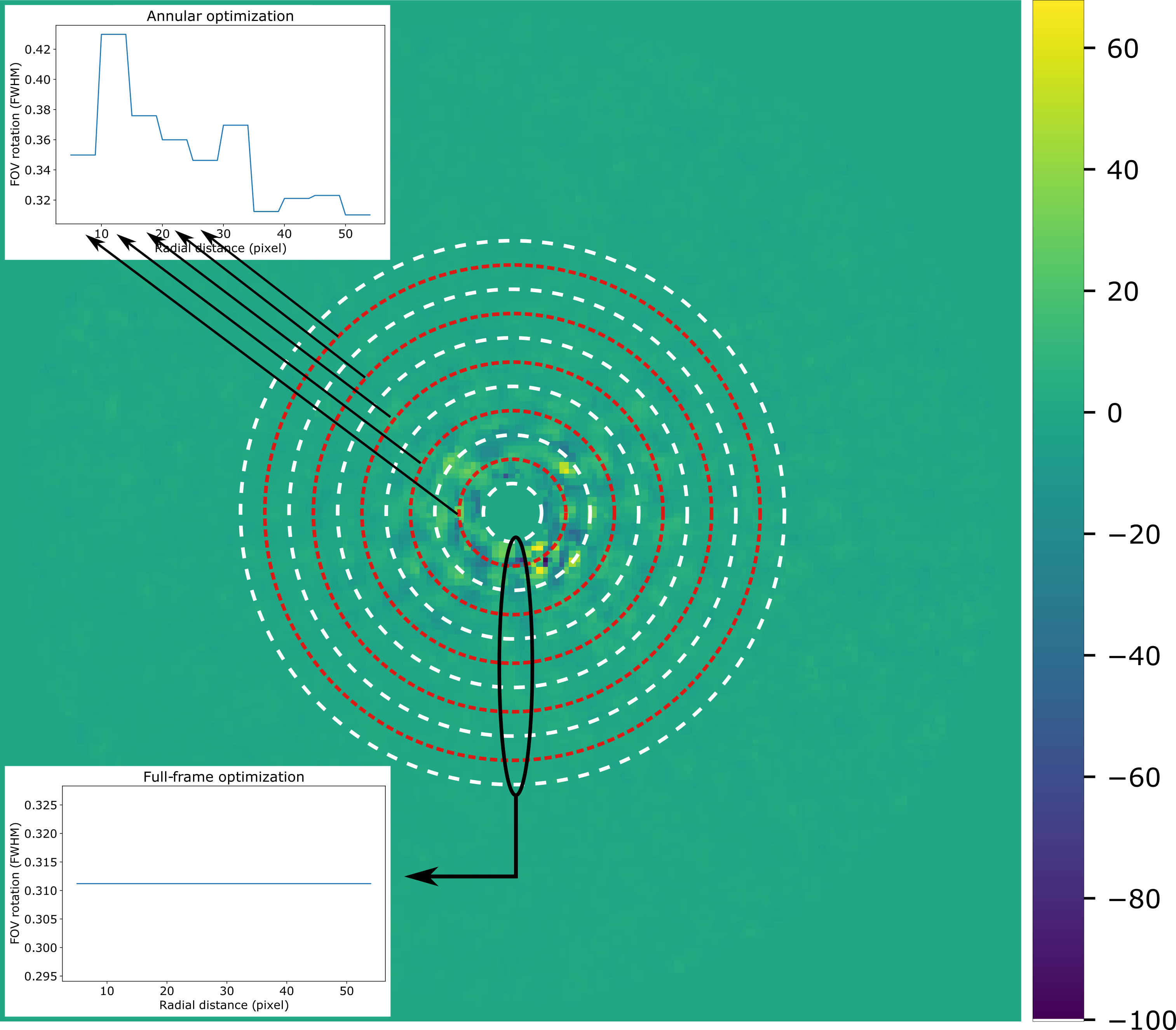}}
 \caption{\label{optimode} Graphical representation of the optimisation of the FOV minimal rotation for the annular PCA with the two modes of the auto-RSM algorithm for the SPHERE 1 data set of the EIDC. The full-frame version (illustrated in the bottom left corner) considers a set of annuli of width equal to one FWHM to provide a single set of optimal parameters. The annular version (top left) considers successive annuli of width equal to one FWHM to provide annulus-wise sets of optimal parameters.}
\end{figure}

After having presented the different steps of the proposed optimisation framework for the RSM map algorithm, these steps can now be merged into a single optimisation procedure, which is implemented in the PyRSM package\footnote{The PyRSM package is available at GitHub: \url{https://github.com/chdahlqvist/RSMmap}.}. Two different modes of this optimisation procedure are proposed: the full-frame mode and the annular mode. The two modes share a common structure but their output dependence on the angular separation is different.  In the full-frame mode, there is no dependence between the optimal set of parameters and the angular separation to the host star, with a single set of parameters being used for every annulus. In the annular mode, the frames are divided into successive annuli of pre-defined width, and a set of optimal parameters is defined for each annulus. As the noise distribution and parameters evolve with the angular separation, this second mode accommodates possible evolutions of the optimal parametrisation with the angular separation to the host star. Figure \ref{optimode} provides a graphical representation of the two different modes in the case of optimisation of the FOV minimum rotation used for the annular PCA estimation. 

As illustrated in Fig.\ref{optimode}, in both cases the frames are divided into successive annuli of one FWHM\footnote{A width equal to one FWHM often provides the best performance, but other widths can be used.} in width. The red dotted circles represent the centres of the selected annuli on which  the apertures for the optimisation of the PSF-subtraction techniques are
centred, and for which probabilities are computed to optimise the parameters of the RSM algorithm and select the optimal set $ \bm{Z}^a$. We do not consider all angular separations but a subset of them separated by one FWHM, as we expect a slow evolution of the parameters. This should give a good representation of the evolution of the parameters or a good overview in the case of the full-frame mode, while reducing the computation time\footnote{A mode considering all angular separations has been tested and provides results close to the other modes while requiring a much longer computation time.}. 

In the case of the full-frame mode, we consider a subset of the annuli of the annular mode, with an increasing distance between the selected annuli as we move away from the host star. This allows us to increase the relative weight of small angular separations, the noisier region being located near the host star, and reduce the estimation time. The proposed annulus selection rule for the full-frame mode can be summarised as

\begin{eqnarray}
\label{ffcase}
\Delta a = 
  \begin{cases}
    \text{ FWHM}  & \quad \text{if } a \in [\text{ FWHM} ,4 \text{ FWHM} ] \\
    2 \text{ FWHM}  & \quad \text{if } a \in ] 4 \text{ FWHM} ,8 \text{ FWHM} ] \\
    4 \text{ FWHM}  & \quad \text{if } a \in ] 8 \text{ FWHM} , a_{max}] \\
  \end{cases}
,\end{eqnarray}
where $\Delta a$ is the separation between two successive annuli used in the optimisation procedure, and $a_{max}$ is the largest annulus to be considered in the RSM map computation. Selection of the optimal parameter set for the PSF-subtraction techniques in the full-frame mode is achieved by comparing the normalised contrasts generated with the different tested parametrisations summed over the selected angular separations. We start by computing contrasts for a common set of parametrisations\footnote{We consider all the tested parametrisations for the NMF and LLSF and the parametrisations tested during the initialisation of the Gaussian process for the other PSF-subtraction techniques.} for each considered angular separation. For a given angular separation, the median of the obtained contrasts is then computed and used to normalise all the contrasts. The normalised contrasts are finally summed over the selected angular separations provided by Eq. \ref{ffcase} for each considered parametrisation. The optimal set of parameters is then the one that minimises the summed normalised contrast\footnote{The inverse of the normalised average contrast summed over the considered angular separation is used as loss function for the Bayesian optimisation.}. As the contrast decreases with the angular separation, the normalisation allows a proper summation of the contrasts generated at the different angular distances. A similar approach is used for optimisation of the RSM algorithm  and selection of the optimal likelihoods, although no normalisation is required according to the definition of the performance metric. As regards the annular mode, no normalisation is required as the optimisation is done separately for each selected annulus.

The complete auto-RSM optimisation procedures for the two considered modes are summarised in Tables \ref{ffmode} and \ref{amode}. As can be seen from both tables, the optimisation procedures can be divided into four main steps, (i) optimisation of the PSF-subtraction techniques, (ii) optimisation of the RSM algorithm, (iii) optimal combination of models and sequences, and (iv) computation of the final RSM probability map (respectively the opti\_model, opti\_RSM, opti\_combination, and opti\_map function of the PyRSM class). In both modes we include the estimation of a background noise threshold for every annulus, by taking, for each annulus, the maximum probability observed in the map generated with the reversed parallactic angles. Following subtraction of the angular separation-dependent thresholds, we set all negative probabilities to zero to generate the final map. The threshold subtraction should help to reduce the noise, especially near the host star where most residual speckles are observed. However, these thresholds should not be used as detection thresholds, as the noise statistics properties of the original ADI sequence are not exactly equivalent to the ADI sequence with sign-flipped parallactic angles. 
                                
Considering the existence of potential bright artefacts in the map generated with the reversed parallactic angles, we rely on a Hampel filter and a polynomial fit to smooth the radial evolution of the thresholds in the full-frame mode. As the parametrisation of the RSM algorithm has a large impact on planetary signals and background noise levels, we do not apply the threshold fit for the annular mode, as the RSM parametrisation evolves with the  angular separations. However, we do apply a smoothing procedure for the parameters of the PSF-subtraction techniques by applying a moving average after a Hampel filter. This helps in smoothing potential discontinuities between annuli in the set of optimal parameters and provides a more consistent final probability map. As computation of the residual cubes is done annulus-wise\footnote{An annular version of the NMF algorithm has been developed for the annular mode of the auto-RSM. The other PSF-subtraction techniques rely already on an annulus-wise estimation of the residuals.}, we need a single set of parameters\footnote{The set of parameters that has been optimised based on the set of apertures centred on the annulus} for a number of angular separations equal to the width of the annulus. However, the RSM map computation requires the definition of a set of parameters for every considered angular separation. A radial basis multiquadric function (RBF) is used to perform an interpolation \citep{Hardy71} of the RSM optimal parameters for the annular mode to provide a set of parameters for each angular distance.

\section{Performance assessment}
\label{pa}

\subsection{Description of the data sets}

As mentioned in the introduction, we base our performance analysis on the data set of the EIDC ADI subchallenge \citep{Cantalloube20}. This data set regroups nine ADI sequences, three for each considered HCI instrument, namely VLT/SPHERE- IRDIS \citep{Beuzit19}, Keck/NIRC2 \citep{Serabyn_2017}, and LBT/LMIRCam \citep{Skrutskie10}. The ADI sequences were obtained in H2-band for SPHERE  and Lp-band for the two other instruments. For each ADI sequence, a set of four fits files is provided: the temporal cube of images, the parallactic angles variation corrected from true north, a non-coronagraphic or non-saturated PSF of the instrument, and the pixel scale of the detector. The ADI sequences are pre-reduced using the dedicated pre-processing pipeline for the three instruments \citep[more details about the reduction are provided by][]{Cantalloube20}. 

As the LMIRCam ADI sequences regroup between 3219 and 4838 frames, we relied on moving averages to reduce this number to around 250 frames in order to limit the computation time. The reduction of the number of frames starts with the de-rotation of the original cube of images using the parallactic angle variations provided. A moving average is then applied on the de-rotated cubes along the time axis with a window and step size of 20 frames for the LMIRCam sequence 1 and 3, and 15 frames for the LMIRCam sequence 2. The same is done on the set of parallactic angle variations. The inverse reduced parallactic angle variations are then used to re-rotate the resulting ADI sequences. In addition to reducing the computation time, the moving average allows part of the noise to be averaged out in advance. More details about the nine ADI sequences are provided in Table~\ref{ADIdesc}.

To assess the performance of the HCI data-processing techniques, fake companions were injected by the EIDC organisers using the VIP package \citep{Gonzalez17}. Between 0 and 5 point sources were injected into each ADI sequence for a total of 20 planetary signals within the entire EIDC ADI subchallenge data set. These point sources were injected using the opposite parallactic angles, avoiding any interference with potential existing planetary signals while keeping the speckle noise statistics\footnote{As the auto-RSM relies on reversed parallactic angles to optimise the model parameters, the optimisation is done on the original ADI sequences in the case of the EIDC data sets. However, the ADI sequences selected for the EIDC do not contain any known planetary candidates. The optimisation should therefore not be affected.}. The separation, the azimuth, and the contrast of the injected fake companions were chosen randomly. The contrasts range between $2\sigma$ and $8\sigma$ based on a contrast curve computed with the regular annular PCA implemented in the VIP package \citep{Gonzalez17}, which is referred to as the `baseline' in the performance analysis presented in \cite{Cantalloube20}. The detection maps of the baseline consist in S/N maps computed using the approach of \cite{Mawet14}. The detection maps generated with the baseline approach are used in our model comparison.

\subsection{Performance metrics}
\label{perfmet}

The performance assessment of HCI data-processing techniques is done via the definition of a classification problem, counting detections and non-detections on a grid of FWHM-sized apertures applied to the detection maps. A true positive (TP) is defined as a value above the threshold provided by the user along with the S/N or probability maps within the FWHM aperture centred on the position of the injected fake companion. Any values above the provided threshold that are not in the set of apertures containing injected fake companions are considered as false positives (FPs). The false negatives (FNs) regroup all the non-detections at the position of injected fake companions, while the true negatives (TNs) are the non-detections at any other position. Different performance metrics are computed using these four categories:

\begin{itemize}
\item True positive rate: $TPR = \frac{TP}{TP + FN}$
\item False positive rate: $FPR =  \frac{FP}{FP + TN}$
\item False discovery rate: $FDR = \frac{FP}{FP + TP}$
\item F1 score: : $F1 =  \frac{2 TP}{FP + FN + 2TP}$
\end{itemize}

In addition to the F1 score computed at the pre-defined threshold, we follow the same approach as in \cite{Cantalloube20} and also consider the area under the curve (AUC) for the TPR, FPR, and FDR as a function of the threshold to classify the different versions of the proposed optimisation procedure. The AUCs of the TPR, FPR, and FDR are preferred to the values of these latter at the provided threshold, as this allows us to mitigate the arbitrariness of the threshold selection by considering their evolution for a range of thresholds. The AUC of the TPR should be as close as possible to 1 and the AUCs of the FPR and FDR as close as possible to zero. The F1 score being the harmonic mean of the recall and precision of the classification problem, it ranges between 0 and 1, with values close to 1 being favoured (perfect recall and precision). 

\subsection{Results}

We have now all the elements to apply the auto-RSM optimisation procedure described in Sects. 2 and 3 to the nine selected ADI sequences. Only PSF-subtraction techniques relying on an off-axis PSF (and not on forward models) are considered during the optimisation procedure in order to reduce the computation time, considering the large set of ADI sequences on which the method is tested, and the numerous parametrisations of the auto-RSM algorithm that we consider. This also allows a fair comparison with the results of the RSM algorithm already used in \cite{Cantalloube20}, which relied only on the PSF-subtraction techniques based on off-axis PSFs. We note however that the PyRSM Python package also accommodates the use of a forward-model version of KLIP and LOCI, where a parameter defines the maximum angular separation above which the forward model is no longer considered. This allows us to take advantage of the higher performance of forward-modeled PSF-subtraction techniques at small angular separations, while limiting their impact on the computation time at larger separations.

For the Bayesian optimisation of APCA and LOCI, the contrast is computed for respectively 80 and 60 points of the parameter space to initialise the Gaussian process, while 60 iterations of the minimum-expectation Bayesian optimisation are used to determined the optimal set of parameters. The smaller number of points for the initialisation of LOCI comes from its smaller set of parameters compared to APCA. The number of points for the initialisation and the number of iterations have been chosen to ensure the convergence to the global optimum\footnote{This parametrisation of the Bayesian optimisation algorithm ensures that the same set of optimal parameters is found when the algorithm is applied several times to the same ADI sequence.}. The ranges of possible values that have been selected to define the parameter space for the PSF-subtraction optimisation are shown in Table \ref{parameterspace}. Most ADI sequences share a common range of possible values. However, differences may be found in the definition of the parameter space boundaries for the NIRC2 ADI sequences due to the reduced number of frames.

\subsubsection{Full-frame and annular auto-RSM parametrisation}

Regarding the parameters of the full-frame version of auto-RSM, the set of selected annuli is truncated at $a_{max}=10 \lambda /D$ to favour small angular separations during the optimisation, the region close to the host star being more noisy\footnote{It also allows us to reduce the computation time, the larger angular separations being computationally more expensive.}. The order of the polynomial fit of the annular threshold is set to three in order to limit the impact of small artefacts appearing in the RSM map generated with inverted parallactic angles, while keeping the main characteristics of the angular evolution of the  noise. The full-frame (FF) auto-RSM was tested with three different parametrisations, allowing the comparison between the bottom-up (BU) and top-down (TD) selection of the optimal cubes of likelihoods, as well as the comparison between the forward (F) and forward-backward (FB) approaches to compute the final probabilities.

The annular version of auto-RSM requires the definition of two additional parameters: the window sizes for the Hampel filter and the moving average used to smooth the PSF-subtraction parameters (see Table \ref{amode}). The window sizes are respectively equal to 3 and 5, and the window is centred on the angular distance for which the filtered value or the moving average is computed. Two different flavours of the annular auto-RSM were tested, one relying on the annular framework for the optimisation of the entire set of parameters (A), and one using the annular framework for the PSF-subtraction parametrisation and the full-frame framework for the RSM parametrisation and the selection of the optimal set of cubes of likelihoods. The hybrid approach mixing full-frame and annular frameworks (AFF) aims to reduce the angular variability of the background residual probabilities, which are mainly affected by the parametrisation of the RSM model.

\subsubsection{Performance metric computation and model comparison}

\begin{figure*}[ht]
\footnotesize
  \centering
  \subfloat[SPHERE-1]{\includegraphics[width=160pt]{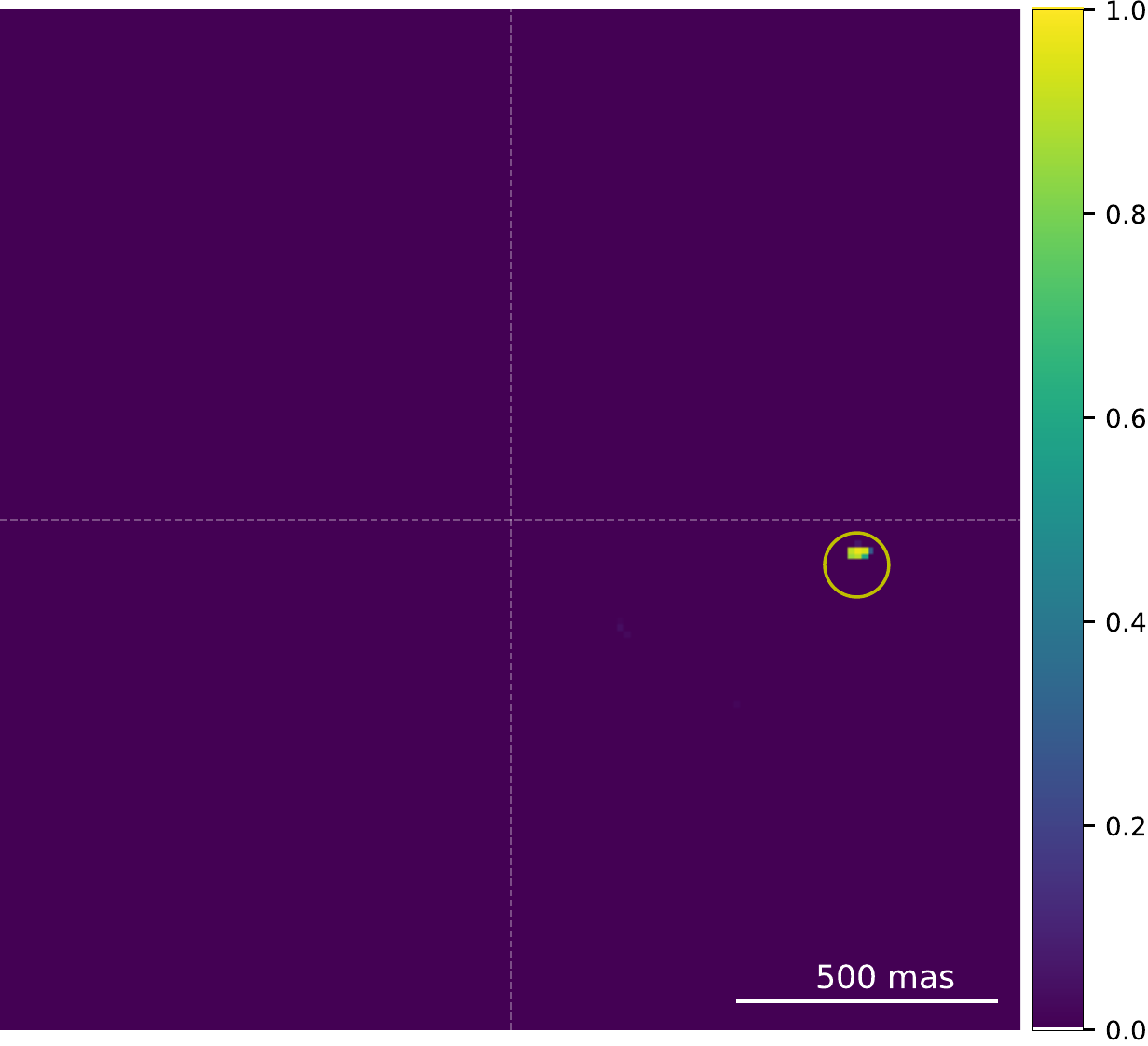}}
  \subfloat[SPHERE-2]{\includegraphics[width=160pt]{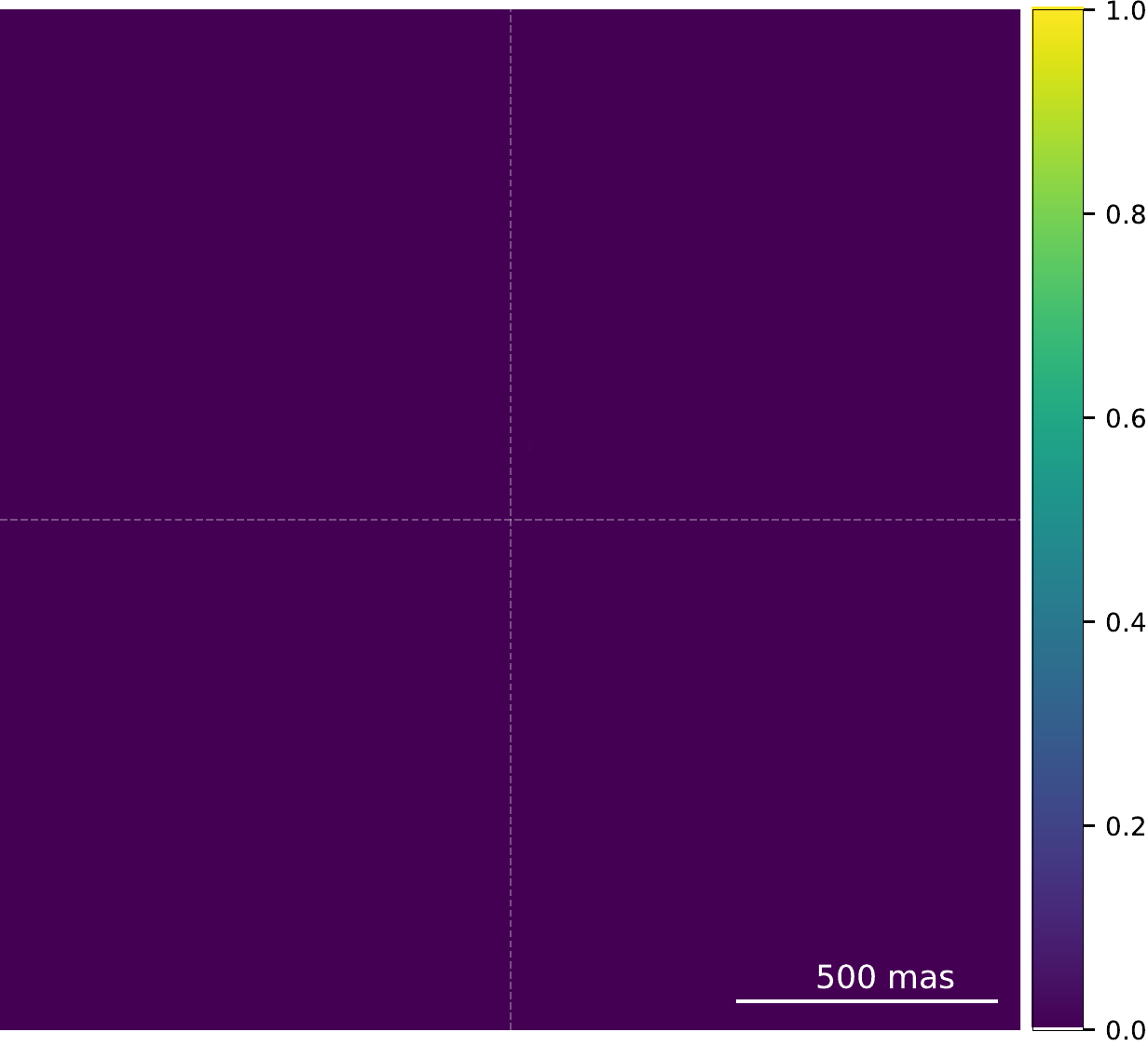}}
  \subfloat[SPHERE-3]{\includegraphics[width=160pt]{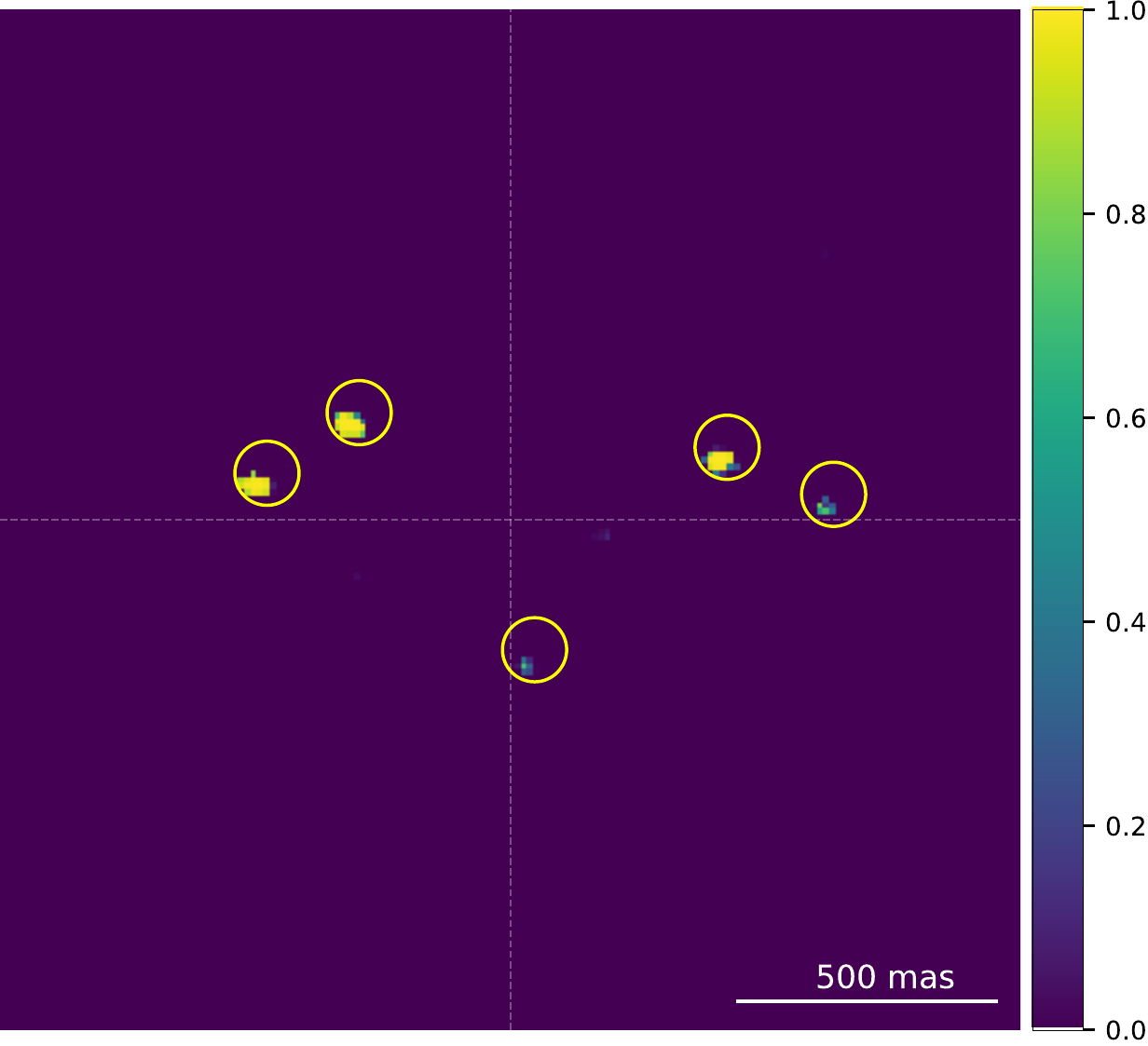}}\\
   \subfloat[NIRC2-1]{\includegraphics[width=160pt]{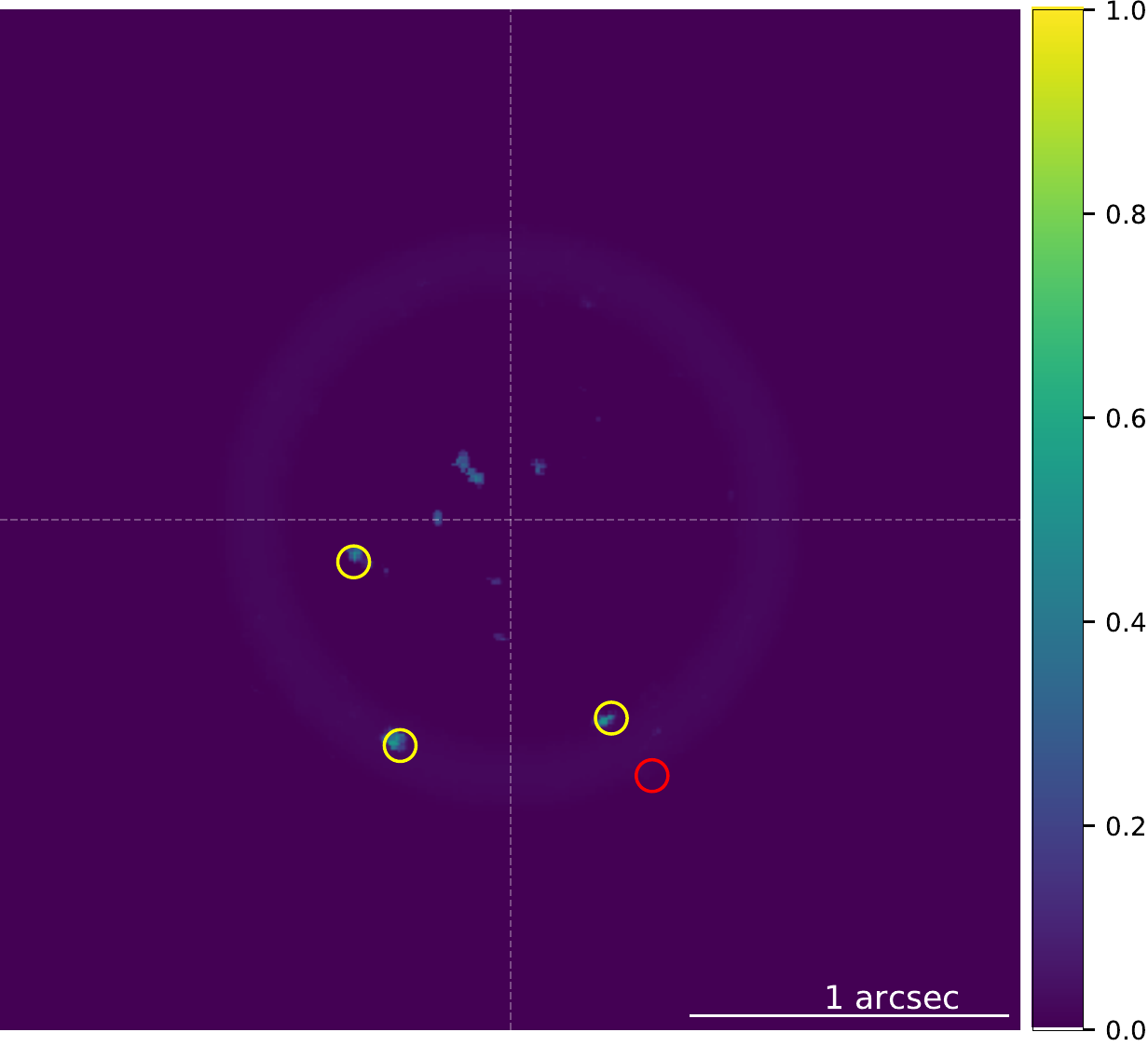}}
  \subfloat[NIRC2-2]{\includegraphics[width=160pt]{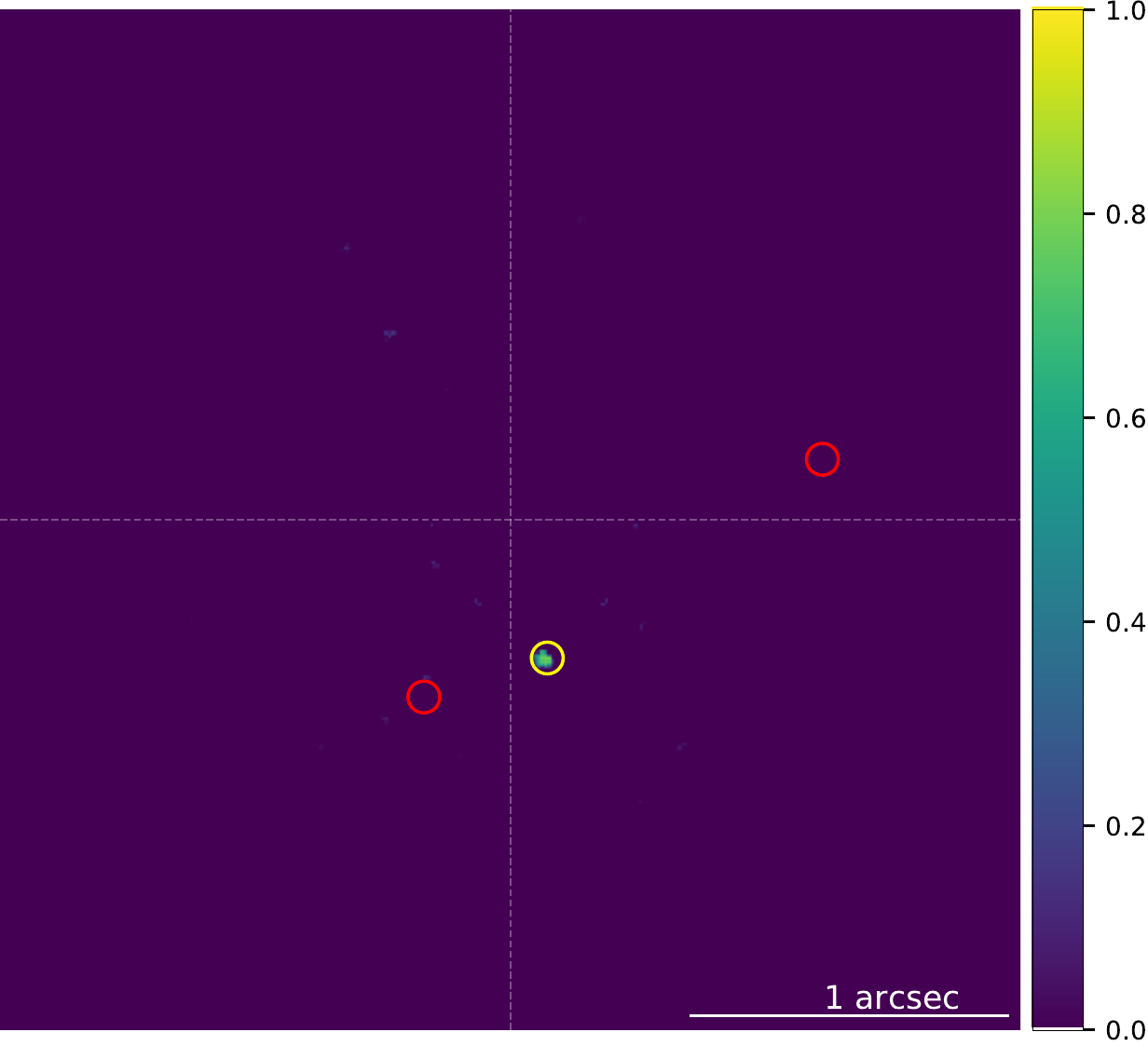}}
  \subfloat[NIRC2-3]{\includegraphics[width=160pt]{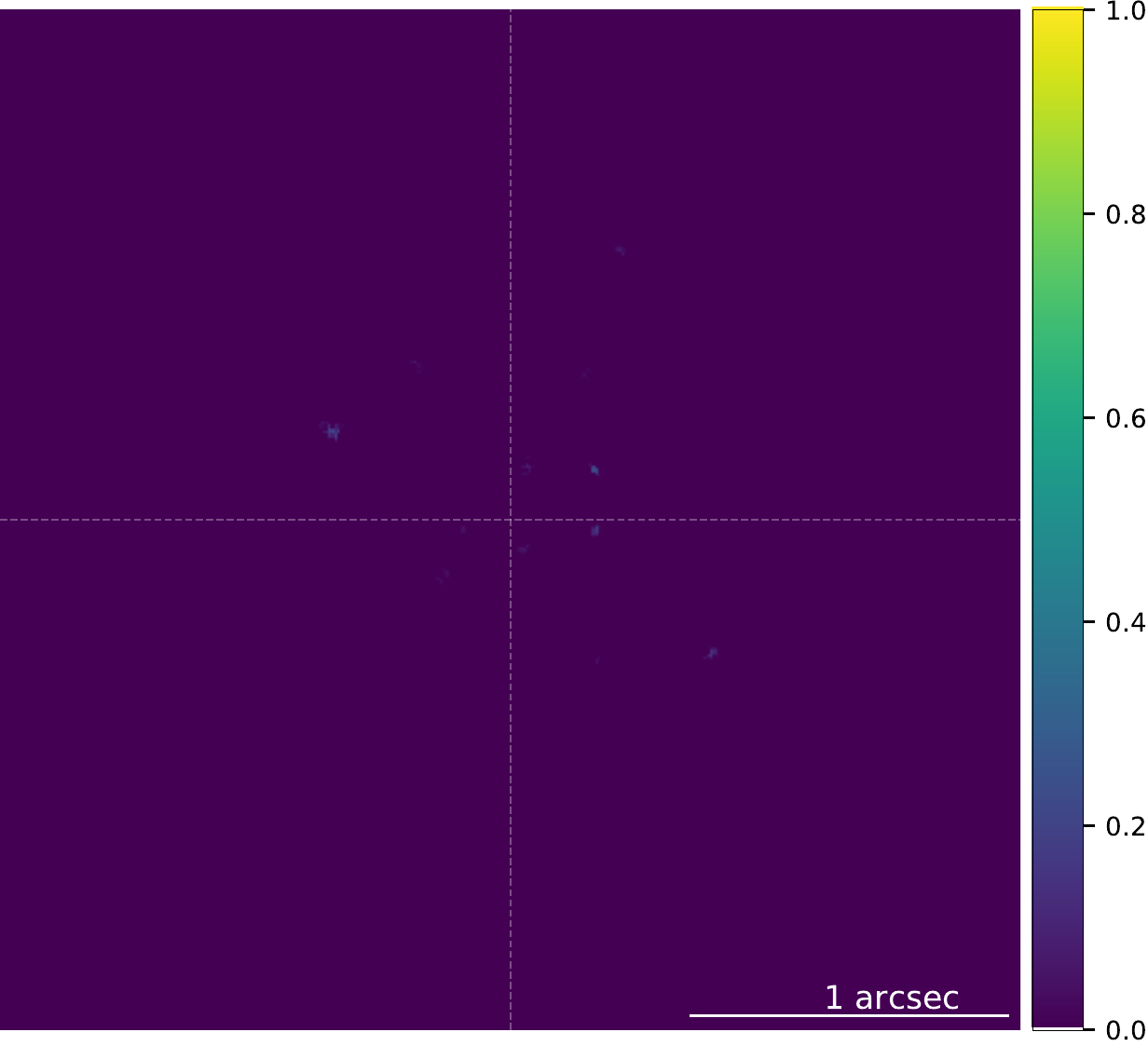}}\\
    \subfloat[LMIRCam-1]{\includegraphics[width=160pt]{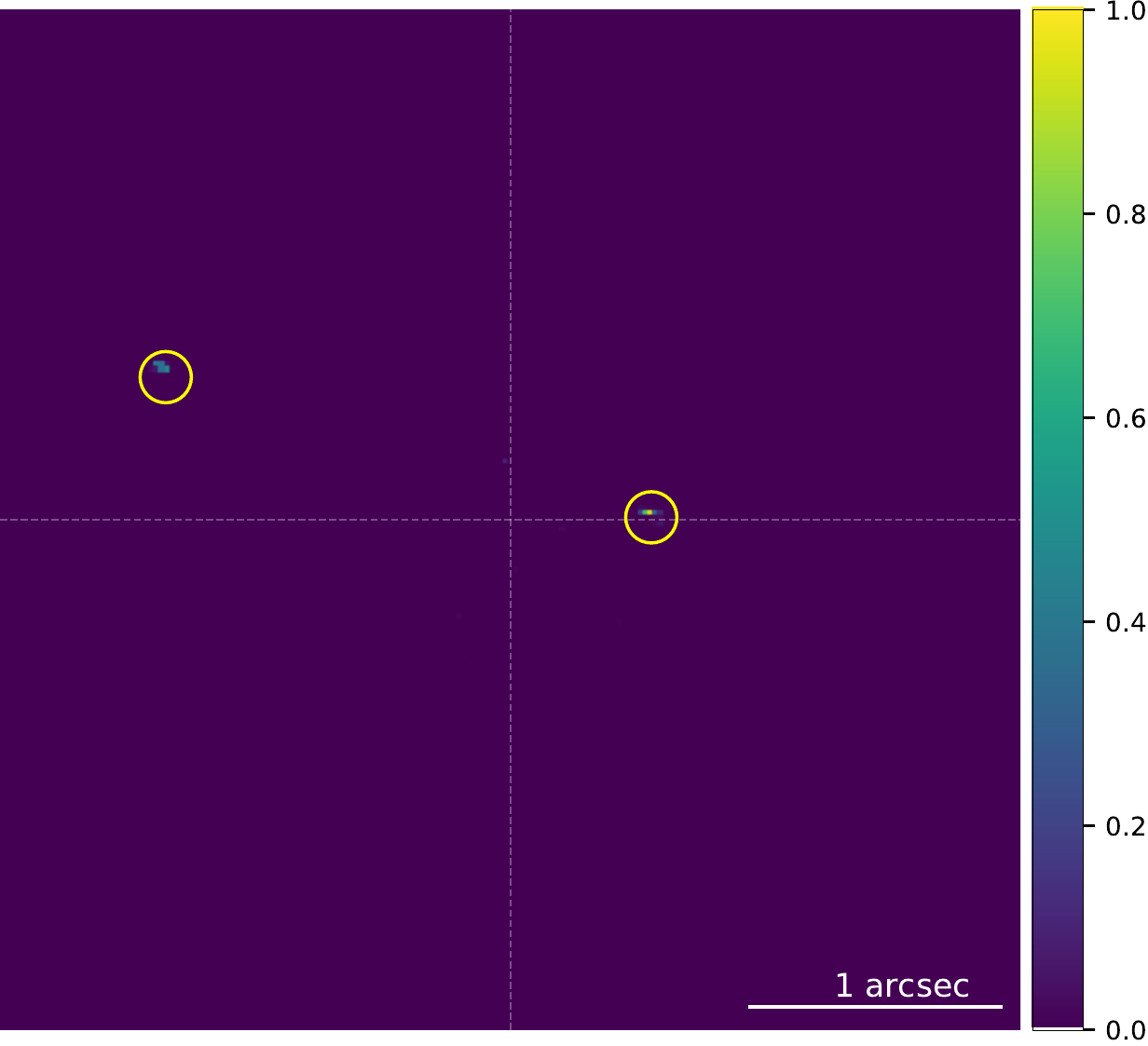}}
  \subfloat[LMIRCam-2]{\includegraphics[width=160pt]{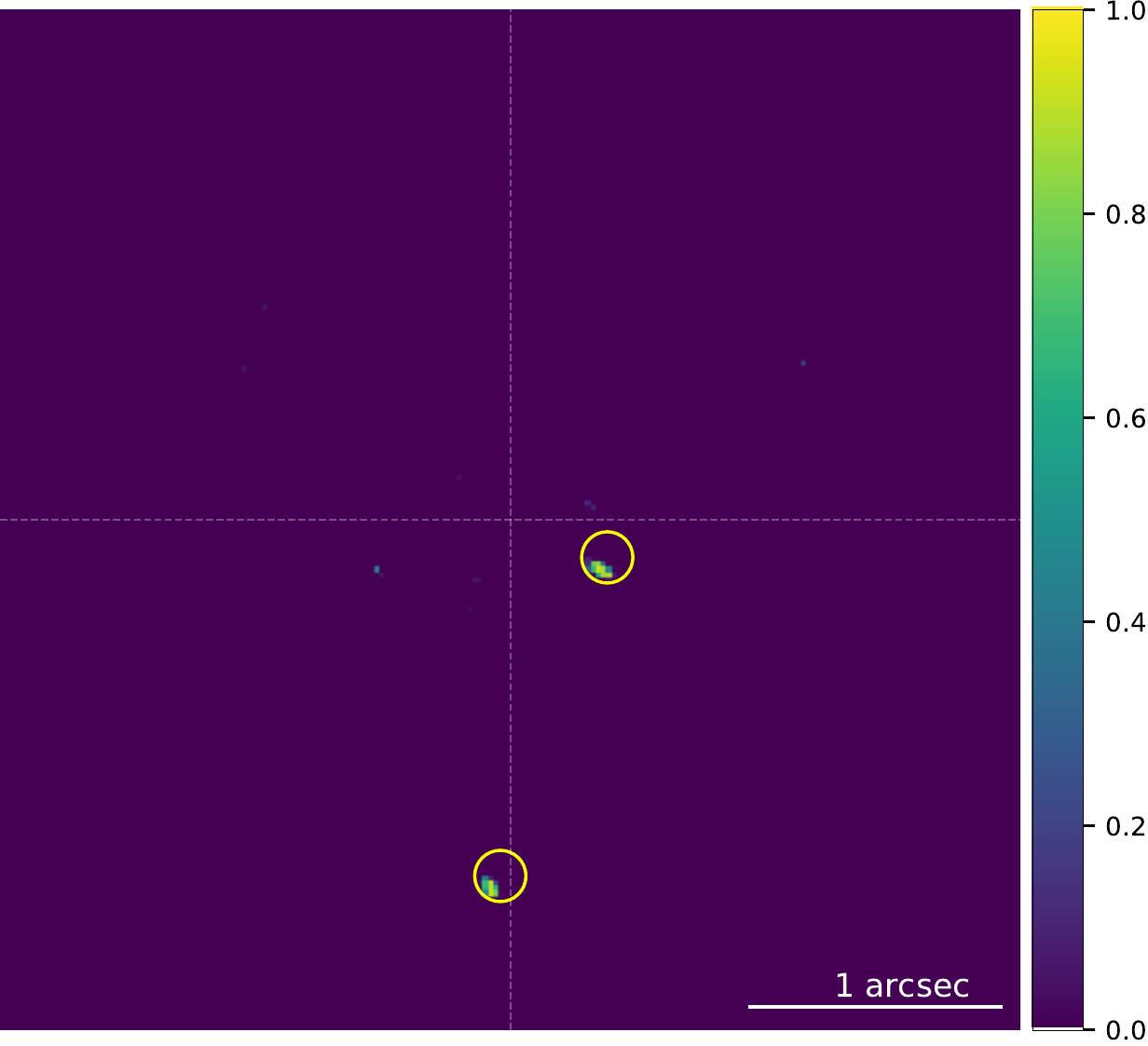}}
  \subfloat[LMIRCam-3]{\includegraphics[width=160pt]{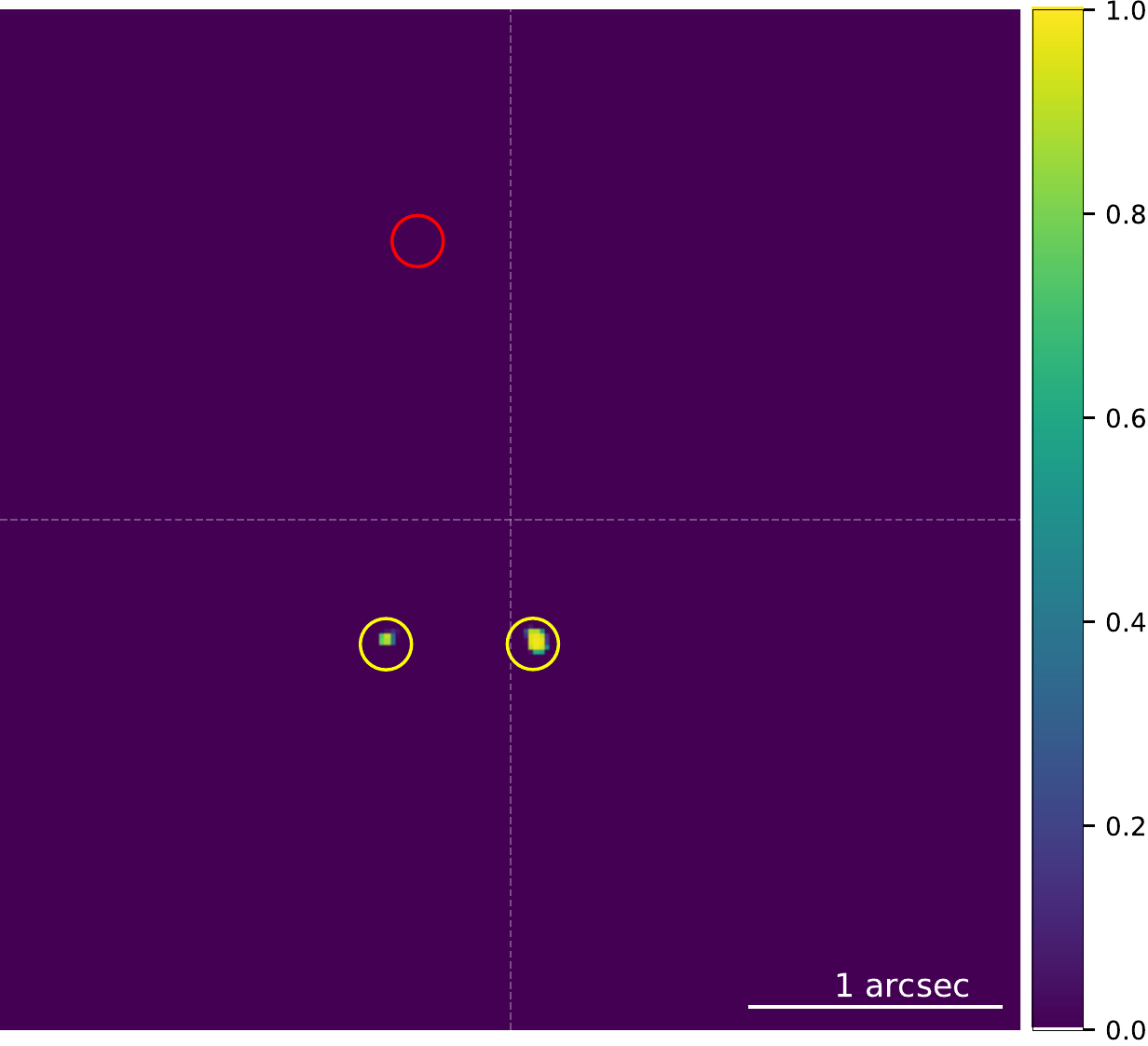}}\\

  \caption{\label{MapFFBUF} Detection maps corresponding to the nine data sets of the EIDC, generated with the full-frame version of auto-RSM using the bottom-up approach for the selection of the optimal set of cubes of likelihoods, as well as the forward approach for the computation of the probabilities. The yellow circles are centred on the true position of the detected targets (TP) and the red circles give the true positions of FNs.}
\end{figure*}

Having presented the five tested parametrisations of the full-frame and annular auto-RSM, we now turn to the estimation of the detection maps and the computation of the performance metrics, which will allow us to rank these parametrisations and compare them with both the original RSM algorithm and the baseline presented in \cite{Cantalloube20}. All parametrisations of the auto-RSM were applied to the nine data sets of the EIDC. Figure \ref{MapFFBUF} presents the detection maps generated with the full-frame auto-RSM using the bottom-up greedy algorithm to select the optimal set of cubes of likelihoods and the forward approach to compute the probabilities (auto-RSM FF\_BU\_F). The detection maps for all five parametrisations of the auto-RSM are provided in Appendix \ref{paramperf}. As can be seen from Fig. \ref{MapFFBUF}, the contrast between detected targets and background residual probabilities is very high compared to standard S/N maps, demonstrating the ability of the proposed approach to easily disentangle planetary signals from residual speckles and ease the selection of a detection threshold. As an illustration, the ratio between the peak probability (or S/N) of the target and the mean of the background probabilities (or S/Ns) in the detection map of the SPHERE 1 data set is larger than 3000 for the auto-RSM FF\_BU\_F, and only 2 for the baseline\footnote{In the case of the baseline S/N map, the minimum S/N value has been added to the S/N map to have only positive values.}. 

\begin{figure*}[h!]
\footnotesize
  \centering
  \subfloat[SPHERE-1]{\includegraphics[width=160pt]{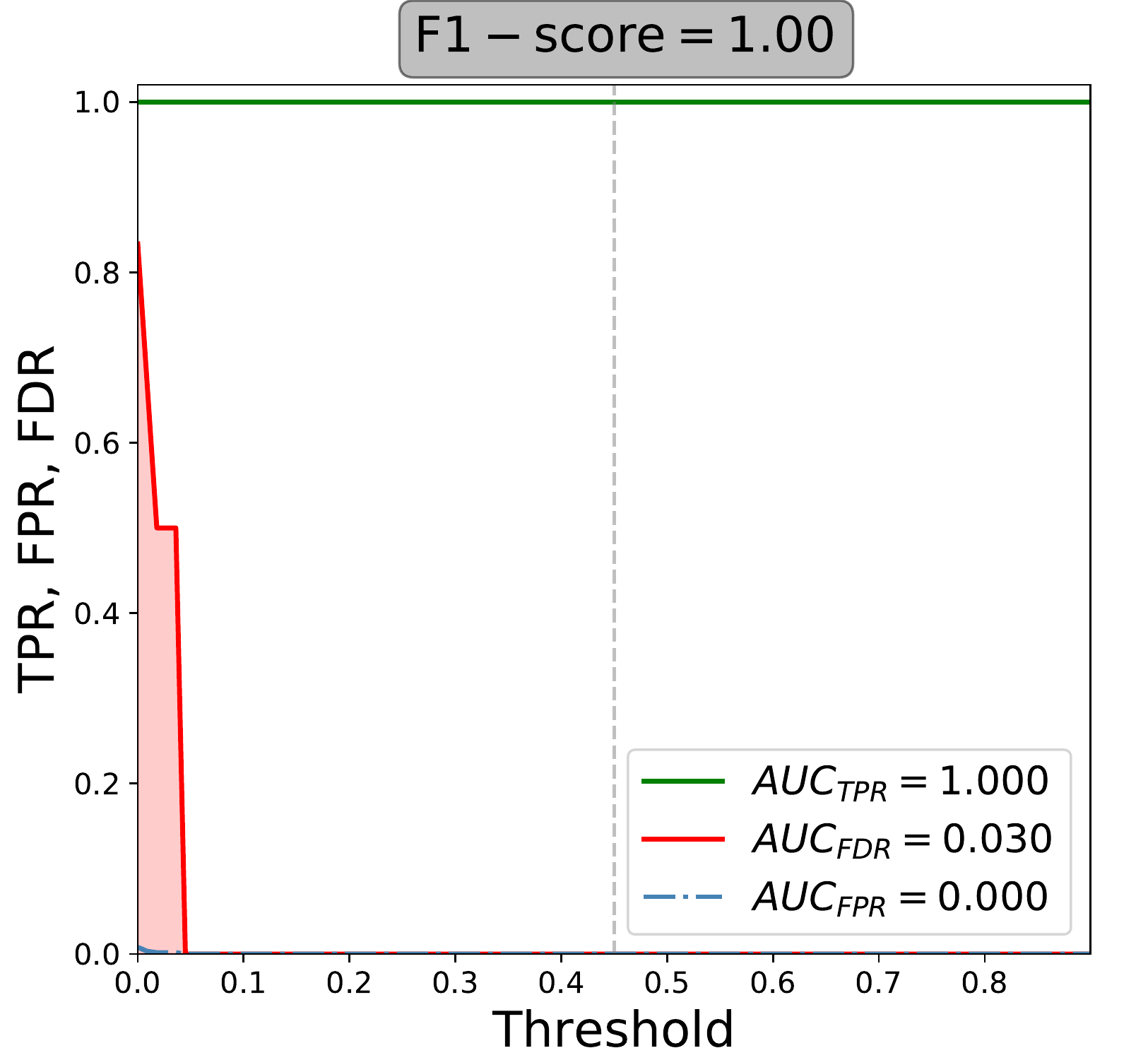}}
  \subfloat[SPHERE-2]{\includegraphics[width=160pt]{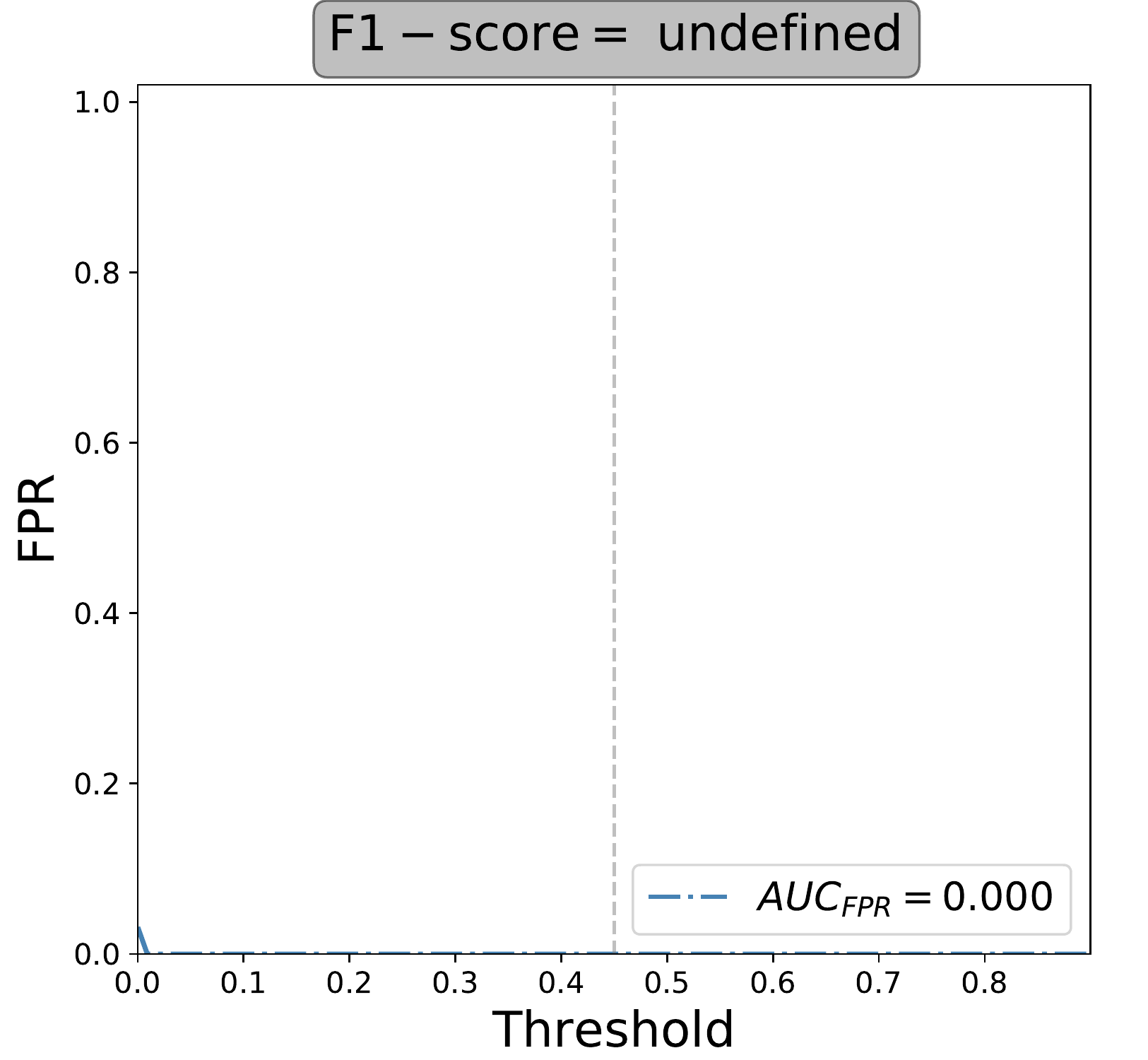}}
  \subfloat[SPHERE-3]{\includegraphics[width=160pt]{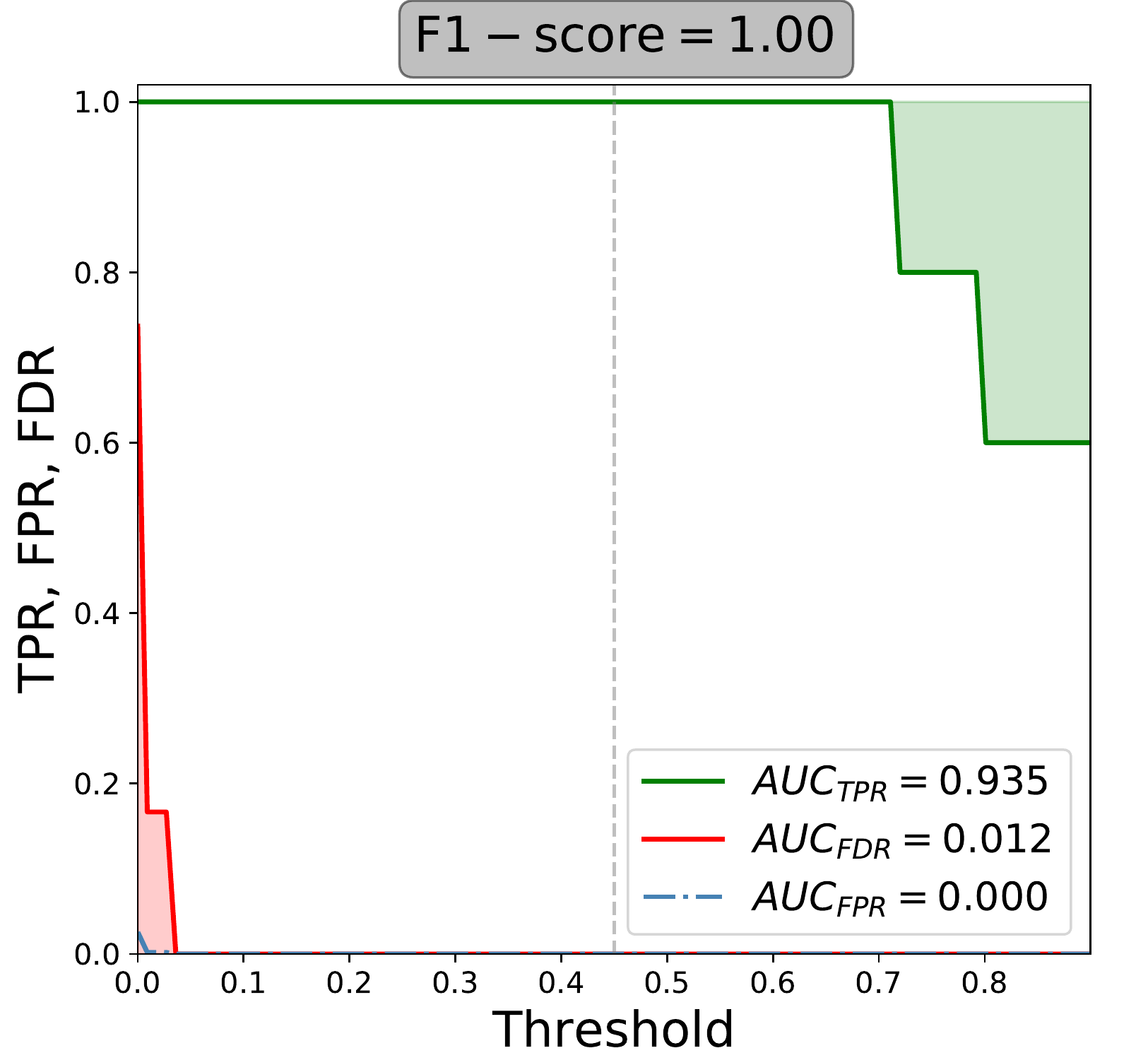}}\\
   \subfloat[NIRC2-1]{\includegraphics[width=160pt]{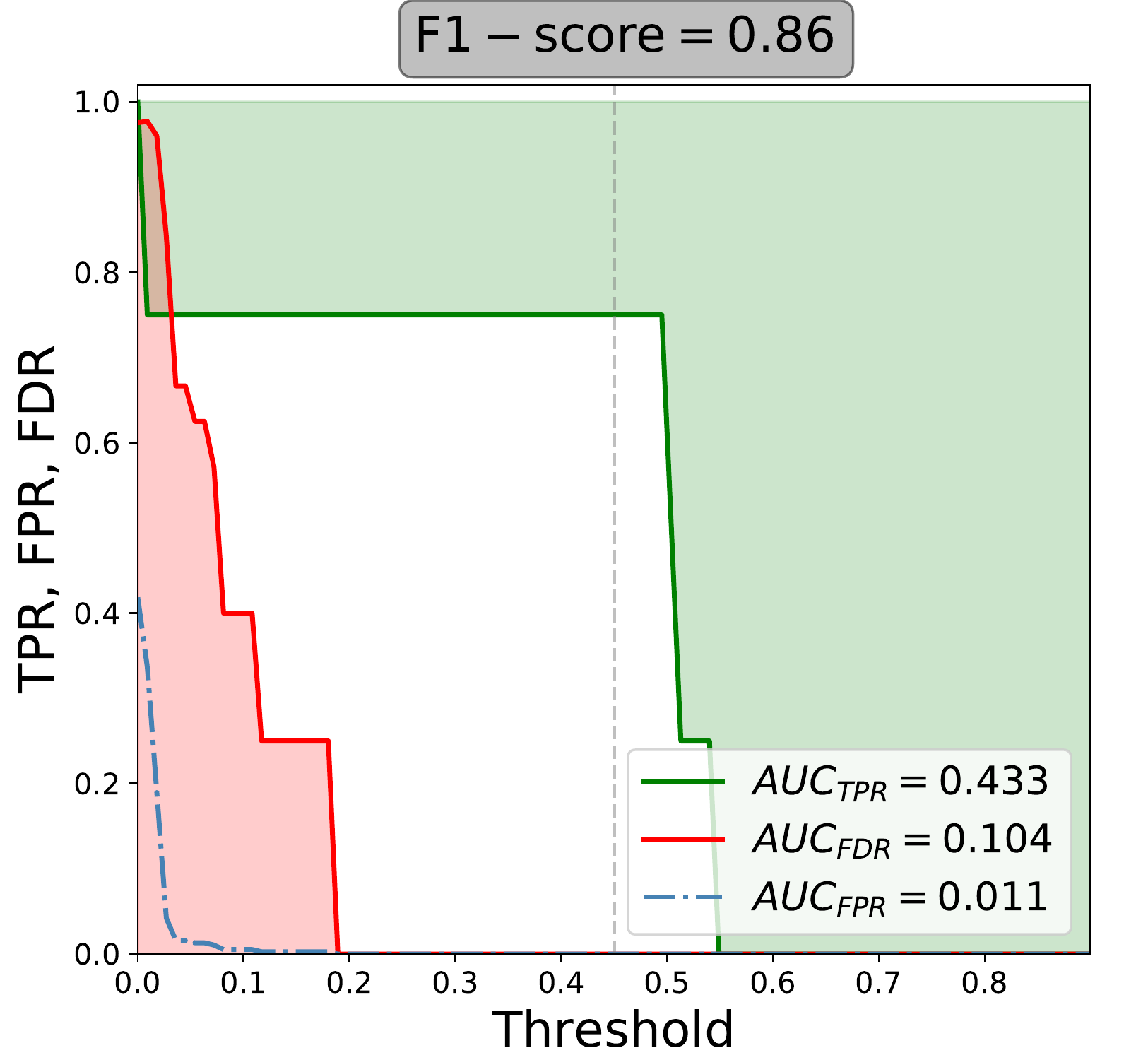}}
  \subfloat[NIRC2-2]{\includegraphics[width=160pt]{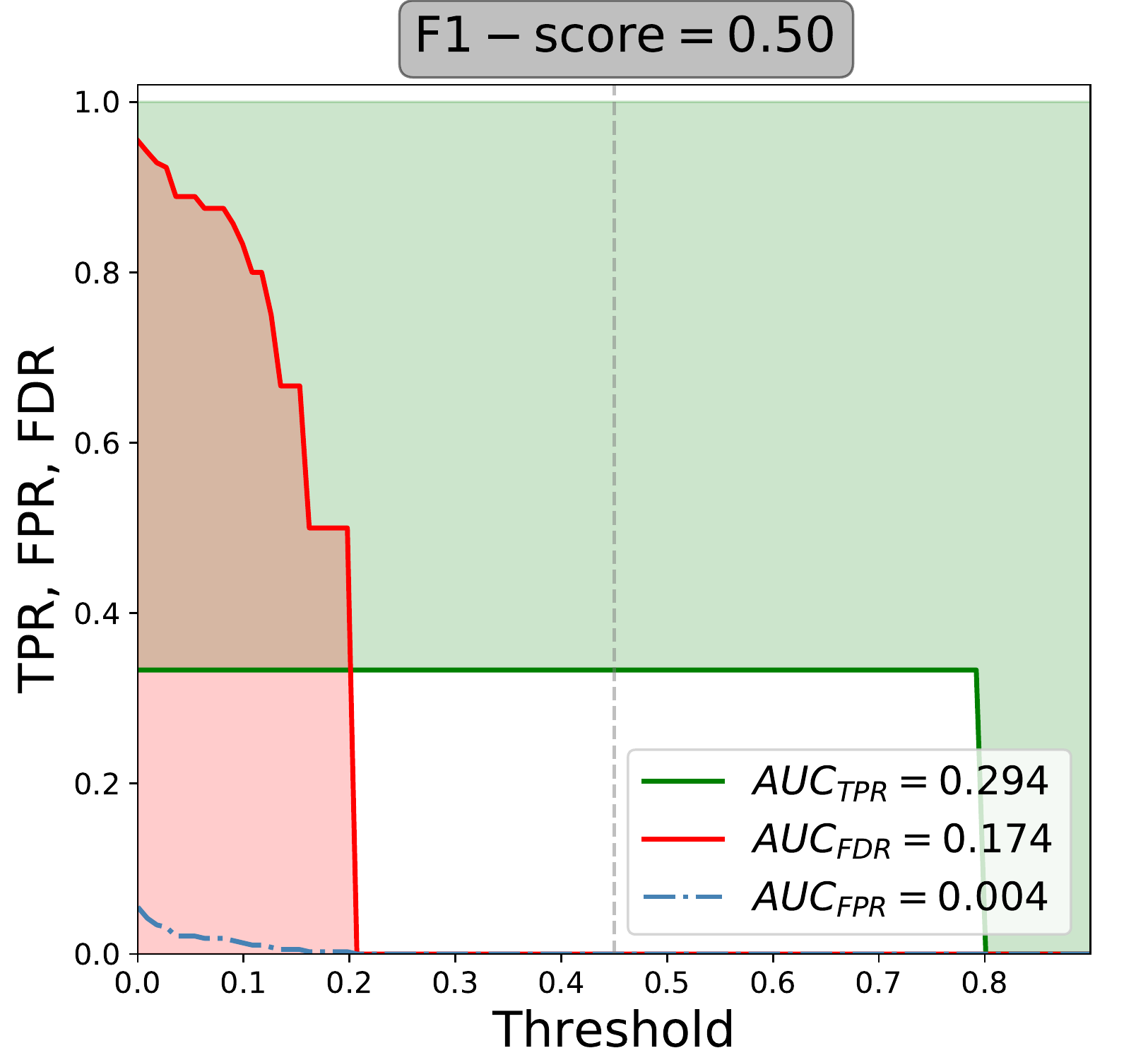}}
  \subfloat[NIRC2-3]{\includegraphics[width=160pt]{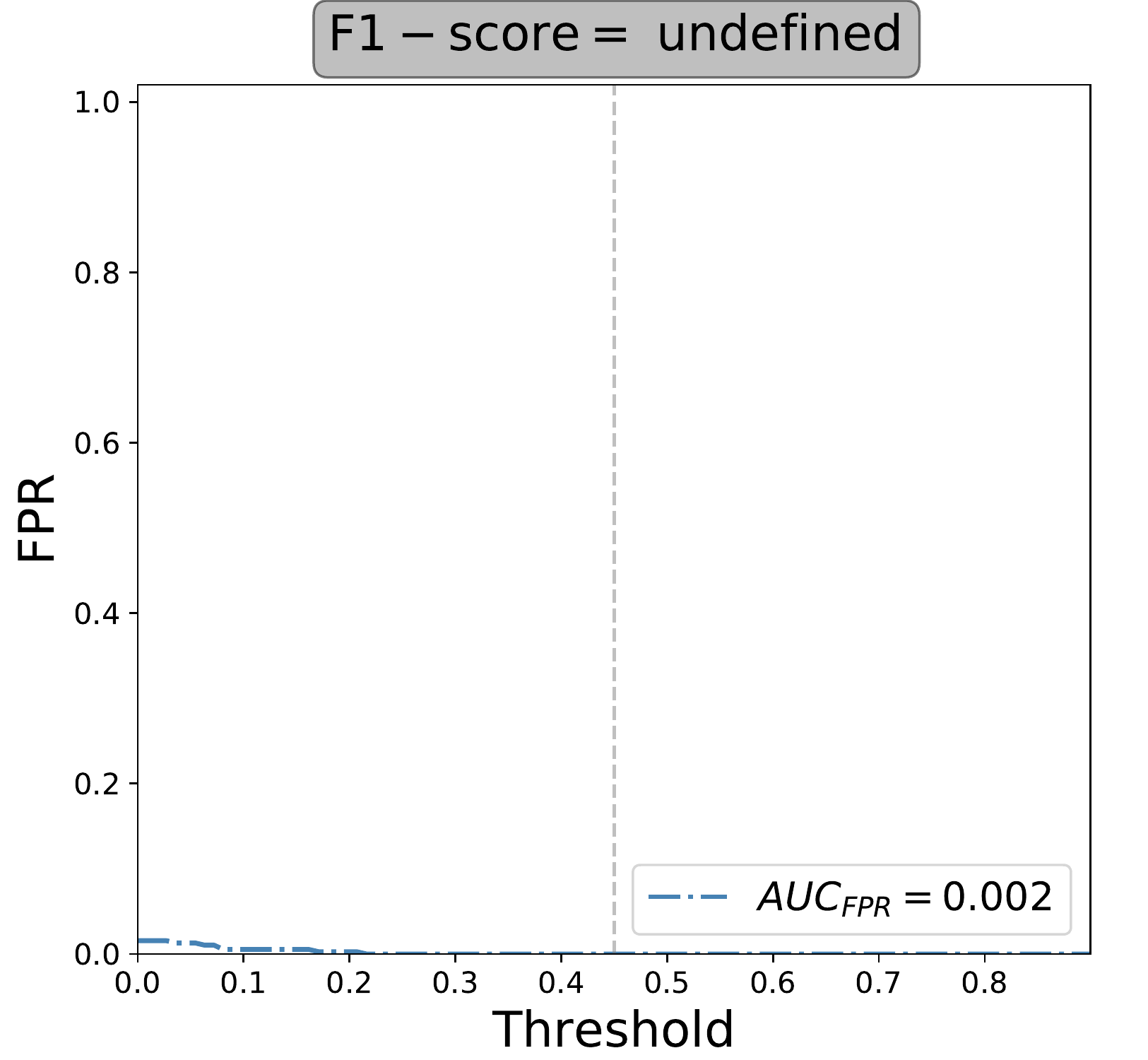}}\\
   \subfloat[LMIRCam-1]{\includegraphics[width=160pt]{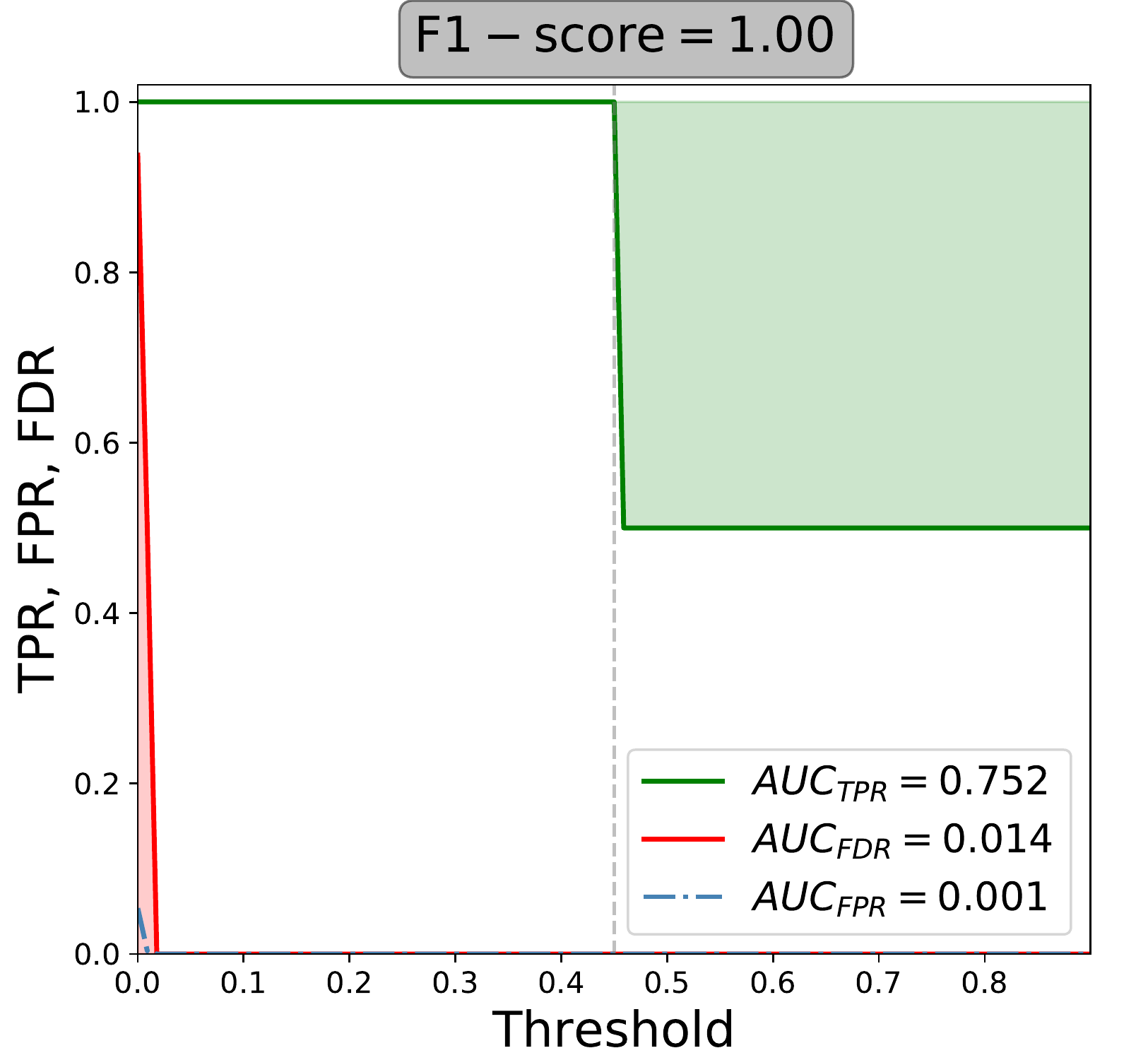}}
  \subfloat[LMIRCam-2]{\includegraphics[width=160pt]{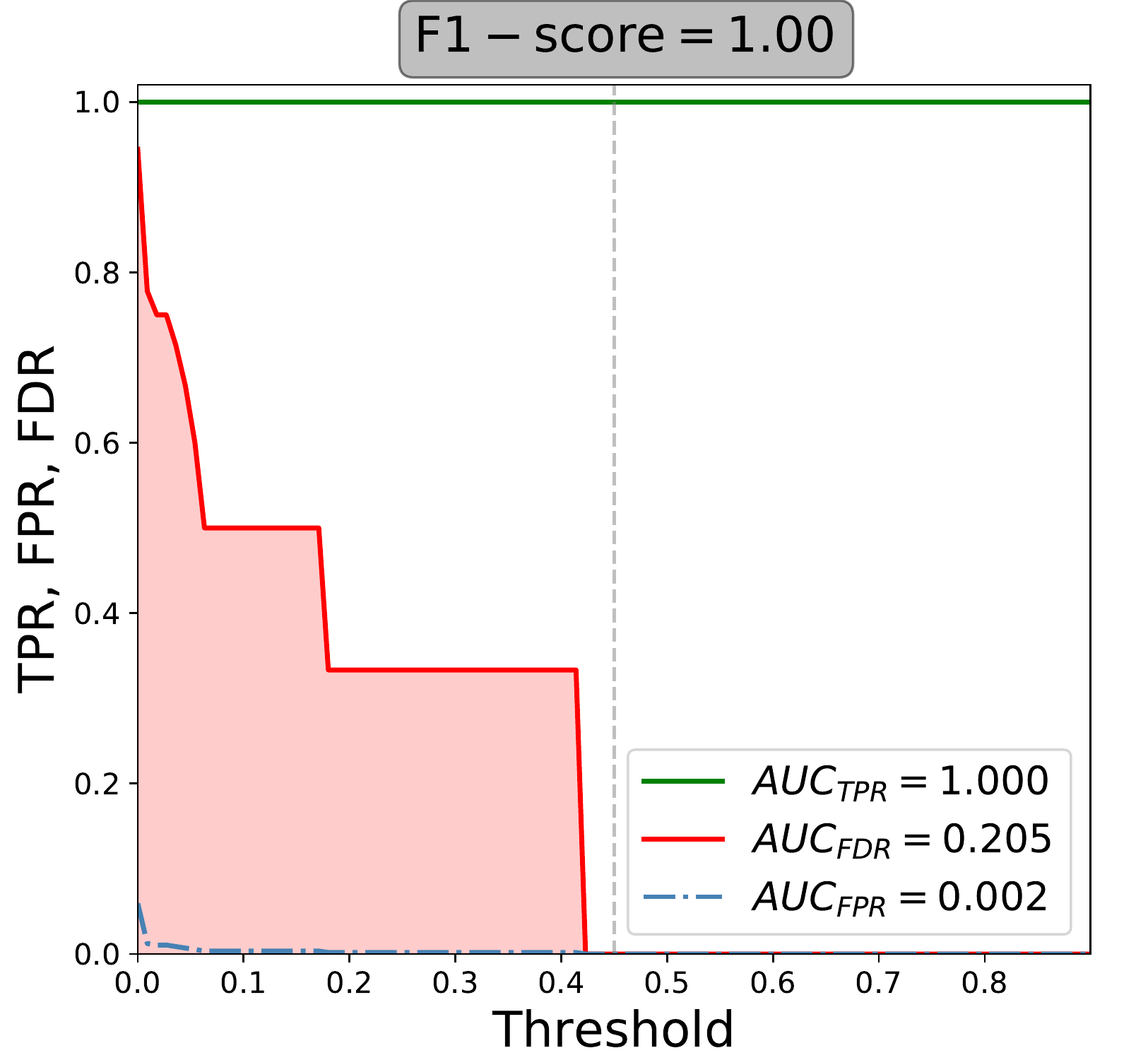}}
  \subfloat[LMIRCam-3]{\includegraphics[width=160pt]{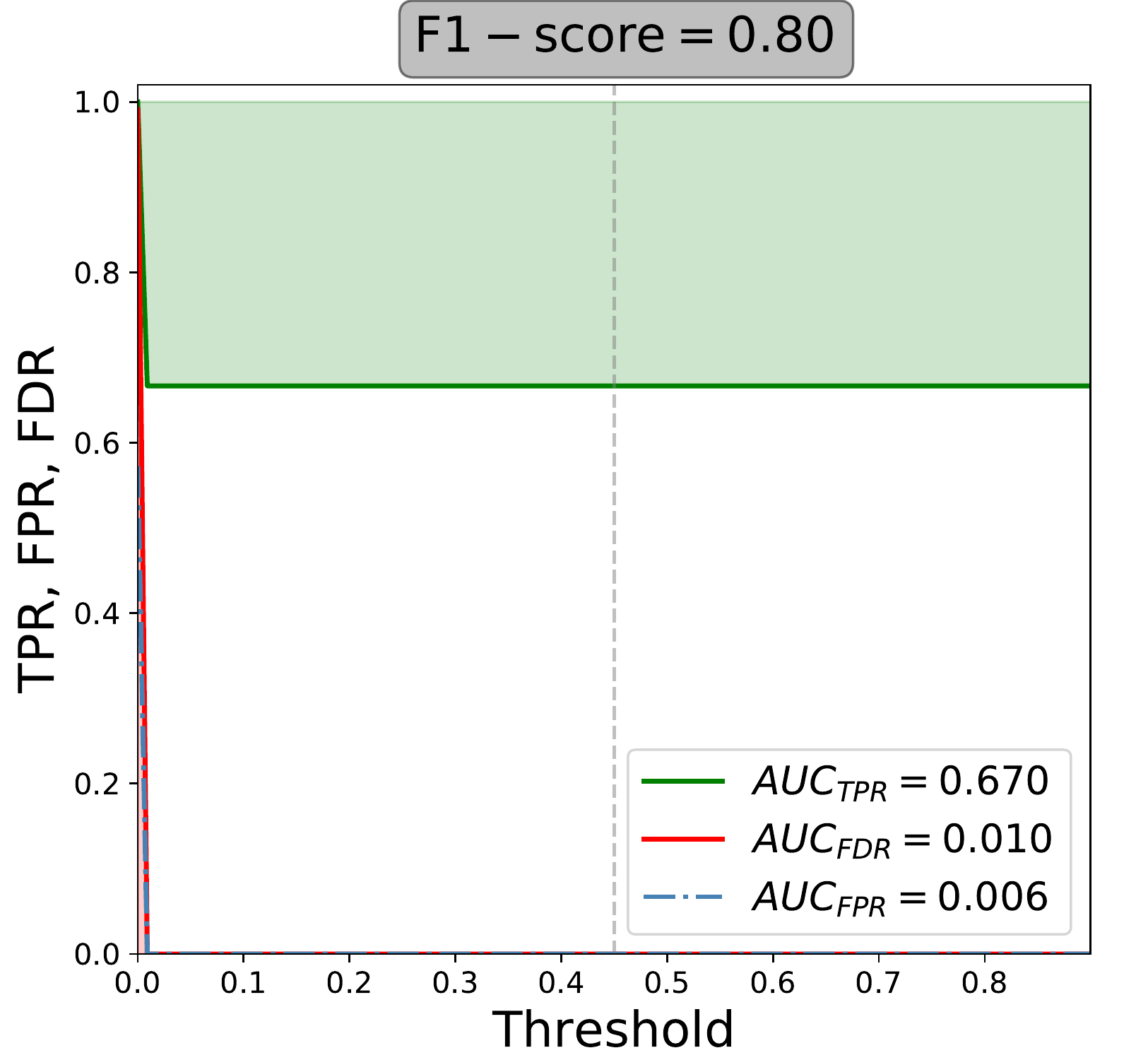}}\\

  \caption{\label{CurveFFBUF} True positive rate (green), false discovery rate (red) and false positive rate (dash-dotted blue line) computed for a range of thresholds varying from zero to twice the selected threshold (represented by a dotted vertical line). These curves are computed for the nine data sets of the EIDC, relying on the detection maps estimated with the full-frame version of auto-RSM using the bottom-up approach for the selection of the optimal set of cubes of likelihoods, as well as the original forward approach for the computation of the probabilities. The green line representing the TPR should be as close as possible to 1 for the entire range of thresholds, while the red and dash-dotted blue line representing respectively the FDR and the FPR, should be as close as possible to zero.}
\end{figure*}

Following the EIDC procedure, a single threshold was selected for all data sets, for each parametrisation of the auto-RSM. This threshold allows estimation of the F1 score. As mentioned in Sect. \ref{perfmet}, in addition to the F1 score, the AUCs of the TPR, FPR, and FDR are also computed. Figure \ref{CurveFFBUF} illustrates the computation of the different performance metrics for the nine data sets, relying on the detection maps generated with the auto-RSM FF\_BU\_F. The TPR, FPR, and FDR are computed for different threshold values ranging from zero to twice the selected threshold. The AUCs of the TPR, FPR, and FDR are computed in this interval. Apart from the NIRC2 data sets and LMIRCam-2, the AUC of the FDR is very small for the remaining data sets compared to the baseline. The AUC of the FPR is close to zero for all data sets, especially for the SPHERE data sets for which the AUC values are below the considered 0.001 limit.

\begin{figure*}[h!]
\footnotesize
  \centering
  \subfloat[(a)]{\includegraphics[trim=0 20 0 6, clip,width=260pt]{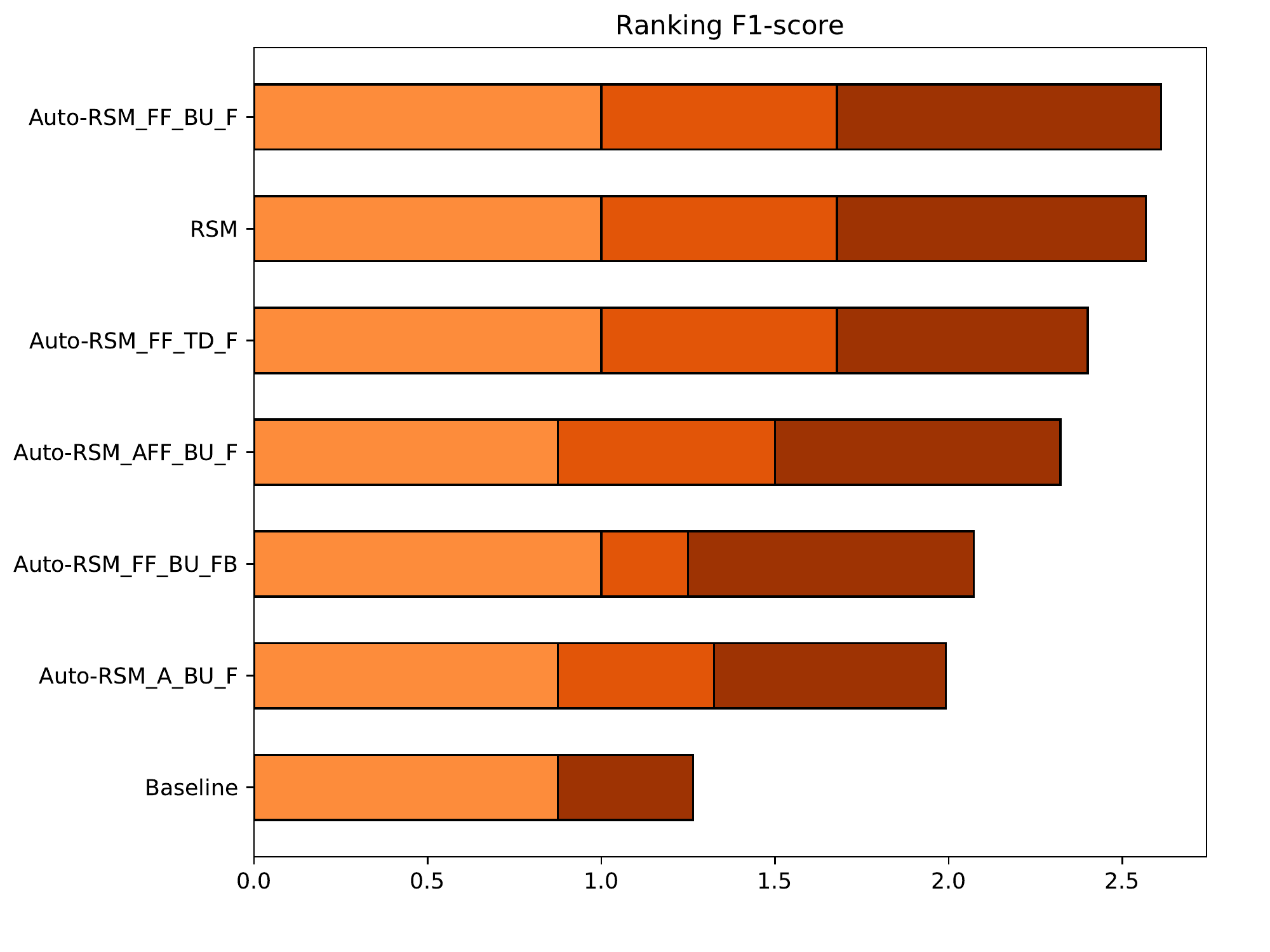}}
  \subfloat[(b)]{\includegraphics[trim=0 20 0 6, clip,width=260pt]{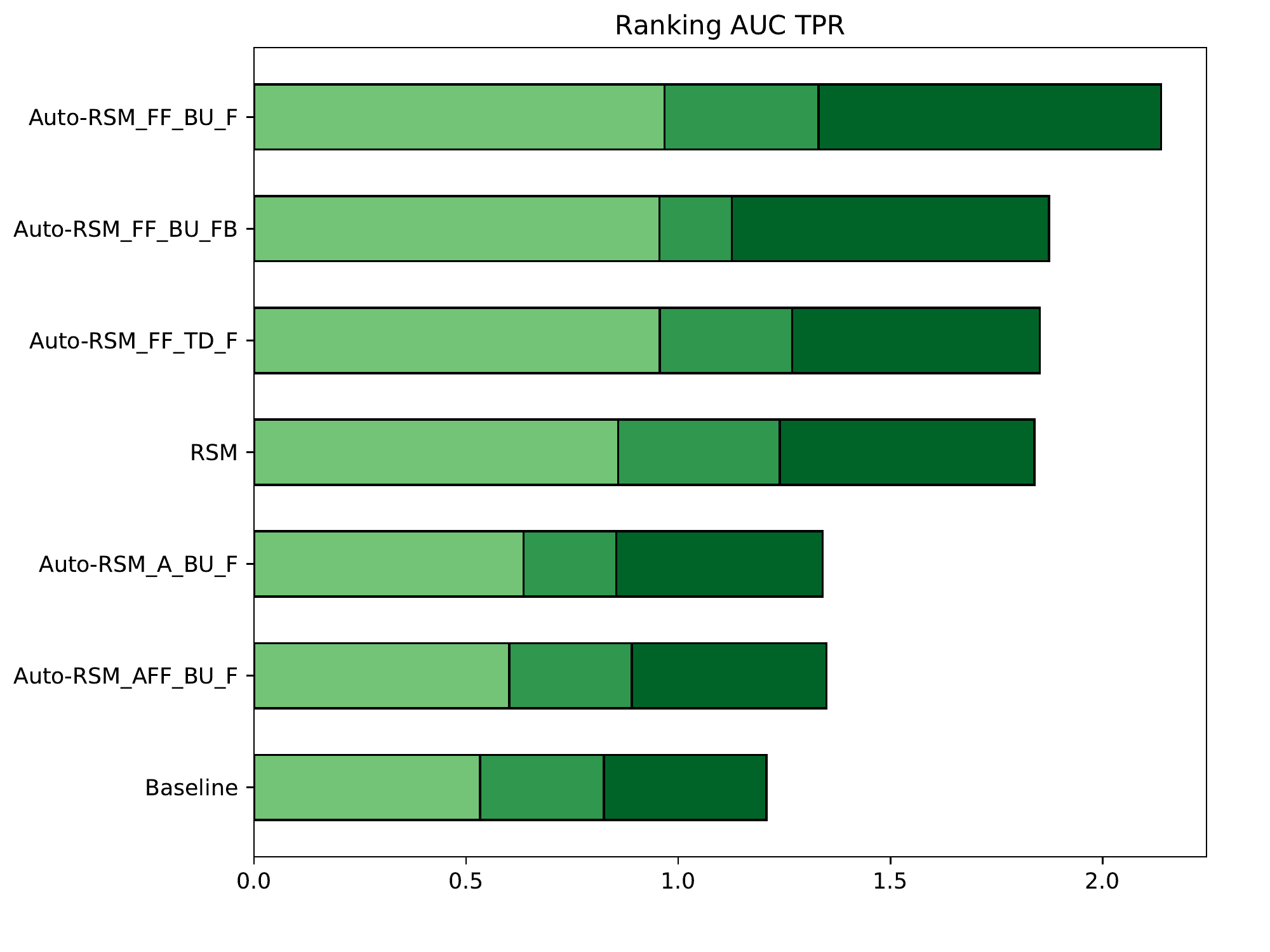}}\\
  \subfloat[(c)]{\includegraphics[trim=0 20 0 6, clip,width=260pt]{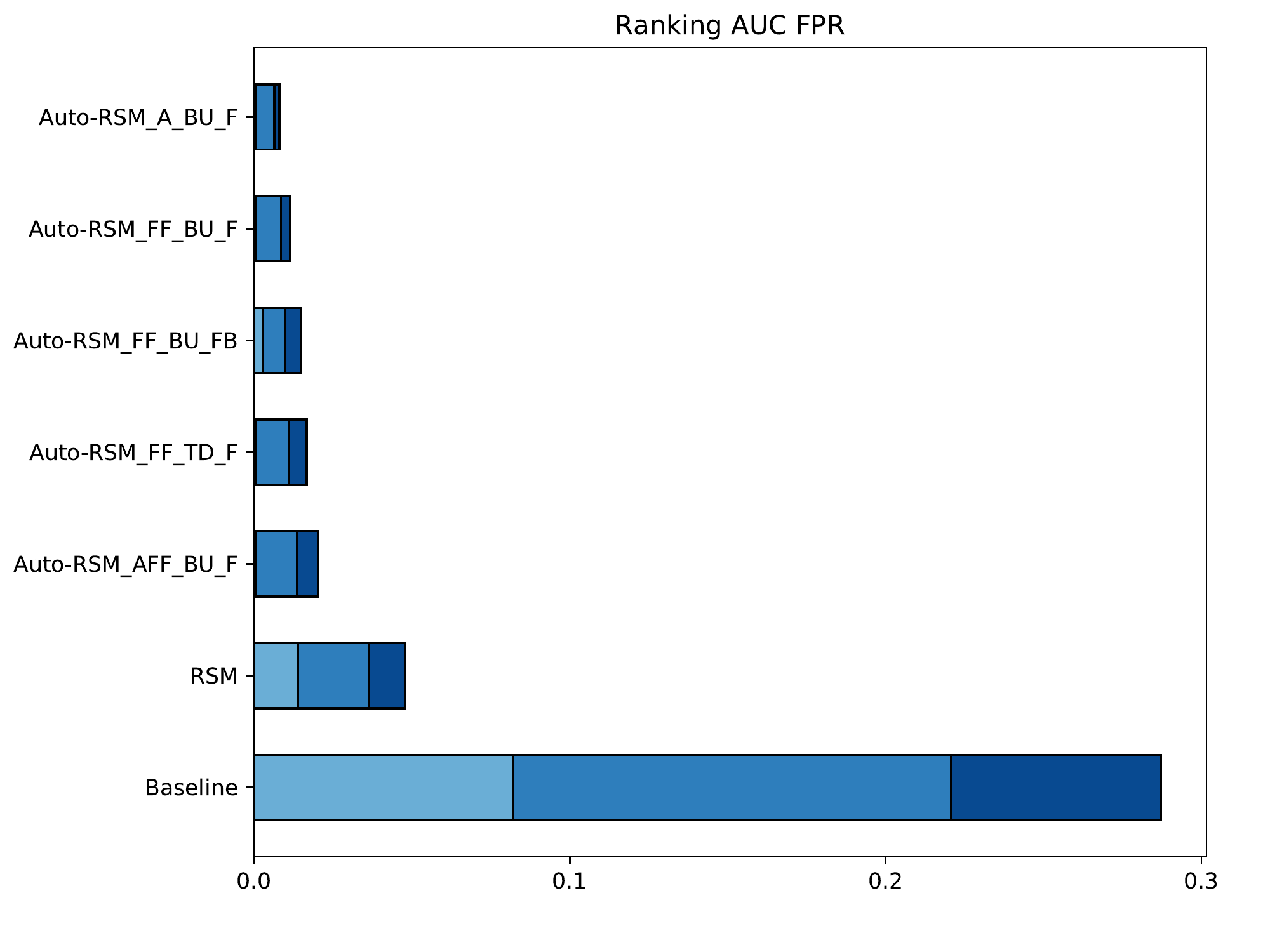}}
  \subfloat[(d)]{\includegraphics[trim=0 20 0 6, clip,width=260pt]{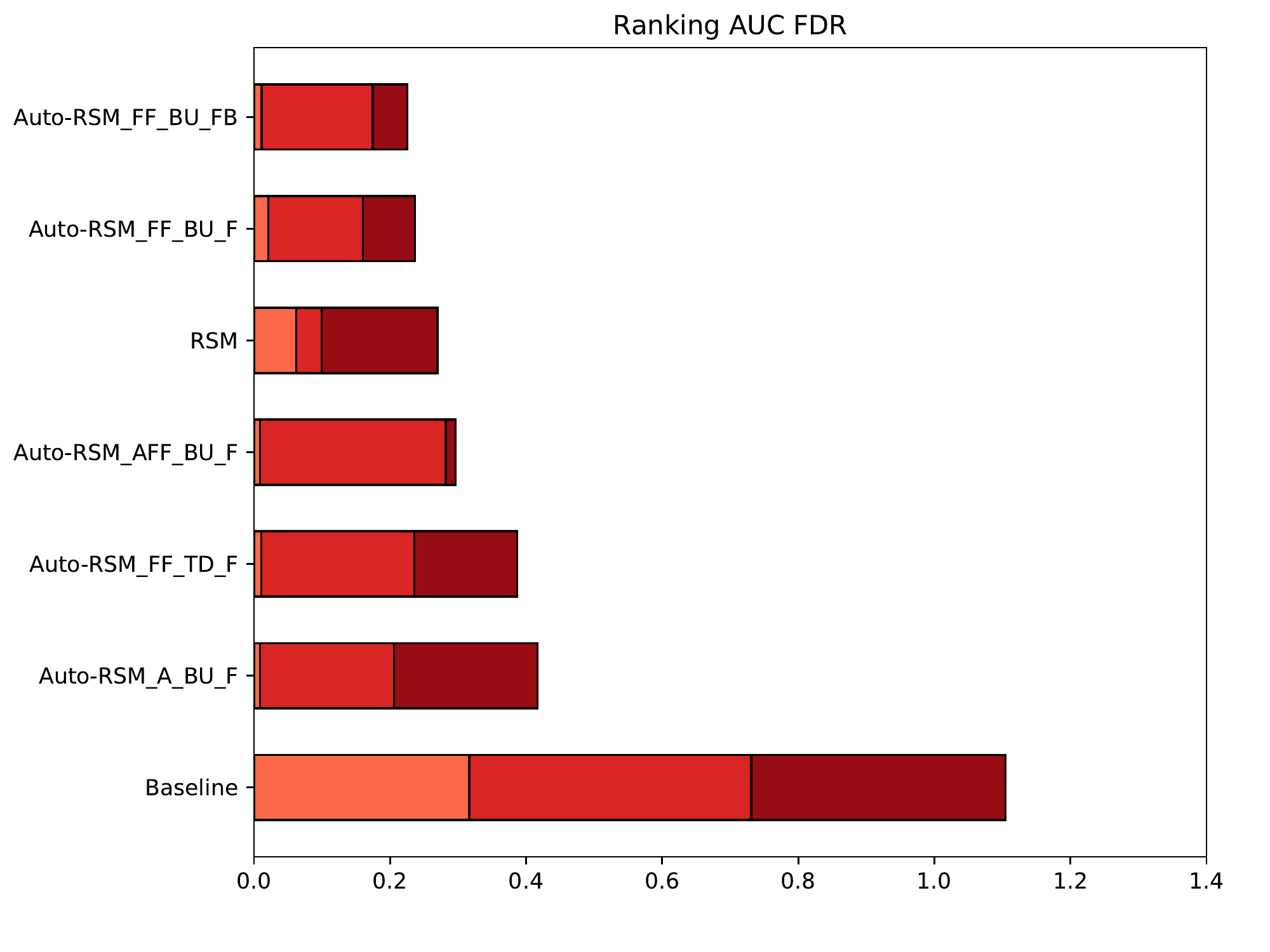}}\\
  \caption{\label{Ranking} Ranking of the different parametrisations of the full-frame and annular versions of the auto-RSM along the original RSM and baseline presented in \cite{Cantalloube20}. Figure (a) provides the ranking based on the F1 score obtained at the selected threshold. Figure (b) gives the ranking based on the AUC of the TPR. Figure (c) gives the ranking based on the AUC of the FPR, while Figure (d) provides the ranking based on the AUC of the FDR. FF stands for full-frame, A for annular, AFF for annular full-frame (annular approach used to optimise the PSF-subtraction parameters and full-frame approach used for the RSM parameter optimisation and the selection of the optimal set of cubes of likelihoods), and BU, TD, F, and FB as explained in Sect. 4.3.1. The light, medium, and dark colours correspond to VLT/SPHERE-IRDIS, Keck/NIRC2, and LBT/LMIRCam data sets, respectively.}
\end{figure*}

Having illustrated the computation of the performance metrics for the different data sets, we now consider aggregated results to compare the performance of the five auto-RSM parametrisations with the baseline and RSM algorithm submission to the EIDC. The different rankings for the four considered performance metrics are shown in Fig. \ref{Ranking}. The light, medium, and dark colours correspond to the three instruments, with the VLT/SPHERE-IRDIS, Keck/NIRC2, and LBT/LMIRCam data sets, respectively. Figure \ref{Ranking} highlights the fact that the RSM-based approaches largely outperform the baseline with much higher F1 scores, a much larger AUC of the TPR, and much lower AUCs of the FDR and FPR.  Regarding the five considered auto-RSM parametrisations, they all present a smaller F1 score compared to the RSM algorithm parametrised manually, except for the auto-RSM FF\_BU\_F, which performs slightly better. However, when considering the other performance metrics, the auto-RSM approach seems to perform better in most cases, especially when considering false positives. These results demonstrate the ability of the auto-RSM approach to better cope with residual speckle noise, while maintaining a high detection rate. This is a key element in reducing arbitrariness in the selection of the detection threshold. The selection of a detection threshold is indeed often a complex task, especially when relying on S/N maps, as the noise probability distribution is often non-Gaussian.

Looking in more detail at the five parametrisations of the auto-RSM, we see clearly that the auto-RSM FF\_BU\_F leads to the best performance metrics in most cases, and should therefore be favoured for detection when using the auto-RSM approach. The results for the annular and hybrid annular full-frame auto-RSM seem to demonstrate that considering the radial evolution of the optimal parameters does not lead to a significant improvement in performance. The slightly degraded performance of the annular mode can be explained by the fact that the auto-RSM optimisation relies on the inverted parallactic angle approach. The noise structure being similar but not equivalent when inverting the parallactic angles, the annular optimisation is more affected by local differences in the noise structure. These local differences in the noise structure prevent the algorithm from improving the overall performance. Considering the longer computation time required for the annular auto-RSM, and its performance, the full-frame version should clearly be preferred. 

As regards the difference between the bottom-up and top-down approaches, the better results obtained with the bottom-up approach may be explained by its ability to select the cube of likelihoods in the right order. Indeed, the probability associated with a planetary signal increases along the temporal axis when computing the RSM detection map. This probability increases faster and stays high for longer when selecting first the cube of likelihoods providing the highest probability ratio between injected fake companion peak probability and background residual probabilities. Sorting the cubes of likelihoods in descending order of quality leads to a higher average probability for the planetary signal, while it should not affect the probability associated with residual noise.

In addition to the automated selection of the optimal parameters, the results obtained with the auto-RSM FF\_BU\_F show a clear performance improvement compared to the set of RSM detection maps originally submitted to the EIDC \citep{Cantalloube20}. We observe an overall reduction of 76\%\ and 33\% compared to the RSM submission for the AUC of the TPR and the AUC of the FDR,  respectively, as well as an increase of 19\%\ and 2\%\ for the AUC of the TPR and the F1 score, respectively.

\subsection{Commonalities in optimal parametrisations}

The proposed auto-RSM optimisation procedure is relatively time consuming even when relying on the full-frame mode, which may potentially preclude its use for very large surveys. In this
section, we therefore investigate the possibility of using a smaller set of ADI sequences to generate an optimal parametrisation, which could then be applied to a larger set of ADI sequences. This requires a homogeneity in the sets of optimal parameters selected for different ADI sequences generated by a given HCI instrument, as surveys generally consider multiple observation sequences generated by a single HCI instrument. Sources of heterogeneities in the parametrisation of ADI sequences  for a common instrument can originate from the number of frames, the observing conditions, the parallactic angle range, or the target position in the sky. 

As the EIDC data set contains multiple ADI sequences generated with different instruments under different observing conditions and with different characteristics (see \ref{ADIdesc} for the frames number, FOV rotation), it should allow us to estimate the homogeneity of the parametrisations for a common instrument, and potential heterogeneity between instruments. As the full-frame version of the optimisation algorithm provides the best performance, we rely on the set of parameters generated by this mode to conduct our analysis. We define a heterogeneity metric, which differs for the PSF-subtraction techniques and the RSM algorithm.  We use  the distance between parametrisations as a metric with which to gauge the performance of the PSF-subtraction techniques. This distance is defined as the difference between the optimal values of all the parameters (see Table \ref{parametersff}) for each pair of ADI sequences. For each parameter, we normalise the absolute value of the distance between the two considered ADI sequences with the mean of the two optimal values used to compute the distance. This allows proper comparison of the relative weight of the different parameters. For the RSM parametrisation, most parameters are non-numerical and we therefore replace the notion of distance by the notion of similarity. For a given pair of ADI sequences,  as a metric for each parameter we use the percentage of dissimilarity within the entire set of PSF-subtraction techniques, i.e. the number of PSF-subtraction techniques for which the parameter values are different divided by the total number of PSF-subtraction techniques.

\begin{figure}[h!]
  \centering
  \subfloat[]{\includegraphics[width=260pt]{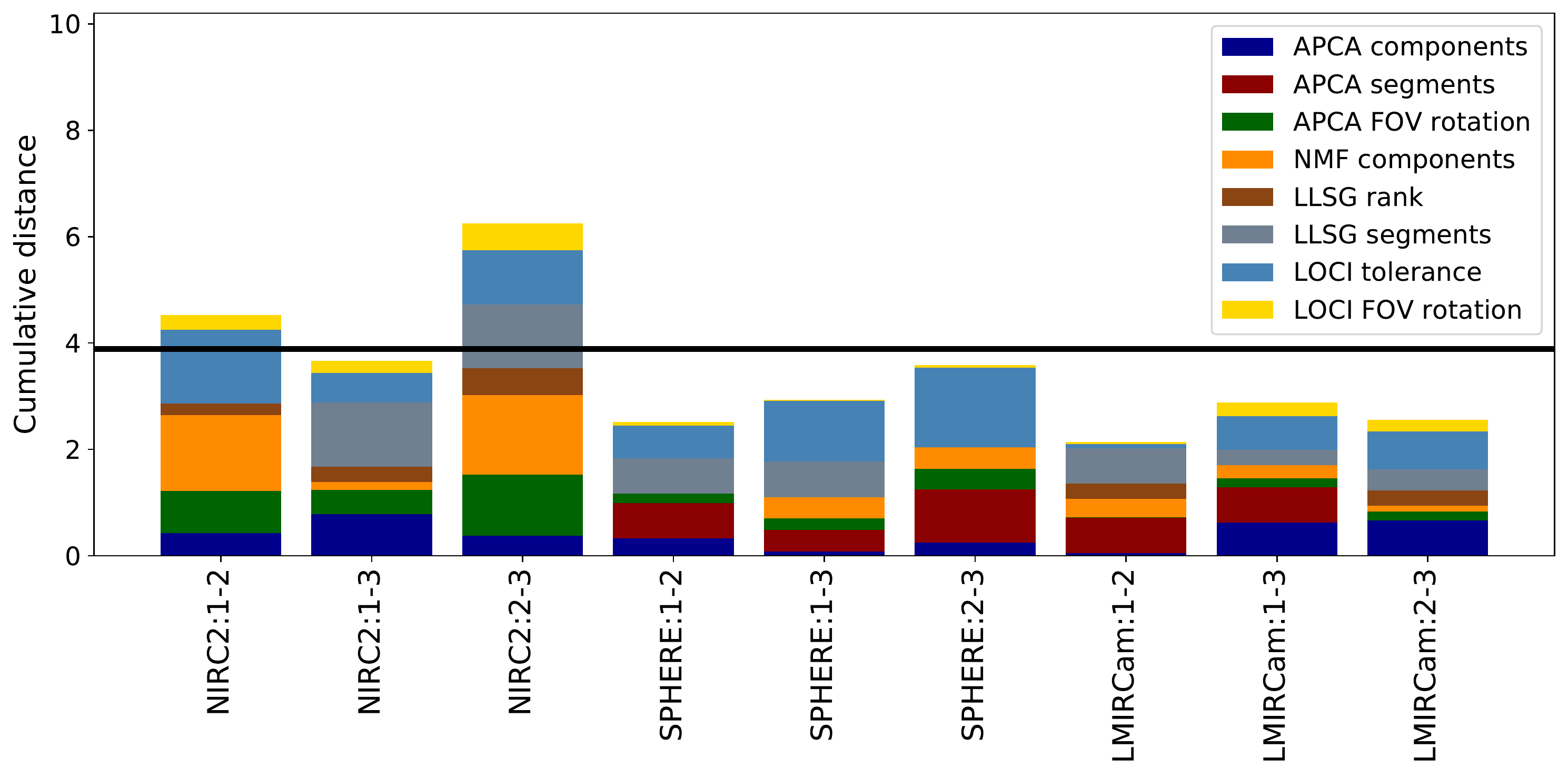}}\\
  %\subfloat[(b) PSF-subtraction techniques]{\includegraphics[width=260pt]{Distance_full_psfsub.pdf}}
 
  \caption{\label{commonalitiespsf} Normalised distances between PSF-subtraction-technique parameter sets for nine pairs of ADI sequences. The different coloured bars provide the contribution of the different parameters to the cumulative normalised distance. The considered pairs of ADI sequences are generated by the same instrument. The black horizontal line represents the normalised distances averaged over the 36 possible pairs of ADI sequences. }
\end{figure}

Figure \ref{commonalitiespsf} shows the cumulative normalised distances for every pair of ADI sequences within each instrument, along with the relative weight of all the parameters. The black line gives the mean distance computed based on the 36 possible pairs of ADI sequences. A cumulative distance below the back line indicates a higher homogeneity of the parameters for the considered pair of ADI sequences. As can be seen from Figure \ref{commonalitiespsf}, the ADI sequences generated by the SPHERE and LMIRCam instruments seem to be characterised by a relatively homogeneous set of parameters, which implies that a common set of parameters could be defined and used for larger surveys. The larger heterogeneity for the NIRC2 samples, which seems to be mainly driven by the NIRC2-2 sequence, may be explained by the ADI sequence characteristics (see Table. \ref{parametersff}), or by differences in terms observing conditions.

\begin{figure}[h!]
  \centering
  \subfloat[]{\includegraphics[width=260pt]{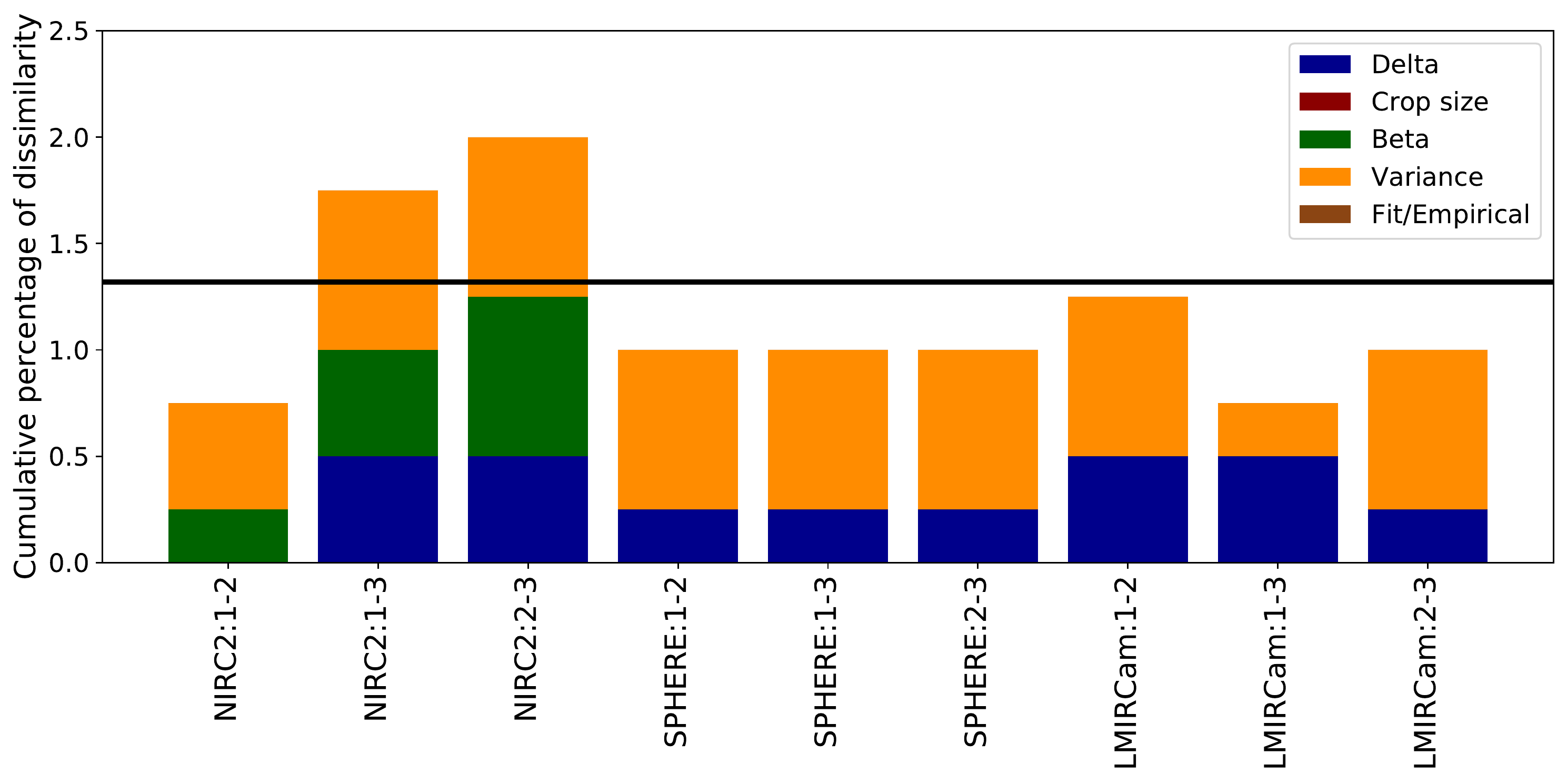}}\\
  %%\subfloat[(b) PSF-subtraction techniques]{\includegraphics[width=260pt]{Distance_full_RSM_2.pdf}}
 
  \caption{\label{commonalitiesrsm} Percentage of dissimilarity between RSM parameter sets for nine pairs of ADI sequences. The different coloured bars provide the contribution of the different parameters to the cumulative dissimilarity. The selected pairs of ADI sequences are generated by the same instrument. The black horizontal line represents the percentage of dissimilarity averaged over the 36 possible pairs of ADI sequences.}
\end{figure}

Considering now the parametrisation of the RSM algorithm, the results presented in Fig. \ref{commonalitiesrsm}, demonstrate again the larger parametric heterogeneity for the NIRC2 samples, with the NIRC2-3 sequence affecting the dissimilarity measures  the most. This confirms the particularity of the NIRC2-3 sequence, which seems to have a different noise structure compared to the other ADI sequences. The crop size of 3 pixels as well as the use of a best-fit approach to estimate the noise properties are common to all ADI sequences (see Table \ref{parametersff}). The heterogeneity is mainly driven by the definition of the region used for computation of the noise properties, which tends to demonstrate the advantage of considering multiple approaches for estimation of the noise properties. We finally consider the set of selected PSF-subtraction techniques used to generate the final RSM map when relying on the full-frame bottom-up auto-RSM. We see from Table \ref{optiset} that the SPHERE ADI sequences share the same set of PSF-subtraction techniques while the set is different for the other ADI sequences. 

\begin{table}[t]
                        \caption{Selected PSF-subtraction techniques for the computation of the final RSM-detection map for the nine ADI sequences in the case of the full-frame version of the auto-RSM procedure using the bottom-up approach. }
                        \label{optiset}
\centering

                        \begin{tabular}{lcccc}
                        
                        \hline
ID/Selected model &\textbf{APCA}  &\textbf{NMF} & \textbf{LLSG}&\textbf{LOCI } \\                            
 \hline
SPHERE 1 &X & X & X &X\\
SPHERE 2 &X & X & X &X\\
SPHERE 3 &X & X & X &X\\
NIRC2 1 & & X &  &X\\
NIRC2 2 &X &  &  &\\
NIRC2 3 & & X & X &\\
LMIRCam 1 & & & X&X\\
LMIRCam 2 &X & X &  &X\\
LMIRCam 3 &X &  & X &X\\
\hline
                        \end{tabular}
                                \end{table}

In addition to the estimation of relative distances and dissimilarity measures, we also applied a K-means clustering algorithm to classify the nine ADI sequences  into three clusters based on the set of parameters used for the PSF-subtraction techniques and the RSM algorithm, as well as the likelihood cubes selected for the detection map computation. Using these 32 parameters to characterise each ADI sequence\footnote{The categorical parameters such as the RSM parameters and the optimal set of likelihood cubes have been binarised. A standard normalisation, using the parameters mean and variance, has been applied on the data before the clustering algorithm. The K-means clustering algorithm relying on euclidean distance, a proper scaling of the parameters is necessary to avoid favouring some parameter.}, the K-mean algorithm classified NIRC2-1 and NIRC2-3 into the first group, NIRC2-2 into the second group, and the remaining sequences into the third group. As expected, the NIRC2-2 is not in the same group as the other sequences generated with the NIRC2 instrument (see Fig.\ref{commonalitiespsf} and Table \ref{optiset}). For the other sequences, we reach the same conclusion as before, apart from the fact that both SPHERE and LMIRCam data sets are regrouped into a single cluster whose centre is close to the SPHERE-1 ADI sequence. Increasing the number of clusters does not lead to clear separation between the SPHERE and LMIRCam instruments, demonstrating the similarity of the ADI sequences generated by both instruments. 

\begin{figure}[h!]
  \centering
  \subfloat[SPHERE-2]{\includegraphics[width=200pt]{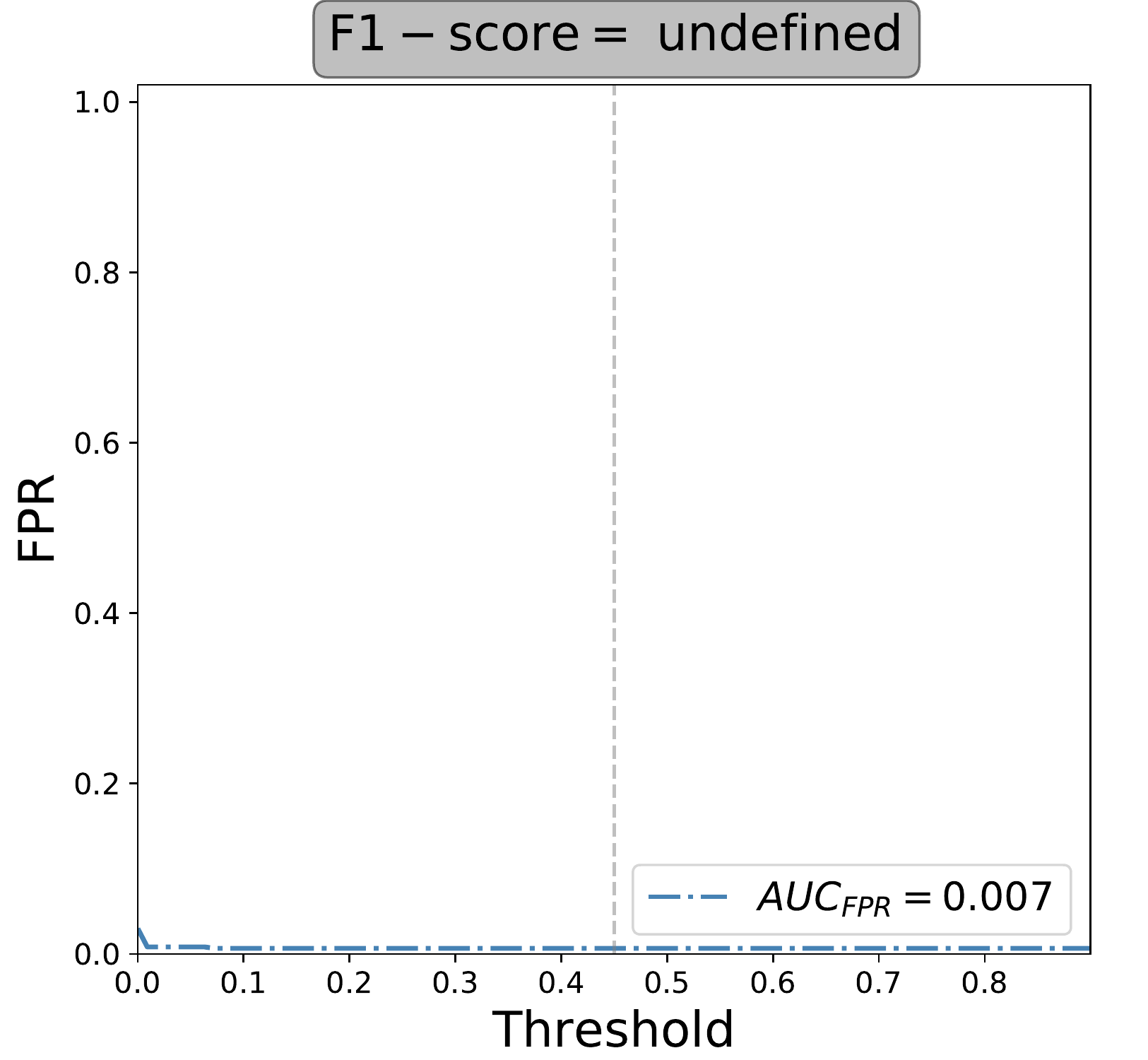}}\\
  \subfloat[SPHERE-3]{\includegraphics[width=200pt]{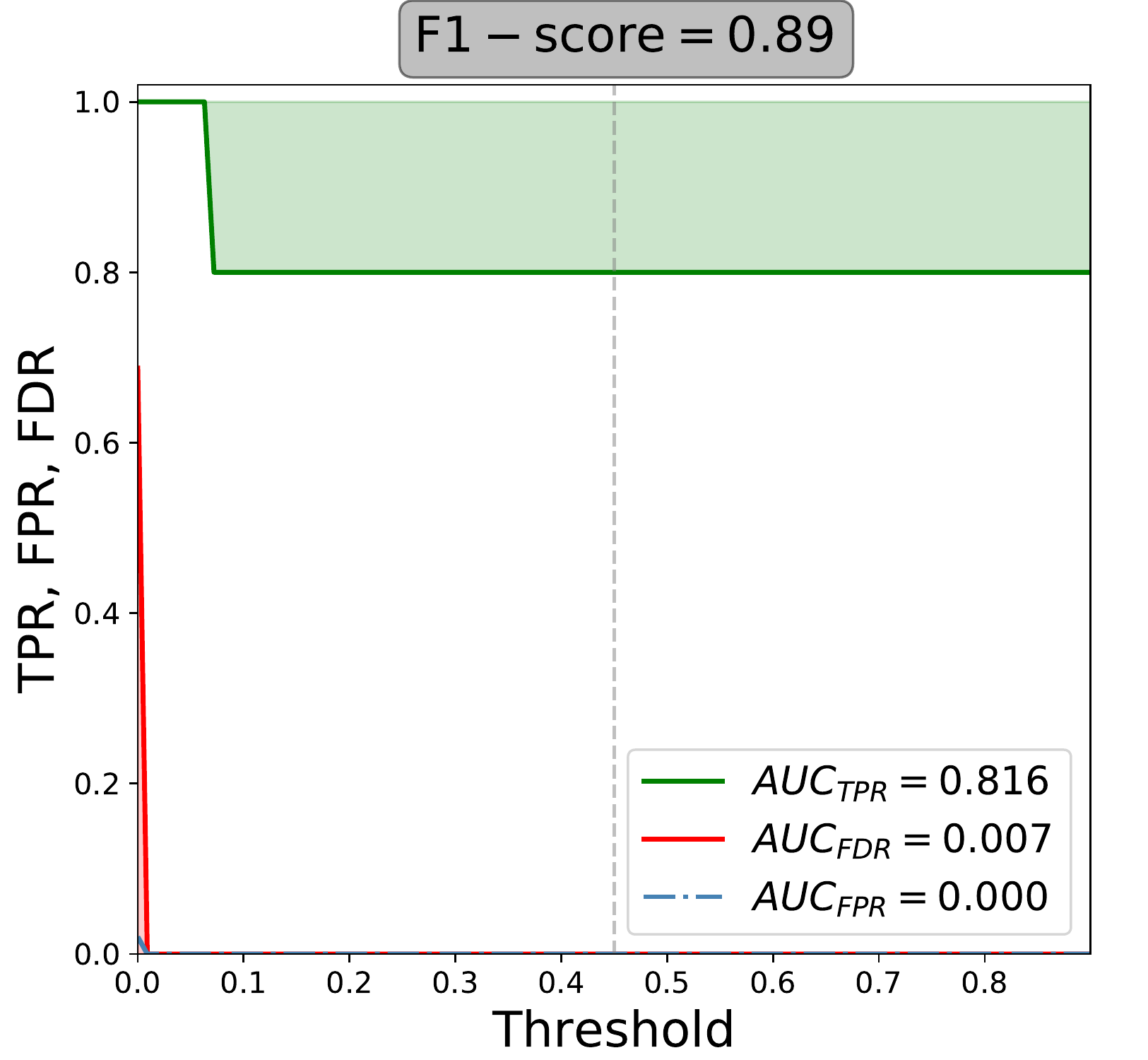}}\\
 
  \caption{\label{res_com} True positive rate (green), False discovery rate (in red), and False positive rate (dash-dotted blue line) computed for a range of thresholds varying from zero to twice the selected 0.45 detection threshold (represented by a dotted vertical line). These curves are computed for the SPHERE-2 and SPHERE-3 data sets of the EIDC, relying on the detection maps estimated with the SPHERE-1 optimal set of parameters along with the full-frame version of auto-RSM using the bottom-up approach . }
\end{figure}

We eventually tested the feasibility of using a single set of parameters for a given instrument, allowing us to investigate the sensitivity of the detection map estimation to the parametrisation. We selected the optimal parametrisation of the SPHERE-1 data set as this parametrisation is the closest to the centre of the SPHERE-LMIRCam cluster, and estimated the detection maps for the SPHERE-2 and SPHERE-3 data sets. Despite the inhomogeneity of the SPHERE data sets in terms of observing conditions and selected wavelengths, the obtained detection maps differed only slightly from the one presented in Fig. \ref{MapFFBUF}, with a reduced probability for one of the five targets in the SPHERE-3 data set but with a similar background noise level. As can be seen from Fig. \ref{res_com}, the change in terms of AUC of the FPR and FDR is negligible, while the F1 score and the AUC of the TPR reduce a little when considering the 0.45 probability threshold but remain similar if the threshold is adapted.

Considering all these results, the use of a single set of parameters for large surveys seems feasible in most cases, especially for the SPHERE and LMIRCam instruments. However, as demonstrated by the ADI sequences generated by NIRC2, when large dissimilarities are observed in terms of background noise level, a more refined subdivision should be considered.

\section{Conclusion}

In this paper, we present a new automated optimisation framework for the RSM approach, called auto-RSM. The proposed automated parameter selection is designed to reduce the complexity and possible arbitrariness of parameter selection when using HCI post-processing techniques and to provide users with a simple framework to compute reliable detection maps. Based on a single or multiple ADI sequences, after parameter optimisation, auto-RSM generates  a single detection map with high contrast between planetary candidates and residual speckles.

The proposed multi-step parameter-optimisation framework can be divided into three main steps, (i) the selection of the optimal set of parameters for the considered PSF-subtraction techniques, (ii) the optimisation of the RSM approach parametrisation, and (iii) the selection of the optimal set of PSF-subtraction techniques and ADI sequences to be considered when generating the final detection map. The selection of the optimal set of parameters for the PSF-subtraction techniques is based on the minimisation of the mean contrast within the selected set of annuli, while the optimisation of the RSM approach and selection of the optimal set of cubes of likelihoods are based on the probability ratio between injected fake companion peak probability and background residual probabilities. As some PSF-subtraction techniques have a continuous parameters space, a Bayesian optimisation framework is proposed to explore the parameter space and select the optimal set of parameters. Two different versions of the auto-RSM algorithm are proposed, a full-frame version where a single set of parameters is selected for all angular separations, and an annular version where the set of optimal parameters evolves with radial distance. Different parametrisations of the full-frame and annular auto-RSM are tested to investigate the added value of different methods to select the optimal set of cubes of likelihoods or to compute the final probabilities.

The data sets of the EIDC and the performance assessment framework proposed in \cite{Cantalloube20} are used to compare the performance of the different versions and parametrisations of the auto-RSM. The performance assessment is performed via the computation of a data set-dependant F1 score at a predefined threshold, as well as the estimation of the AUC of the TPR, FPR, and FDR. The auto-RSM results demonstrate the interest of the approach: in most cases, it provides better performance than the original RSM-detection map submitted to the EIDC, while the original RSM map was already at or close to the top of the ranking for all performance metrics in the EIDC. The full-frame auto-RSM using the bottom-up approach to select the optimal set of cubes of likelihood and the forward approach to compute the RSM probabilities provides the best overall performance in terms of detection. Considering the longer computation time and lower performance of the annular version, the full-frame auto-RSM should be preferred.

As auto-RSM is computationally expensive even when using the full-frame version, we investigate the possibility of using a common set of parameters for each instrument. We studied the commonalities existing between the parametrisations of the nine data sets of the EIDC, and found that the distance between the parametrisations for a common instrument is smaller than the distance between the parametrisations of different instruments. Potential differences between the noise characteristics of different data sets generated with a common instrument should nevertheless be taken into account, as illustrated by the NIRC2 data sets. However, the use of a single  parametrisation (or a limited number of them) for large surveys seems possible.

The auto-RSM framework is not limited to the RSM algorithm and the first step of the algorithm may be used separately to optimise the parametrisation of PSF-subtraction techniques and generate S/N maps. A S/N version of the proposed optimisation framework, called auto-S/N, was developed. Despite the degraded performance compared to auto-RSM, auto-S/N is characterised by a reduced computation time and can sometimes be a good complement to auto-RSM. All these versions of the proposed optimisation framework are available in a single Python package called PyRSM. This package offers a parameter-free detection map computation algorithm with a very low level of residual speckles, especially for the auto-RSM, allowing a simple detection threshold selection.

\begin{acknowledgements}
This work was supported by the Fonds de la Recherche Scientifique - FNRS under Grant n$^{\circ}$ F.4504.18 and by the European Research Council (ERC) under the European Union's Horizon 2020 research and innovation program (grant agreement n$^{\circ}$ 819155). 
\end{acknowledgements}

\bibliographystyle{aa}
\bibliography{AutoRSM.bib}

\begin{appendix}

\section{Mathematical notations for auto-RSM}
\label{paramdesc}    

This Appendix regroups in Table~\ref{variable}, all the mathematical notations used throughout the paper.
                           
\begin{table*}[!htbp]
                        \caption{Description of the mathematical notations for the variables used in the Bayesian optimisation algorithm, the RSM detection map and Auto-RSM optimisation framework.}
                        \label{variable}
\centering

                        \begin{tabular}{lll}
                        \hline
                        \hline
Symbol  & Dimension & Comments\\                        
 \hline \\
Bayesian optimisation & &\\
\hline
$\mathcal{O}_{1:t}$&$t$& Set of observations of the loss function\\
$t$&$1$& Number of tested sets of parameters\\
$\bm{p}_{1:t}$&$n_p \times t$ & Tested sets of parameters\\
$f(\bm{p})$&$1$& Loss function evaluated with the set of parameters $\bm{p}$\\
$\mathcal{GP}\left( m(\bm{p}_{1:t}),\bm{K}\right)$&1& Gaussian process returning the mean and variance of a Gaussian distribution\\
&& over the possible values of $f$ at $\bm{p}$\\
$m(\bm{p})$&$1$& Mean of the Gaussian distribution of the loss function at $\bm{p}$\\
$\bm{K}$&$1$&Covariance function of the tested set of parameters $\bm{p}_{1:t}$\\
$\mu(\bm{p}_{t+1})$&$1$& Mean of the Gaussian posterior distribution at $\bm{p}_{t+1}$\\
$\sigma(\bm{p}_{t+1})$ &$1$& Variance of the Gaussian posterior distribution at $\bm{p}_{t+1}$\\
$\widehat{\bm{p}}$ &$1$& Set of parameters providing the current known maximum value of the\\
&& loss function $f(\widehat{\bm{p}})$\\
\hline
RSM map & &\\
\hline
$\bm{x}_{i_a}$ &$\theta \times \theta \times T L_a$ & Patch of residuals centred on pixel $i_a$\\
$F_{i_a}$& $T L_a$ & Realisation of a two-state Markov chain representing the state of\\
&& the system for pixel $i_a$\\
$ \bm{m}$& $\theta \times \theta$ & Cropped planetary signal (off-axis PSF)\\
$\bm{\varepsilon_{s,i_a}}$& $2 \theta\times \theta \times T L_a$ & Error terms associated with the two regimes\\
$S_{i_a}$& $T L_a$ & State of the system for every pixel $i_a$\\
$\xi_{s,i_a}$& $2 \times T L_a$& Probability associated with state $s$ for every pixel $i_a$\\
$\eta_{s,i_a}$ & $2 \times T L_a$ & Likelihood of being in each state for every pixel $i_a$\\
$p_{q,s}$& $2 \times 2 $& Transition probabilities between the regimes\\
$\mu$& $1$ & Mean of the residuals contained in an annulus $a$, with width equal to $\theta$\\
$\sigma$& $1$ & Standard deviation of the residuals contained in an annulus $a$, with width\\
 &&equal to $\theta$ \\
$\beta$& $1$ &  Intensity of the planetary signal in the cube of residuals \\
$a$& $1$ & Annulus index\\
$L_a$& $1$ & Number of pixels included in the annulus $a$\\
$T$& $1$ & Number of frames in the cube of residuals\\
$i_a$&$1$& Index associated with every pixel from every frame in the annulus $a$\\
&& (ranges from $1$ to $T L_a$)\\
$\theta$& $1$ & Angular size of the considered planetary signal ( set to $1\lambda/D$)\\
$\delta$&$1$& Multiplicative factor of the noise standard deviation providing the \\
&&intensity parameter $\beta$ \\
\hline 
Auto-RSM & &\\
\hline
$\bm{i}$ &$N^{pix}$ & Vectorised science image/annulus before speckle subtraction \\
$ \bm{m}$& $N^{pix}$ & Vectorised planetary signal (off-axis PSF) inside the selected annulus\\
$a_{max}$ &$1$& Largest angular separation considered for the detection map computation\\
$C_{a,m,c}$&$L_a \times N_{technique} \times N_{sequence}$& Average contrast obtained with the optimal set of parameters\\
$M_{a,m,c}$&$L_a \times N_{technique} \times N_{sequence}$& Position of the median contrast aperture within the selected annulus\\
$P^{PSF}_{a,m,c}$&$L_a \times N_{technique} \times N_{sequence}$& Optimal set of parameters for the reference PSF computation \\
$P^{PSF,*}_{a,m,c}$&$L_a \times N_{technique} \times N_{sequence}$& Optimal set of parameters for the reference PSF smoothed via moving average\\
$P^{RSM}_{a,m,c}$&$L_a \times N_{technique} \times N_{sequence}$& Optimal set of parameters for the computation of the RSM map \\
$P^{RSM,*}_{a,m,c}$&$L_a \times N_{technique} \times N_{sequence}$& Optimal set of parameters for the computation of the RSM map\\
&& interpolated via RBF\\
$\bm{Y}^a$&$L_a$& Set of likelihood time series available (one per couple of ADI sequence\\ 
&& and PSF-subtraction technique) for the computation of the RSM map\\
$Y^a_{c,m}$&$L_a \times N_{technique} \times N_{sequence}$& Likelihood time series\\
$Y^a_{h*} $&$L_a \times T$& Likelihood time series maximising the RSM performance index at a\\
&& given iteration \\
$\bm{Z}^a$& $L_a$& Selected likelihood time series for the computation of the final detection map\\
$T_{a}$& $L_a $& Annulus-wise thresholds computed with flipped parallactic angles\\
$T^*_{a}$& $L_a $& Smoothed annulus-wise thresholds computed with flipped parallactic angles\\
\hline
                        \end{tabular}
                                \end{table*}
                                
\section{Computation of the expected improvement and update of posterior probability moments}
\label{gpmodel}

The objective of the expected improvement approach is to estimate the magnitude of the improvement that a set of parameters can potentially yield in terms of loss function value. As the true maximum value of the loss function $f(\bm{p}^*)$ is not known, \cite{Mockus78} propose maximising the expected improvement with respect to a known maximum $f(\widehat{\bm{p}})$. They define the improvement function as 

\begin{eqnarray}
I(\bm{p}_{t+1})=\max \left\lbrace 0, f(\bm{p}_{t+1})-f(\widehat{\bm{p}})\right\rbrace 
,\end{eqnarray}          
where $I(\bm{p}_{t+1})$ is positive when the prediction is higher than the current maximum loss function value and zero otherwise. The new set of parameters $\bm{p}_{t+1}$ is found by maximising the expected improvement, as follows:

\begin{eqnarray}
\bm{p}_{t+1}=argmax_{\bm{p}_{t+1}} \mathbb{E} \left(  \left\lbrace 0, f(\bm{p}_{t+1})-f(\widehat{\bm{p}})\right\rbrace \mid \mathcal{O}_{1:t} \right) 
.\end{eqnarray}          

The likelihood of improvement $I$ when considering the Gaussian process giving the posterior distribution is then

\begin{eqnarray}
\frac{1}{\sqrt{2 \pi} \sigma(\bm{p}_{t+1})} \exp \left( -\frac{(\mu(\bm{p}_{t+1})-f(\widehat{\bm{p}})-I)^2}{2\sigma^2(\bm{p}_{t+1})}\right) ,
\end{eqnarray}  
with $\mu(\bm{p}_{t+1})$ and $\sigma(\bm{p}_{t+1})$ being  the mean and standard deviation, respectively, of the posterior probability $f(\bm{p}_{1:t}) \sim \mathcal{N}( 0,\bm{K})$ for the new set of parameters $\bm{p}_{t+1}$. The expected improvement is then simply the integral over this likelihood function:

\small
\begin{eqnarray}
\mathbb{E}(I(\bm{p}_{t+1})) = \int^{\infty}_{0} \frac{I}{\sqrt{2 \pi} \sigma(\bm{p}_{t+1})} \exp \left( -\frac{(\mu(\bm{p}_{t+1})-f(\widehat{\bm{p}})-I)^2}{2\sigma^2(\bm{p}_{t+1})}\right) dI,
\end{eqnarray}  
\normalsize
which gives after integration by part
\small
\begin{eqnarray}
&&\mathbb{E}(I(\bm{p}_{t+1}))  = \sigma(\bm{p}_{t+1}) \nonumber \\
&&\left[  \frac{\mu(\bm{p}_{t+1})-f(\widehat{\bm{p}})}{\sigma(\bm{p}_{t+1})} \Phi \left( \frac{\mu(\bm{p}_{t+1})-f(\widehat{\bm{p}})}{\sigma(\bm{p}_{t+1})} \right)  +  \phi\left(  \frac{\mu(\bm{p}_{t+1})-f(\widehat{\bm{p}})}{\sigma(\bm{p}_{t+1}} \right] \right)  .
\end{eqnarray}  
\normalsize
Considering the improvement function definition, we obtain the expression of Eq. \ref{bayesian6},

\begin{eqnarray}
\label{bayesian7}
&&\text{EI}(\bm{p}_{t+1}) = \nonumber \\
  &&\begin{cases}
    (\mu(\bm{p}_{t+1})-f(\widehat{\bm{p}}))\Phi(Z) +\sigma(\bm{p}_{t+1})\phi(Z)  & \quad \text{if } \sigma(\bm{p}_{t+1})>0\\
    0      & \quad \text{if } \sigma(\bm{p}_{t+1})=0\\
  \end{cases},
\end{eqnarray}
with $Z=\frac{\mu(\bm{p}_{t+1})-f(\widehat{\bm{p}})}{\sigma(\bm{p}_{t+1})}$. 

We see that the computation of EI requires an estimation of the mean $\mu(\bm{p}_{t+1})$ and standard deviation $\sigma(\bm{p}_{t+1})$ of the posterior probability of $f(\bm{p}_{1:t})$. Starting from the posterior probability $f(\bm{p}_{1:t}) \sim \mathcal{N}( 0,\bm{K})$ and taking into account the new set of parameters $\bm{p}_{t+1}$ we get

\begin{eqnarray}
\begin{pmatrix}
f(\bm{p}_{1:t})  \\
f(\bm{p}_{t+1}) 
\end{pmatrix}=
\mathcal{N} \left( 0, \begin{pmatrix}
\bm{K} & \bm{k}  \\
\bm{k}^T & k(\bm{p}_{t+1},\bm{p}_{t+1})
\end{pmatrix} \right) ,
\end{eqnarray}  
where $\bm{k}=\left\lbrace k(\bm{p}_{1},\bm{p}_{t+1}),k(\bm{p}_{2},\bm{p}_{t+1}), \dots, k(\bm{p}_{t},\bm{p}_{t+1}) \right\rbrace $.

We then get the following expression for the posterior distribution using the Sherman-Morrison-Woodbury formula:

\begin{eqnarray}
P(f(\bm{p}_{t+1}) \mid\mathcal{O}_{1:t},\bm{p}_{t+1})=\mathcal{N} (\mu(\bm{p}_{t+1}),\sigma^2(\bm{p}_{t+1}))
,\end{eqnarray}  
with the mean and variance given by

\begin{eqnarray}
\mu(\bm{p}_{t+1})&=&\bm{k}^T\bm{K}^{-1} f(\bm{p}_{1:t}) ,\\
\sigma^2(\bm{p}_{t+1})&=&k(\bm{p}_{t+1},\bm{p}_{t+1})-\bm{k}^T\bm{K}^{-1}\bm{k}
.\end{eqnarray}

\section{Average contrast computation via multiple fake companion injections}
\label{multifc}
\begin{figure}[h!]
  \centering
  \subfloat[(a) Sequential injections]{\includegraphics[width=130pt]{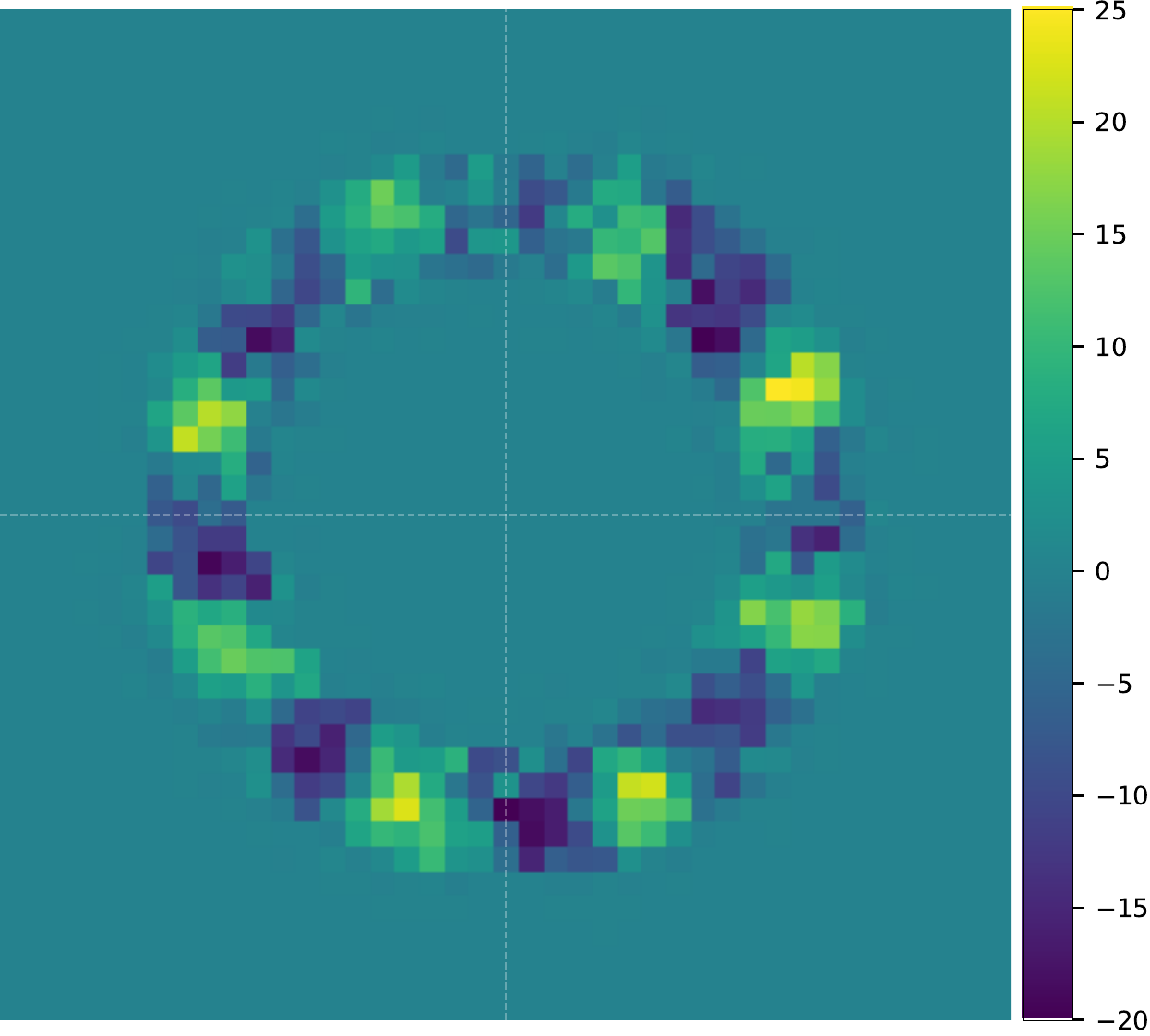}}
  \subfloat[(b) Multiple injections]{\includegraphics[width=130pt]{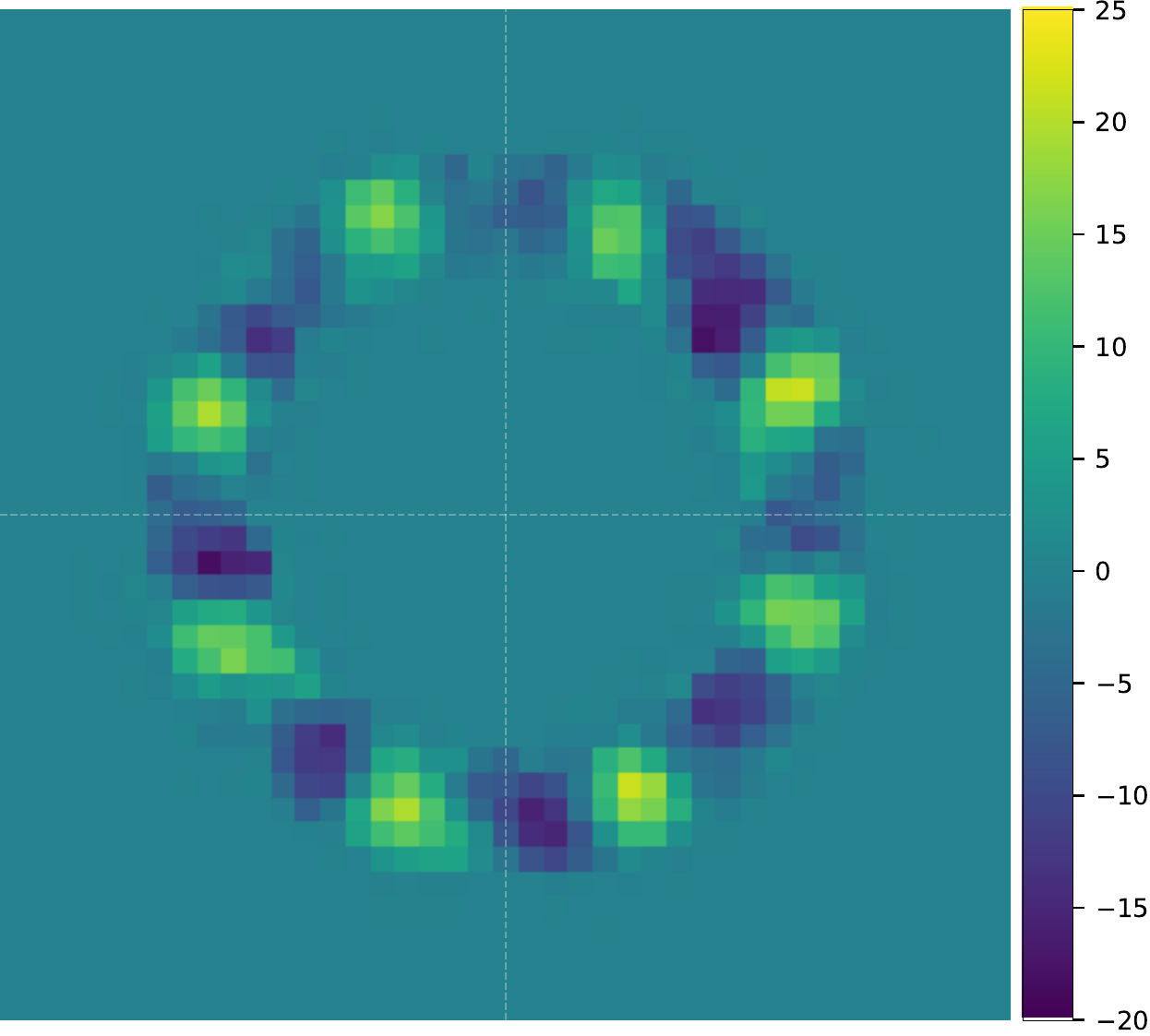}}
 
  \caption{\label{multi1} Comparison of the recovered intensities of eight fake companions injected sequentially or at once, at a radial distance of 2 $\lambda/D$, using the SPHERE 1 data set of the EIDC, and relying on annular PCA to generate the reference PSF with a number of principal components equal to 20.}
\end{figure}

In this section we aim to assess the validity of our approximation when computing the average contrast by considering the agreement between the average contrasts generated using multiple injections and sequential injections. When relying on multiple injection, the self- and over-subtraction associated with an injected fake companion may affect neighbouring apertures, especially at small angular separations. We impose, for multiple injections, a minimal separation of two FWHMs between the positions of two injected fake companions in order to reduce the impact of these interferences on the estimation of the average contrast. The number of injected fake companions is therefore limited to half the number of apertures contained in a given annulus with a maximum of eight fake companions, which should provide a reliable estimate of the speckle field within the annulus while limiting the interference between fake companions. As can be seen from Fig. \ref{multi1}, the intensity patterns for multiple injections are similar to the ones observed for sequential injections, with the companions having the smallest or largest flux positioned at the same azimuth.

\begin{figure}[h!]
  \centering
  \subfloat[(a) 2 $\lambda/D$]{\includegraphics[width=240pt]{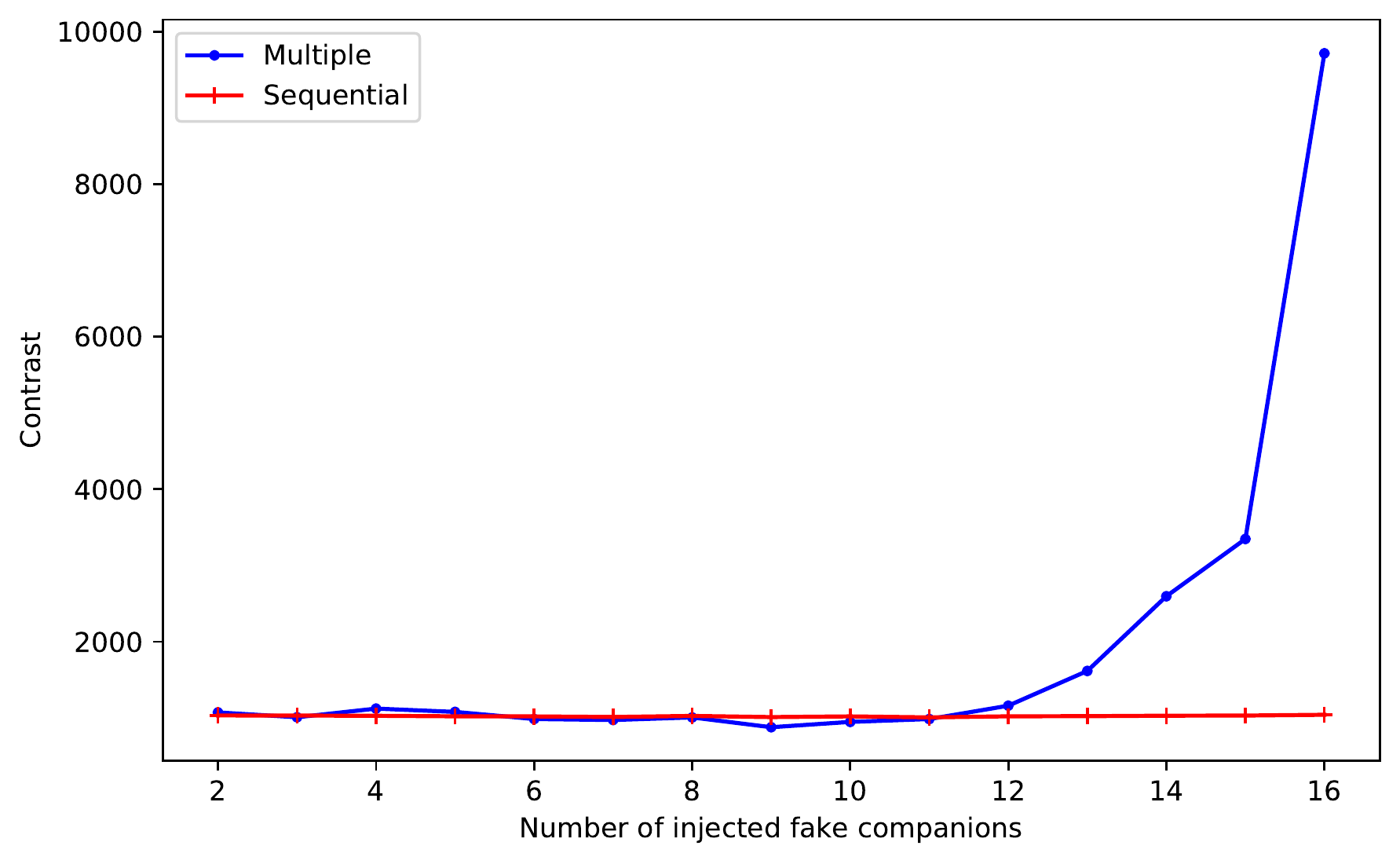}}\\
  \subfloat[(b) 4 $\lambda/D$]{\includegraphics[width=240pt]{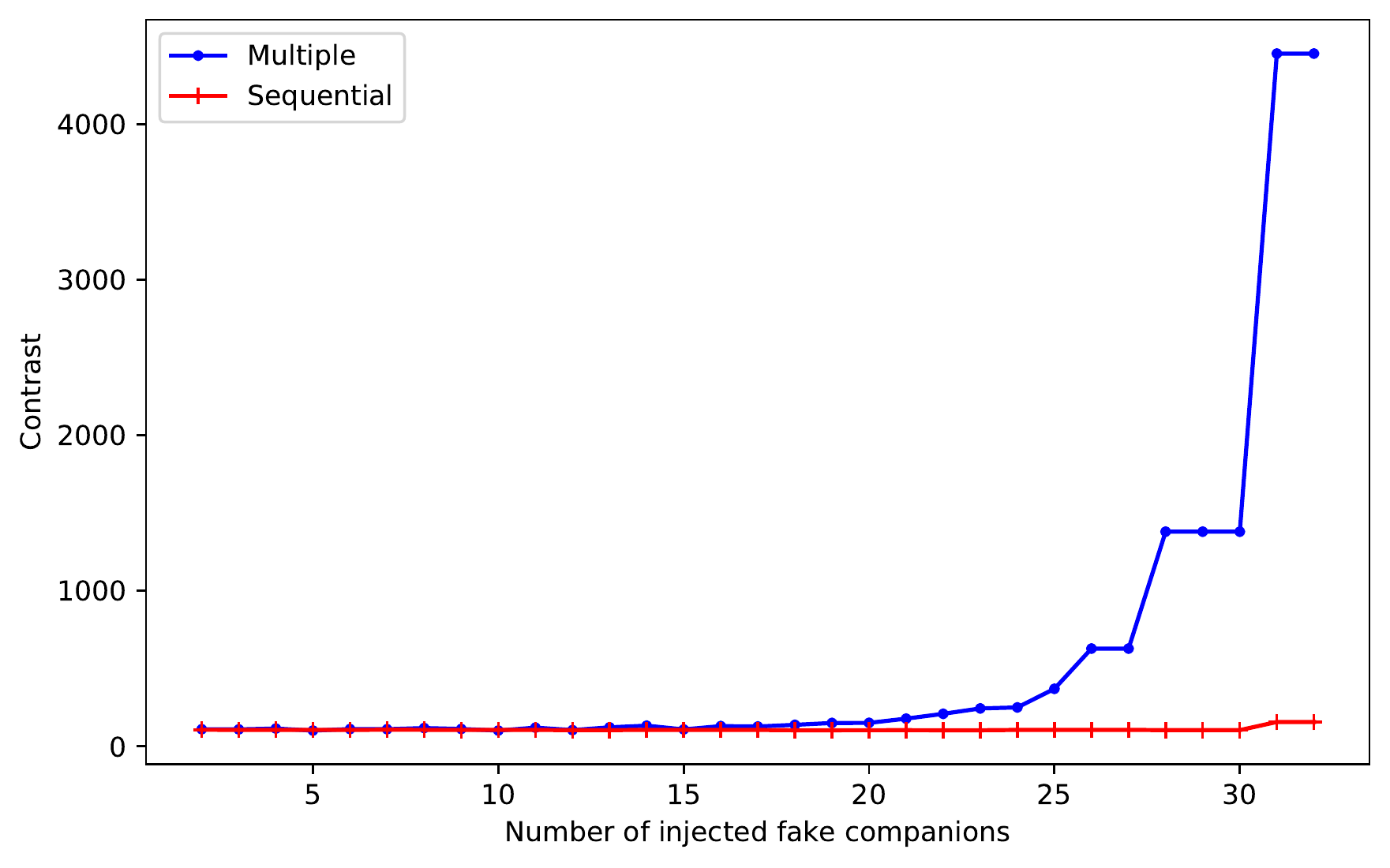}}\\
  \subfloat[(b) 8 $\lambda/D$]{\includegraphics[width=240pt]{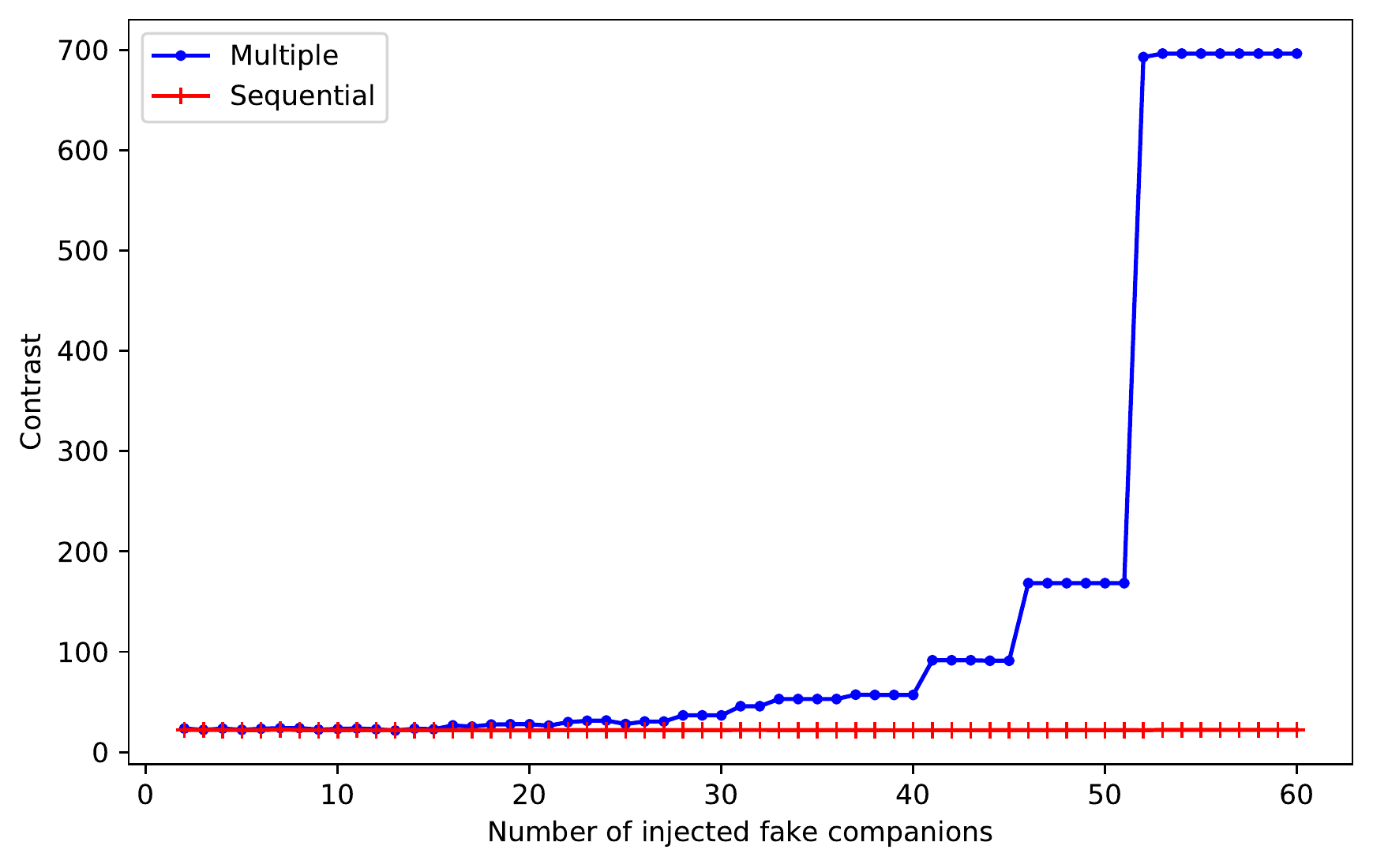}}
 
  \caption{\label{multi2} Comparison of the average contrasts obtained with sequential and multiple injections for an increasing number of injected fake companions. The curves have been computed at three different angular separations (2, 4, and 8 $\lambda/D$), using the SPHERE 1 data set of the EIDC, and relying on annular PCA to generate the reference PSF with a number of principal components equal to 20. }
\end{figure}

Figure \ref{multi2} provides the evolution of the average contrast computed with the sequential and multiple injections for an increasing number of fake companions. As expected the average contrast does not vary significantly for the sequential injections. However, for multiple injections, the average contrast starts to strongly diverge for distances between the injected positions of neighbouring companions below two FWHMs. A distance of two FWHMs corresponds to 9, 18, and 33 companions for an angular distance of  2, 4, and 8 $\lambda/D$,  respectively. Looking at Fig. \ref{multi2}, we see that for eight injected fake companions, the average contrasts generated with the multiple and sequential injections are very similar.

\begin{figure}[h!]
  \centering
  \subfloat[(a) 2 $\lambda/D$]{\includegraphics[width=240pt]{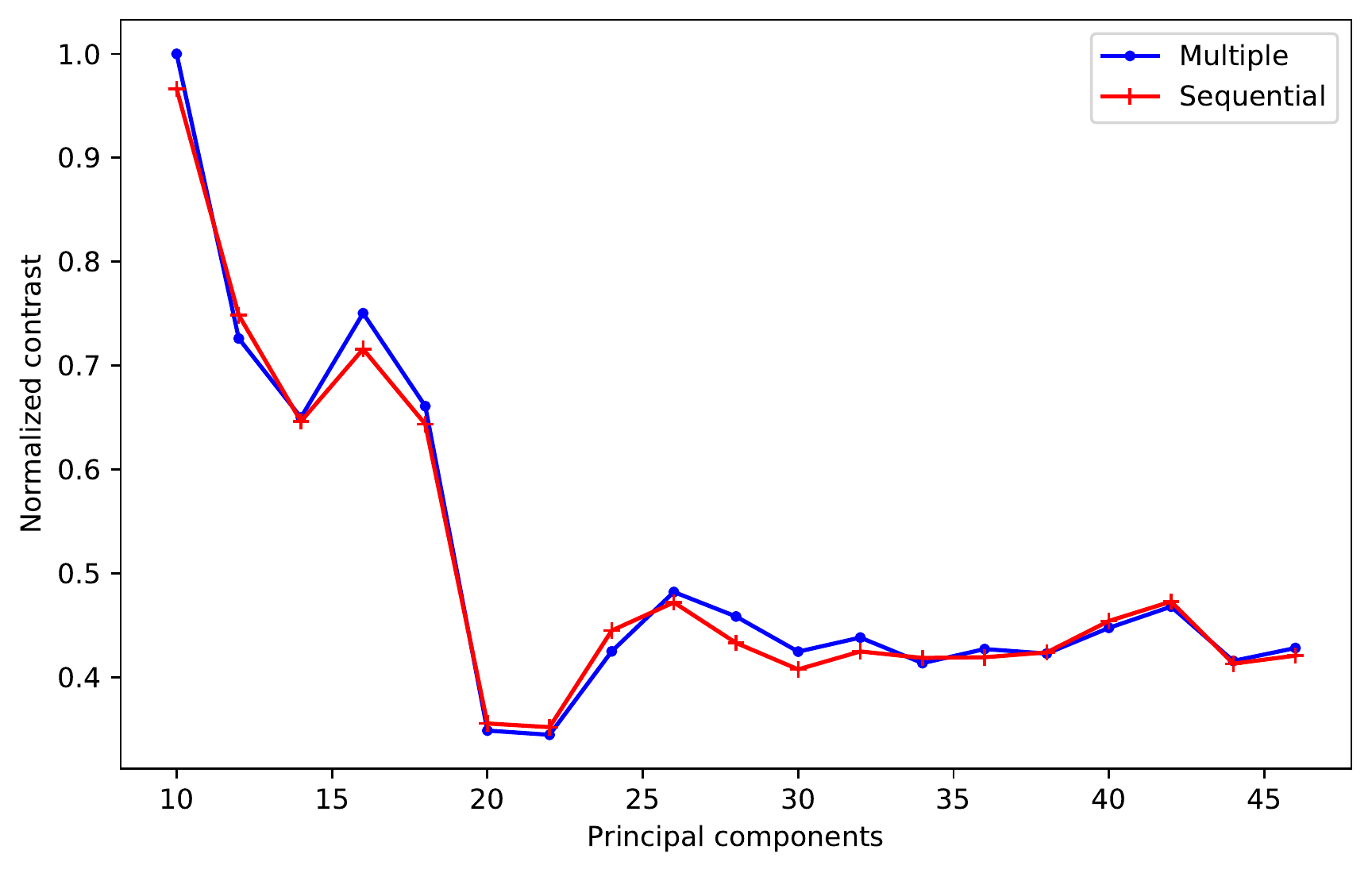}}\\
  \subfloat[(b) 4 $\lambda/D$]{\includegraphics[width=240pt]{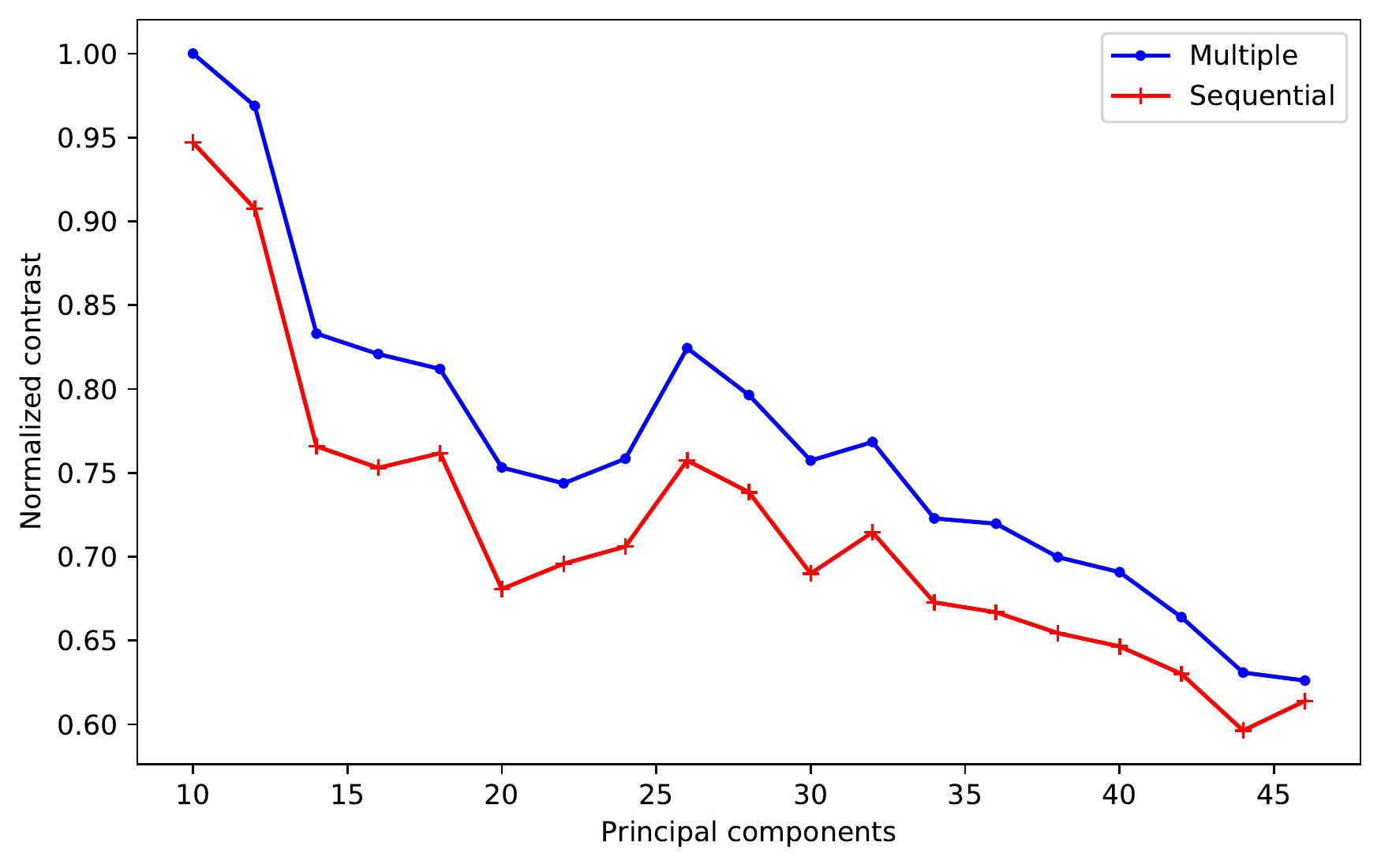}}\\
  \subfloat[(c) 8 $\lambda/D$]{\includegraphics[width=240pt]{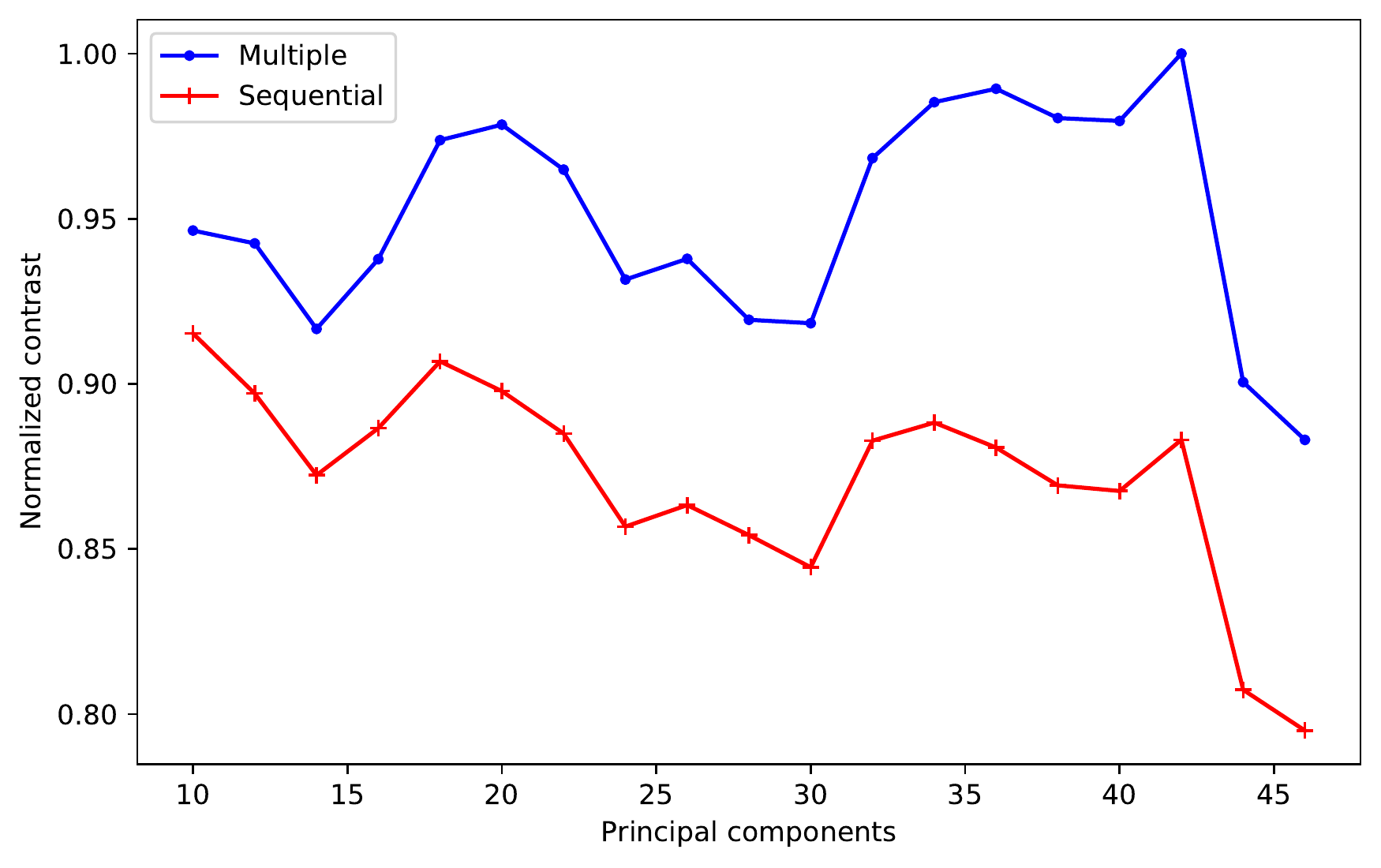}}
 
  \caption{\label{multi3} Comparison of the average contrasts obtained with sequential and multiple injections for a range of principal components used by the annular PCA to generate the reference PSF (between 10 and 45 principal components). The curves were computed at three different angular separations (2, 4, and 8 $\lambda/D$) using the SPHERE 1 data set of the EIDC. }
\end{figure}

Besides the distance between the average contrasts generated by the two approaches, the behaviour of these average contrasts when modifying the parameters of the PSF-subtraction techniques is the most important element, as it drives the optimal parameter selection. We computed the average contrasts for several different numbers of principal components in the case of annular PCA to determine if the behaviour of the contrast curves generated with multiple and sequential injections was similar. Figure \ref{multi3} shows the evolution of the average contrast with the number of principal components used for the reference PSF computation for different angular separations. Although there exists a gap between both curves, the two curves seem to evolve in parallel, especially for smaller angular distances. We observe high Pearson correlations between the curves generated with multiple and sequential injections, with a correlation of 0.996, 0.992, and 0.704 for an angular distance of respectively 2, 4, and 8 $\lambda/D$. This seems to indicate that the same set of parameters will minimise the average contrast and confirm the validity of our approximation when relying on multiple injections to compute the average contrast. 

\section{Auto-S/N}
\label{autosn}
\subsection{Algorithm definition}

We define in this section, the auto-S/N algorithm which is derived from the auto-RSM framework. The first step of the auto-RSM algorithm, i.e. the parameter optimisation of the PSF-subtraction techniques, is used to generate an optimised cube of residuals for every considered PSF-subtraction technique. As in the case of auto-RSM, the auto-S/N aims to combine the obtained cubes of residuals to generate a final detection map. As the cubes of residuals generated by the different PSF-subtraction techniques have their own noise distribution, a simple mean-combination is not possible. A simple way to overcome this limitation is to mean-combine the S/N maps instead of the cubes of residuals. As the dissimilarities in the noise structure of the different cubes of residuals are reflected in their respective S/N maps, part of the residual speckle noise should average out. The main difficulty pertains to the proper definition of the throughput to estimate the contrast used for the optimal selection, as we are combining S/N maps. 

Considering the detection map obtained by averaging $N$ different S/N maps, each pixels S/N is defined as:

\begin{eqnarray}
\text{S/N}_{i_a}&=& \frac{1}{N} \sum^N_{j=1} \frac{\text{F}_{i_a,j}}{\text{N}_{a,j}} \\
&=&\frac{1}{N} \frac{\sum^N_{j=1}\text{F}_{i_a,j}\prod^N_{k\neq j} {N}_{a,k}}{\prod^N_{k=1} {N}_{a,k}} ,
\end{eqnarray}
with $\text{F}_{i_a,j}$ the flux associated with the aperture centred on pixel $i_a$, where $a$ is the considered annulus in the mean-combined de-rotated cube of residuals generated with the PSF-subtraction technique $j$, and $\text{N}_{a,j}$ is the associated noise computed in annulus $a$. Following this expression, the throughput obtained from the injection of a fake companion at pixel $i_a$ is given in the case of a combination of $N$ S/N maps:

\begin{eqnarray}
\text{throughput}_{i_a}&=& \frac{\sum^N_{j=1} \text{IF}_{i_a,j}\prod^N_{k\neq j} {N}_{a,k}}{\sum^N_{j=1} \text{RF}_{i_a,j}\prod^N_{k\neq j} {N}_{a,k}}  ,
\end{eqnarray}
where $\text{IF}$ stands for injected flux and $\text{RF}$ for retrieved flux. The throughput becomes simply a sum of fluxes weighted by the noises of the other considered PSF-subtraction techniques. This implies that the throughput associated with a less noisy mean-combined de-rotated cube of residuals has a higher weight as both the injected flux and retrieved flux are multiplied by larger noise values than the others. The noise appearing in the expression of the contrast (see eq. \ref{contrast}) is then computed as the noise averaged over the different S/N maps for the considered angular separation $a$. Similarly to the parameter optimisation for the PSF-subtraction techniques, fake companions are injected at different azimuths to obtain an average contrast. The obtained average contrast is then used to select the optimal set of S/N maps either via the bottom-up or the top-down approach described respectively in Tables \ref{bu} and \ref{td}. As for the auto-RSM, the auto-S/N can also use either the full-frame or the annular optimisation mode. 

\subsection{Performance assessment}

\begin{figure*}[h!]
\footnotesize
  \centering
  \subfloat[SPHERE-1 Baseline]{\includegraphics[width=115pt]{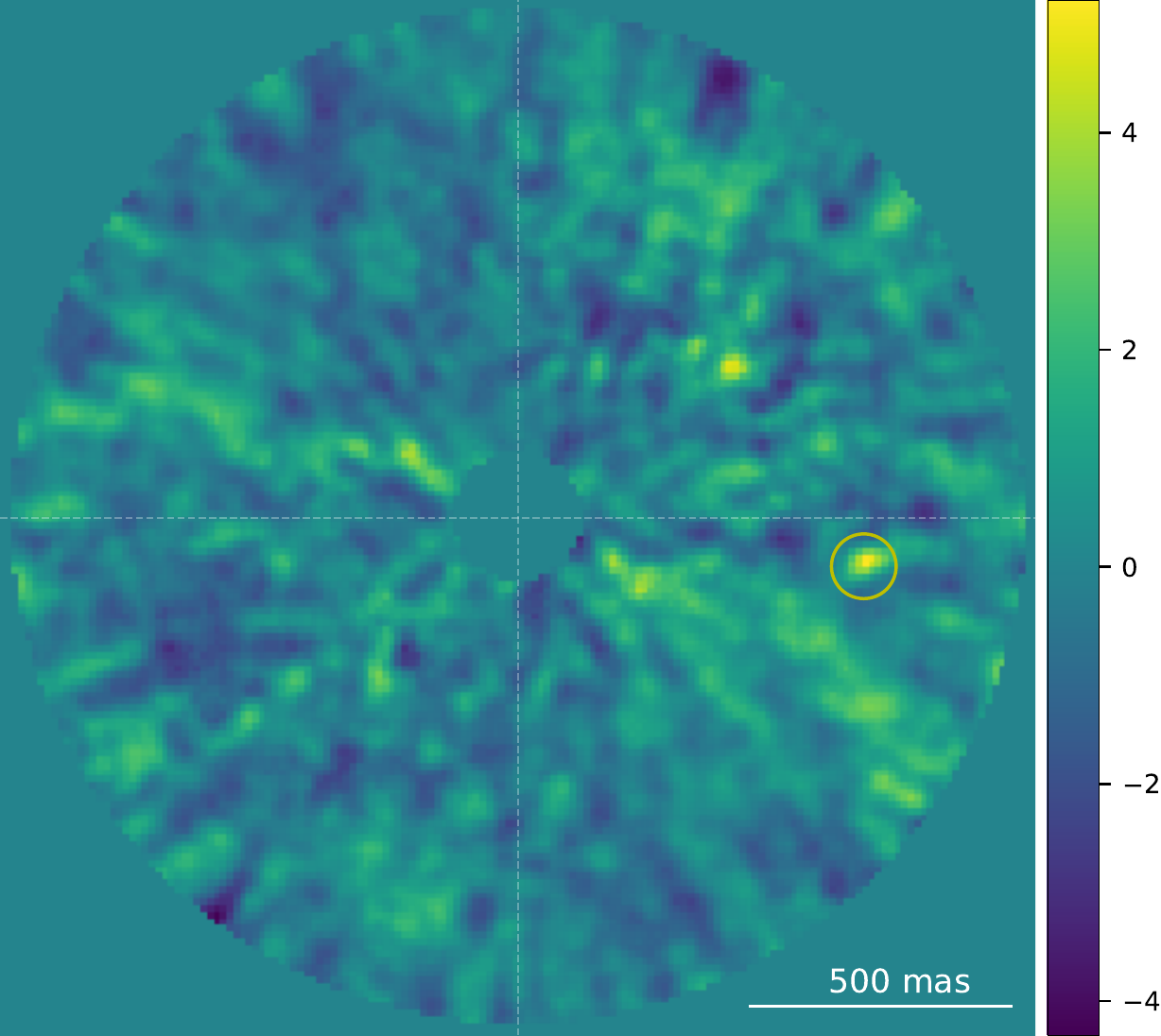}}
   \subfloat[SPHERE-3 Baseline]{\includegraphics[width=115pt]{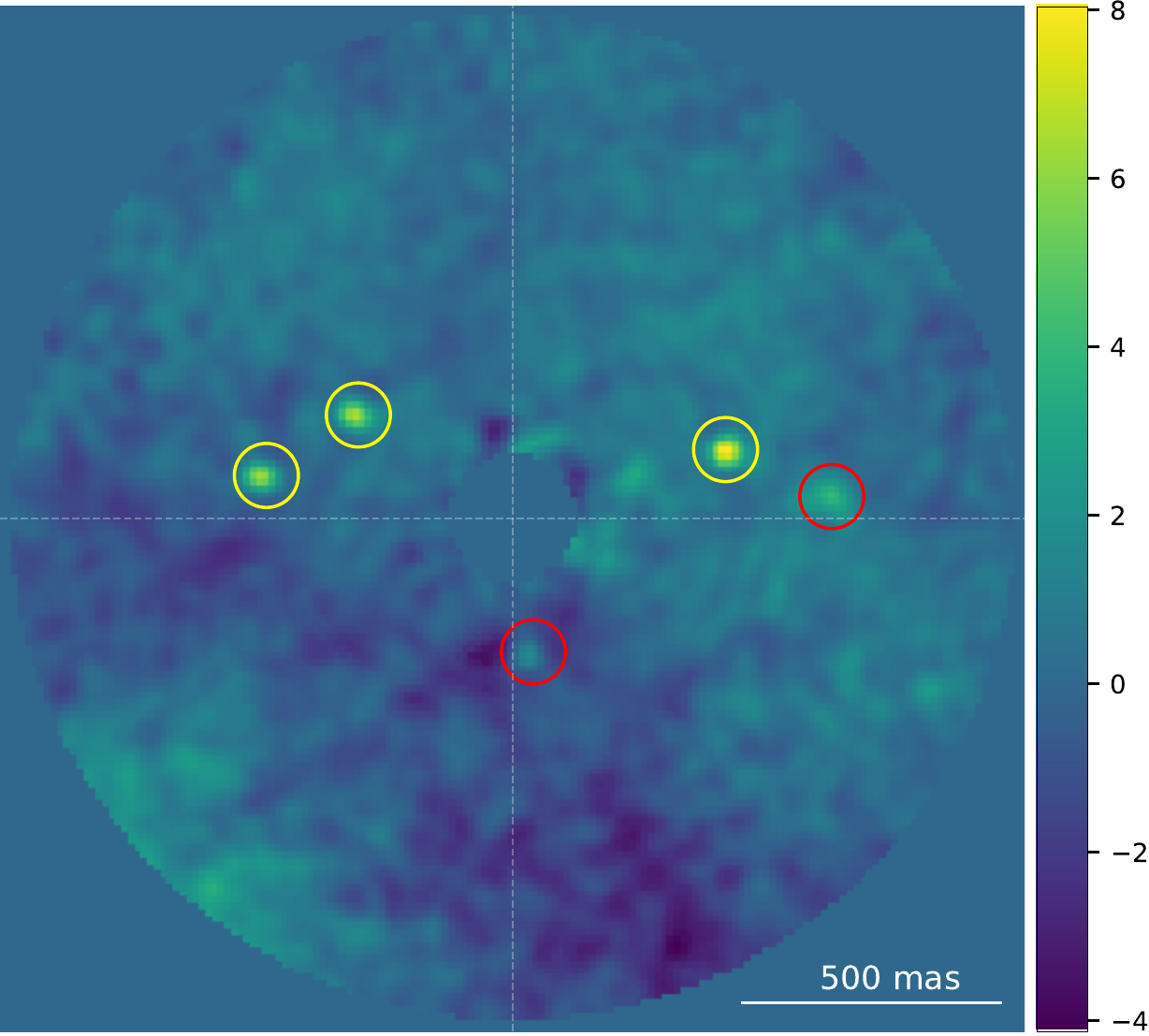}}
     \subfloat[NIRC2-1 Baseline]{\includegraphics[width=115pt]{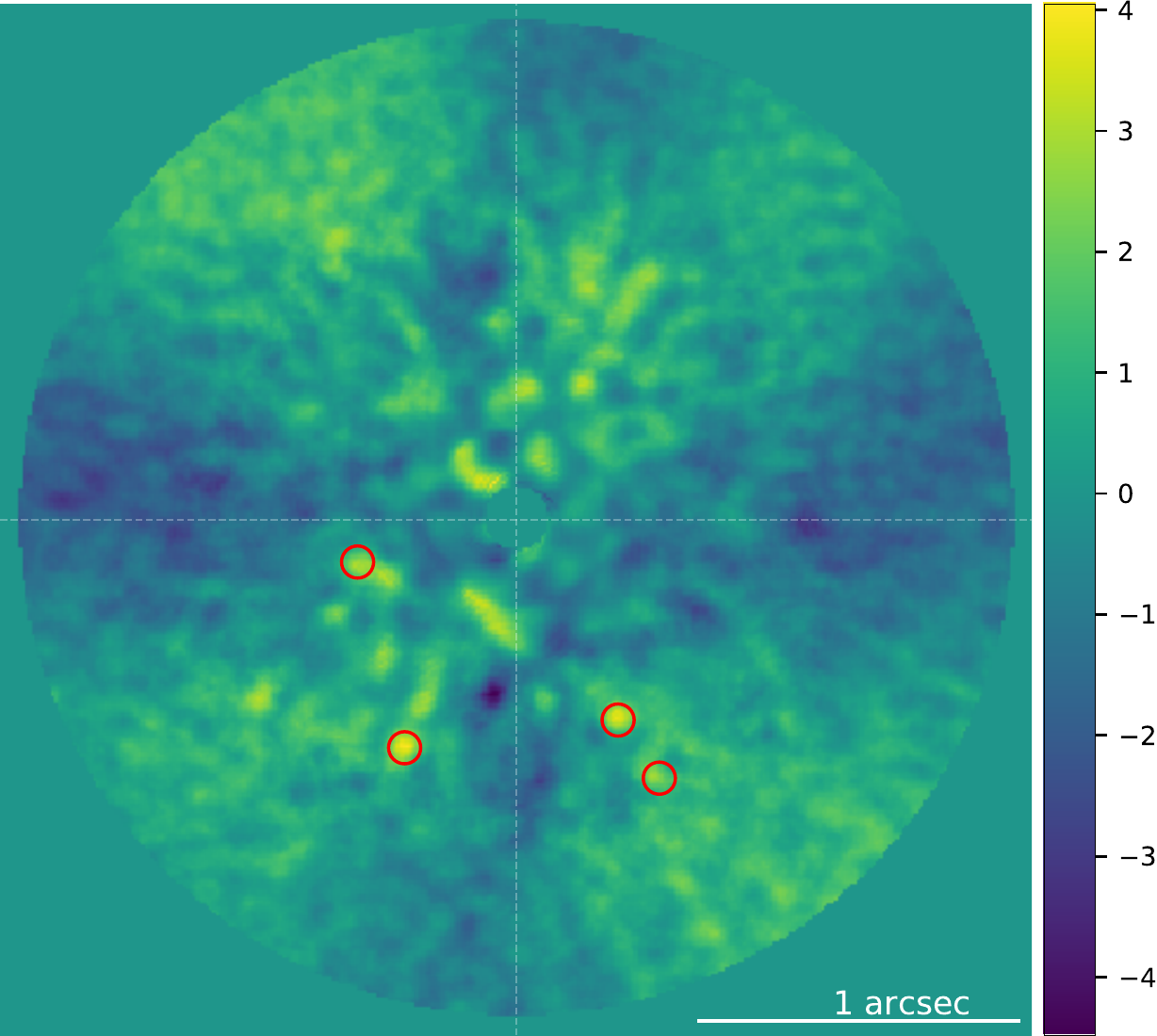}}
       \subfloat[NIRC2-2 Baseline]{\includegraphics[width=115pt]{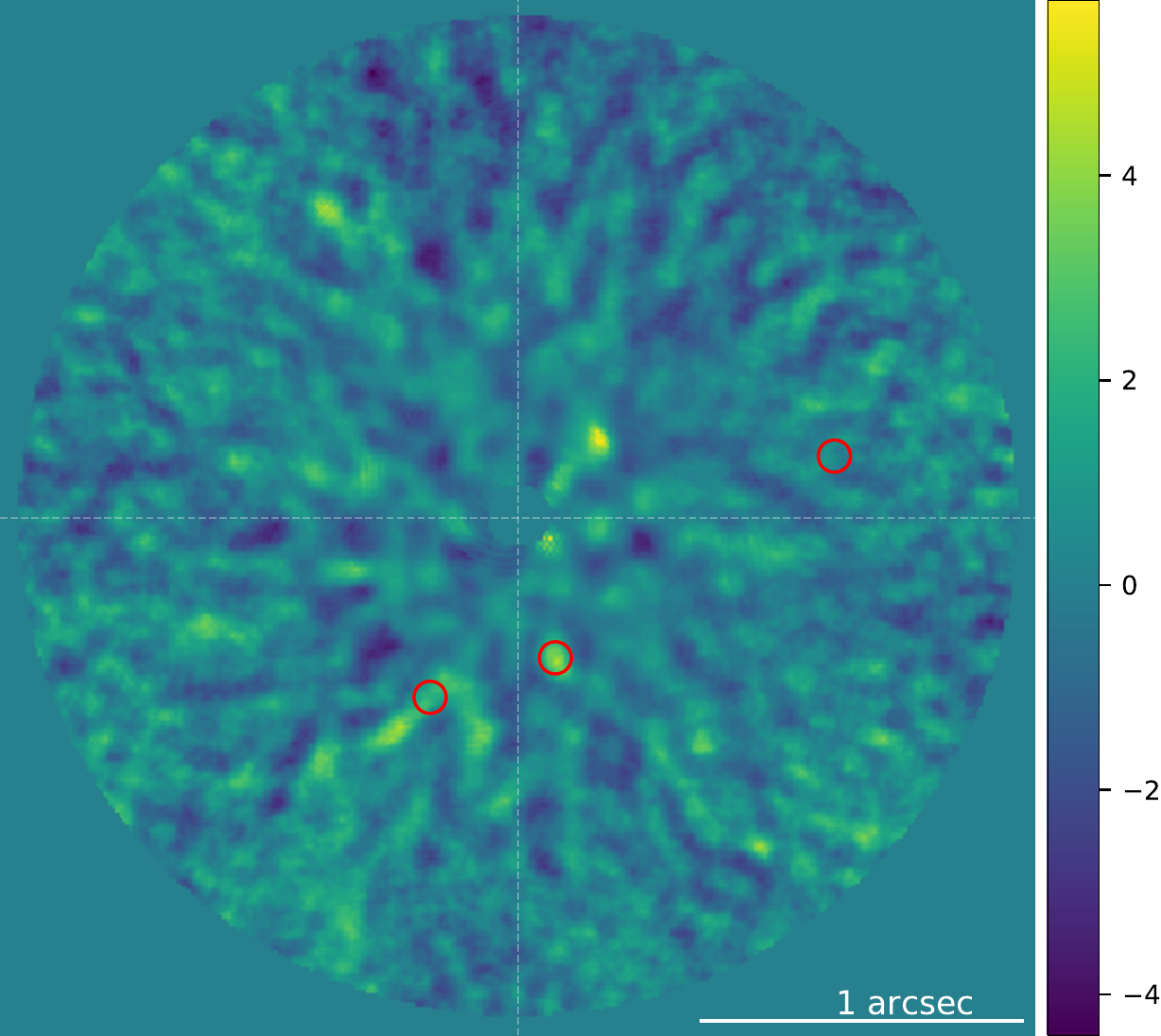}}\\
  \subfloat[SPHERE-1 S/N FF BU]{\includegraphics[width=115pt]{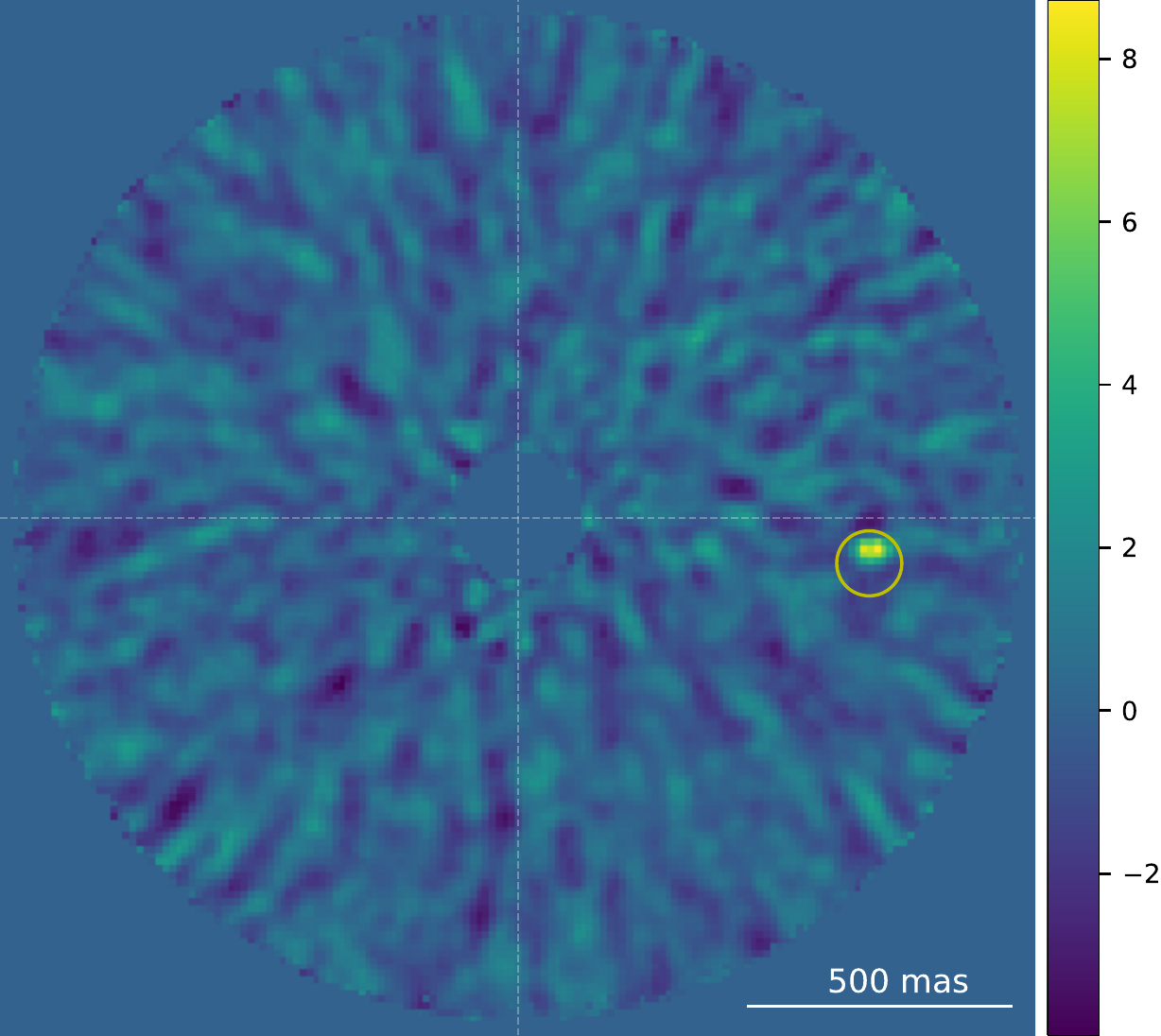}}
    \subfloat[SPHERE-3 S/N FF BU]{\includegraphics[width=115pt]{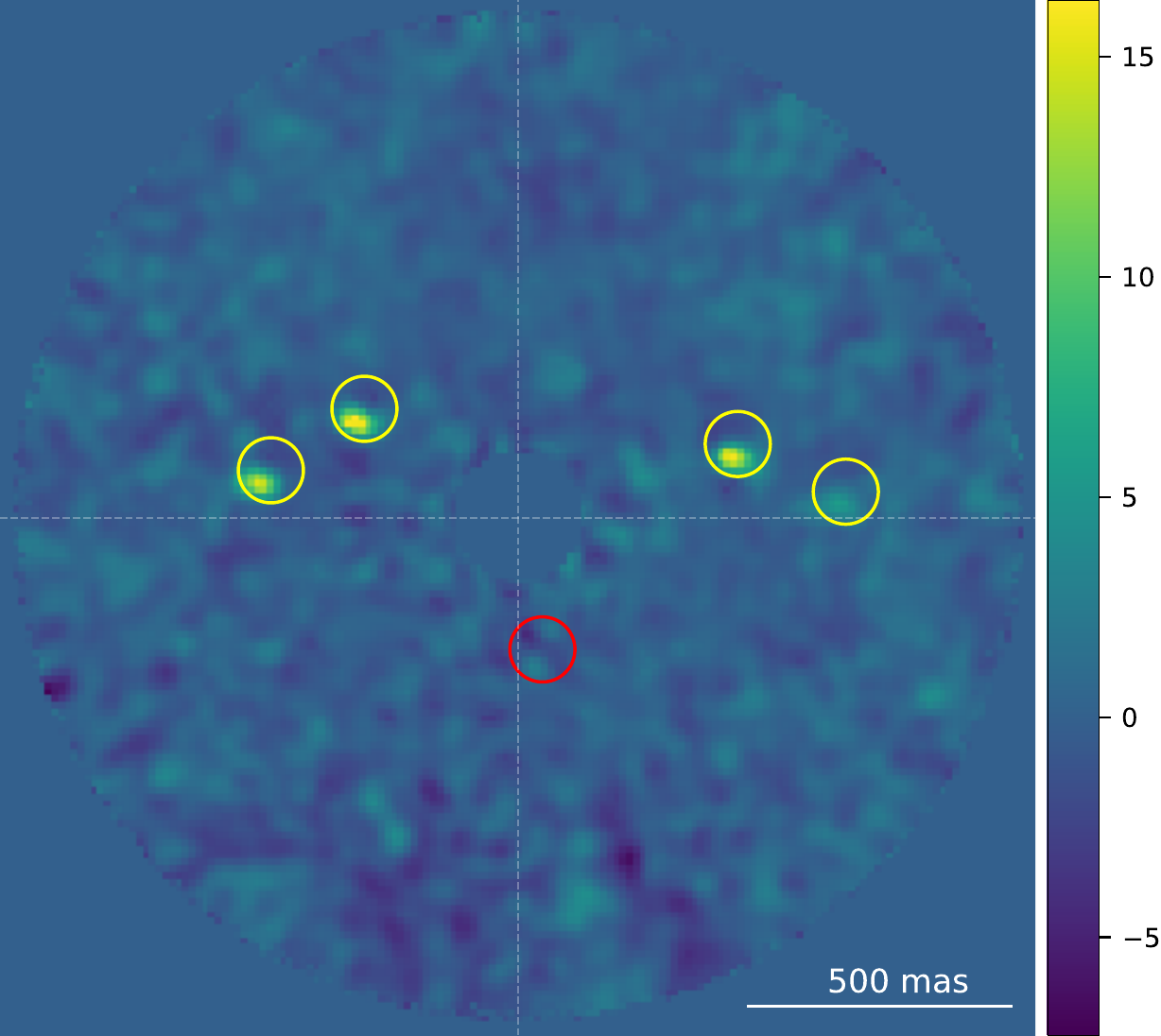}}
      \subfloat[NIRC2-1 S/N FF BU]{\includegraphics[width=115pt]{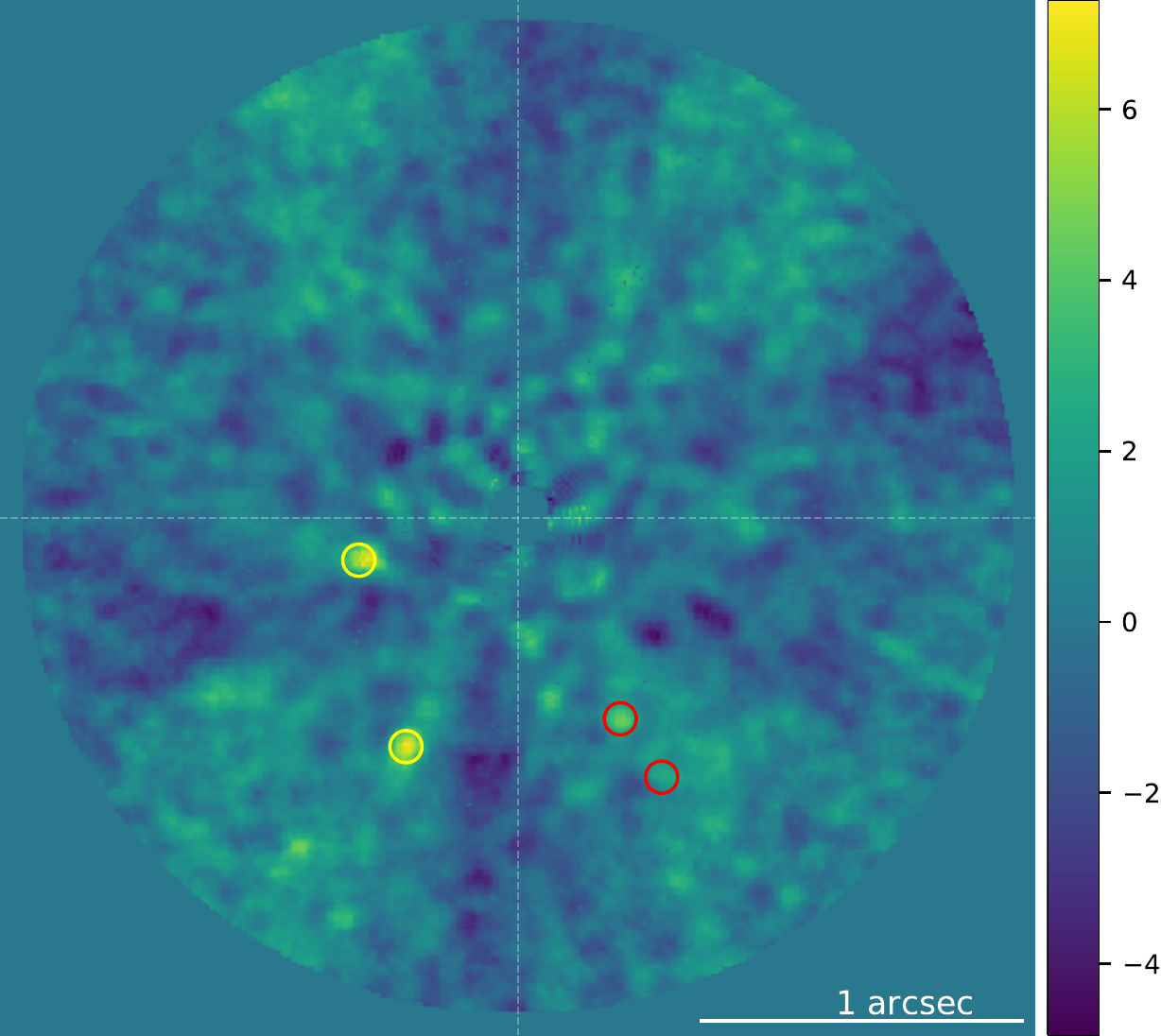}}
        \subfloat[NIRC2-2 S/N FF BU]{\includegraphics[width=115pt]{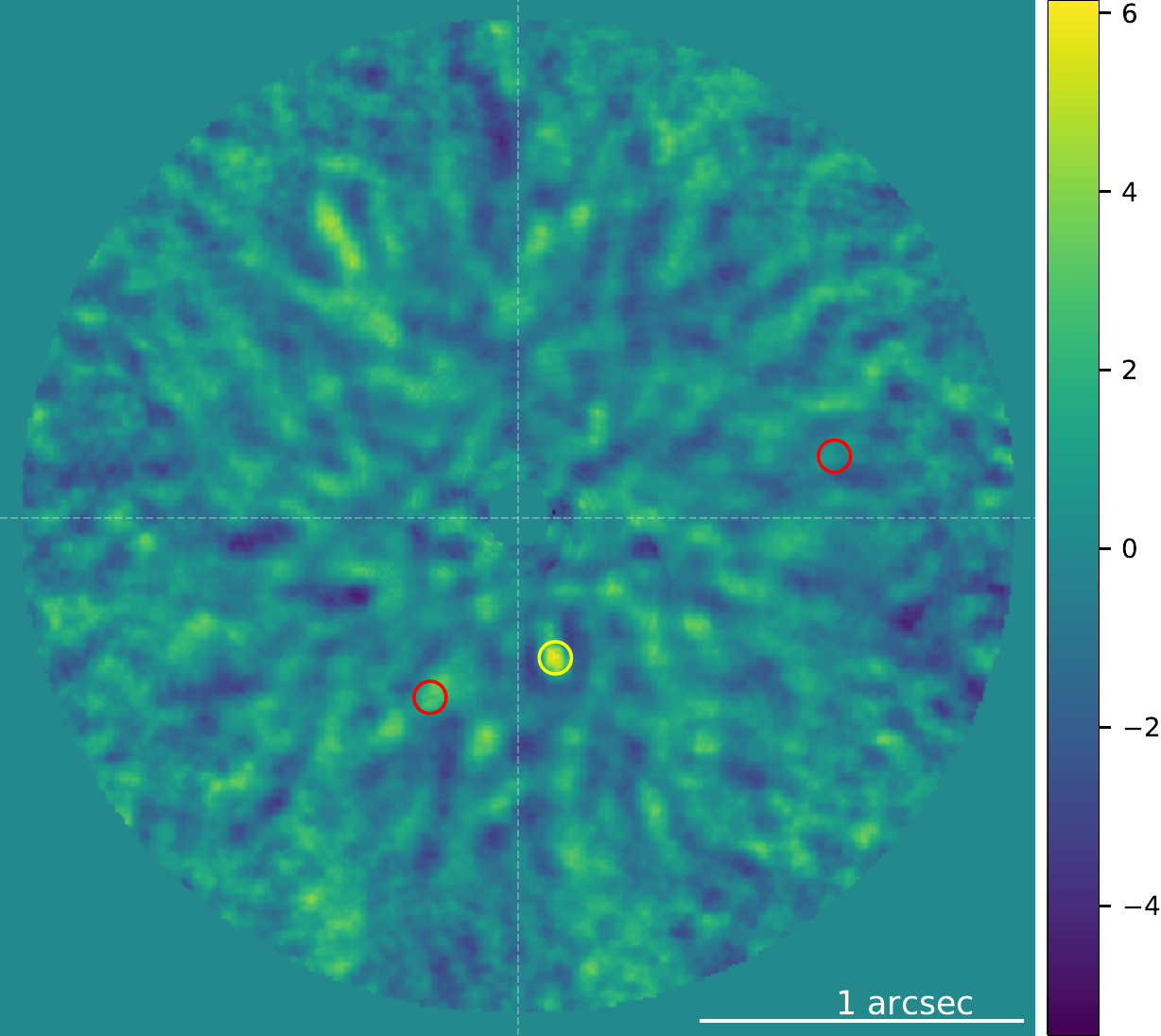}}\\
  \subfloat[SPHERE-1 S/N FF TD]{\includegraphics[width=115pt]{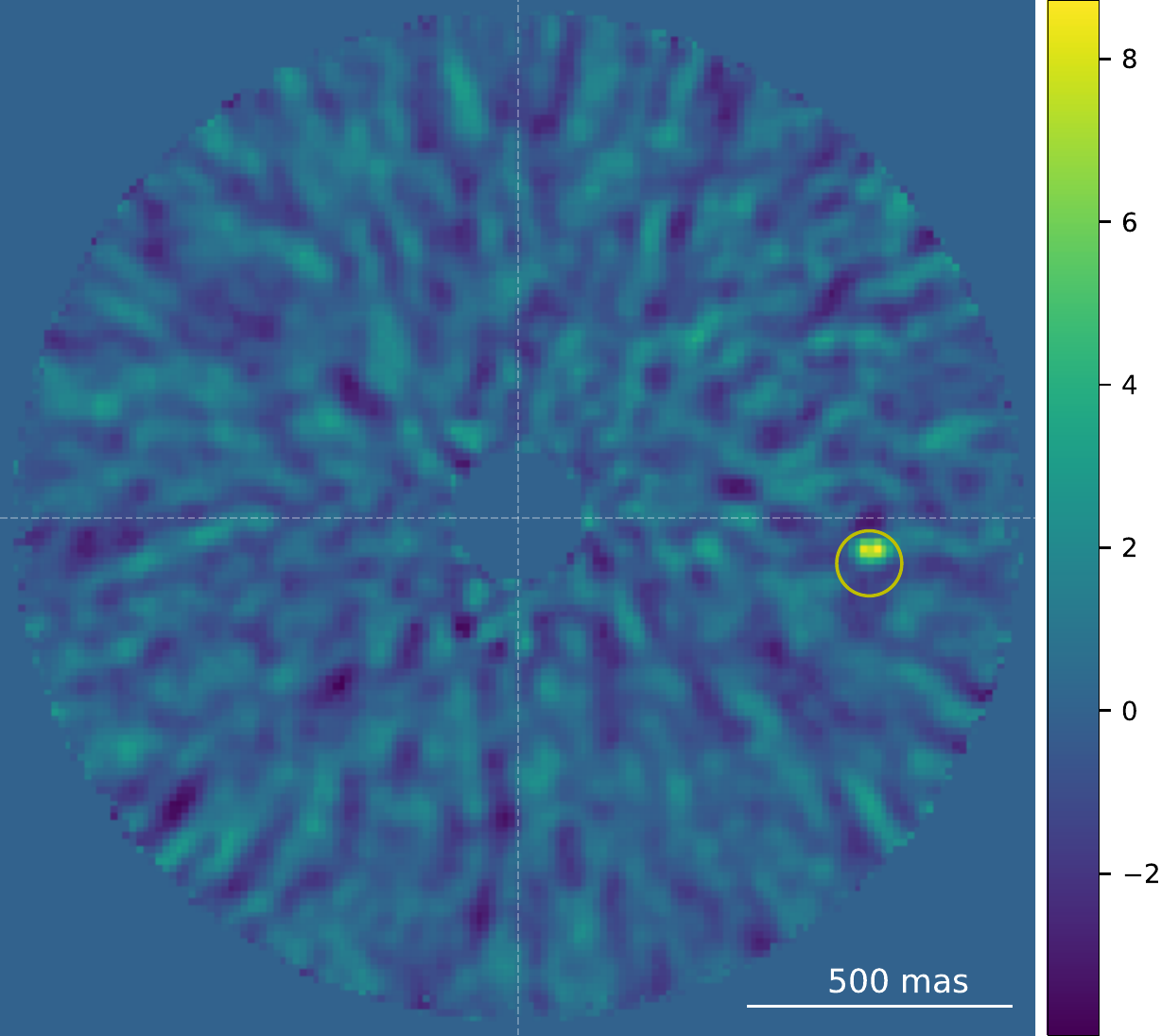}}
    \subfloat[SPHERE-3 S/N FF TD]{\includegraphics[width=115pt]{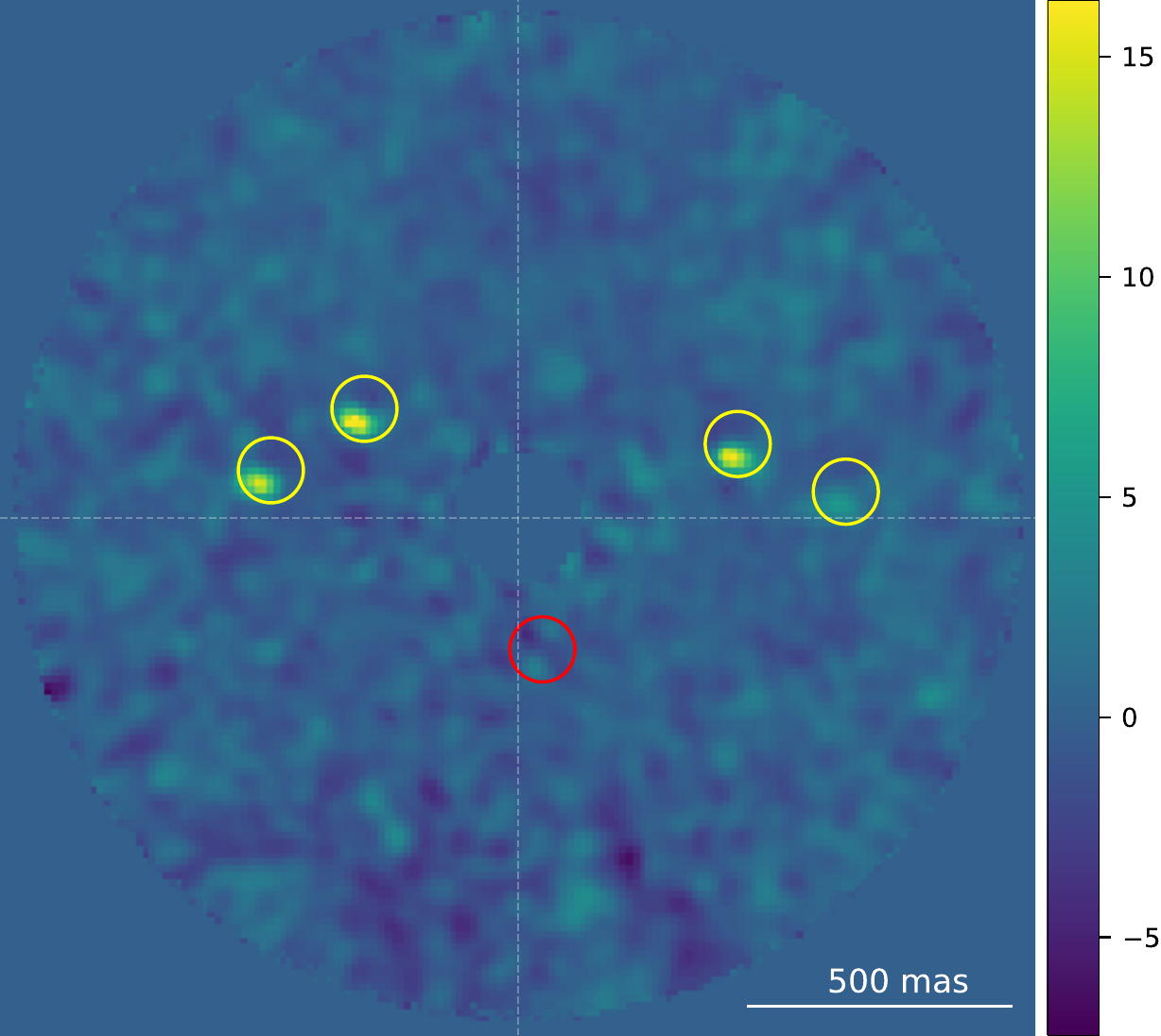}}
      \subfloat[NIRC2-1 S/N FF TD]{\includegraphics[width=115pt]{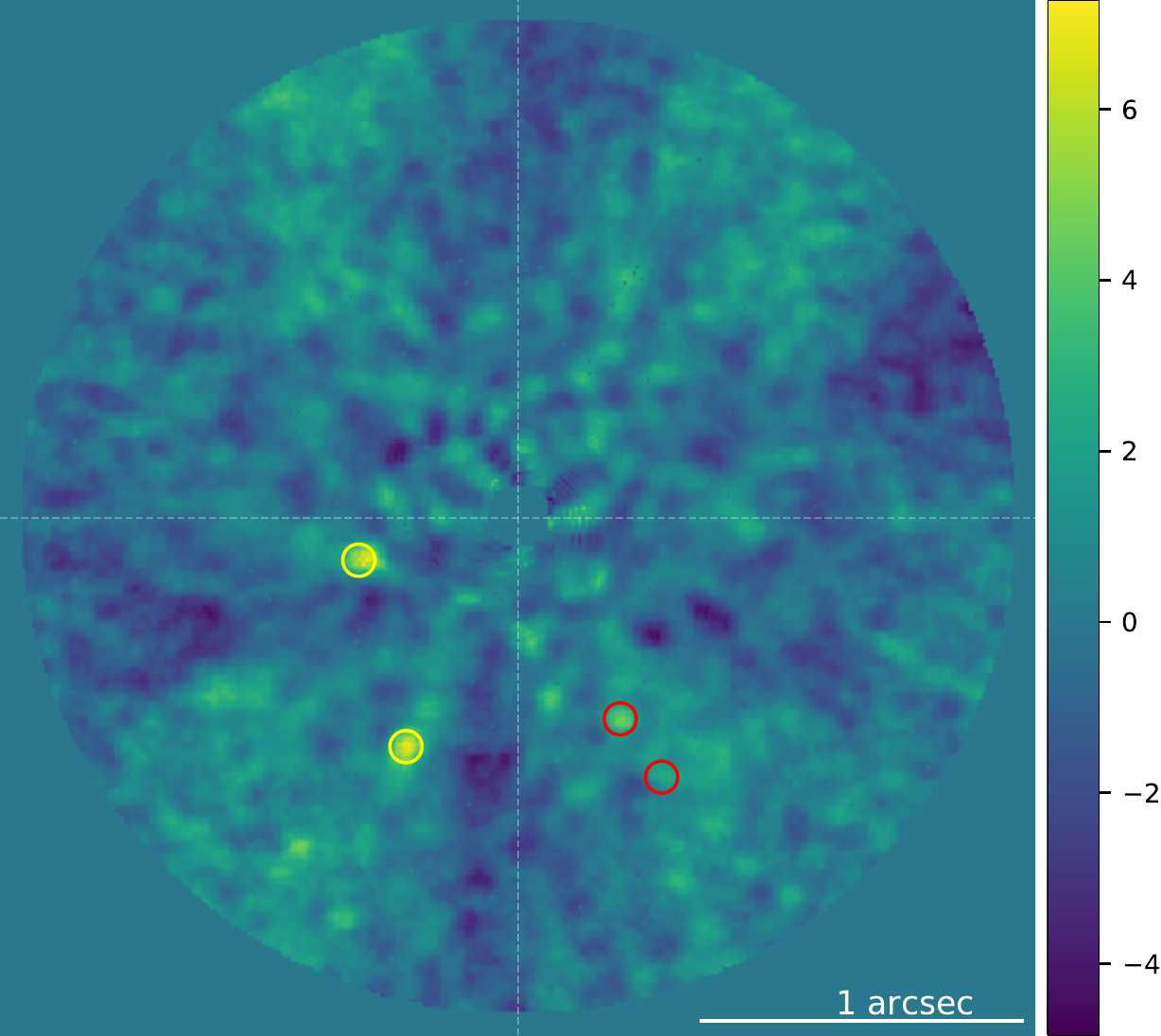}}
       \subfloat[NIRC2-2 S/N FF TD]{\includegraphics[width=115pt]{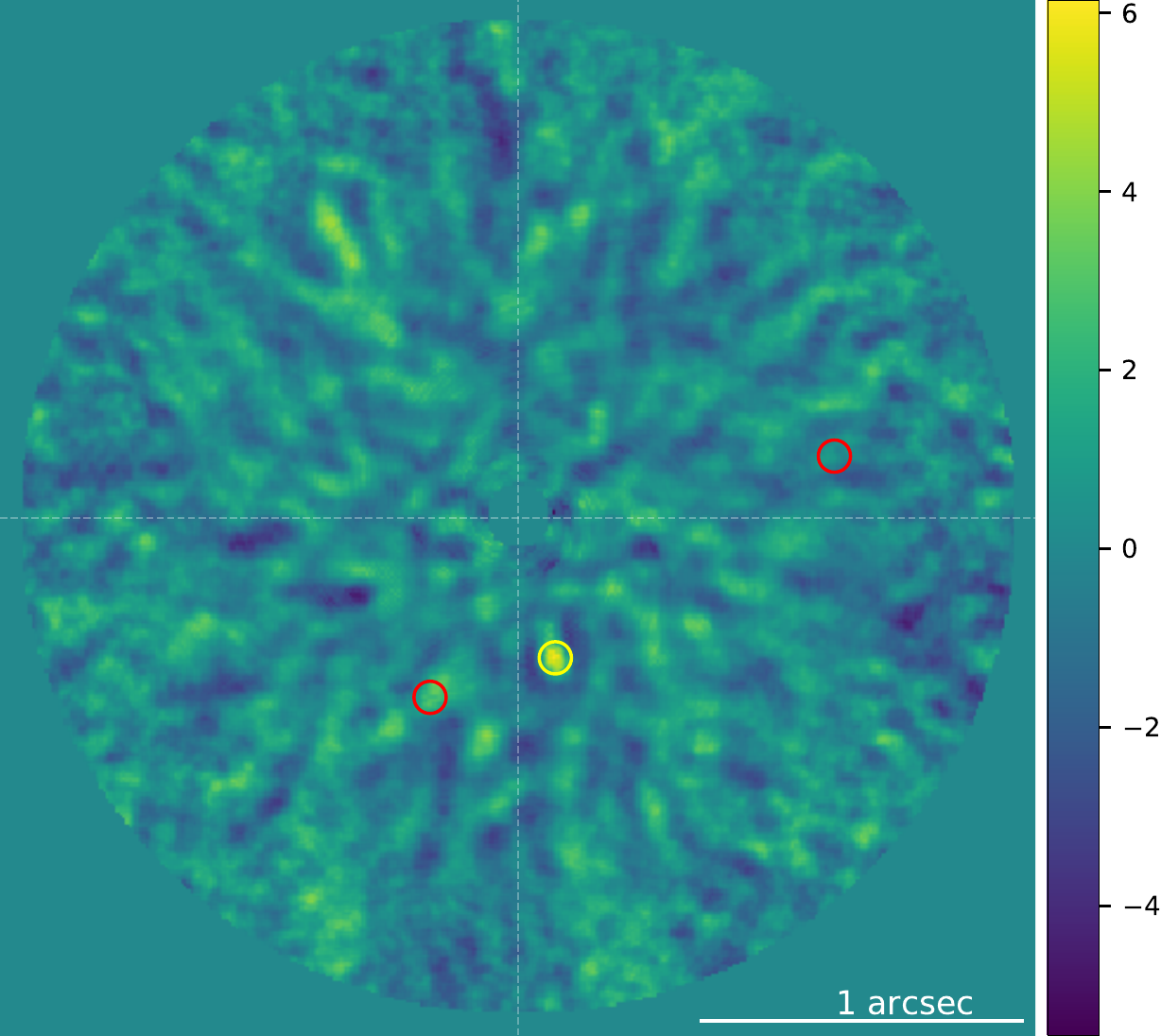}}\\
    \subfloat[SPHERE-1 S/N A BU]{\includegraphics[width=115pt]{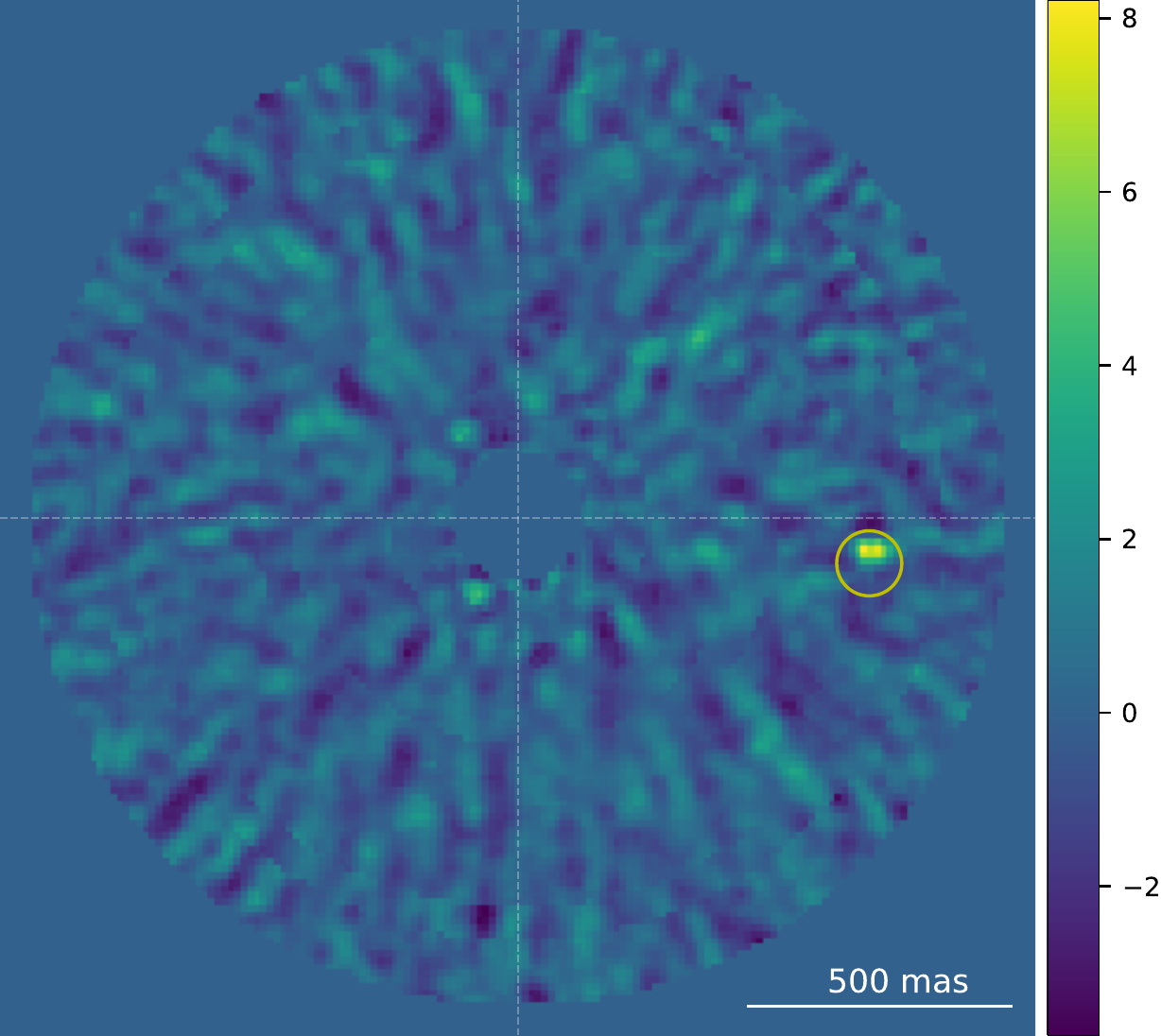}}
        \subfloat[SPHERE-3 S/N A BU]{\includegraphics[width=115pt]{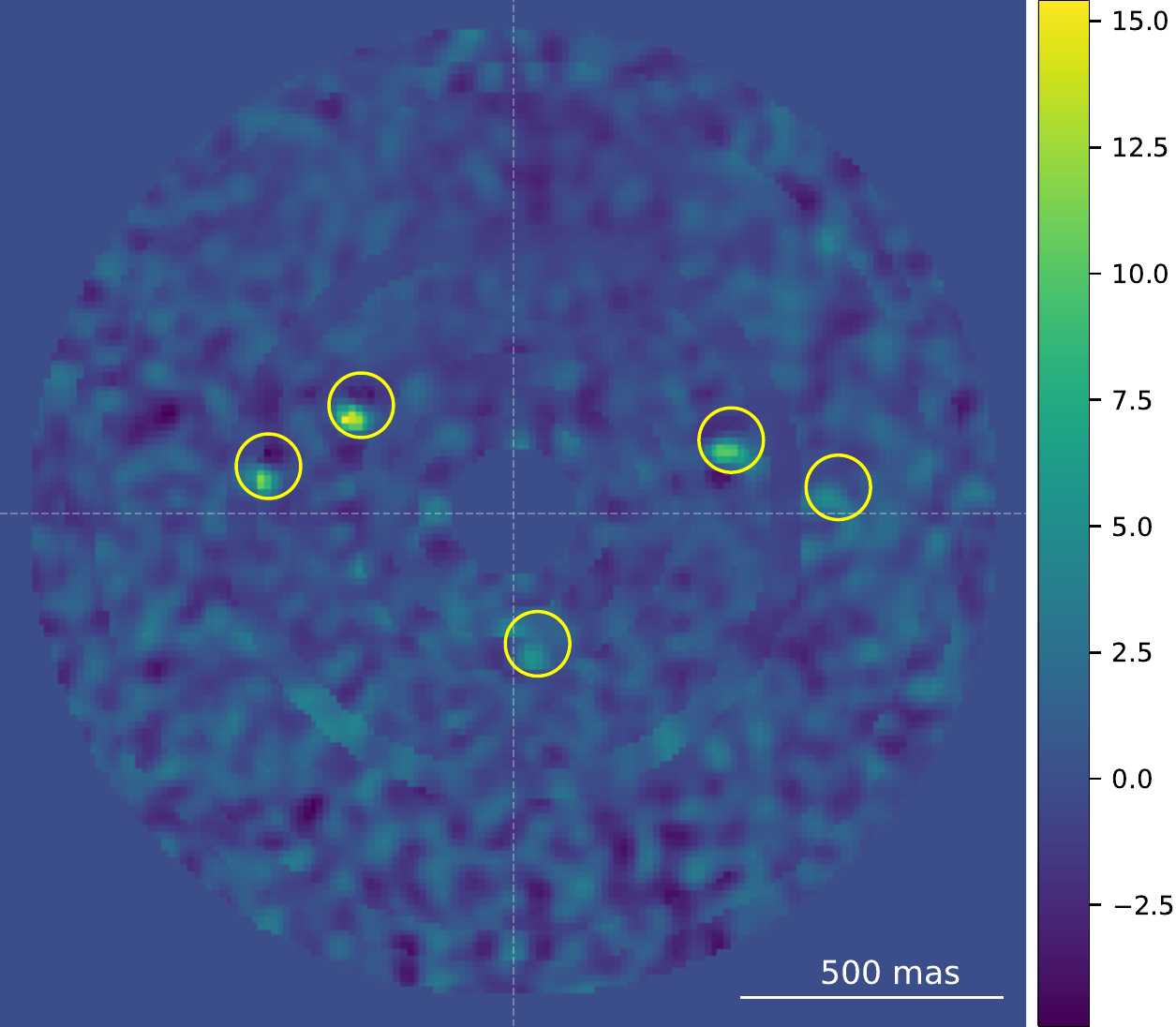}}
           \subfloat[NIRC2-1 S/N A BU]{\includegraphics[width=115pt]{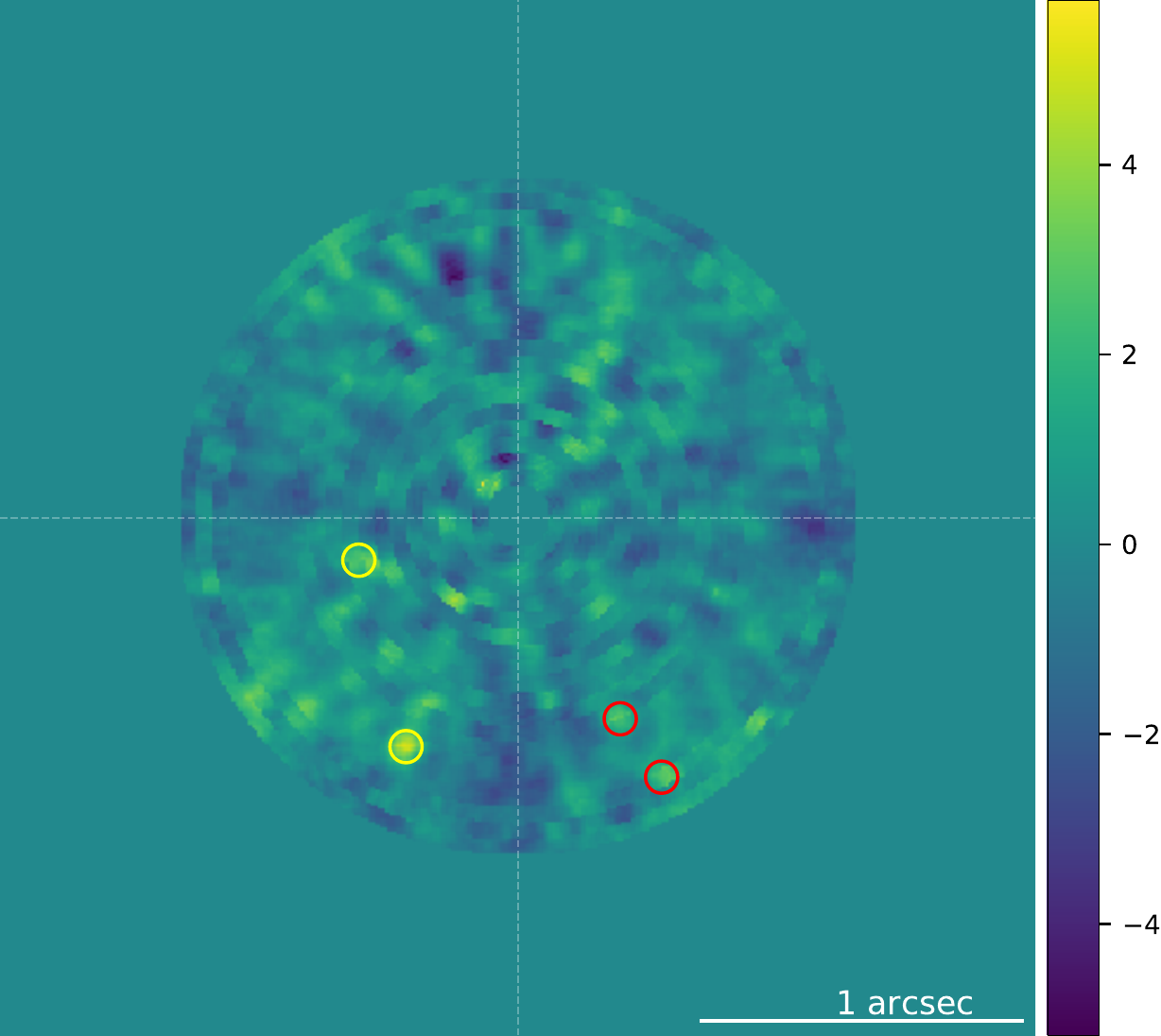}}
              \subfloat[NIRC2-2 S/N A BU]{\includegraphics[width=115pt]{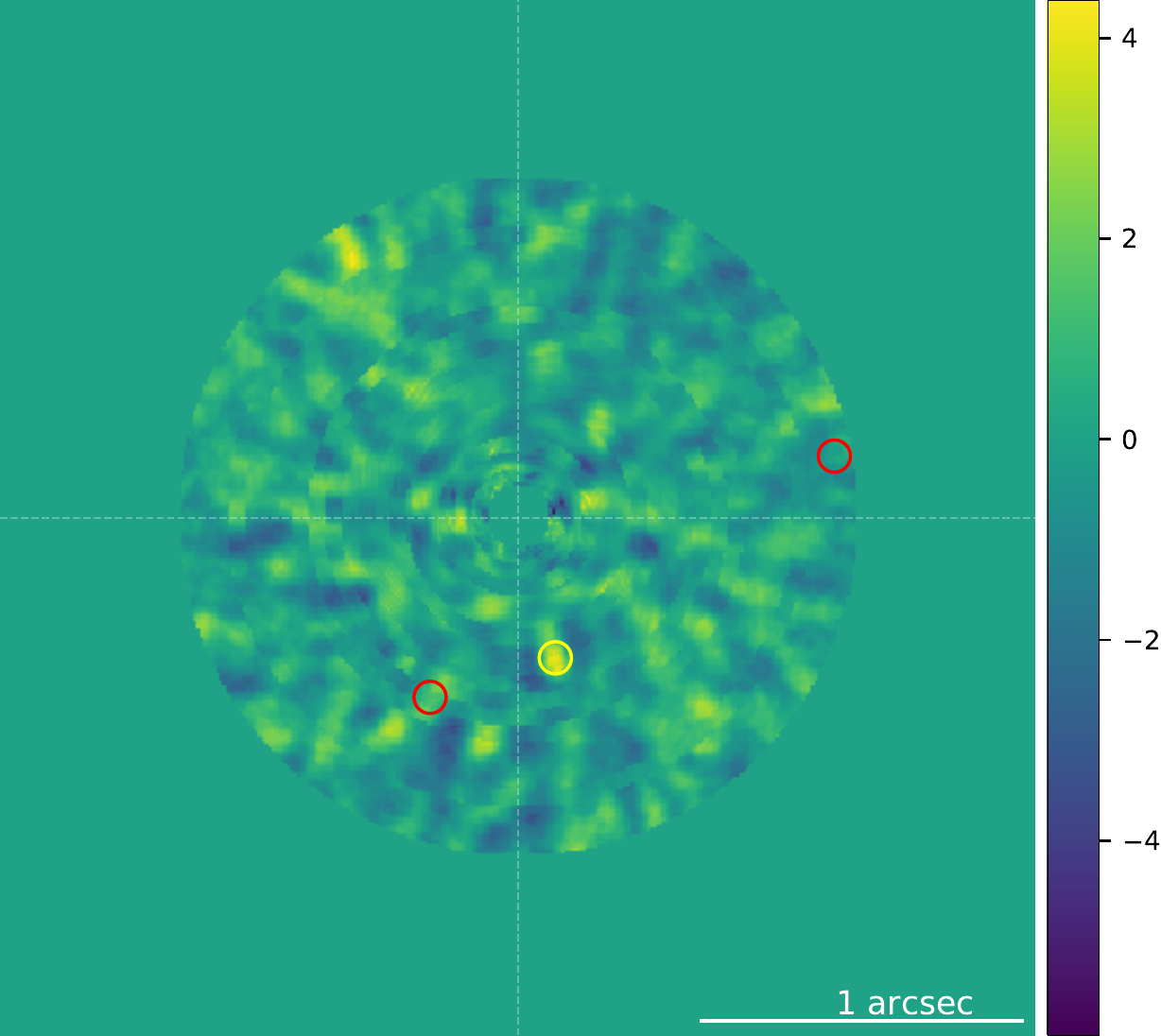}}\\
     \subfloat[SPHERE-1 S/N AFF BU]{\includegraphics[width=115pt]{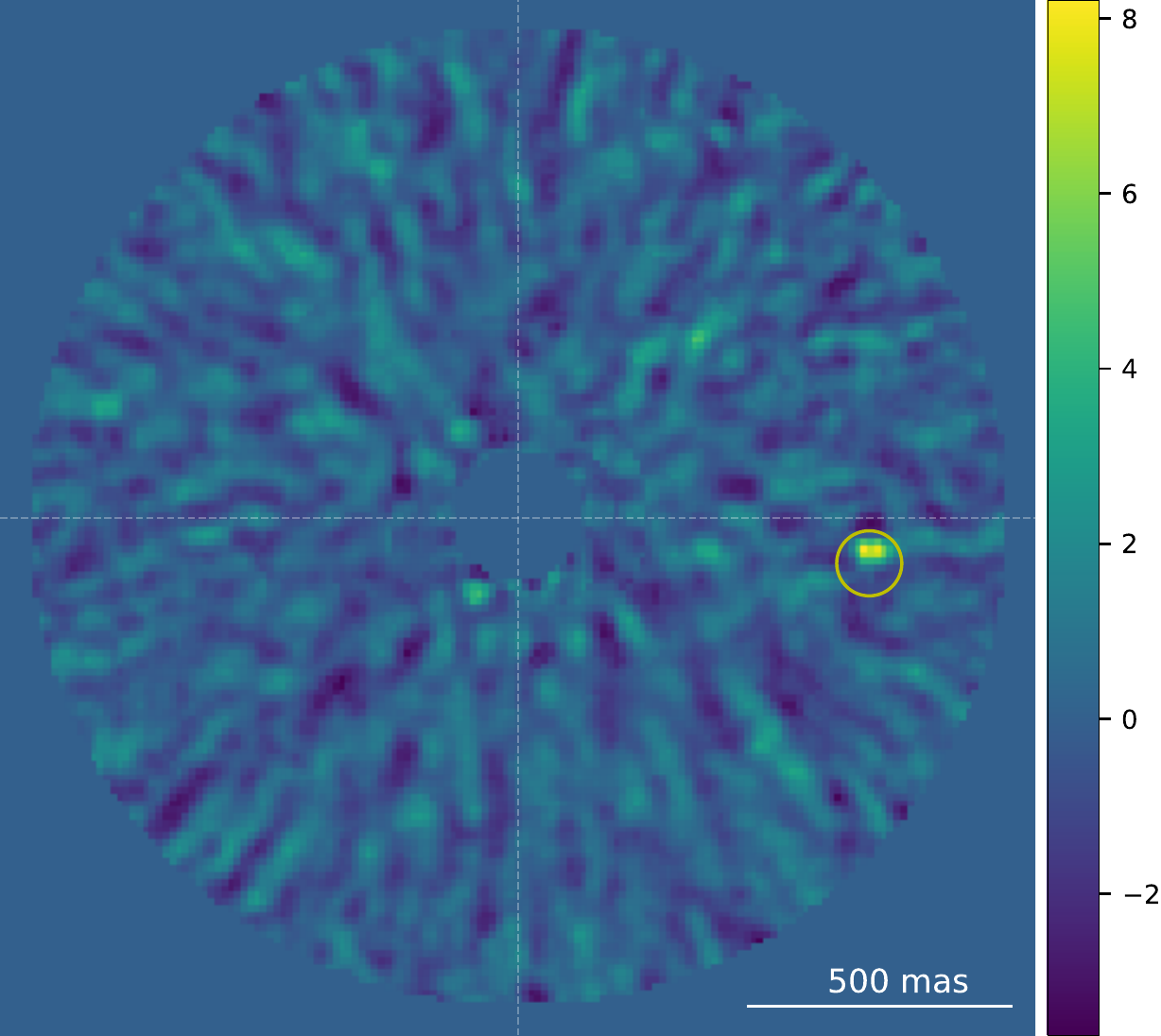}}
      \subfloat[SPHERE-3 S/N AFF BU]{\includegraphics[width=115pt]{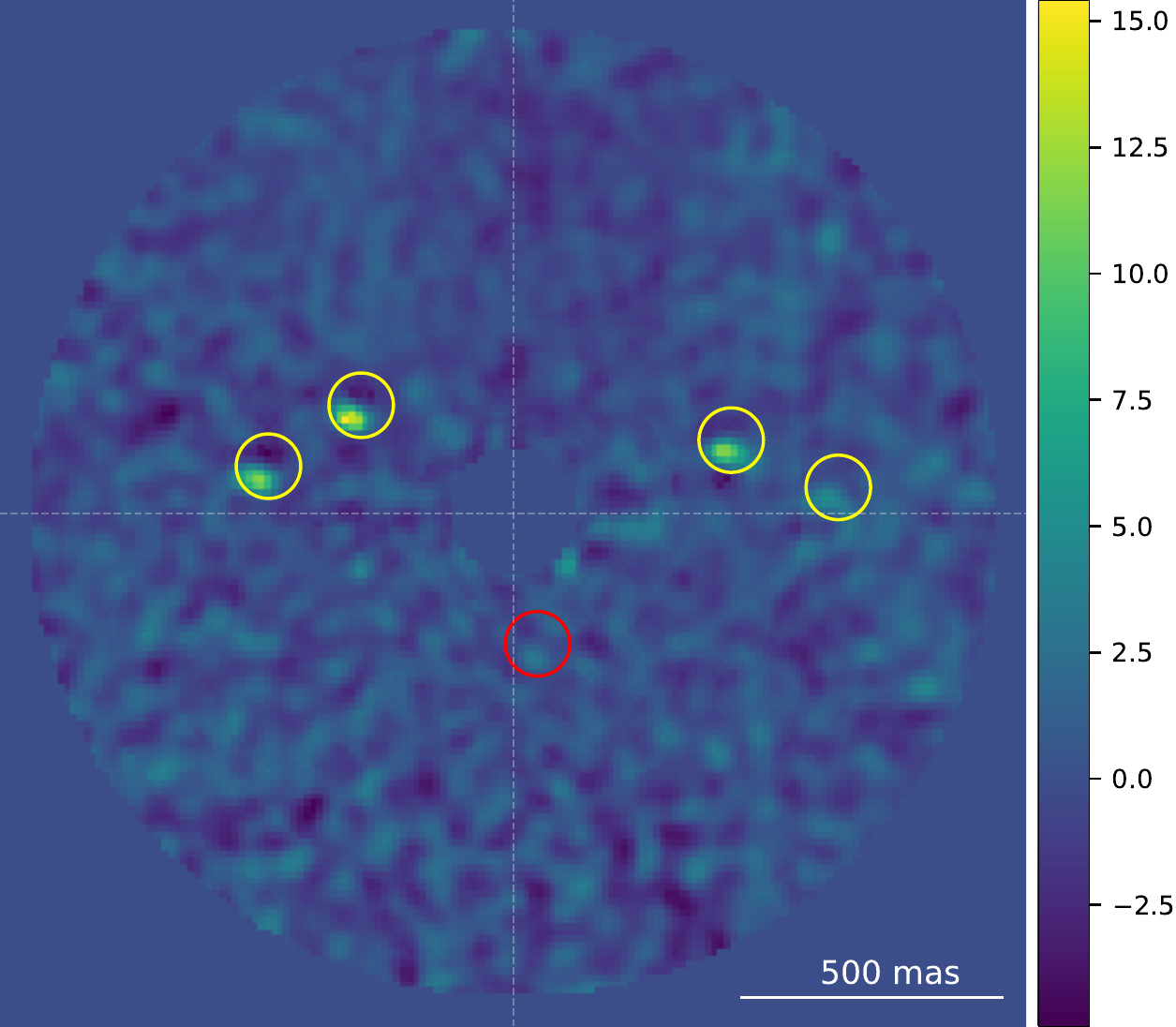}}
       \subfloat[NIRC2-1 S/N AFF BU]{\includegraphics[width=115pt]{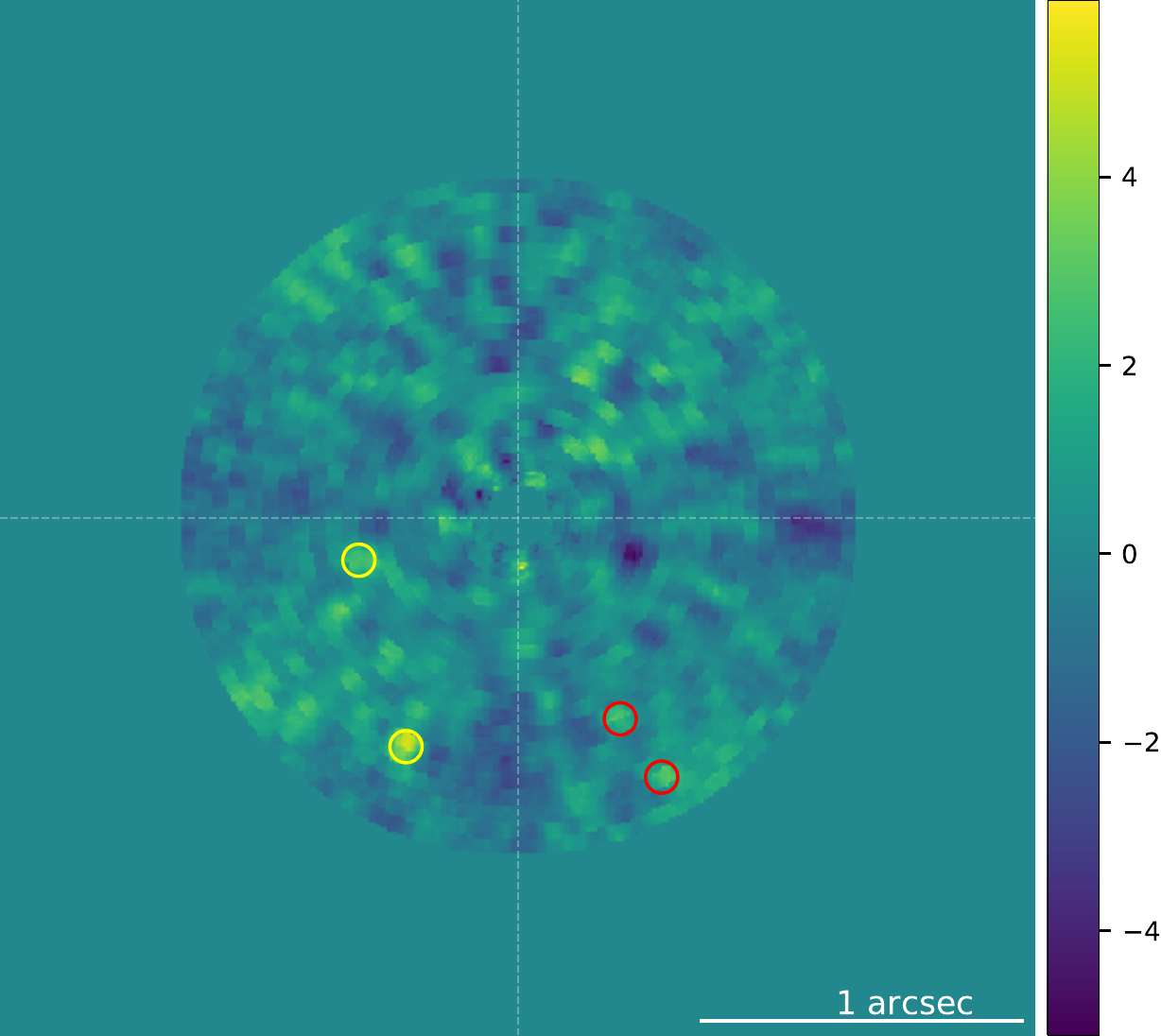}}
         \subfloat[NIRC2-2 S/N AFF BU]{\includegraphics[width=115pt]{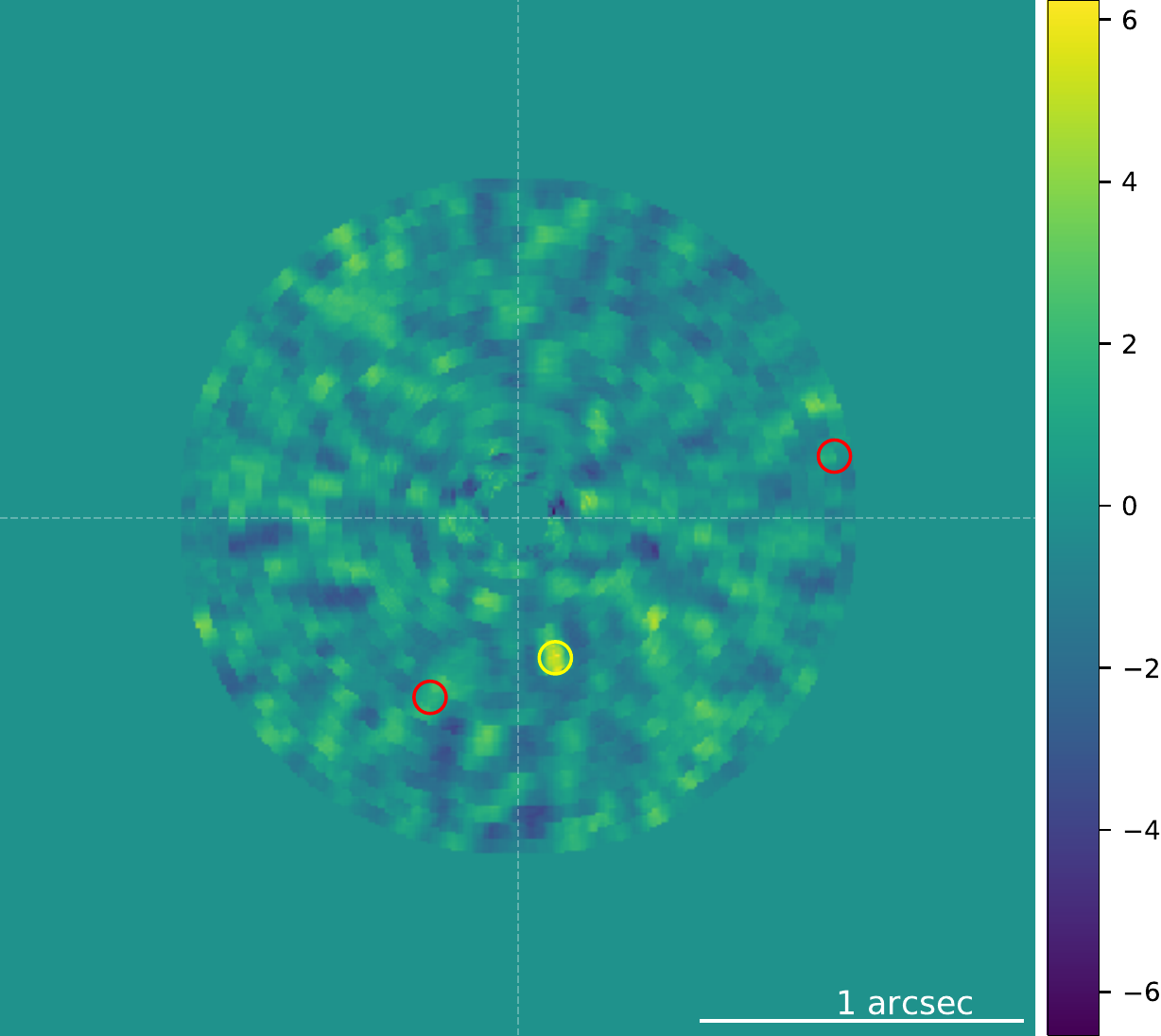}}\\

  \caption{\label{ResSNR1} Detection maps corresponding to the SPHERE and NIRC2 data sets generated with different parametrisations of the full-frame and annular auto-S/N along the baseline model presented in \citep{Cantalloube20}. The SPHERE-2 and NIRC2-3 detection maps are not shown, as no fake companions were injected in these two data sets. See Sect. 4.3.1. for the definition of the acronyms used to characterise the auto-RSM versions. The yellow circles are centred on the true position of the detected targets (TP) and the red circles give the true positions of FNs. }
\end{figure*}

\begin{figure*}[h!]
\footnotesize
  \centering
  \subfloat[LMIRCam-1 Baseline]{\includegraphics[width=115pt]{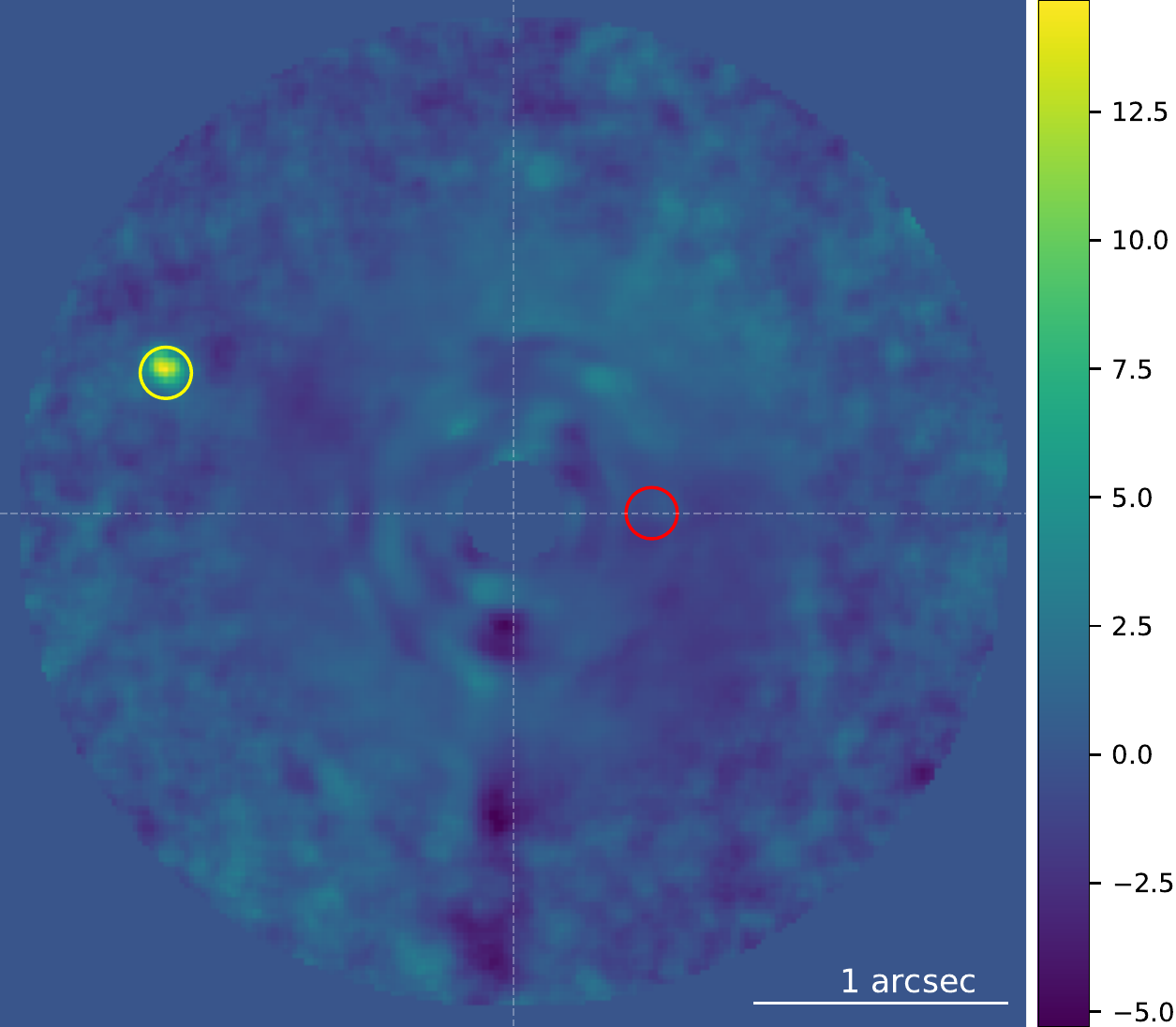}}
   \subfloat[LMIRCam-2 Baseline]{\includegraphics[width=115pt]{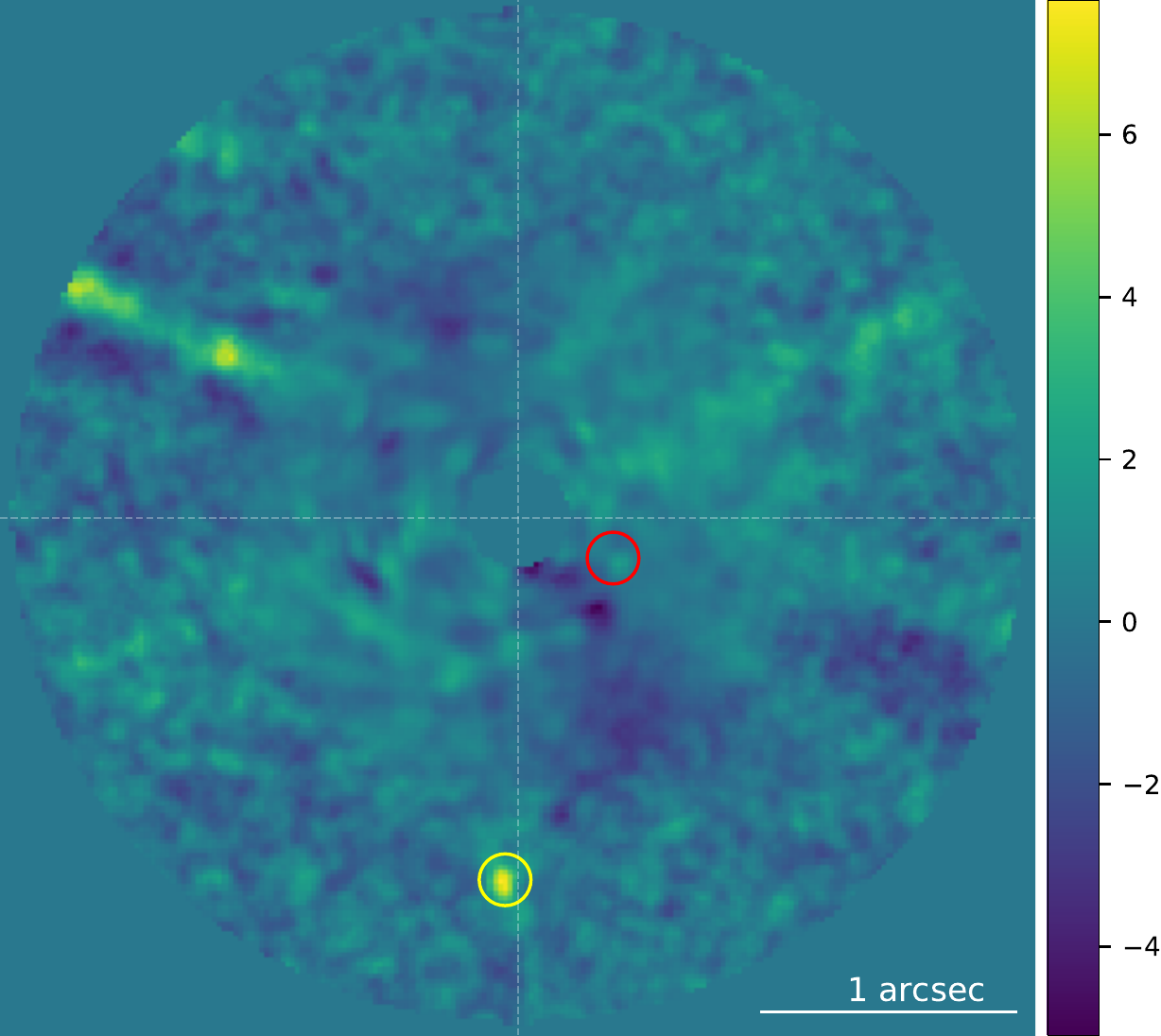}}
     \subfloat[LMIRCam-3 Baseline]{\includegraphics[width=115pt]{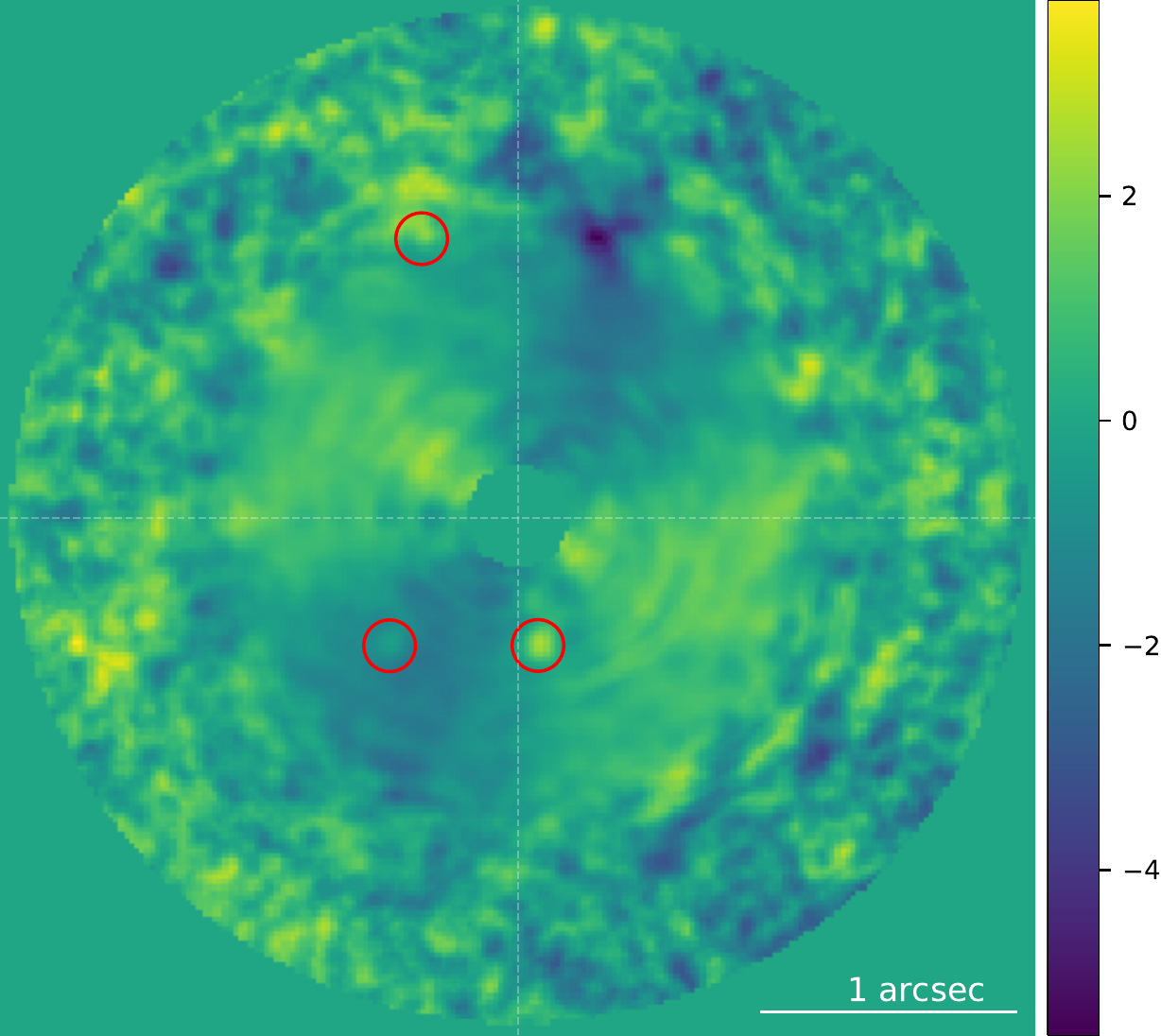}}\\
  \subfloat[LMIRCam-1 S/N FF BU]{\includegraphics[width=115pt]{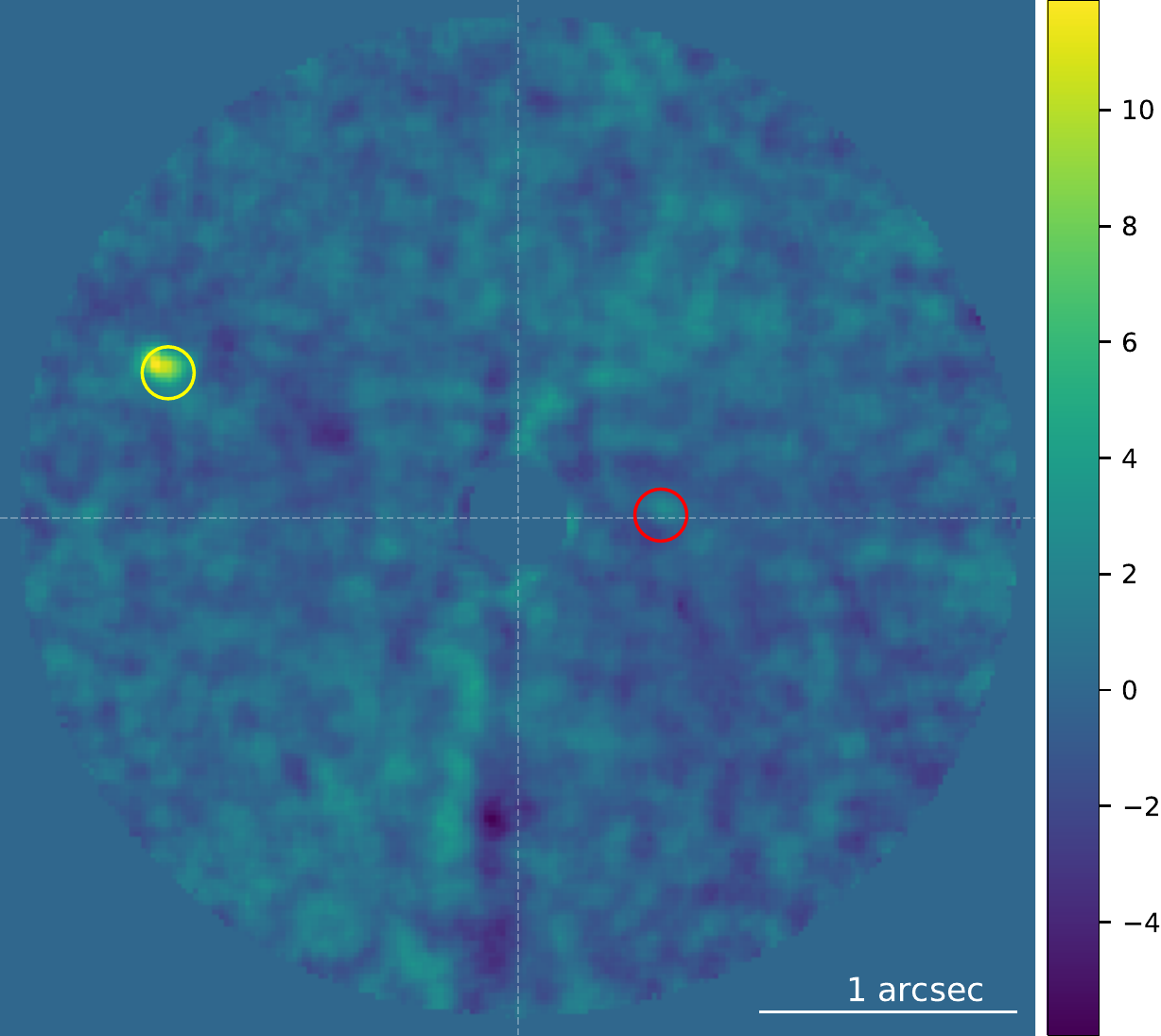}}
   \subfloat[LMIRCam-2 S/N FF BU]{\includegraphics[width=115pt]{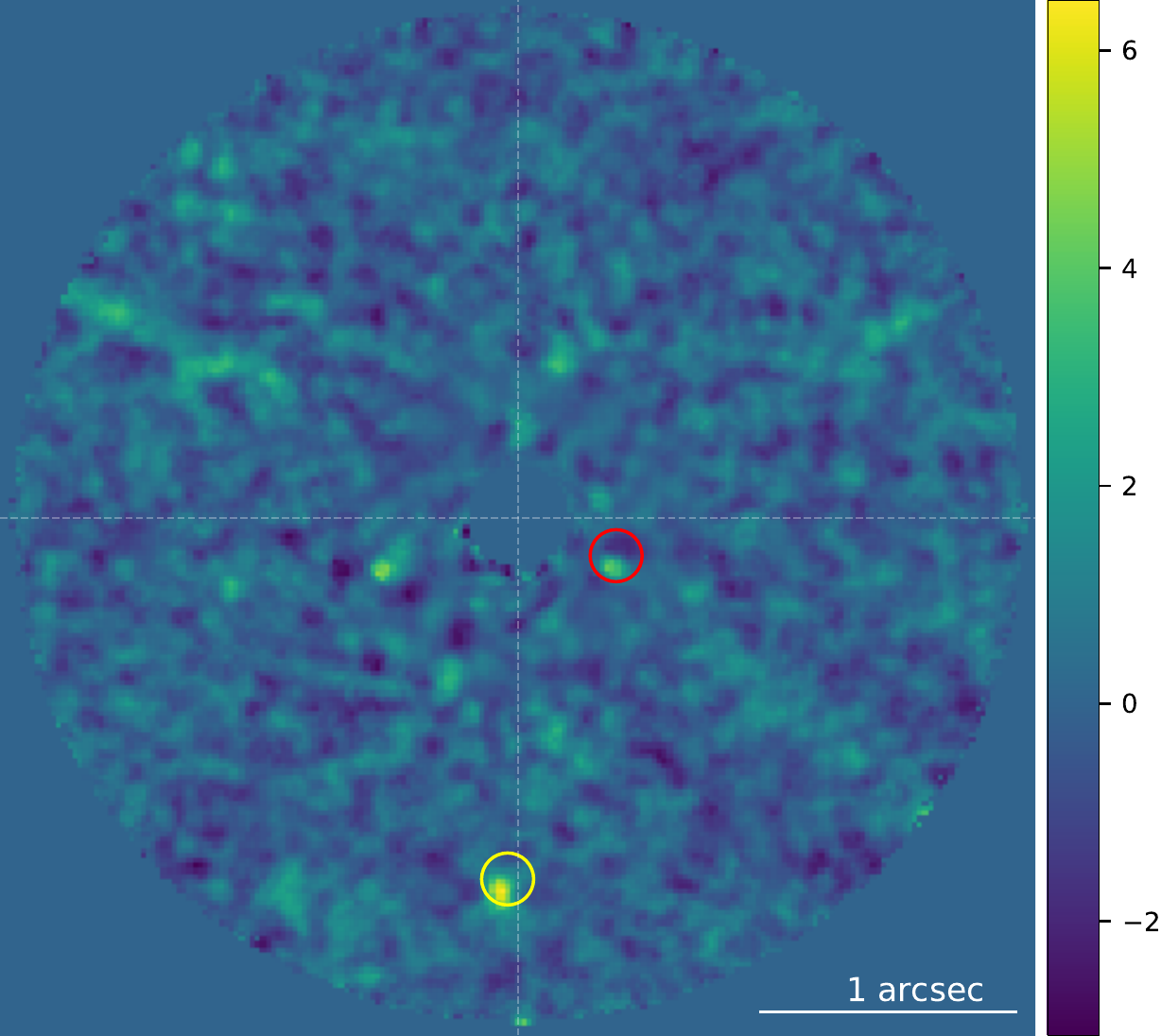}}
    \subfloat[LMIRCam-3 S/N FF BU]{\includegraphics[width=115pt]{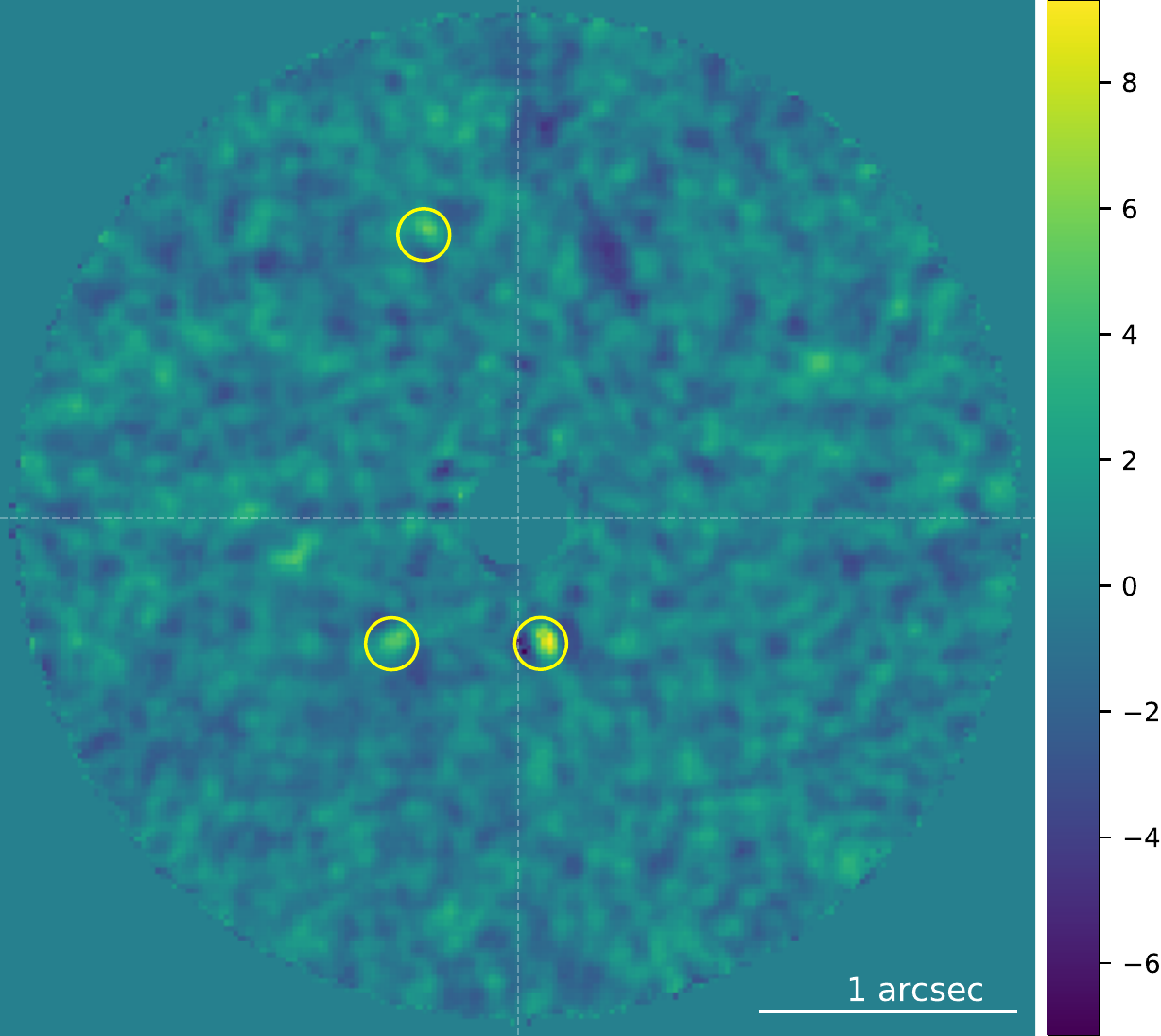}}\\
  \subfloat[LMIRCam-1 S/N FF TD]{\includegraphics[width=115pt]{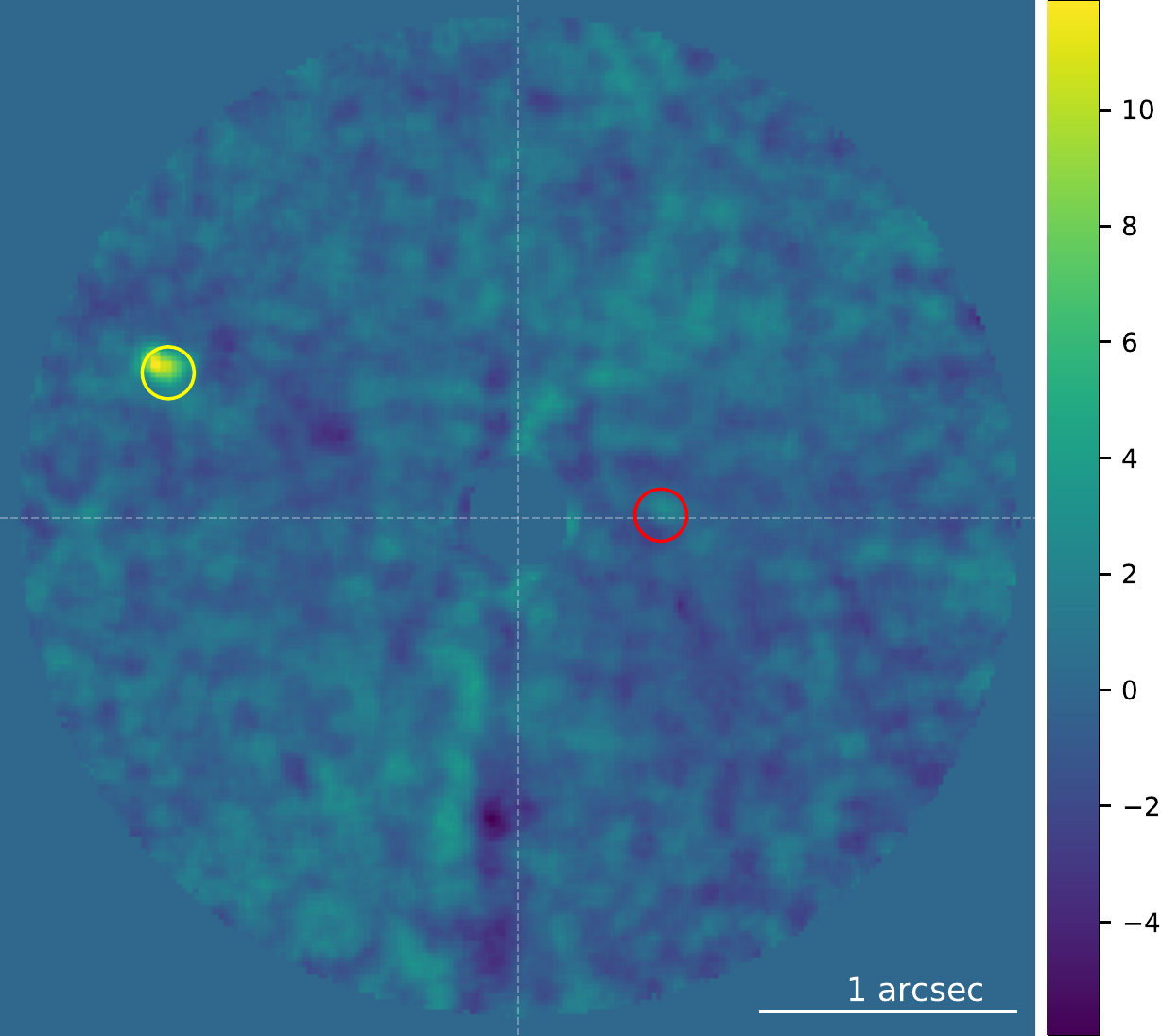}}
   \subfloat[LMIRCam-2 S/N FF TD]{\includegraphics[width=115pt]{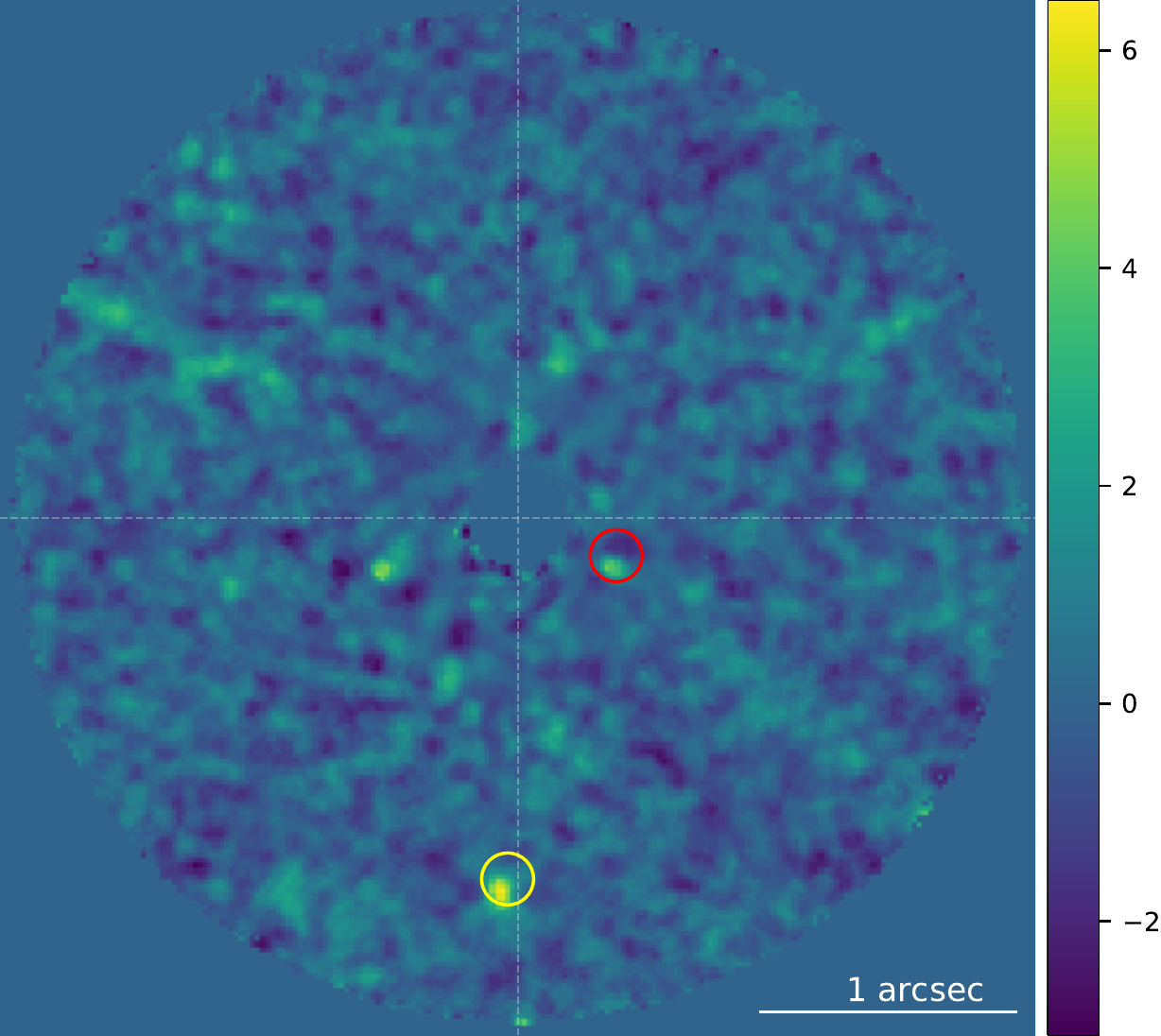}}
    \subfloat[LMIRCam-3 S/N FF TD]{\includegraphics[width=115pt]{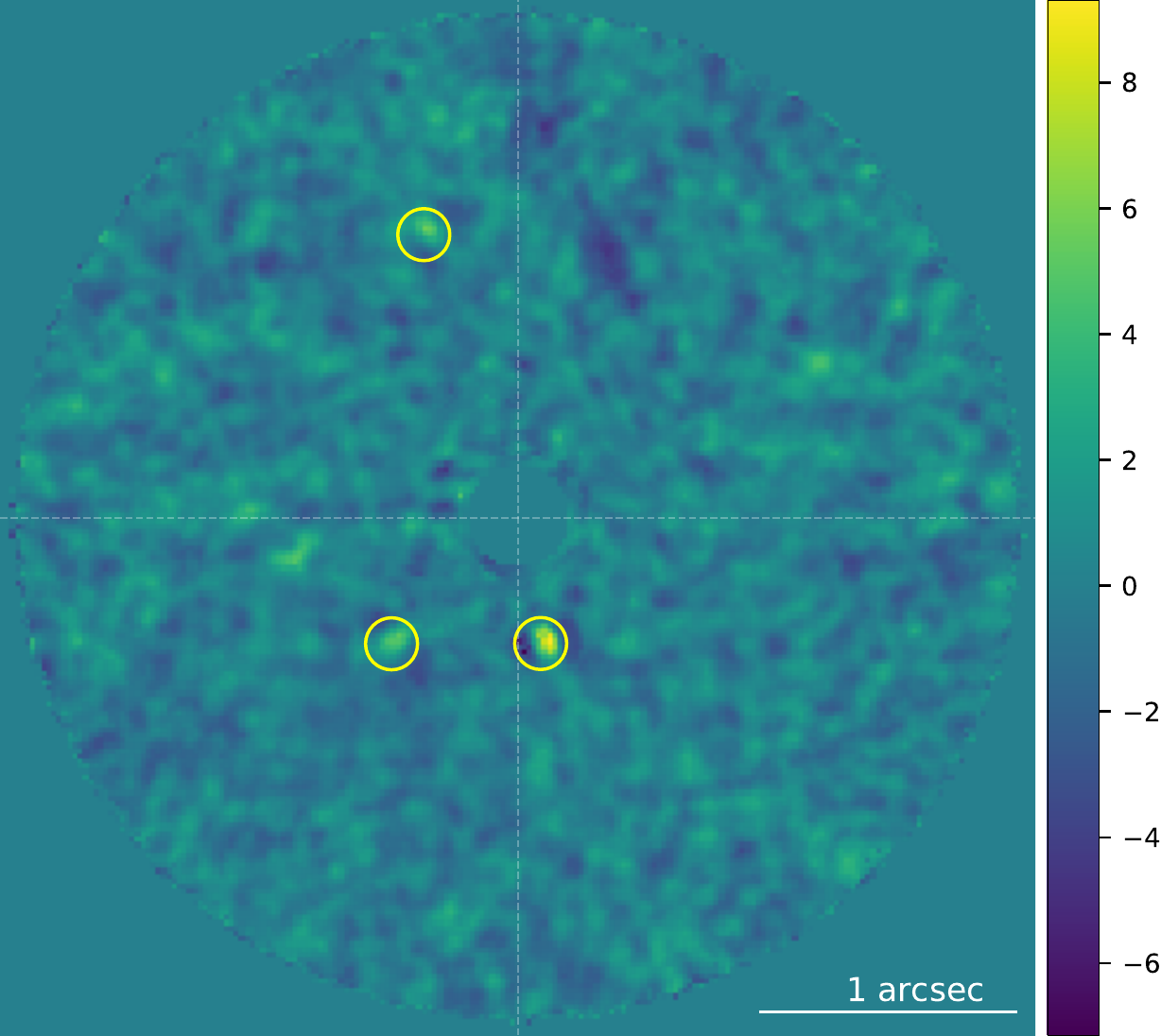}}\\
    \subfloat[LMIRCam-1 S/N A BU]{\includegraphics[width=115pt]{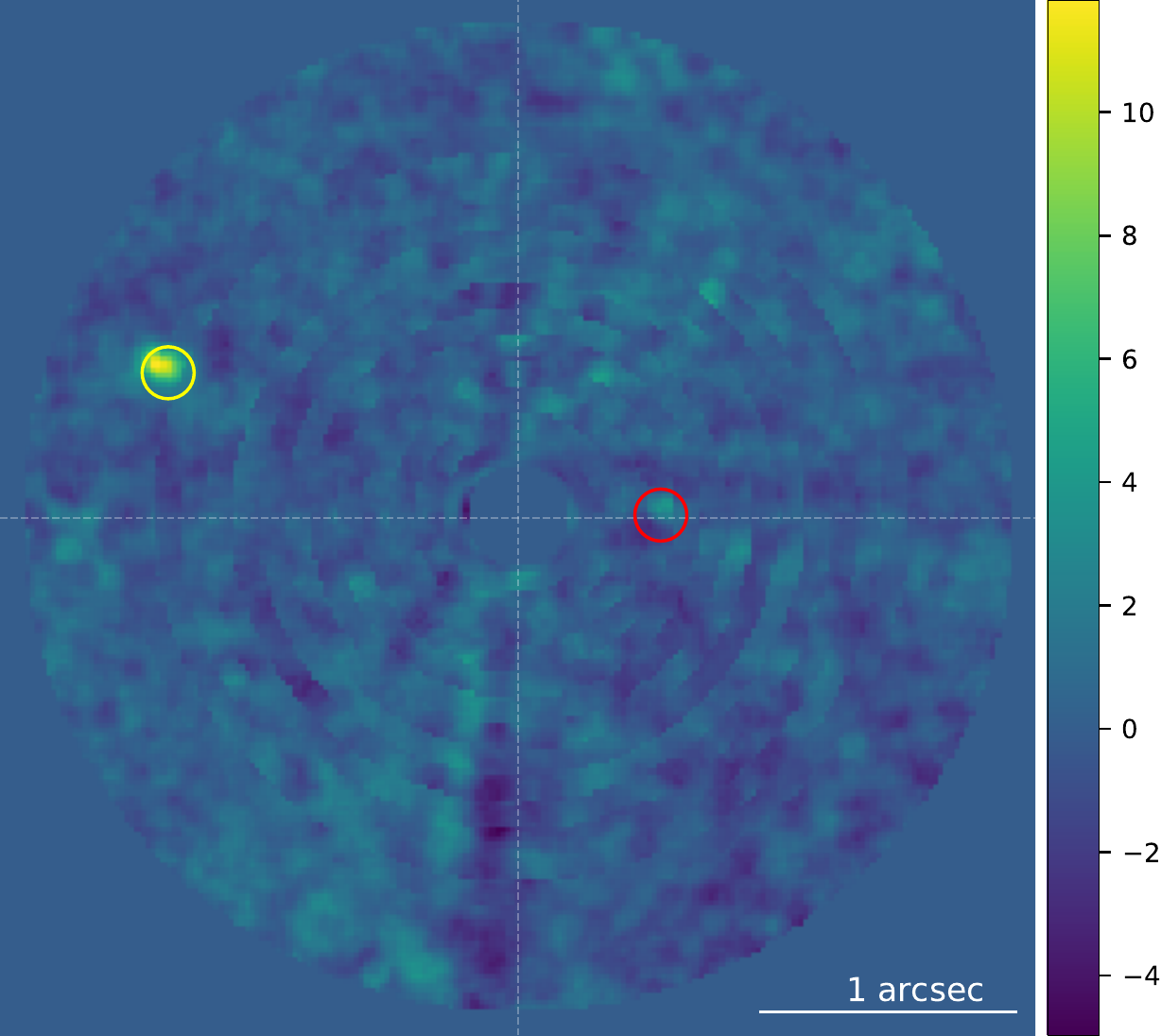}}
      \subfloat[LMIRCam-2 S/N A BU]{\includegraphics[width=115pt]{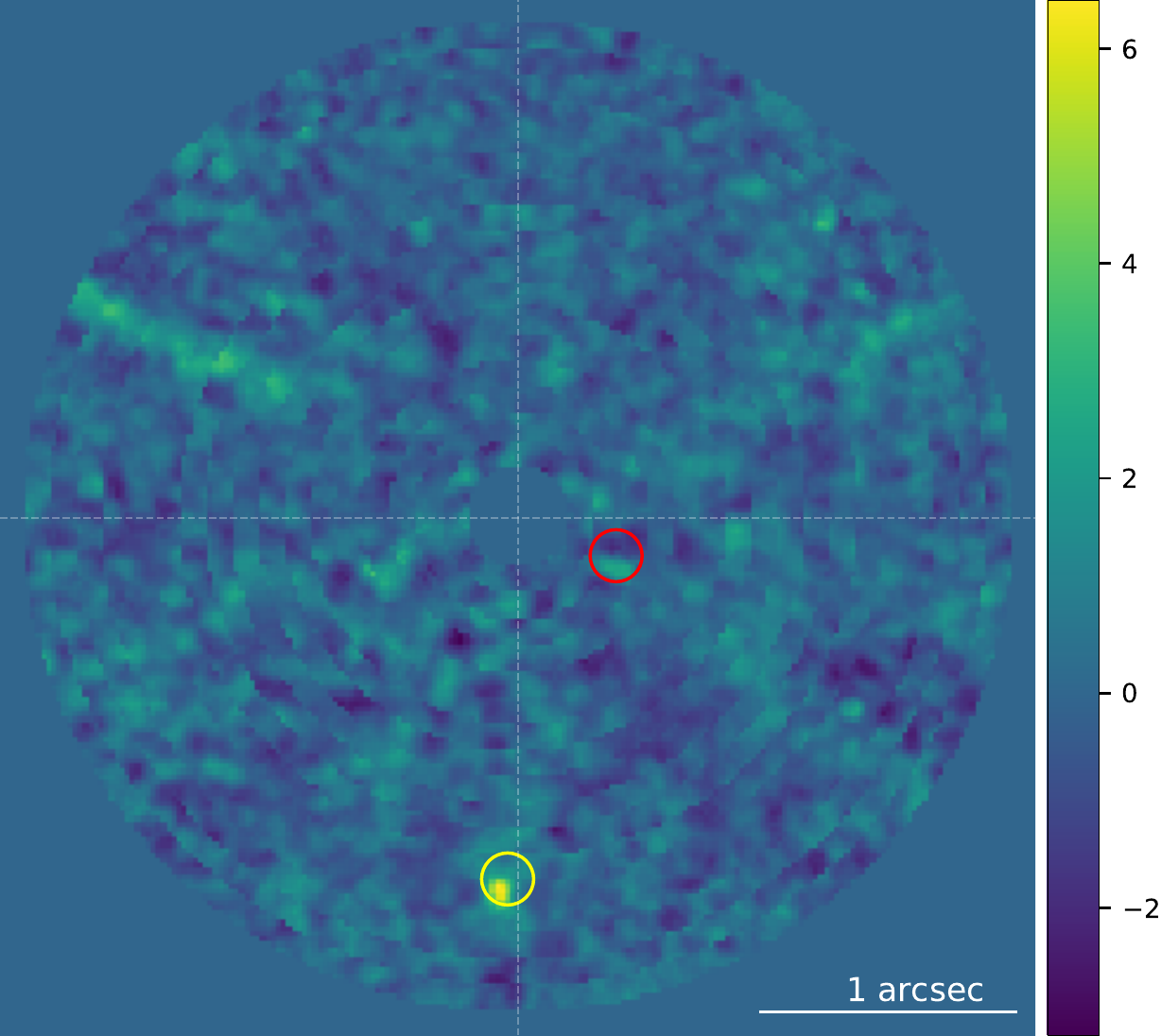}}
         \subfloat[LMIRCam-3 S/N A BU]{\includegraphics[width=115pt]{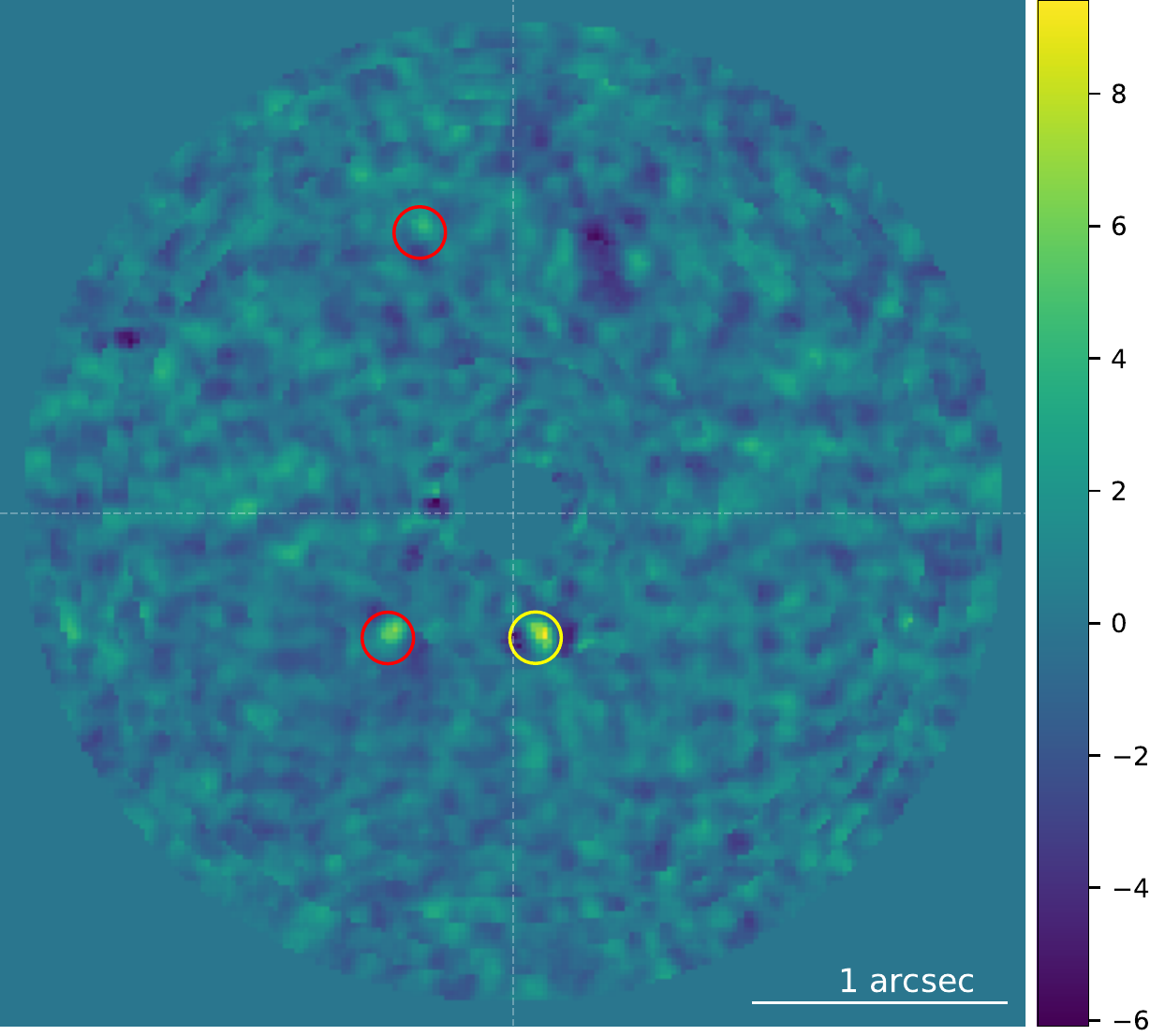}}\\
     \subfloat[LMIRCam-1 S/N AFF BU]{\includegraphics[width=115pt]{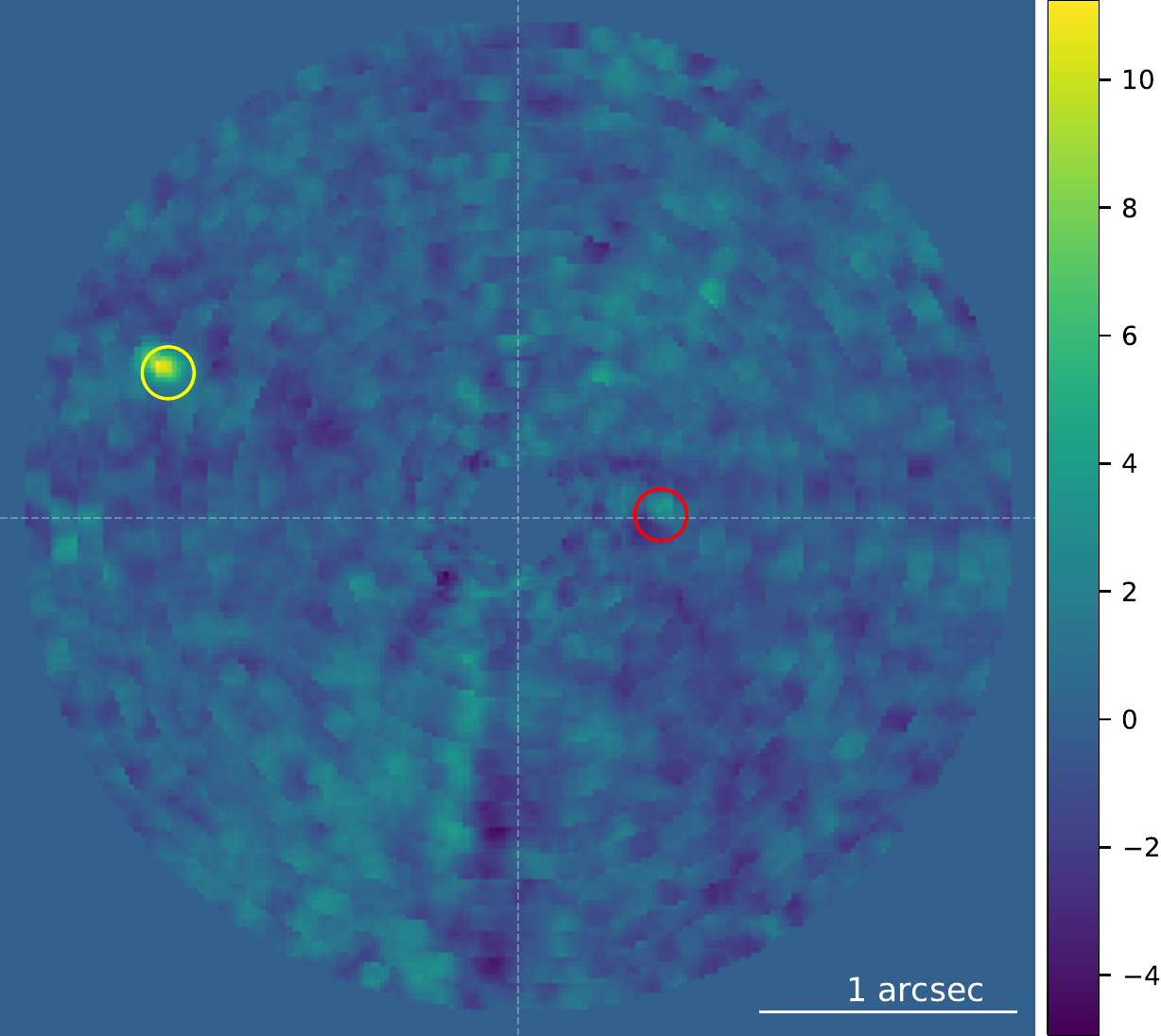}}
      \subfloat[LMIRCam-2 S/N AFF BU]{\includegraphics[width=115pt]{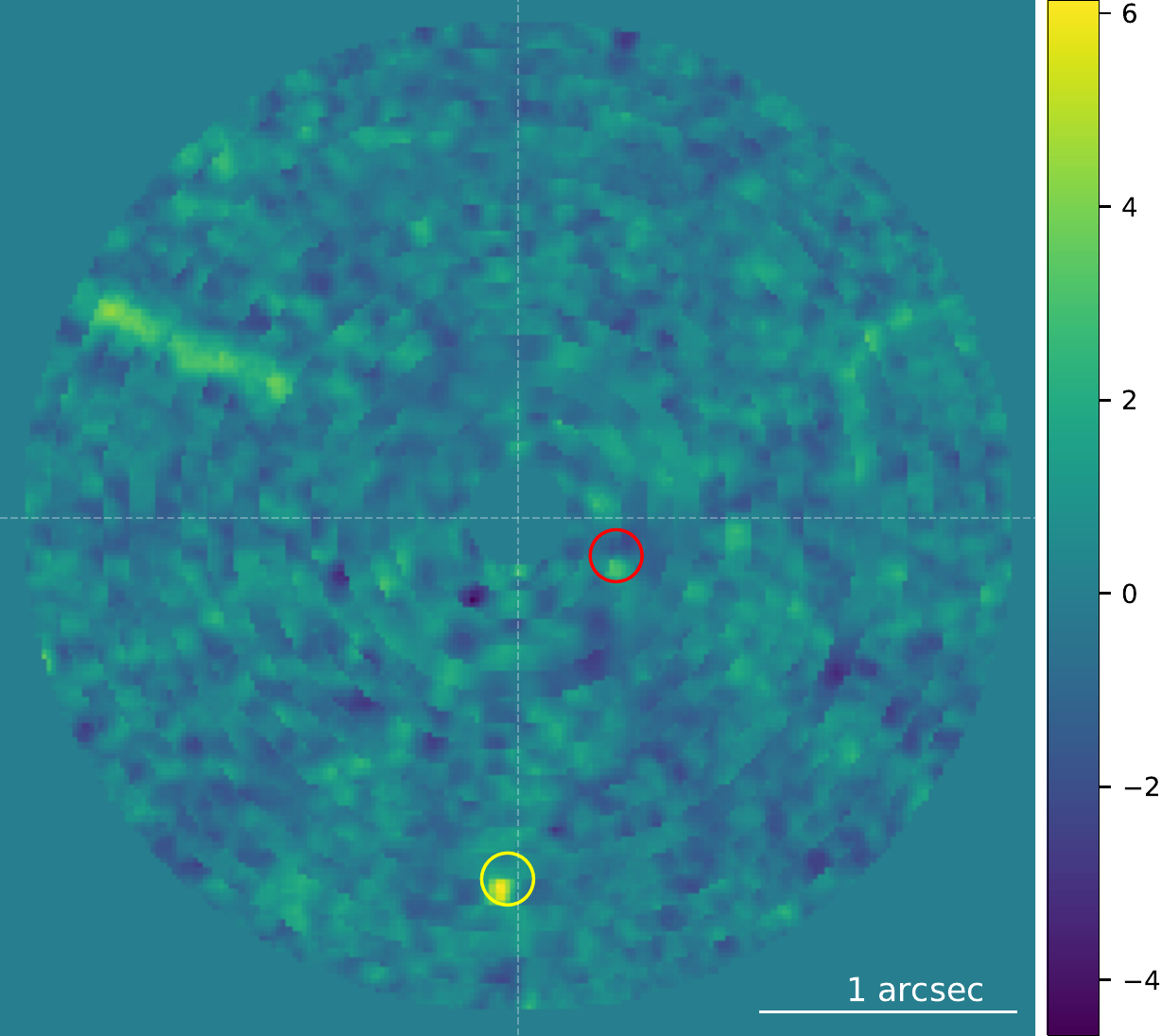}}
         \subfloat[LMIRCam-3 S/N AFF BU]{\includegraphics[width=115pt]{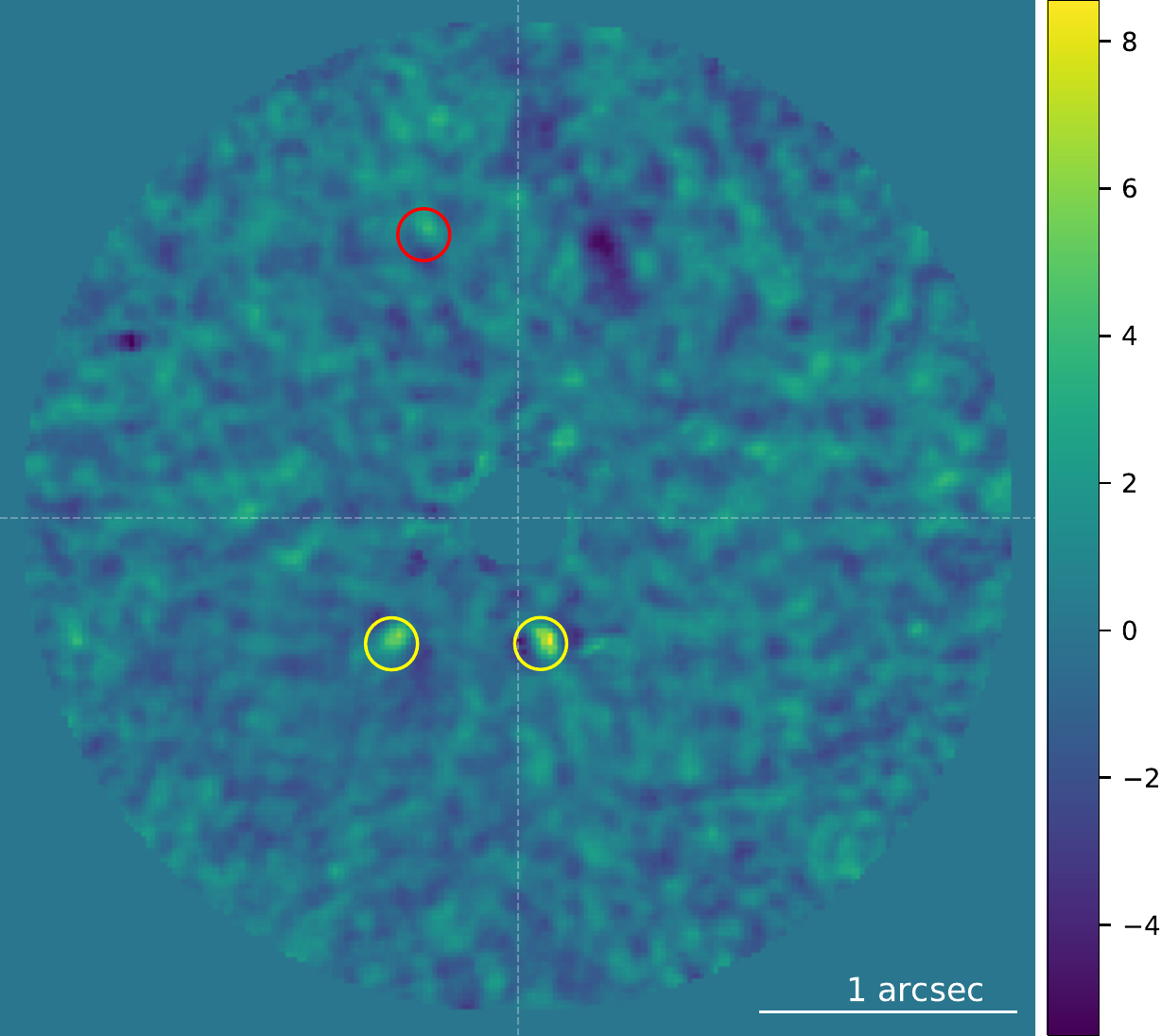}}\\

  \caption{\label{ResSNR2} S/N maps corresponding to the LMIRCam data sets generated with different parametrisations of the full-frame  and annular auto-S/N along the baseline model presented in \citep{Cantalloube20}. See Sect. 4.3.1. for the definition of the acronyms used to characterise the auto-RSM versions. The yellow circles are centred on the true position of the detected targets (TP) and the red circles give the true positions of FNs. }
\end{figure*}

We follow the same procedure as the one proposed in Section \ref{pa} to assess the performance of different parametrisations of the auto-S/N. We consider the full-frame case as well as the annular and annular full-frame optimisation mode along the bottom-up and top-down approach for the PSF-subtraction techniques selection, similarly to the auto-RSM performance assessment in Section \ref{pa}. The S/N maps combinations generated by the four parametrisations of the auto-S/N may be found in Fig. \ref{ResSNR1} and \ref{ResSNR2}. As can be seen from these graphs, the auto-S/N clearly performs better than the baseline proposed in \citep{Cantalloube20}, although the results are degraded compared to the ones obtained with the auto-RSM (see Fig. \ref{MapFFBUF}). This degraded performance was expected, considering the higher performance of RSM probability maps compared to standard S/N maps as demonstrated in \cite{Dahlqvist20}.

\begin{figure*}[h!]
\footnotesize
  \centering
  \subfloat[(a)]{\includegraphics[trim=0 20 0 6, clip,width=260pt]{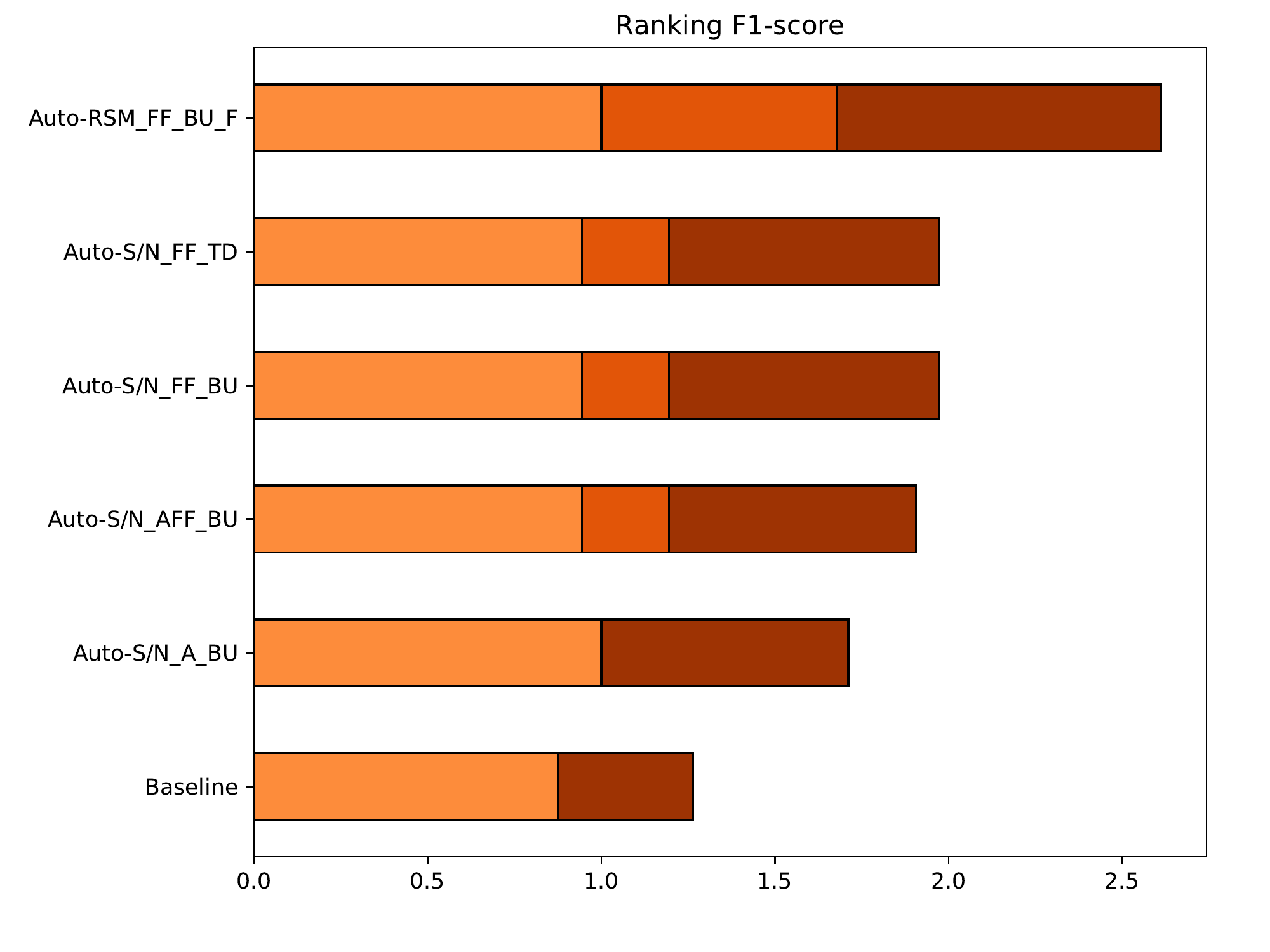}}
  \subfloat[(b)]{\includegraphics[trim=0 20 0 6, clip,width=260pt]{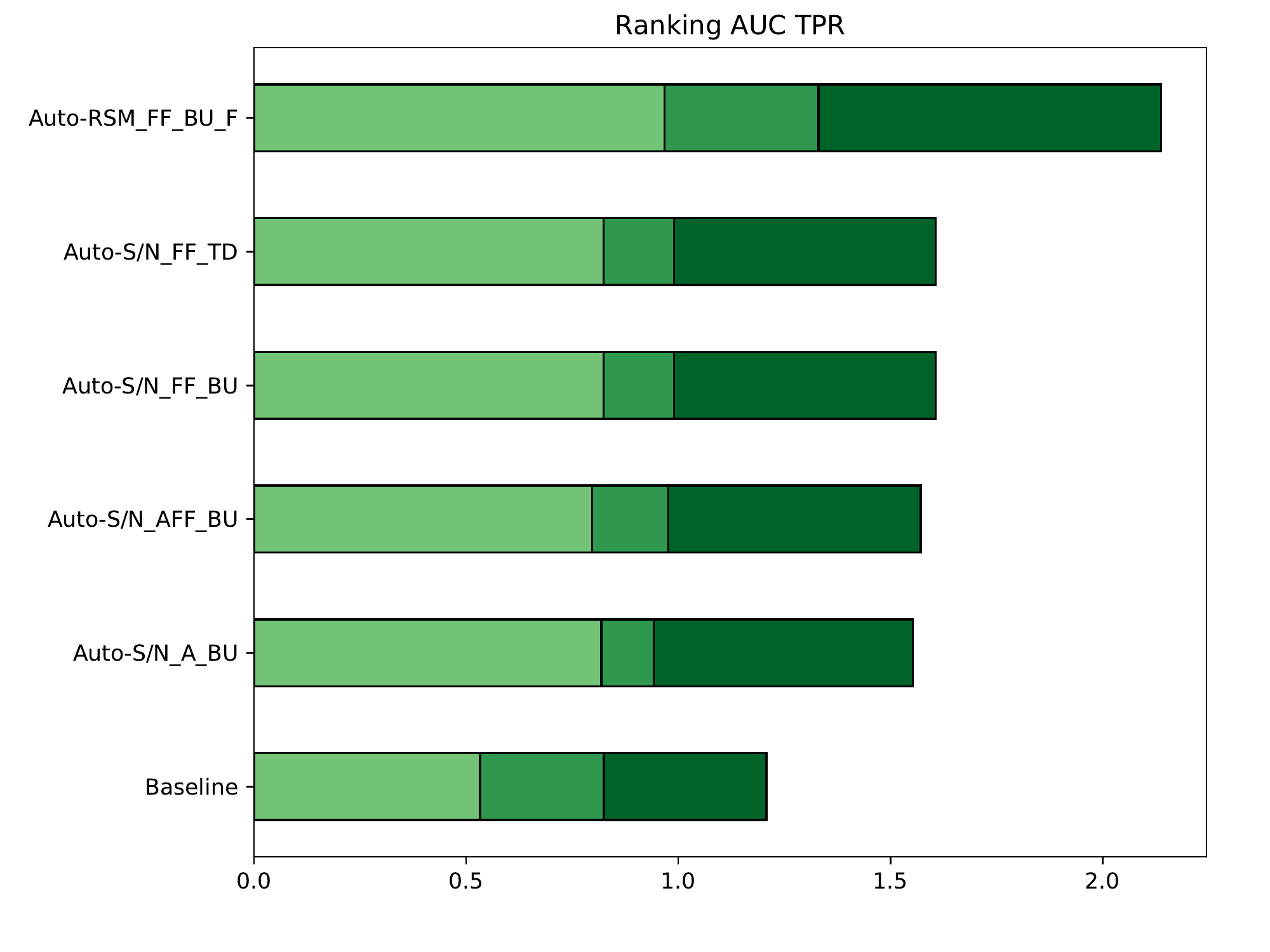}}\\
  \subfloat[(c)]{\includegraphics[trim=0 20 0 6, clip,width=260pt]{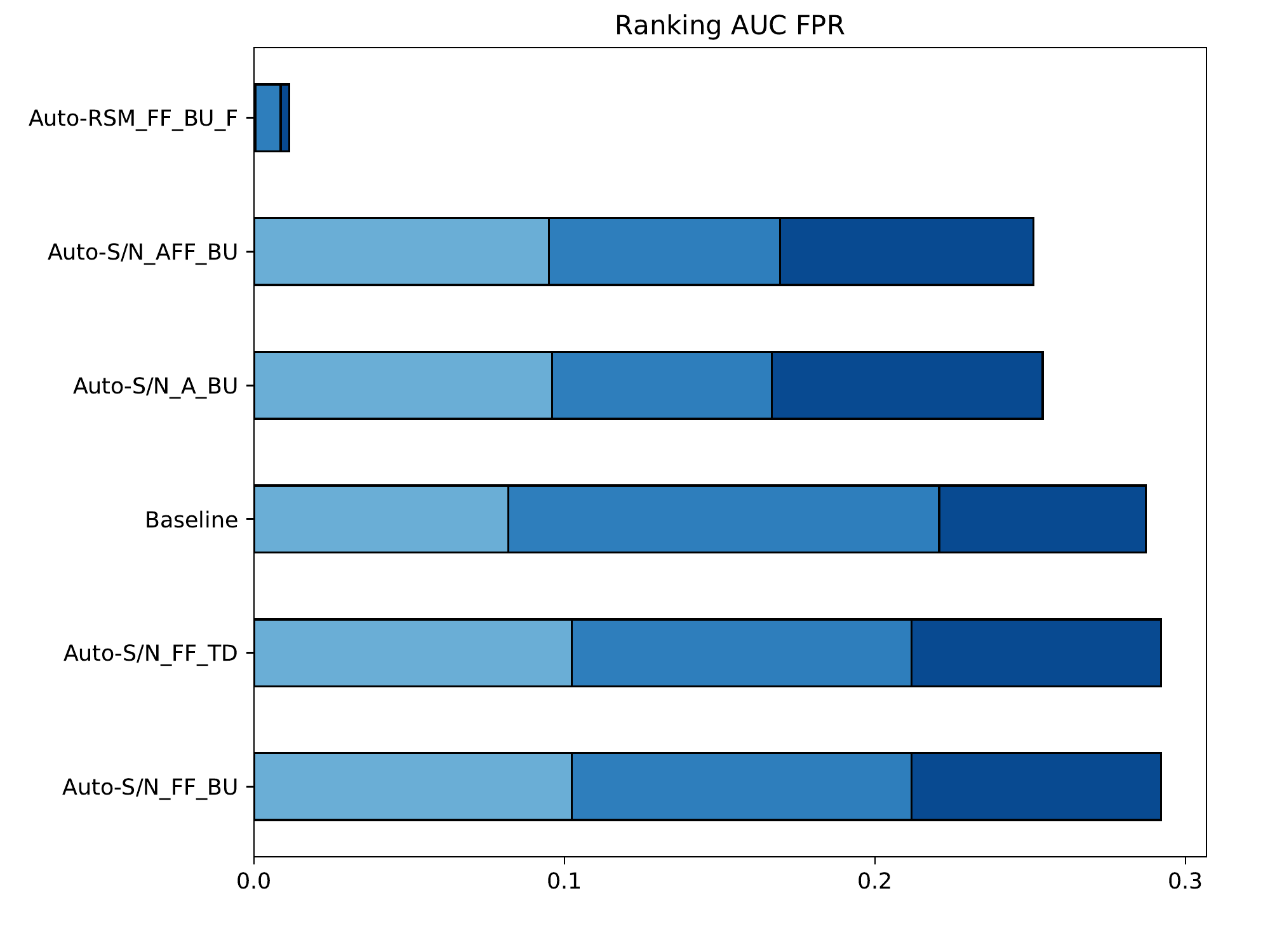}}
  \subfloat[(d)]{\includegraphics[trim=0 20 0 6, clip,width=260pt]{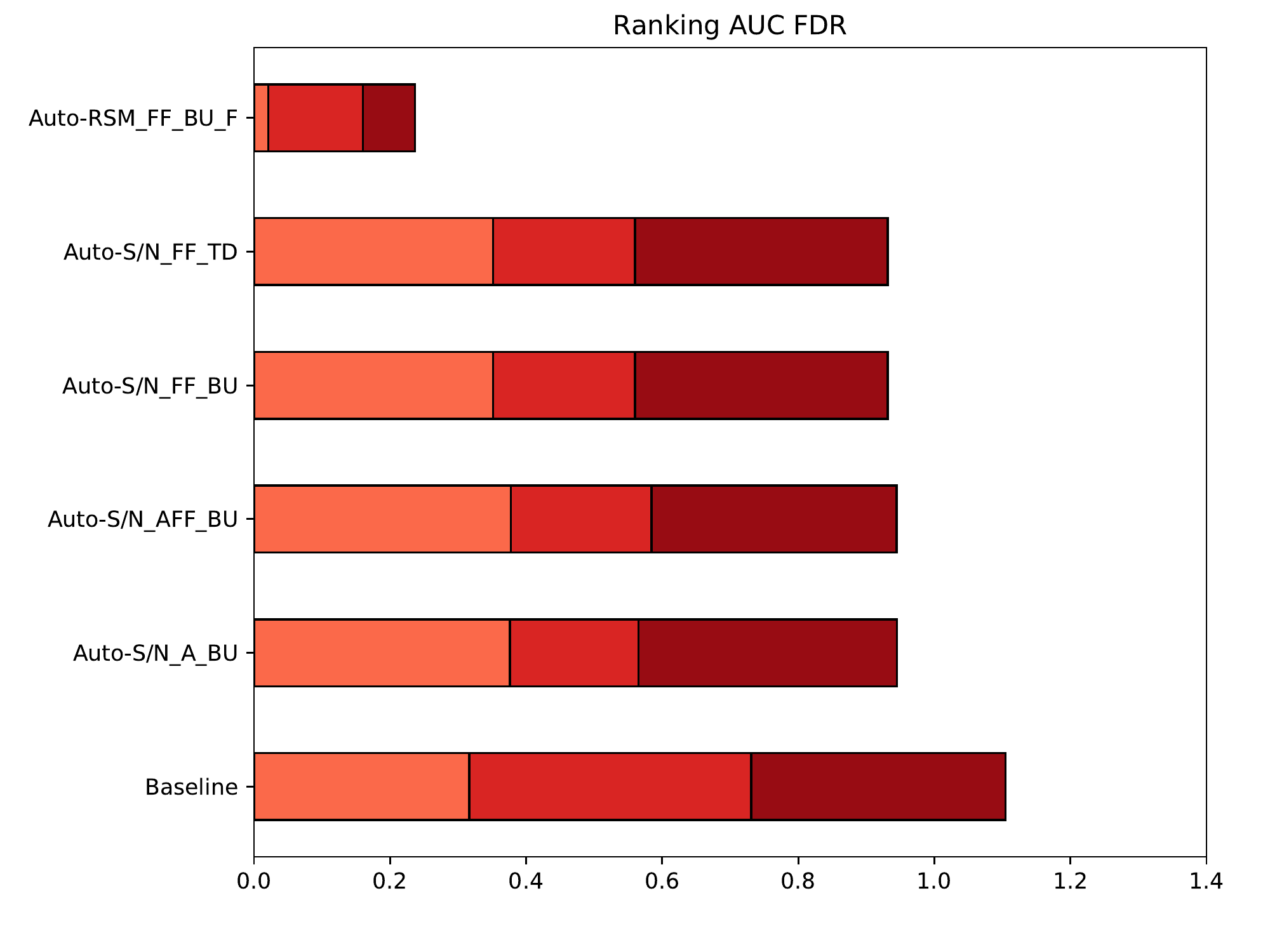}}\\
  \caption{\label{RankingSNR} Ranking of the different parametrisations of the full-frame and annular versions of the auto-S/N along the full-frame bottom-up forward auto-RSM and the baseline presented in \citep{Cantalloube20}. Figure (a) provides the ranking based on the F1 score obtained at the selected threshold. Figure (b) gives the ranking based on the AUC of the TPR. Figure (c) gives the ranking based on the AUC of the FPR, while Figure (d) provides the ranking based on the AUC of the FDR. See Sect. 4.3.1. for the definition of the acronyms used to characterise the auto-RSM versions. The light, medium, and dark colours correspond to the VLT/SPHERE-IRDIS, Keck/NIRC2, and LBT/LMIRCam data sets, respectively.}
\end{figure*}

These results are confirmed by the performance metrics shown in Figs. \ref{RankingSNR}, with both a lower TPR and a higher FPR for the auto-S/N versions compared to the full-frame-bottom-up forward auto-RSM. As in the case of the auto-RSM, the full-frame auto-RSM versions perform better than the annular and hybrid annular full-frame versions.

Considering the shorter computation time compared to auto-RSM and the better performance compared to standard S/N based PSF-subtraction techniques, the auto-S/N can be considered as an interesting alternative to the auto-RSM for large surveys. The auto-S/N may also represent a good complement to the auto-RSM\footnote{ The computation time is further reduced when the auto-RSM has already been applied to a data set.}, as it may lead to the identification of planetary signals missed by the auto-RSM as illustrated by the LMIRCam-3 data set (to be compared with Fig. \ref{MapFFBUF}).

\newpage

\section{Auto-RSM pseudo-code}
\label{pseudocode}

This Appendix presents first the pseudo-codes of the greedy algorithms used to select the optimal set of likelihood cubes generating the final RSM detection map (see Table~\ref{bu} and \ref{td} for respectively the bottom-up and top-down approaches). Tables~\ref{ffmode} and \ref{amode} then provide  the pseudo-codes for the auto-RSM optimisation algorithm in the full-frame mode and annular mode.

\begin{center}
       \begin{table}[h!]
        \caption{Pseudo-code for the bottom-up greedy selection algorithm. The PI symbol represents the RSM performance metrics.}
        \label{bu}
                        \footnotesize
                        \begin{tabular}{l}
                        \hline
1:\hspace*{0.6cm}$\bm{Y}^a=\left\lbrace Y^a_{c,m}, \forall c \in [0,N_{sequence}], m \in [0,N_{technique}] \right\rbrace $\\
2:\hspace*{0.6cm}$\bm{Z}^a = \emptyset$\\
3:\hspace*{0.6cm}$PI^a_p=0$\\
4:\hspace*{0.6cm}\textbf{While} $\bm{Y}^{a} \neq \emptyset$ \textbf{do}\\
5:\hspace*{0.9cm}\textbf{For} $h = 1$ to $length(\bm{Y}^{a})$ \textbf{do}\\
6:\hspace*{1.2cm}$PI^a_h=PI([\bm{Z}^a,\bm{Y}^{a}_{h}]) $\\
7:\hspace*{0.9cm}\textbf{End for}\\
8:\hspace*{0.9cm}\textbf{If} $max(PI^a_{\bm{h}})>PI^a_p$ \\
9:\hspace*{1.2cm}$Y^a_{c*,m*} =Y^a_{h*} = argmax(PI^a_{\bm{h}})$ \\
10:\hspace*{1cm}$\bm{Y}^{a}= \bm{Y}^{a} \setminus \left\lbrace Y^a_{h*} ,Y^a_{h}  \forall (PI^a_h-PI^a_p)<0 \right\rbrace $ \\
11:\hspace*{1cm}$\bm{Z}^a= \bm{Z}^a \cup Y^a_{h*} $ \\
12:\hspace*{1cm}$PI^a_p=max(PI^a_{\bm{h}})$\\
13:\hspace*{0.7cm}\textbf{Else}\\
14:\hspace*{1cm}$\bm{Y}^{a}= \emptyset$\\
15:\hspace*{0.7cm}\textbf{End If}\\
16:\hspace*{0.5cm}\textbf{End While}\\
\hline
                        \end{tabular}
                          \end{table}
                                \end{center}

\begin{center}
       \begin{table}[h!]
        \caption{Pseudo-code for the top-down greedy selection algorithm. The PI symbol represents the RSM performance metrics.}
        \label{td}
                        \footnotesize
                        \begin{tabular}{l}
                        \hline
1:\hspace*{0.6cm}$\bm{Z}^a =\left\lbrace  Y^a_{c,m}, \forall c \in [0,N_{sequence}], m \in [0,N_{technique}] \right\rbrace$\\
2:\hspace*{0.6cm}$PI^a_c=PI(\bm{Z}^a)$\\
3:\hspace*{0.6cm}$PI^a_p=0$\\
4:\hspace*{0.6cm}\textbf{While} $PI^a_c>PI^a_p$ \textbf{do}\\
5:\hspace*{0.9cm}$PI^a_p=PI^a_c$\\
6:\hspace*{0.9cm}\textbf{For} $h = 1$ to $length(\bm{Z}^a)$ \textbf{do}\\
7:\hspace*{1.2cm}$PI^a_h=PI(\bm{Z}^a \setminus Y^a_{h}) $\\
8:\hspace*{0.9cm}\textbf{End for}\\
9:\hspace*{0.9cm}\textbf{If} $max(PI^a_h-PI^a_p)>0$\\
10:\hspace*{1cm}$Y^a_{c*,m*} =Y^a_{h*} = argmax(PI^a_h-PI^a_p)$ \\
11:\hspace*{1cm}$\bm{Z}^a= \bm{Z}^a \setminus Y^a_{h*} $ \\
12:\hspace*{1cm}$PI^a_c=max(PI^a_h-PI^a_p)$\\
13:\hspace*{0.7cm}\textbf{End If}\\
14:\hspace*{0.5cm}\textbf{End While}\\
\hline
                        \end{tabular}
                          \end{table}
                                \end{center}

\begin{center}
       \begin{table}[!htbp]
       
        \caption{Pseudo-code for the auto-RSM algorithm in full-frame mode.The centre of the selected set of annuli $Range_{sel}$ is computed based on the rule provided in eq.\ref{ffcase}, starting at 1.5 FWHM and ending at $a_{max}$.}
        \label{ffmode}
                        \footnotesize
                        \begin{tabular}{l}
                        \hline
1:\hspace*{0.2cm}Flipping parallactic angle sign: $PA=-PA$\\
2:\hspace*{0.2cm}\textbf{For} $c = 1$ to $N_{sequence}$ \textbf{do}\\
3:\hspace*{0.5cm}\textbf{For} $m = 1$ to $N_{technique}$ \textbf{do}\\
4:\hspace*{0.8cm}\textbf{PSF-subtraction technique parameters optimisation}\\
5:\hspace*{1.2cm}\textbf{For} $a$ in $Range_{sel}$ \textbf{do}\\
6:\hspace*{1.5cm}Contrast computation via fake companions injection\\
7:\hspace*{1.5cm}Contrast normalisation\\
8:\hspace*{1.2cm}\textbf{End For}\\
9:\hspace*{1.2cm}Optimal parameters and contrast selection based on \\
 \hspace*{1.5cm}summed normalised contrast: $[C_{a,m,c},P^{PSF}_{m,c}]$\\
10:\hspace*{0.6cm}\textbf{Optimisation of RSM algorithm parameters}\\
11:\hspace*{1cm}\textbf{For} $a$ in $Range_{sel}$ \textbf{do}\\
12:\hspace*{1.3cm}Median position computation: $M_{a,m,c}$\\
13:\hspace*{1.3cm}Performance metric computation via fake companion\\
 \hspace*{1.7cm} injection using $[M_{a,m,c},C_{a,m,c},P^{PSF}_{m,c}]$\\
14:\hspace*{1cm}\textbf{End For}\\
15:\hspace*{1cm}Optimal parameters selection: $P^{RSM}_{m,c}$\\
16:\hspace*{0.5cm}\textbf{End For}\\
17:\hspace*{0.2cm}\textbf{End For}\\
18:\hspace*{0.2cm}\textbf{Optimal combination} $\bm{Z}$\textbf{ selection} via bottom-up or\\
  \hspace*{0.8cm}top-down approach using $[P^{RSM}_{m,c},M_{a,m,c},C_{a,m,c},P^{PSF}_{m,c}]$\\
19:\hspace*{0.2cm}\textbf{For} $a = FWHM$ to $a_{max}$ \textbf{do}\\
20:\hspace*{0.5cm}\textbf{Threshold} $T_{a}$ \textbf{computation} using $[P^{RSM}_{m,c},P^{PSF}_{m,c},\bm{Z}]$\\
21:\hspace*{0.2cm}\textbf{End For}\\
22:\hspace*{0.2cm}Threshold smoothing via polynomial fit: $T^*_{a}$\\
23:\hspace*{0.2cm}Flipping back parallactic angle sign: $PA=-PA$\\
24:\hspace*{0.2cm}\textbf{Final RSM map computation} using $[P^{RSM}_{m,c},P^{PSF}_{m,c},\bm{Z},T^*_{a}]$\\
25:\hspace*{0.2cm}Threshold subtraction from final RSM map\\
\hline
                        \end{tabular}
                          \end{table}
                                \end{center}
                                                                
\begin{center}
       \begin{table}[!htbp]
       
        \caption{Pseudo-code for the auto-RSM algorithm in annular mode.The centre of the selected set of annuli $Range_{sel}$ starts at 1,5 FWHM and end at $a_{max}$ with the centre of every selected annulus separated by one FWHM.}
         \label{amode}
                        \footnotesize
                        \begin{tabular}{l}
                        \hline
1:\hspace*{0.2cm}Flipping parallactic angle sign: $PA=-PA$\\
2:\hspace*{0.2cm}\textbf{For} $c = 1$ to $N_{sequence}$ \textbf{do}\\
3:\hspace*{0.5cm}\textbf{For} $m = 1$ to $N_{technique}$ \textbf{do}\\
4:\hspace*{0.8cm}\textbf{For} $a$ in $Range_{sel}$ \textbf{do}\\
5:\hspace*{1.2cm}\textbf{PSF-subtraction technique parameters optimisation}\\
6:\hspace*{1.5cm}Contrast computation via fake companions injection\\
7:\hspace*{1.5cm}Optimal parameters and contrast $[C_{a,m,c},P^{PSF}_{a,m,c}]$\\
8:\hspace*{0.8cm}\textbf{End For}\\
9:\hspace*{0.8cm}Outliers suppression in $P^{PSF}_{a,m,c}$ via Hampel Filter \\
10:\hspace*{0.8cm}Parameters smoothing via moving average: $P^{PSF,*}_{a,m,c}$\\
11:\hspace*{0.6cm}\textbf{For} $a$ in $Range_{sel}$ \textbf{do}\\
12:\hspace*{0.8cm}\textbf{Optimisation of RSM algorithm parameters}\\
13:\hspace*{1.1cm}Median position computation $M_{a,m,c}$\\
14:\hspace*{1.1cm}Performance metric computation via fake companion\\
 \hspace*{1.6cm} injection using $[M_{a,m,c},C_{a,m,c},P^{PSF,*}_{a,m,c}]$\\
15:\hspace*{1.1cm}Optimal parameters selection: $P^{RSM}_{a,m,c}$\\
16:\hspace*{0.6cm}\textbf{End For}\\
17:\hspace*{0.6cm}Outliers suppression in $P^{RSM}_{a,m,c}$ via Hampel Filter \\
18:\hspace*{0.6cm}Interpolation of optimal parameters via RBF: $P^{RSM,*}_{a,m,c}$\\
19:\hspace*{0.5cm}\textbf{End For}\\
20:\hspace*{0.2cm}\textbf{End For}\\
21:\hspace*{0.2cm}\textbf{Optimal combination} $\bm{Z}^a$\textbf{ selection} via bottom-up or\\
  \hspace*{0.5cm}top-down approach using $[P^{RSM,*}_{a,m,c},M_{a,m,c},C_{a,m,c},P^{PSF,*}_{a,m,c}]$\\
22:\hspace*{0.2cm}\textbf{For} $a = FWHM$ to $a_{max}$ \textbf{do}\\
23:\hspace*{0.5cm}\textbf{Threshold} $T_{a}$ \textbf{computation} using $[P^{RSM,*}_{a,m,c},P^{PSF,*}_{a,m,c},\bm{Z}^a]$\\
24:\hspace*{0.2cm}\textbf{End For}\\
25:\hspace*{0.2cm}Flipping back parallactic angle sign: $PA=-PA$\\
26:\hspace*{0.2cm}\textbf{Final RSM map computation} using $[P^{RSM}_{a,m,c},P^{PSF}_{a,m,c},\bm{Z}^a,T_{a}]$\\
27:\hspace*{0.2cm}Threshold subtraction from final RSM map\\
\hline
                        \end{tabular}
                          \end{table}
                                \end{center}
                                
\newpage

\section{Description of the ADI sequences}
\label{descadi}

This Appendix provides a description of the data sets of the EIDC ADI subchallenge used in the performance assessment of the different modes of the auto-RSM optimisation algorithm (Table~\ref{ADIdesc}).

\begin{table*}[!htbp]
                        \caption{Characteristics of the nine ADI sequences included in the EIDC ADI subchallenge. The original number of frames for the LMIRCam ADI sequences was reduced to limit the computation time, relying on a moving average applied along the time axis on the de-rotated ADI cubes. The step and window sizes have been set to 20 frames for LMIRCam 1 and 3, and 15 frames for LMIRCam 2. }
                        \label{ADIdesc}
\centering

                        \begin{tabular}{lcccc}
                        
                        \hline
Instruments/ID &Number of frames  &Frame size & Plate-scale (mas/pixel) & FOV rotation  \\                                
 \hline
SPHERE 1 &$252$ & $160\times160$ & $12.255$ & $40.3^{\circ}$\\
SPHERE 2 &$80$ & $160\times160$ & $12.255$ & $31.5^{\circ}$\\
SPHERE 3 &$228$& $160\times160$ & $12.255$ & $80.5^{\circ}$\\
NIRC2 1 &$29$ & $321\times 321$ & $20$ & $53.0^{\circ}$\\
NIRC2 2 &$40$ & $321\times 321$ & $20$ & $37.3^{\circ}$\\
NIRC2 3 &$50$ & $321\times 321$ & $20.2$ & $166.9^{\circ}$\\
LMIRCam 1 &$241$ & $200\times200$ & $10.7$ & $153.4^{\circ}$\\
LMIRCam 2 &$214$ & $200\times200$ & $10.7$ & $60.6^{\circ}$\\
LMIRCam 3 &$231$ & $200\times200$ & $10.7$ & $91.0^{\circ}$\\
\hline
                        \end{tabular}
                                \end{table*}
                                
\section{Definition of the parameter ranges}    
\label{paramrange}

This Appendix provides the boundaries of the parameter space for the different data sets of the EIDC ADI subchallenge used for the performance assessment of the auto-RSM optimisation algorithm (Table~\ref{parameterspace}).                     
                                
\begin{table*}[!htbp]
                        \caption{A range of values is provided for each parameter and each ADI sequence in order to define the size of the parameters space considered during the PSF-subtraction techniques optimisation.}
                        \label{parameterspace}
\centering
                        \scriptsize
                        \begin{tabular}{llllllllll}

                        \hline
Parameters/ID &NIRC2-1&NIRC2-2&NIRC2-3&SPHERE-1&SPHERE-2&SPHERE-3&LMIRCam-1&LMIRCam2&LMIRCam-3  \\                             
 \hline         
APCA components& $[5,25]$       &        $[5,25]$       &        $[5,25]$       &        $[15,45]$       &       $[15,45]$       &       $[15,45]$       &       $[15,45]$       &       $[15,45]$       &       $[15,45]$       \\
APCA segments&$[1,4]$   &       $[1,4]$ &       $[1,4]$ &       $[1,4]$ &       $[1,4]$ &       $[1,4]$ &       $[1,4]$ &       $[1,4]$ &       $[1,4]$ \\
APCA FOV rotation&$[0.25,1]$    &       $[0.25,1]$              &       $[0.25,1]$              &       $[0.25,1]$              &       $[0.25,1]$              &$[0.25,1]$             &       $[0.25,1]$              &       $[0.25,1]$              &       $[0.25,1]$              \\
NMF components&$[2,15]$         &       $[2,15]$                &       $[2,15]$                &       $[2,20]$        &       $[2,20]$                &       $[2,20]$                &       $[2,20]$                &       $[2,20]$                &       $[2,20]$                \\
LLSG rank& $[1,5]$              &       $[1,5]$ &       $[1,5]$ &       $[1,10]$        &       $[1,10]$        &       $[1,10]$        &       $[1,10]$        &       $[1,10]$        &       $[1,10]$        \\
LLSG segments & $[1,4]$ &       $[1,4]$ &       $[1,4]$ &       $[1,4]$ &       $[1,4]$ &       $[1,4]$ &       $[1,4]$ &       $[1,4]$ &       $[1,4]$ \\
LOCI tolerance & $[10^{-3},10^{-2}]$    &       $[10^{-3},10^{-2}]$     &       $[10^{-3},10^{-2}]$&    $[10^{-3},10^{-2}]$     &       $[10^{-3},10^{-2}]$&    $[10^{-3},10^{-2}]$     &       $[10^{-3},10^{-2}]$&    $[10^{-3},10^{-2}]$&    $[10^{-3},10^{-2}]$     \\
LOCI FOV rotation &$[0.25,1]$   &       $[0.25,1]$              &       $[0.25,1]$              &       $[0.25,1]$              &       $[0.25,1]$              &$[0.25,1]$             &       $[0.25,1]$              &       $[0.25,1]$              &       $[0.25,1]$              \\
\hline
                        \end{tabular}
                                \end{table*}
                                
\section{Detection maps for auto-RSM parametrisations}  
\label{paramperf}

\begin{figure*}[!htbp]
\footnotesize
  \centering
  
  \subfloat[SPHERE-1 FF BU F]{\includegraphics[width=115pt]{SPHERE_1_FF_BU_F.pdf}}
    \subfloat[SPHERE-3 FF BU F]{\includegraphics[width=115pt]{SPHERE_3_FF_BU_F.pdf}}
      \subfloat[NIRC2-1 FF BU F]{\includegraphics[width=115pt]{NIRC2_1_FF_BU_F.pdf}}
        \subfloat[NIRC2-2 FF BU F]{\includegraphics[width=115pt]{NIRC2_2_FF_BU_F.pdf}}\\
  \subfloat[SPHERE-1 FF TD F]{\includegraphics[width=115pt]{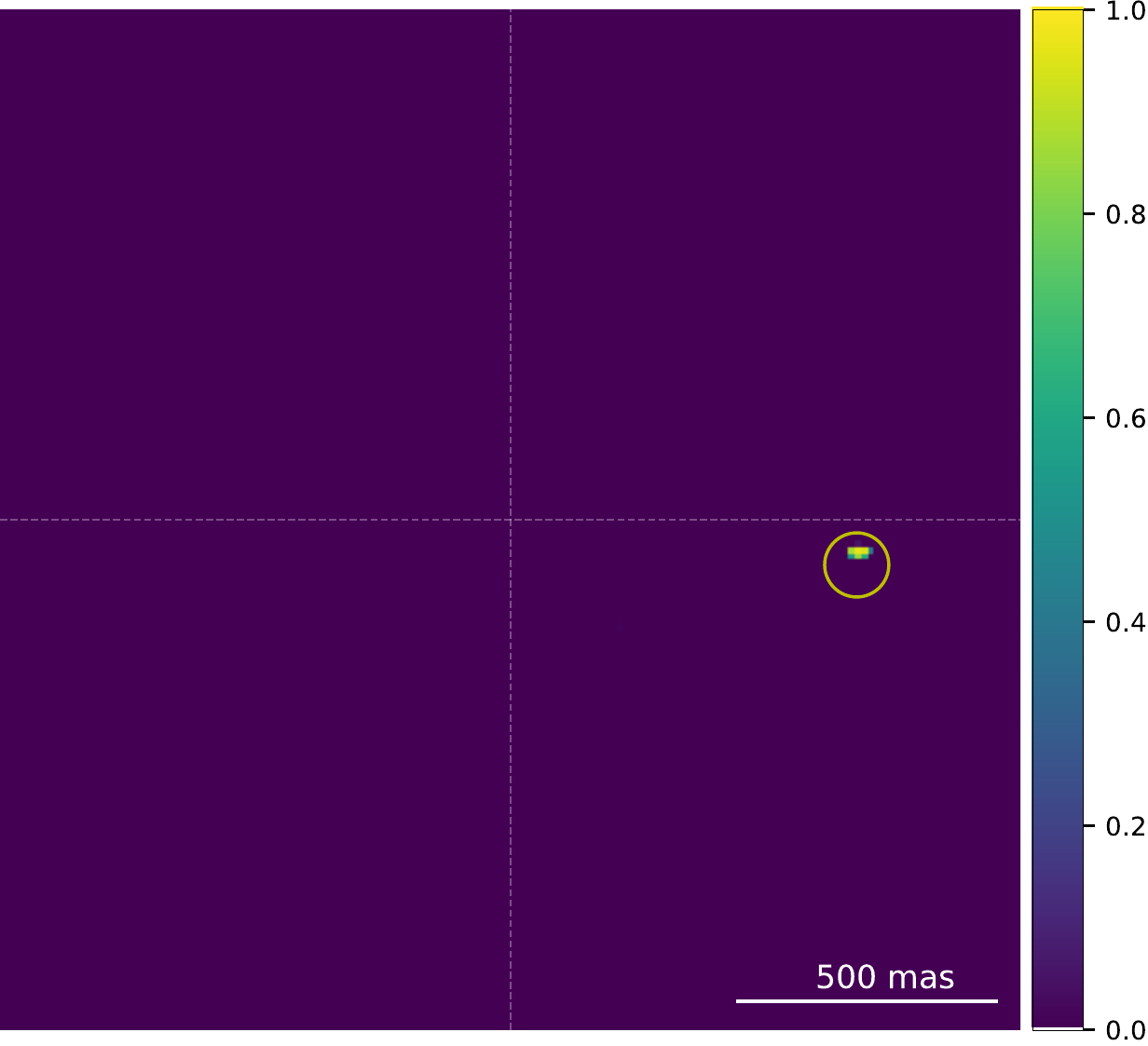}}
   \subfloat[SPHERE-3 FF TD F]{\includegraphics[width=115pt]{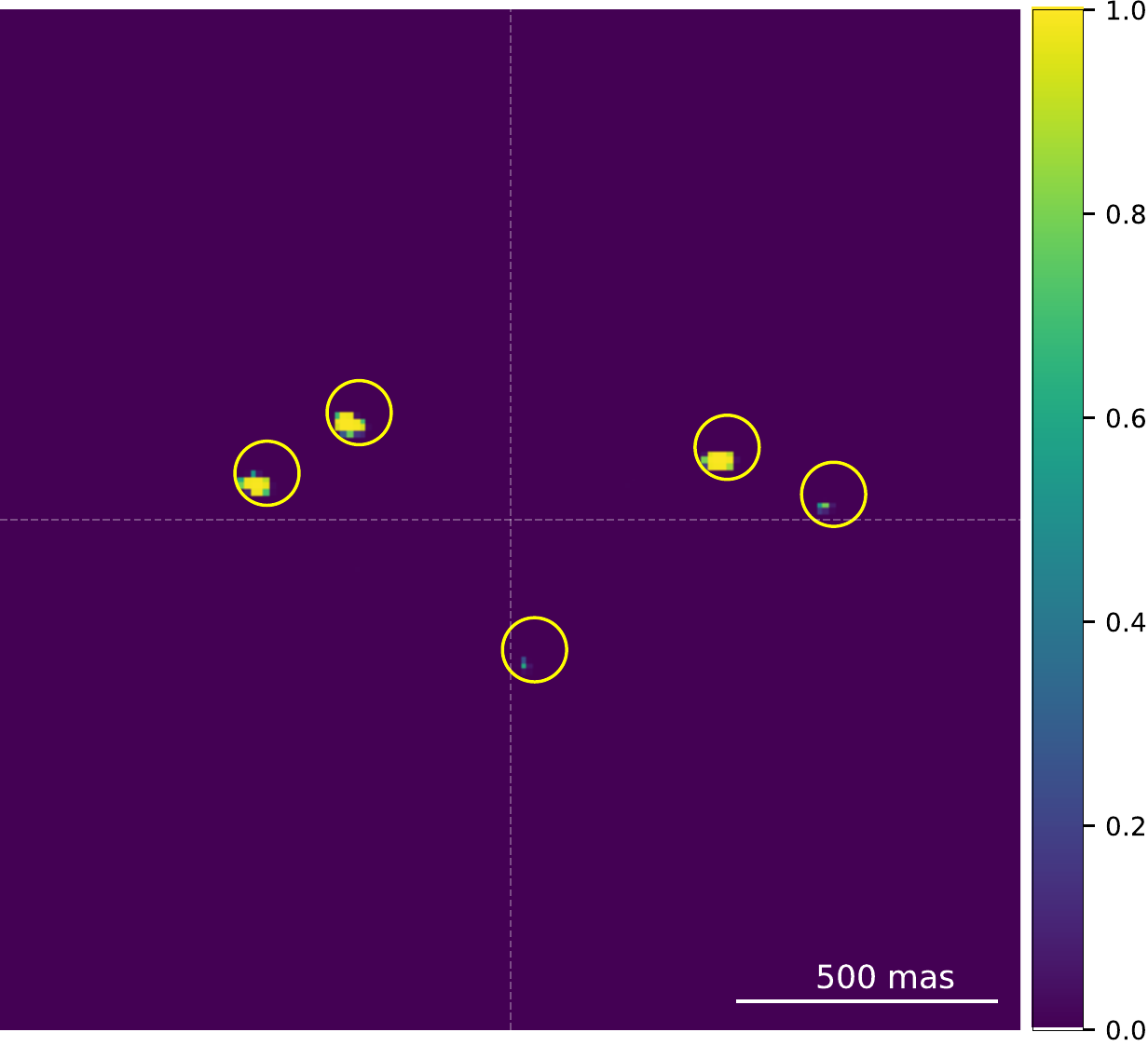}}
    \subfloat[NIRC2-1 FF TD F]{\includegraphics[width=115pt]{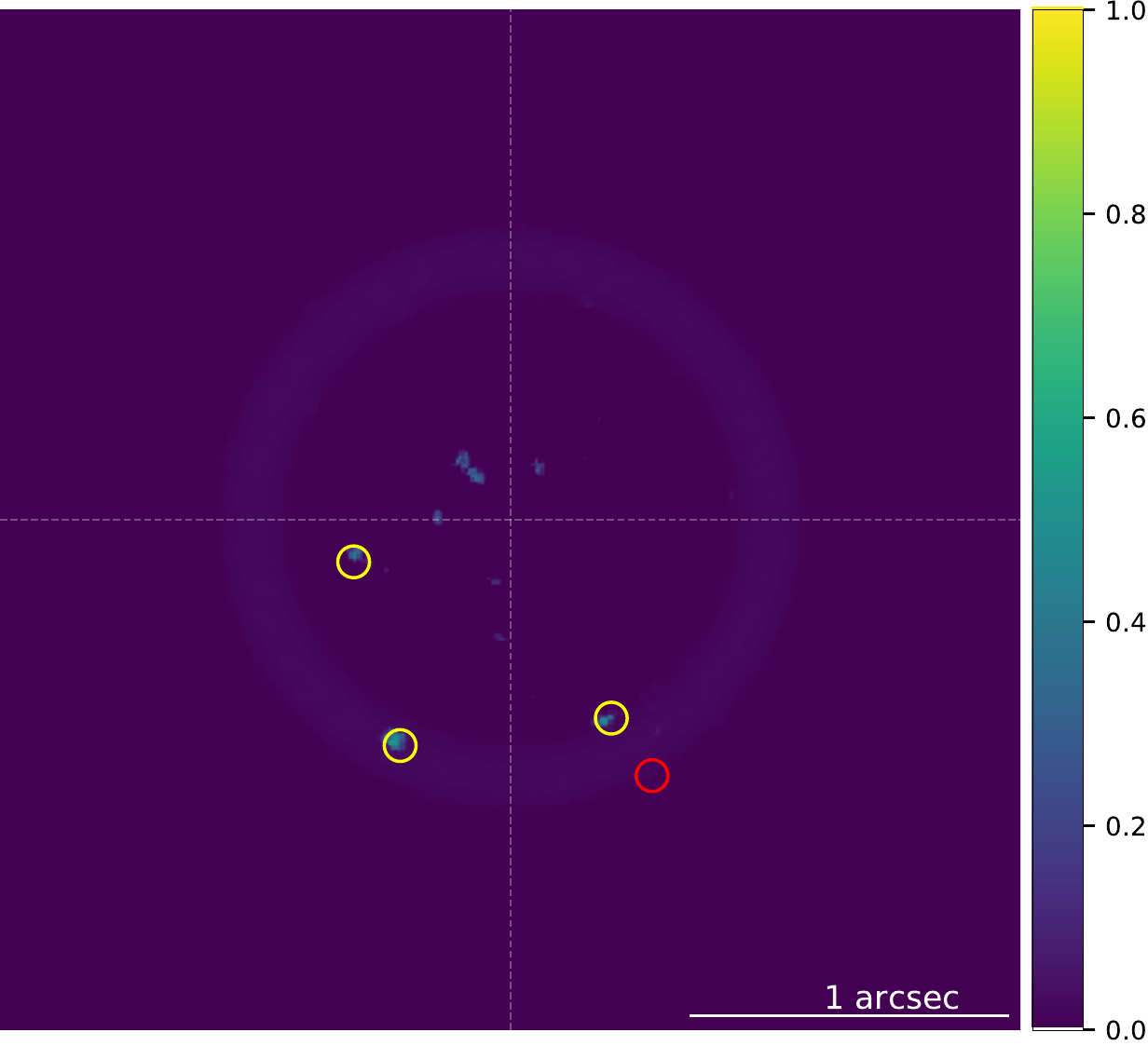}}
     \subfloat[NIRC2-2 FF TD F]{\includegraphics[width=115pt]{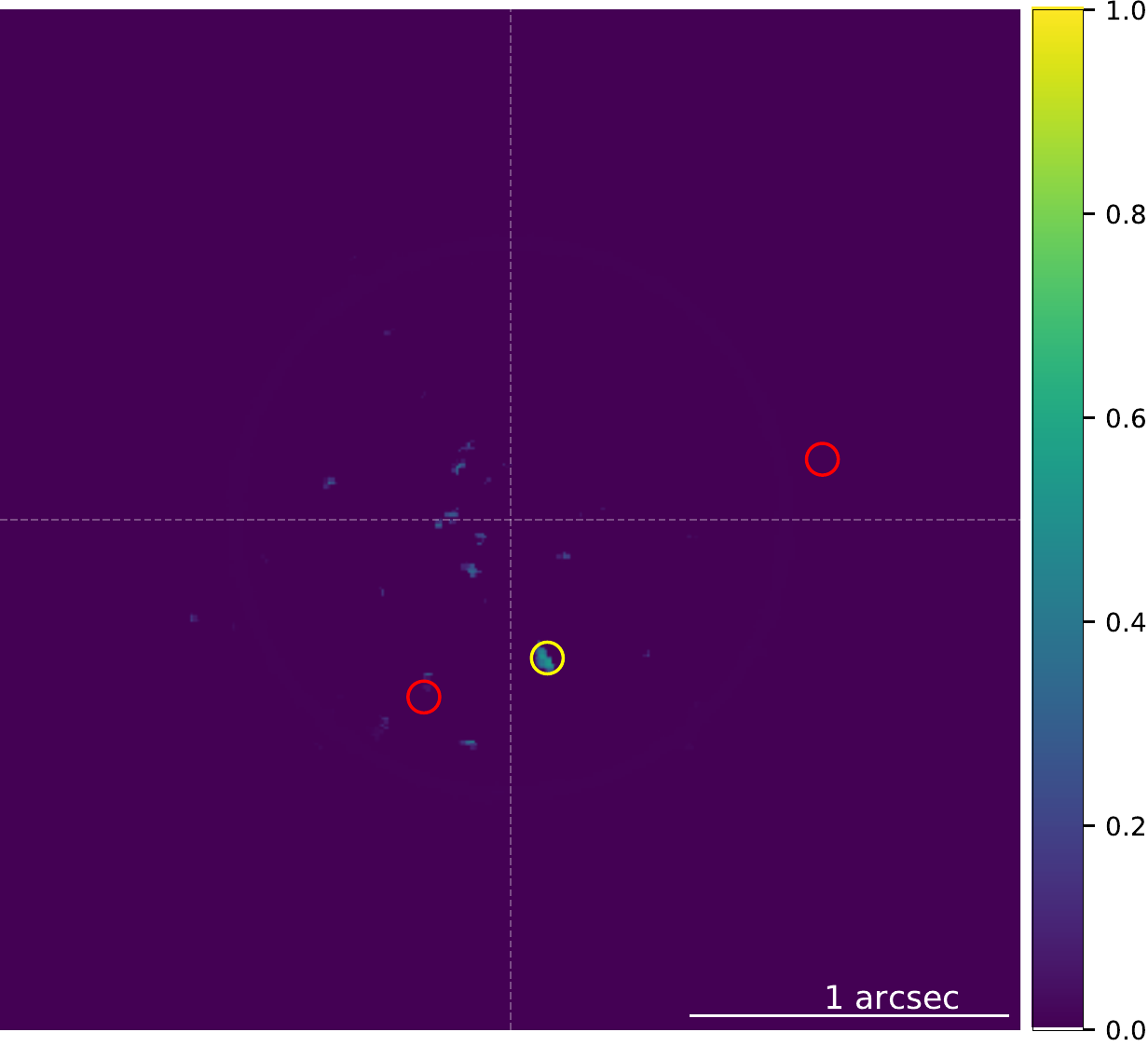}}\\
      \subfloat[SPHERE-1 FF BU FB]{\includegraphics[width=115pt]{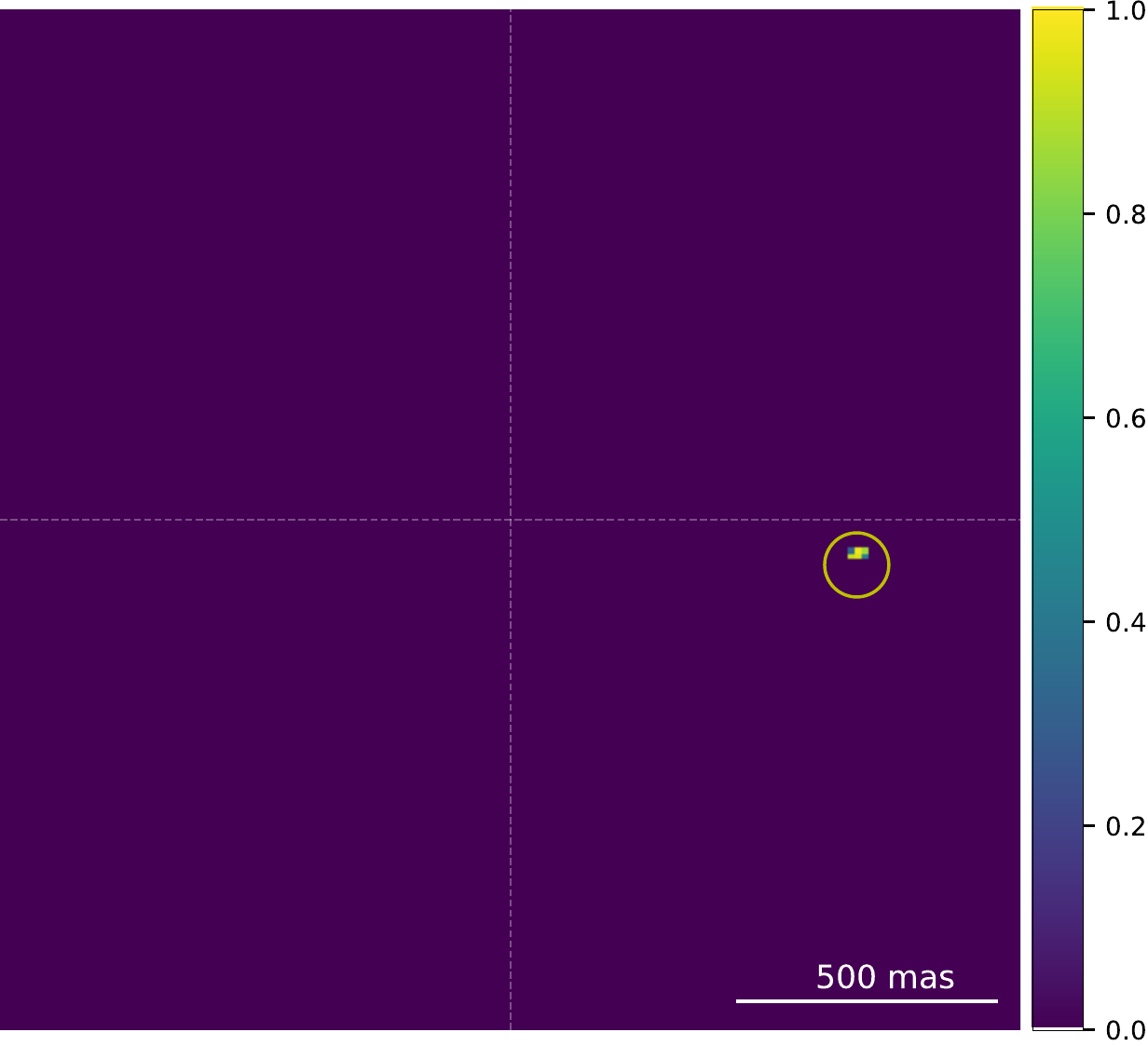}}
    \subfloat[SPHERE-3 FF BU FB]{\includegraphics[width=115pt]{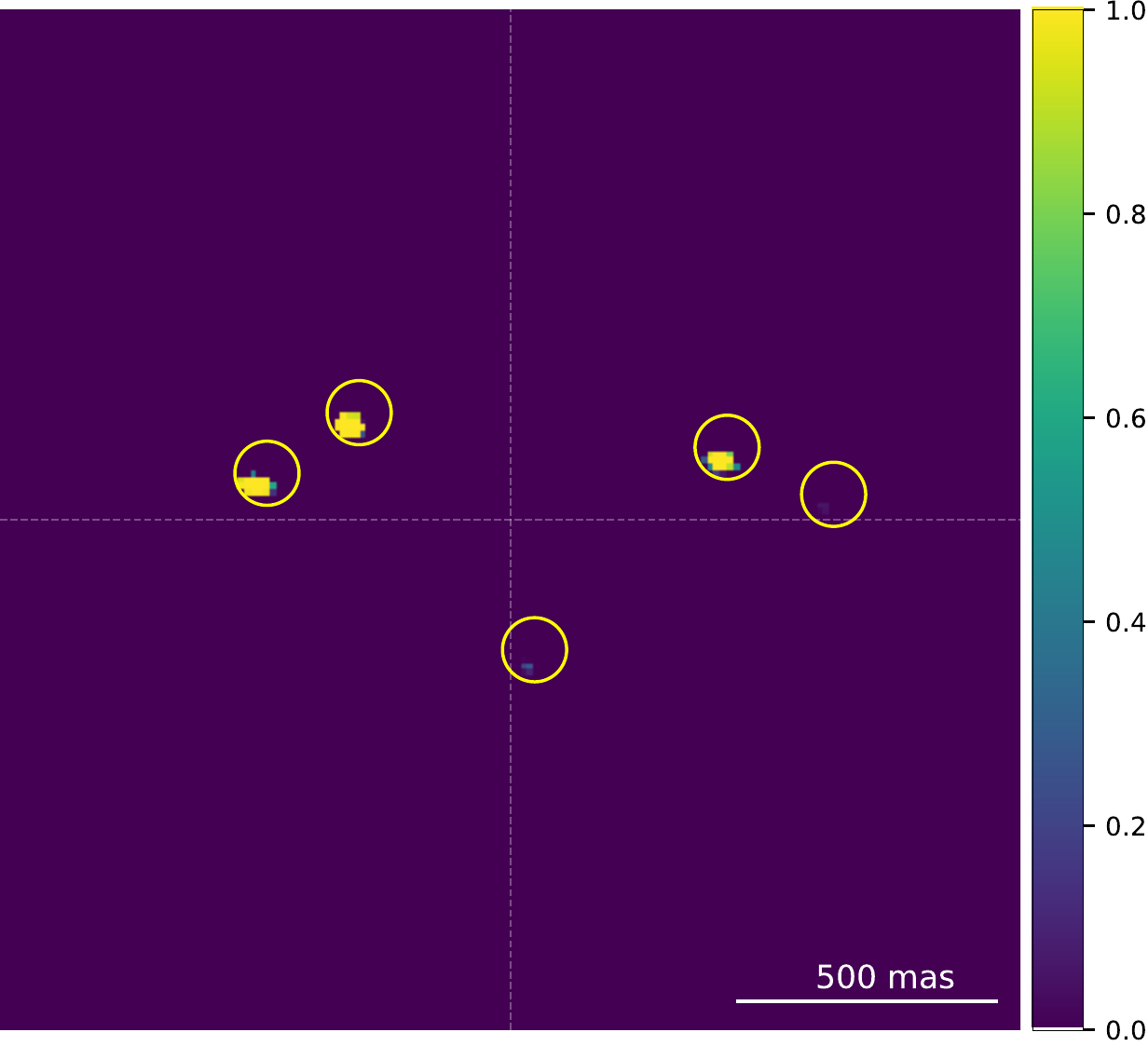}}
      \subfloat[NIRC2-1 FF BU FB]{\includegraphics[width=115pt]{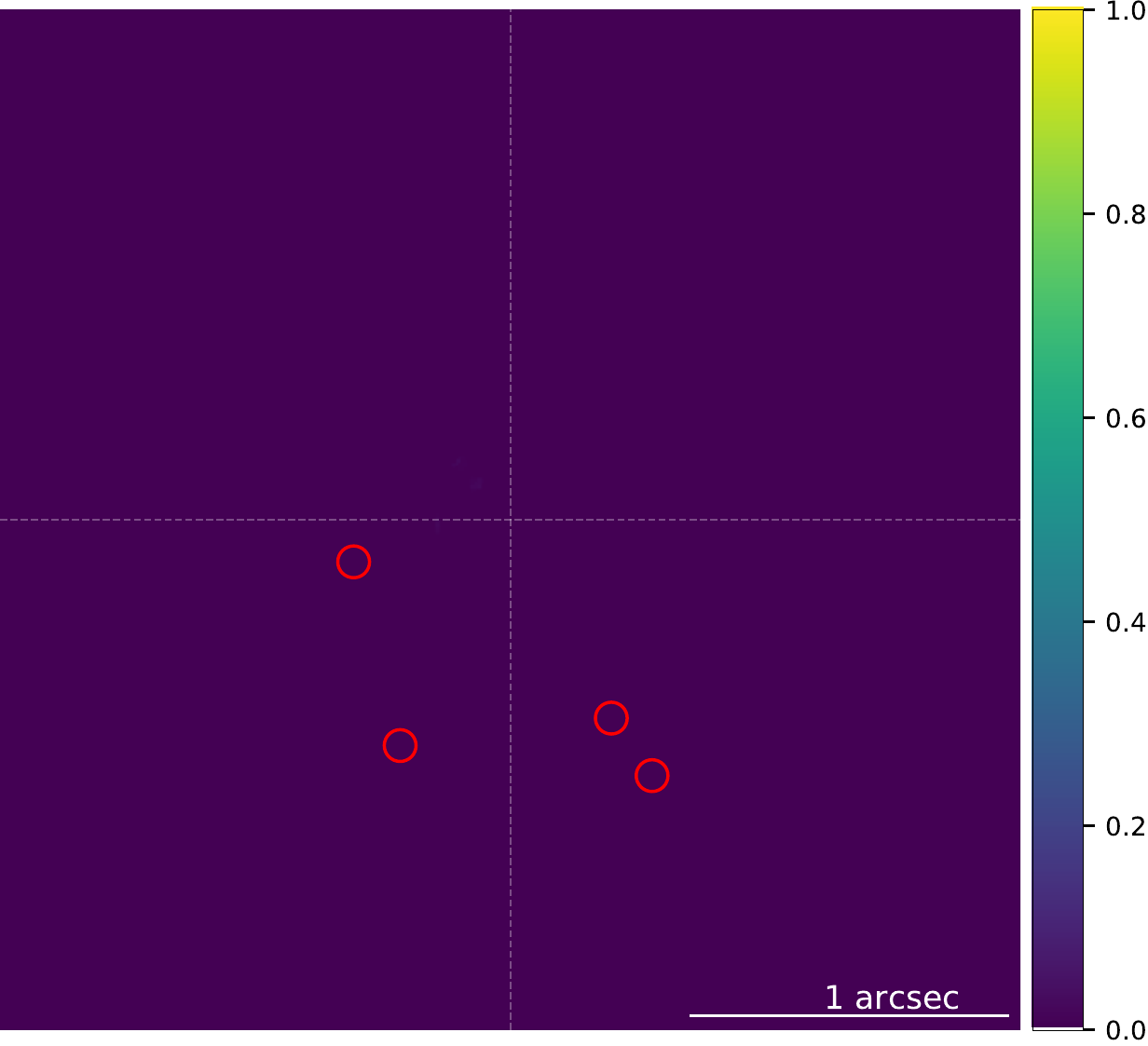}}
        \subfloat[NIRC2-2 FF BU FB]{\includegraphics[width=115pt]{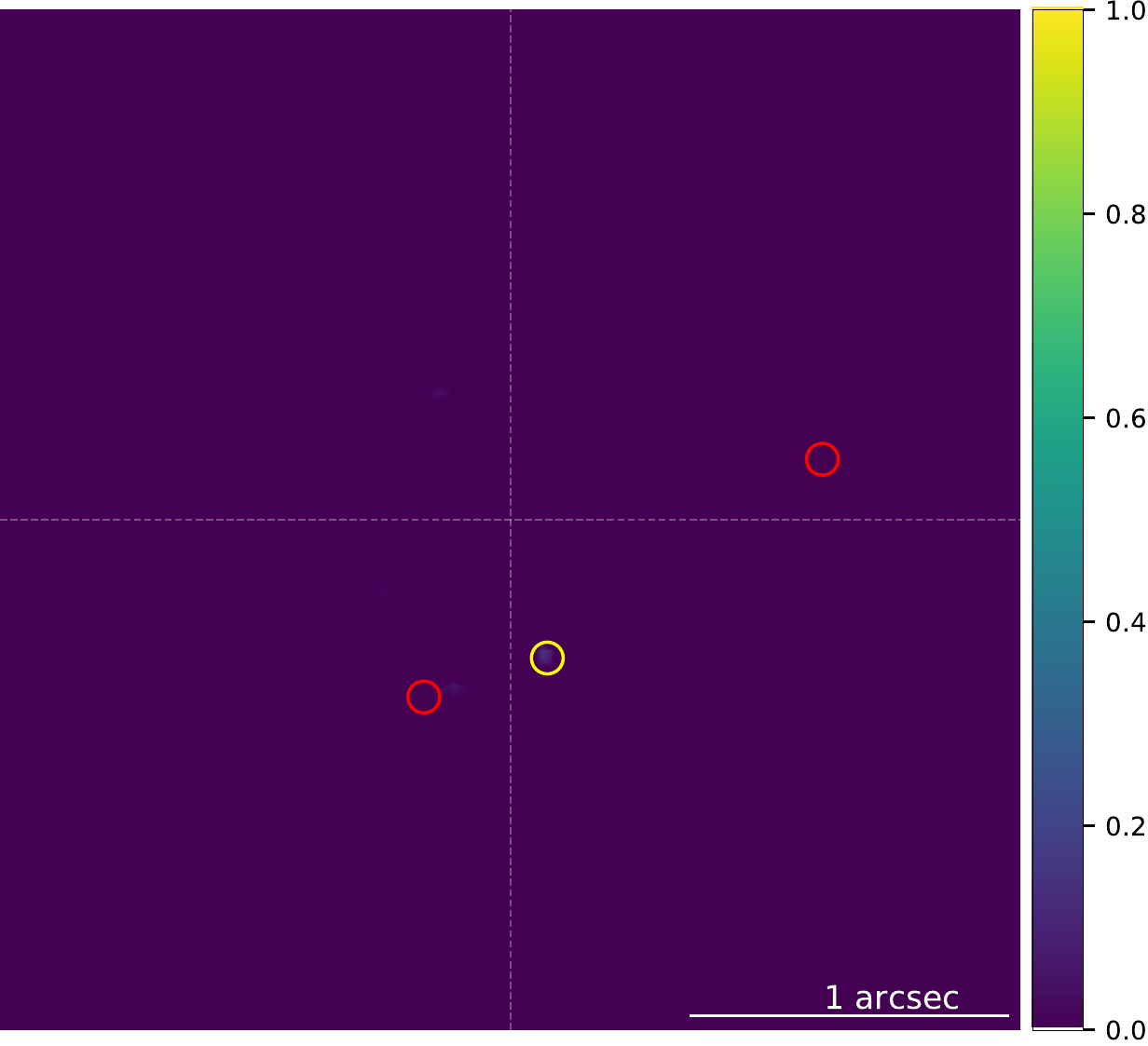}}\\
    \subfloat[SPHERE-1 A BU F]{\includegraphics[width=115pt]{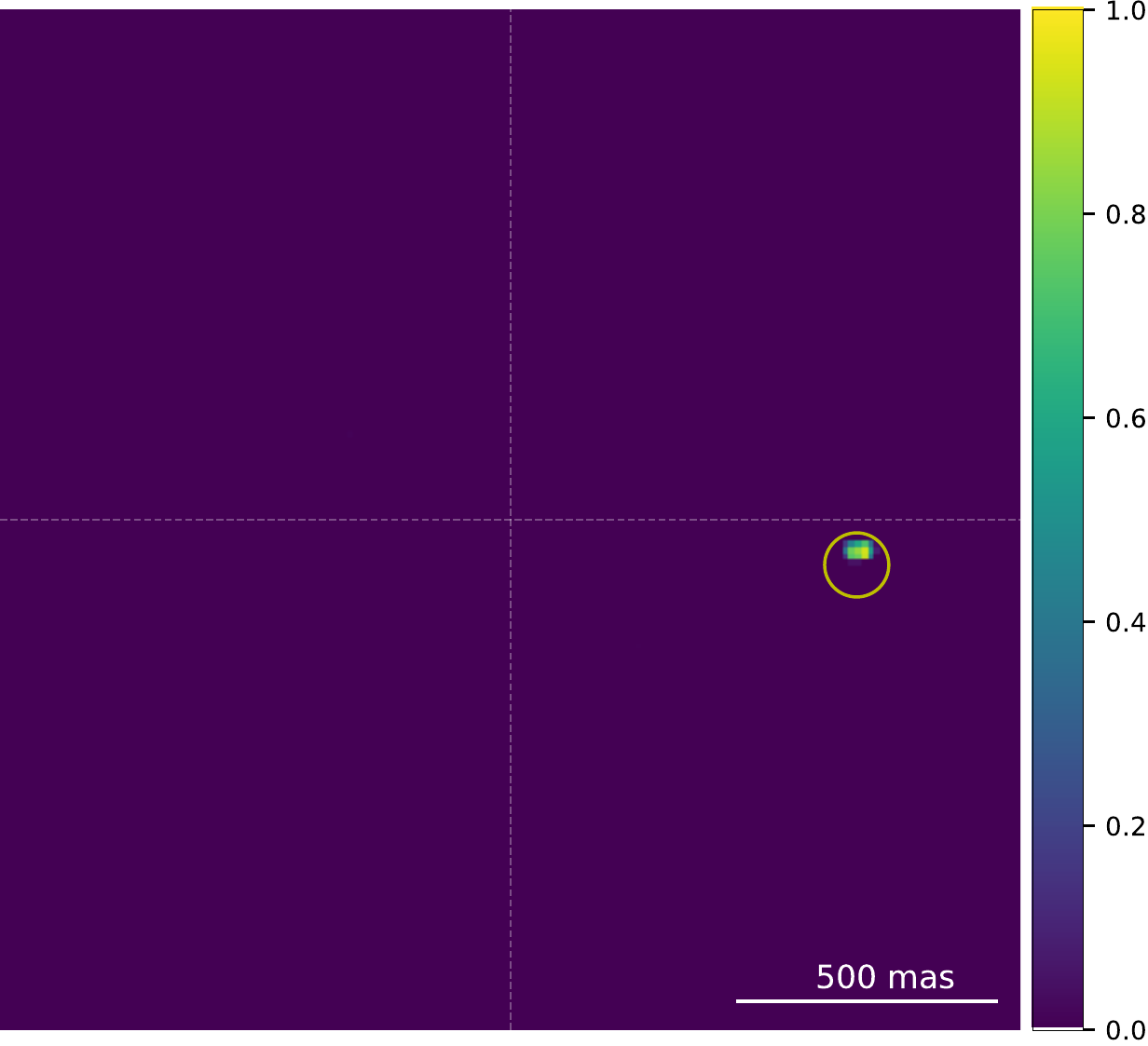}}
      \subfloat[SPHERE-3 A BU F]{\includegraphics[width=115pt]{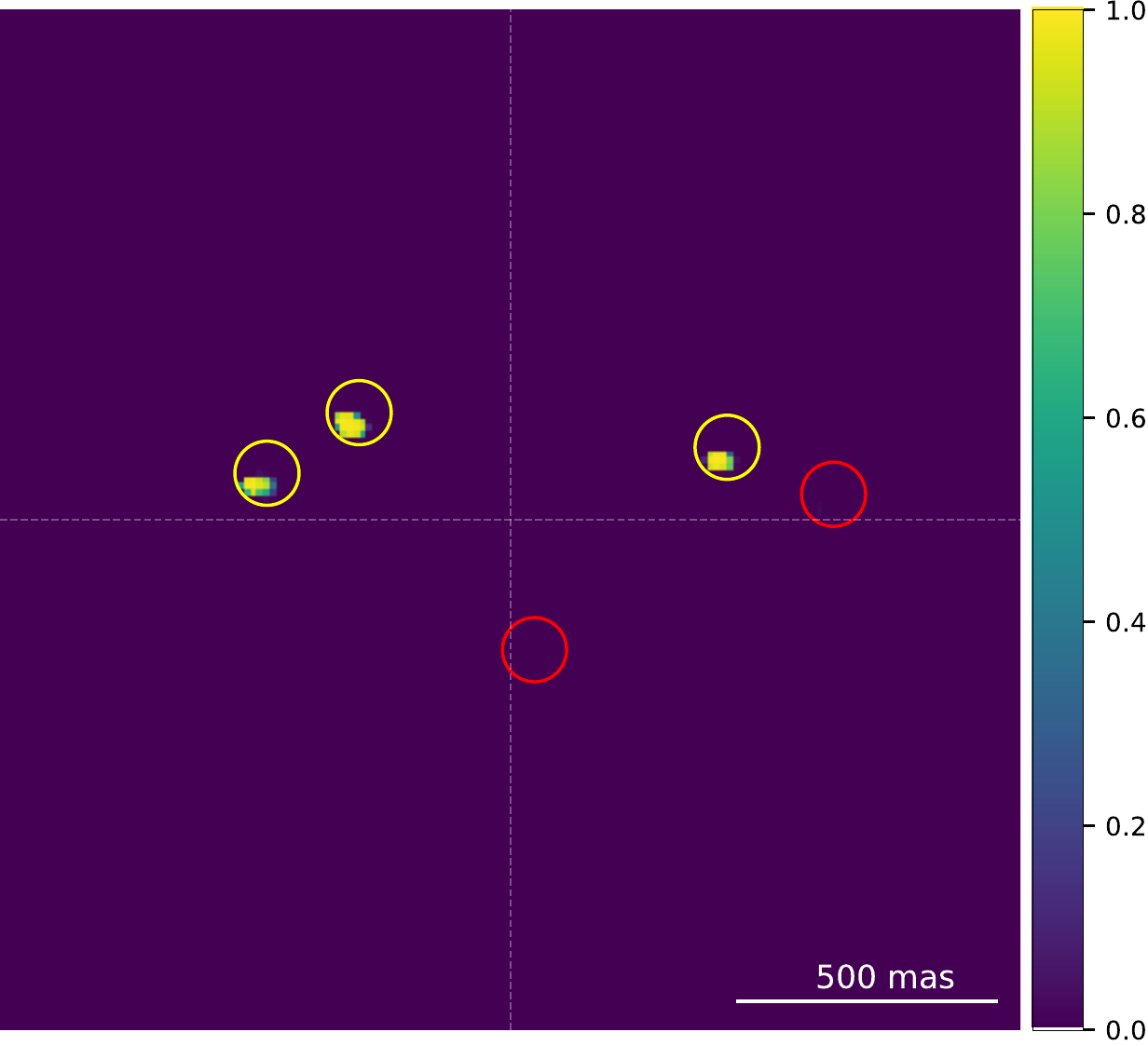}}
        \subfloat[NIRC2-1 A BU F]{\includegraphics[width=115pt]{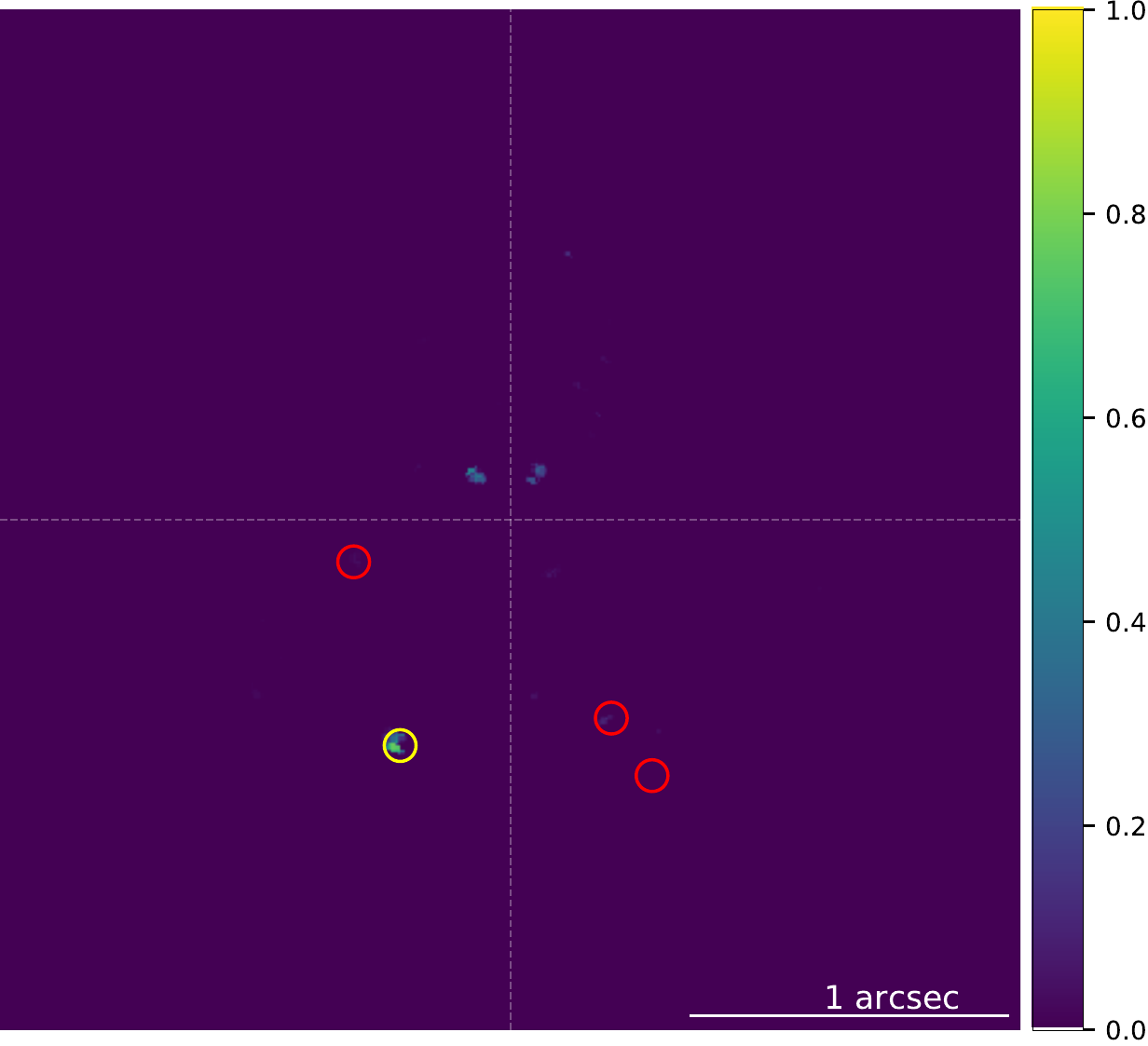}}
             \subfloat[NIRC2-2 A BU F]{\includegraphics[width=115pt]{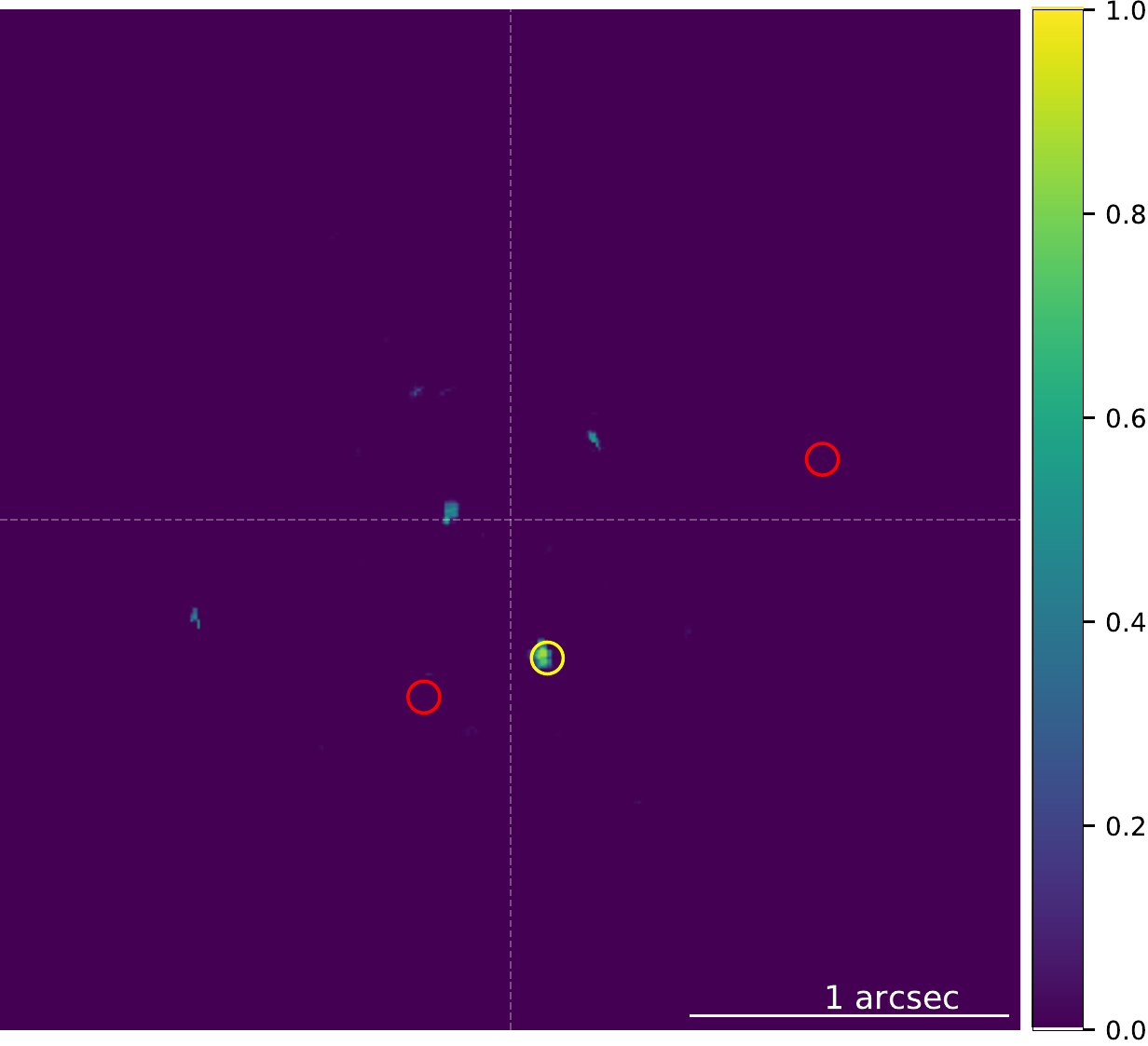}}\\
     \subfloat[SPHERE-1 AFF BU F]{\includegraphics[width=115pt]{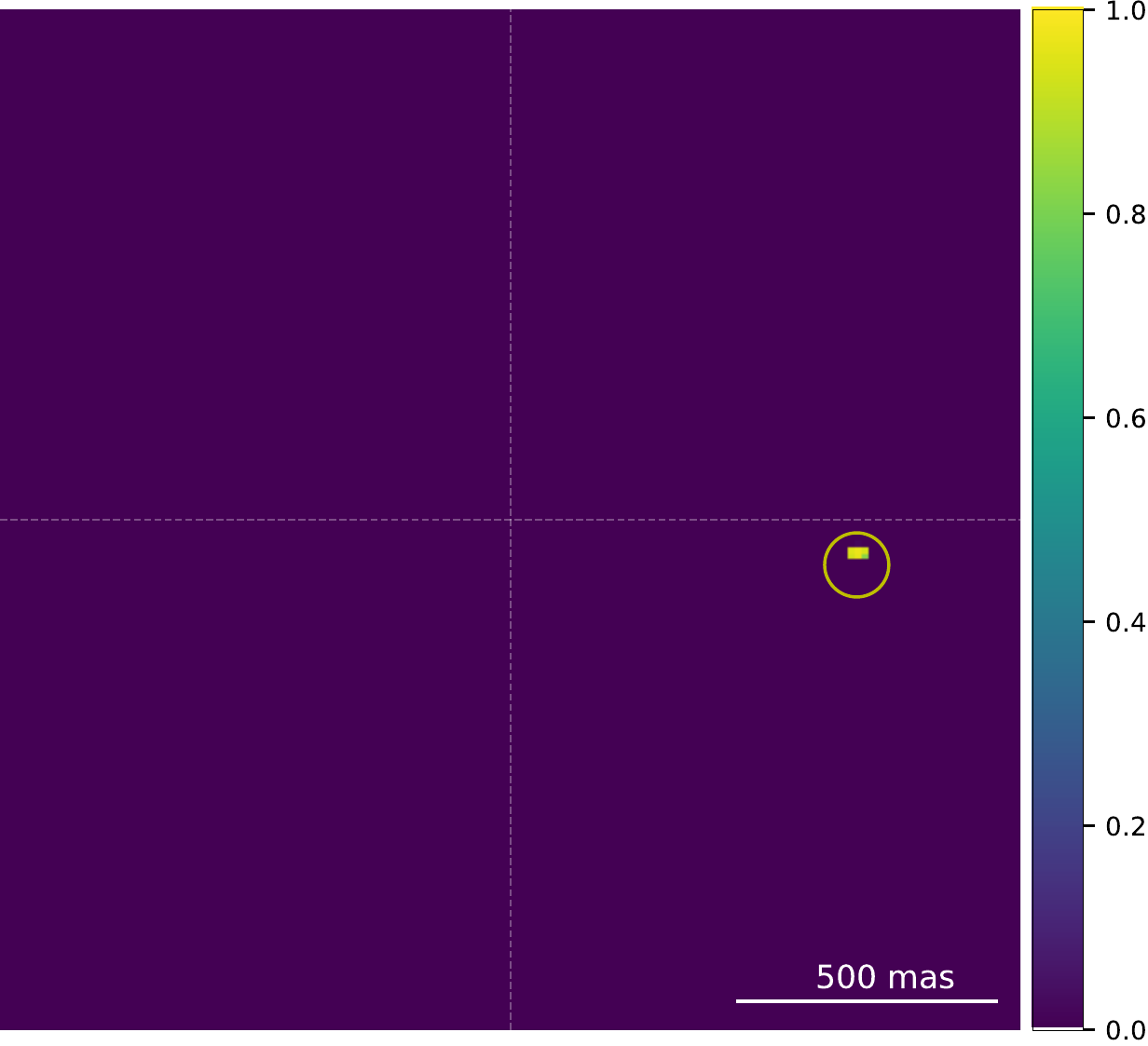}}
      \subfloat[SPHERE-3 AFF BU F]{\includegraphics[width=115pt]{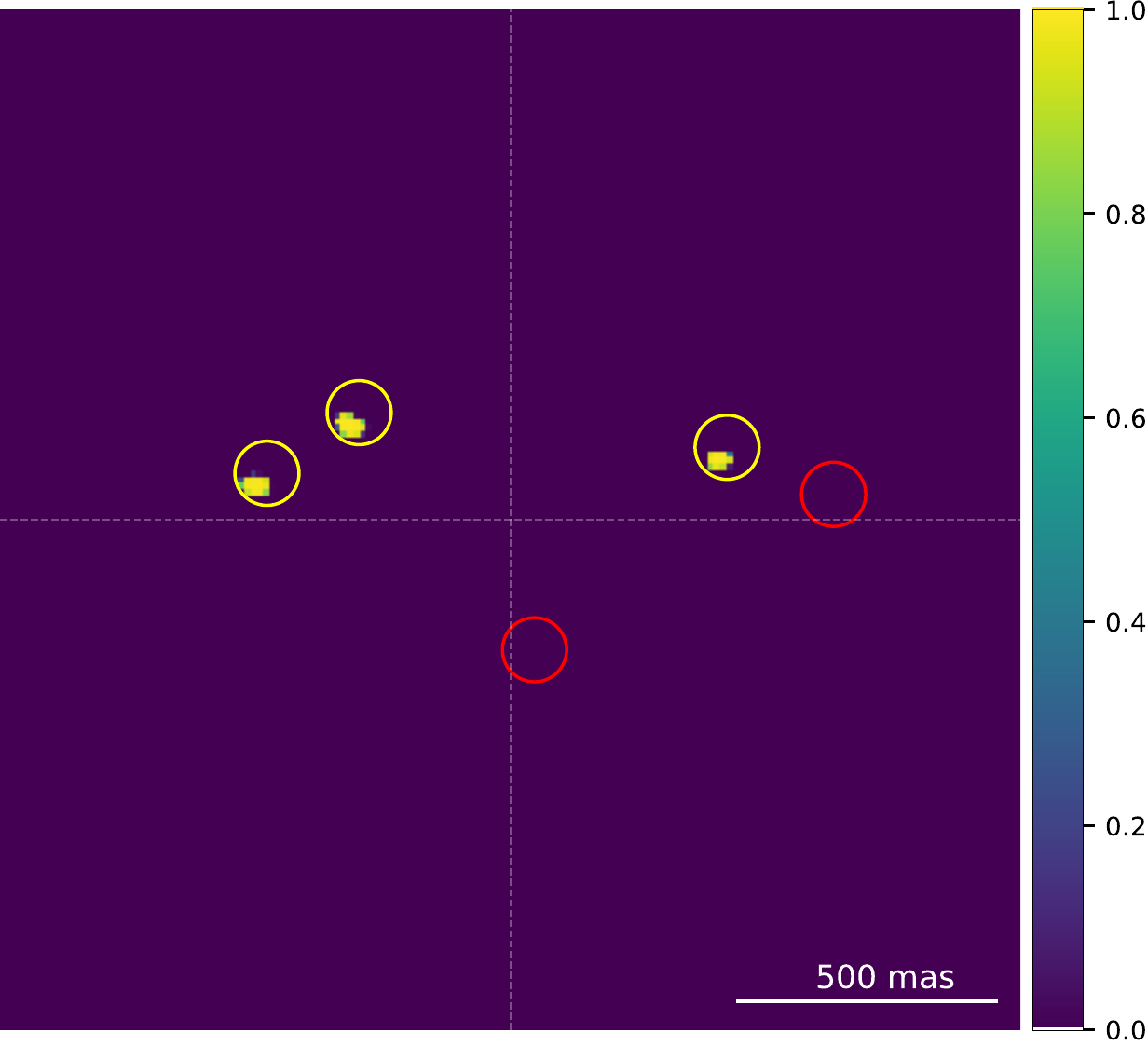}}
       \subfloat[NIRC2-1 AFF BU F]{\includegraphics[width=115pt]{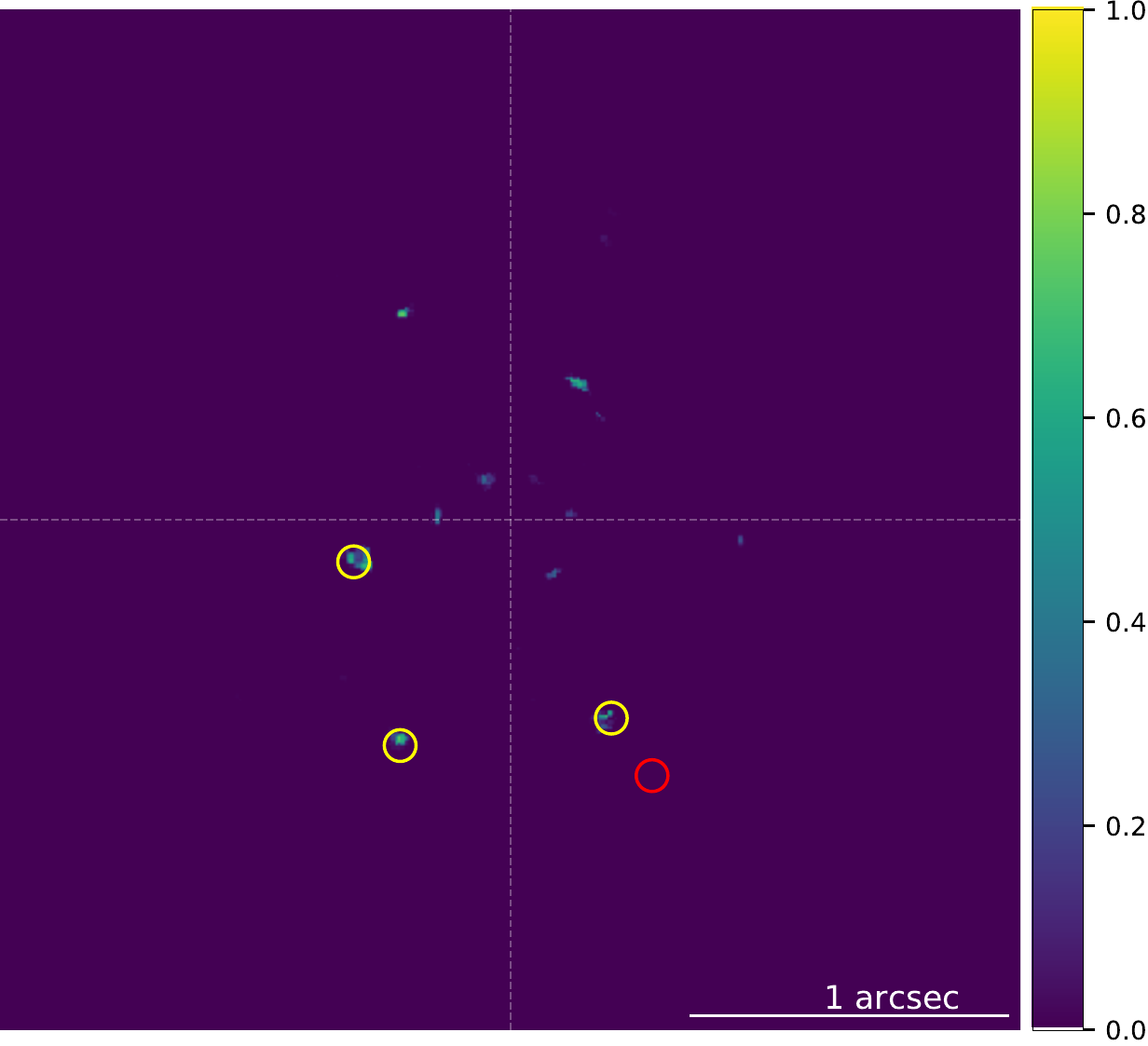}}
          \subfloat[NIRC2-2 AFF BU F]{\includegraphics[width=115pt]{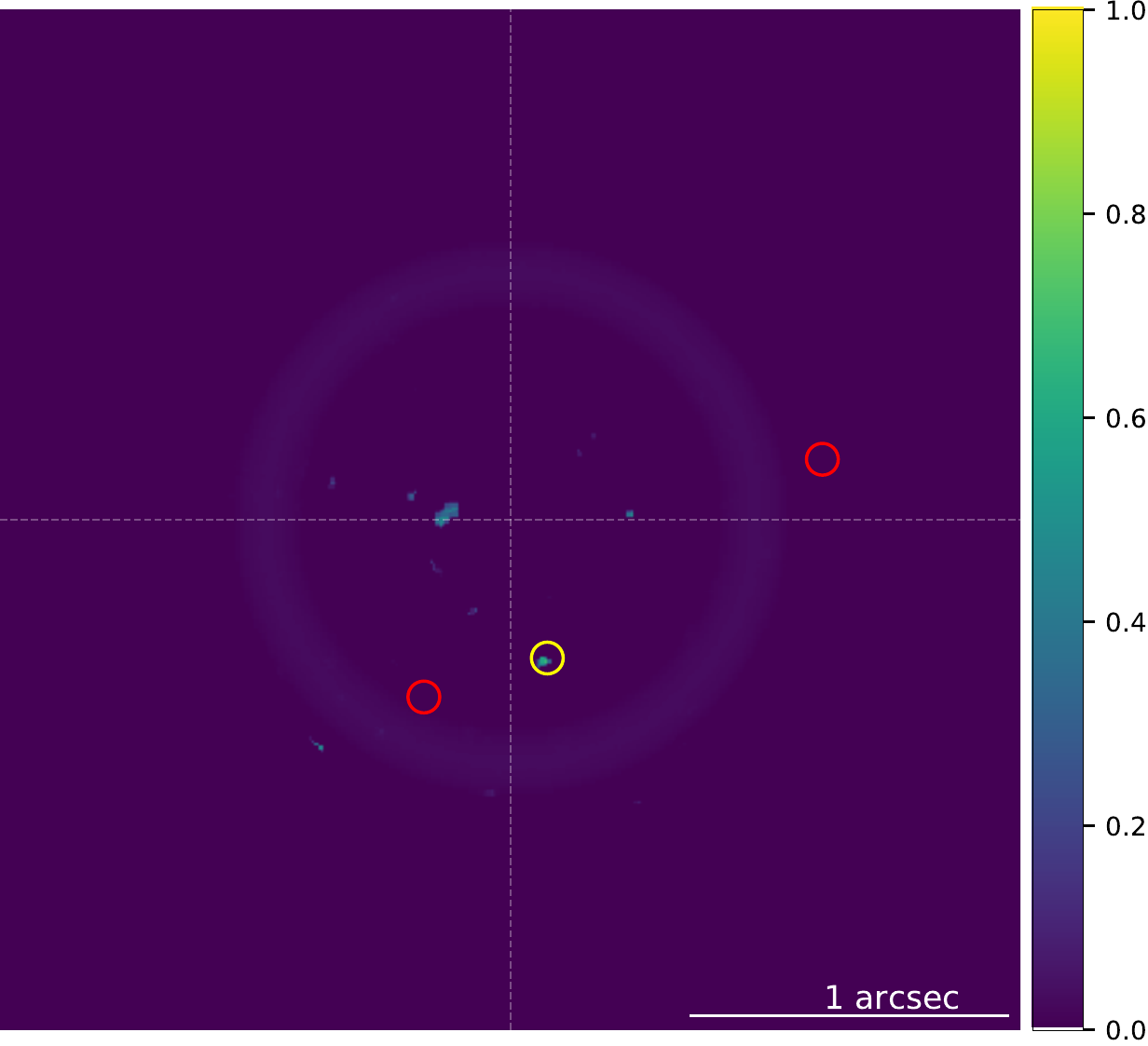}}\\

  \caption{\label{ResRSM1} Detection maps corresponding to the SPHERE and NIRC2 data sets generated with different parametrisations of the full-frame  and annular auto-RSM. The SPHERE-2 and NIRC2-3 detection maps are not shown, as no fake companions were injected in these two data sets. See Sect. 4.3.1. for the definition of the acronyms used to characterise the auto-RSM versions. The yellow circles are centred on the true position of the detected targets (TP) and the red circles give the true positions of FNs. }
\end{figure*}

\begin{figure*}[!htbp]
\footnotesize
  \centering

  \subfloat[LMIRCam-1 FF BU F]{\includegraphics[width=115pt]{LMIRCam_1_FF_BU_F.pdf}}
    \subfloat[LMIRCam-2 FF BU F]{\includegraphics[width=115pt]{LMIRCam_2_FF_BU_F.pdf}}
        \subfloat[LMIRCam-3 FF BU F]{\includegraphics[width=115pt]{LMIRCam_3_FF_BU_F.pdf}}\\
  \subfloat[LMIRCam-1 FF TD F]{\includegraphics[width=115pt]{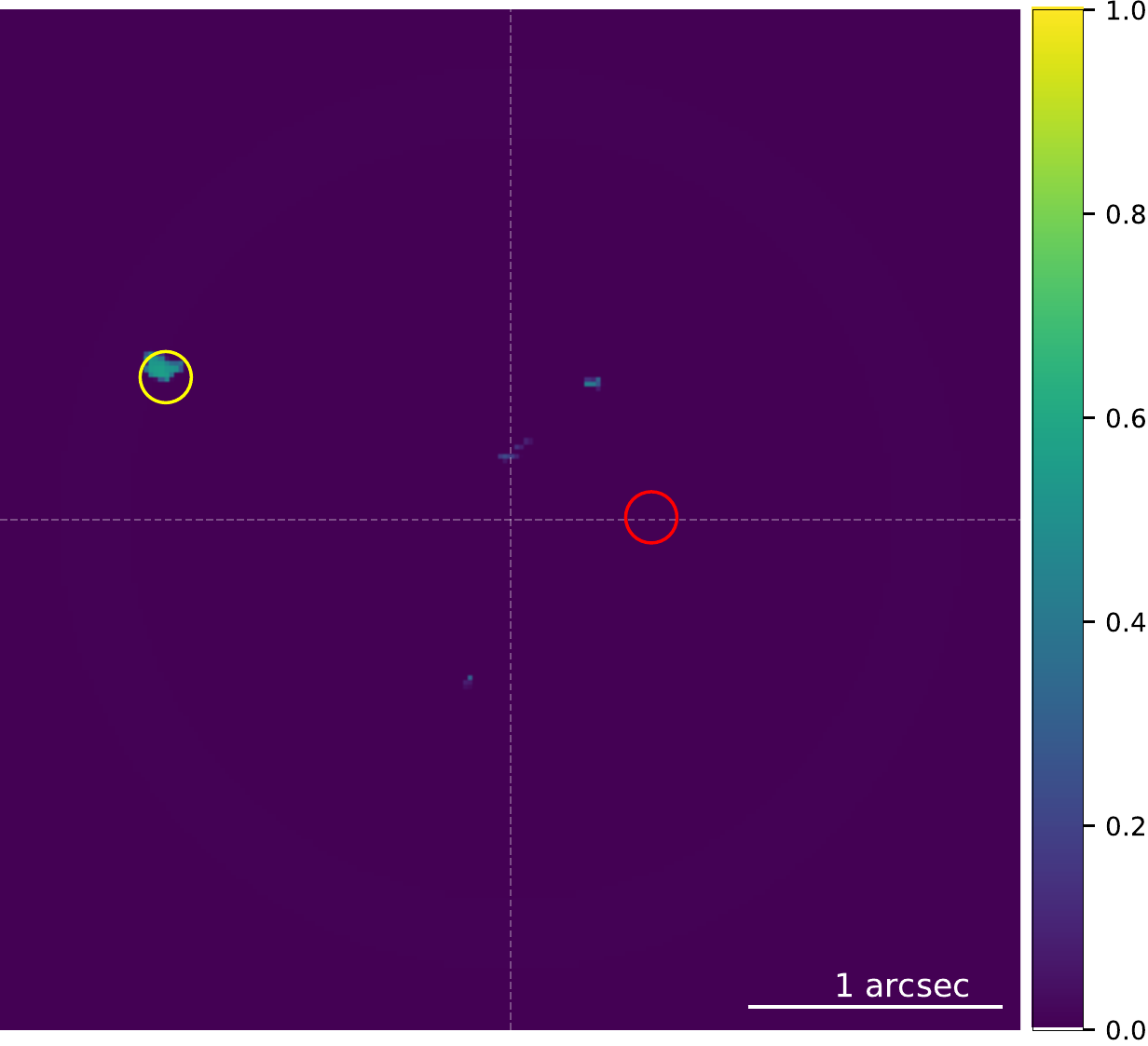}}
    \subfloat[LMIRCam-2 FF TD F]{\includegraphics[width=115pt]{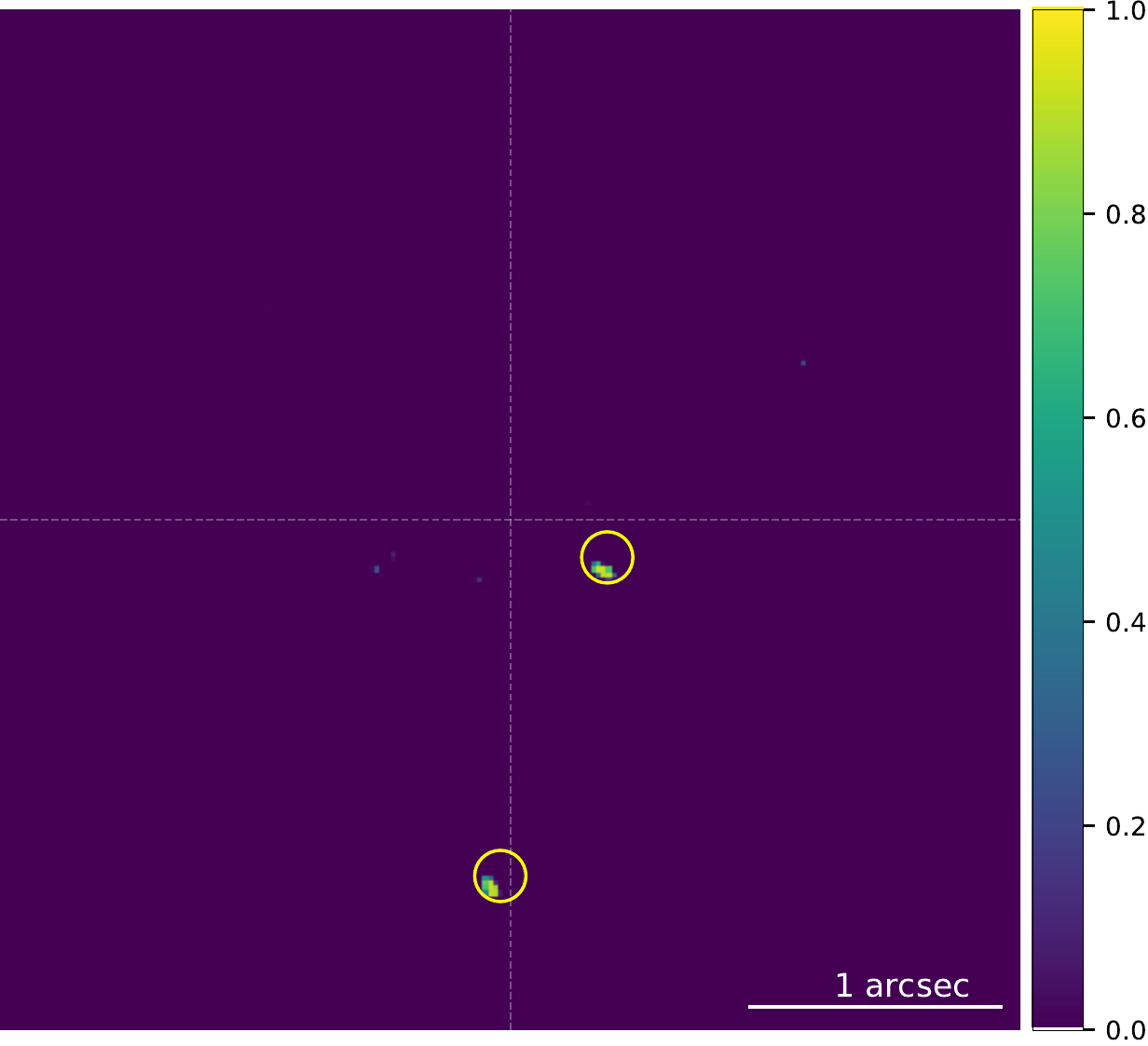}}
       \subfloat[LMIRCam-3 FF TD F]{\includegraphics[width=115pt]{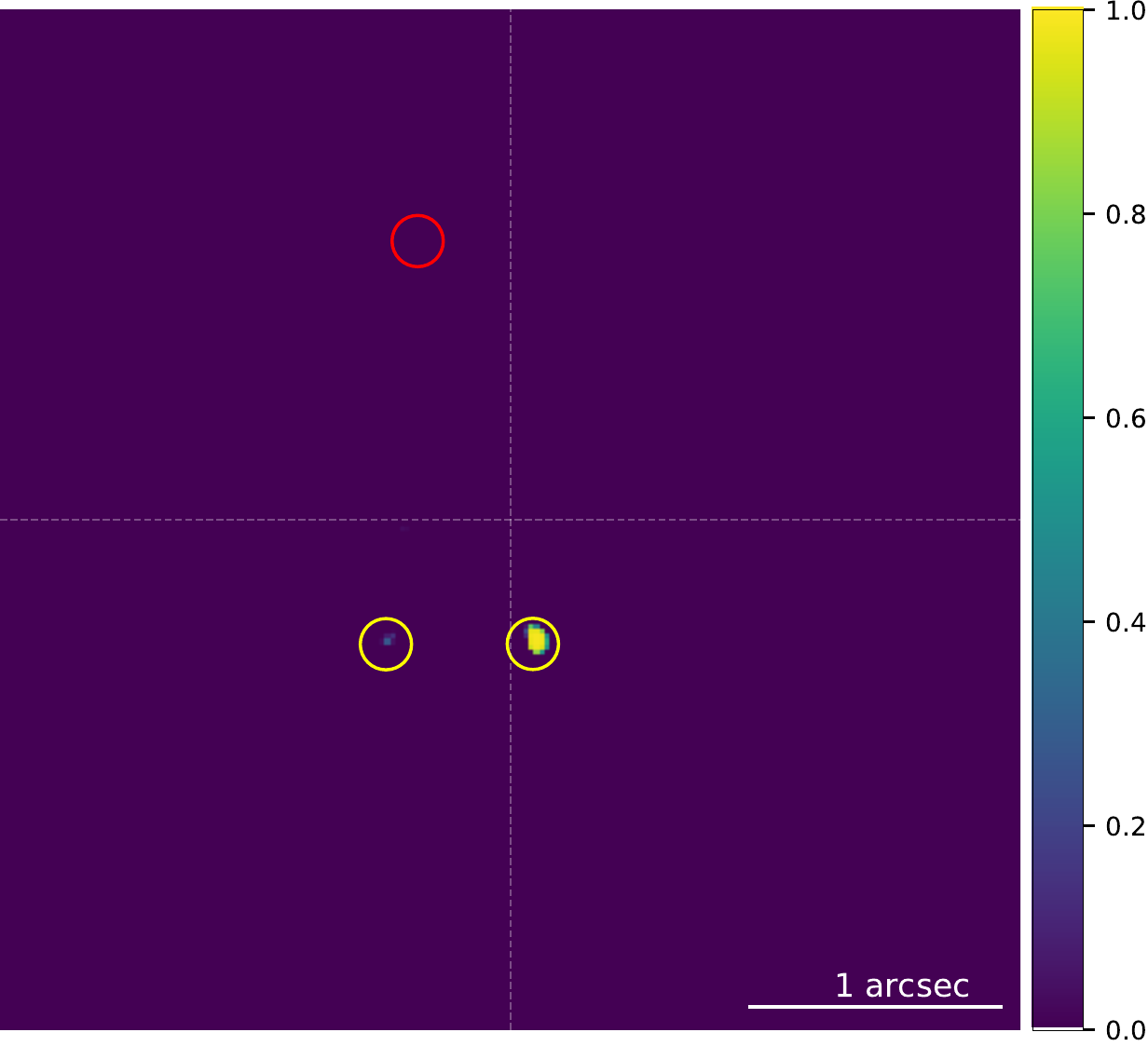}}\\
         \subfloat[LMIRCam-1 FF BU FB]{\includegraphics[width=115pt]{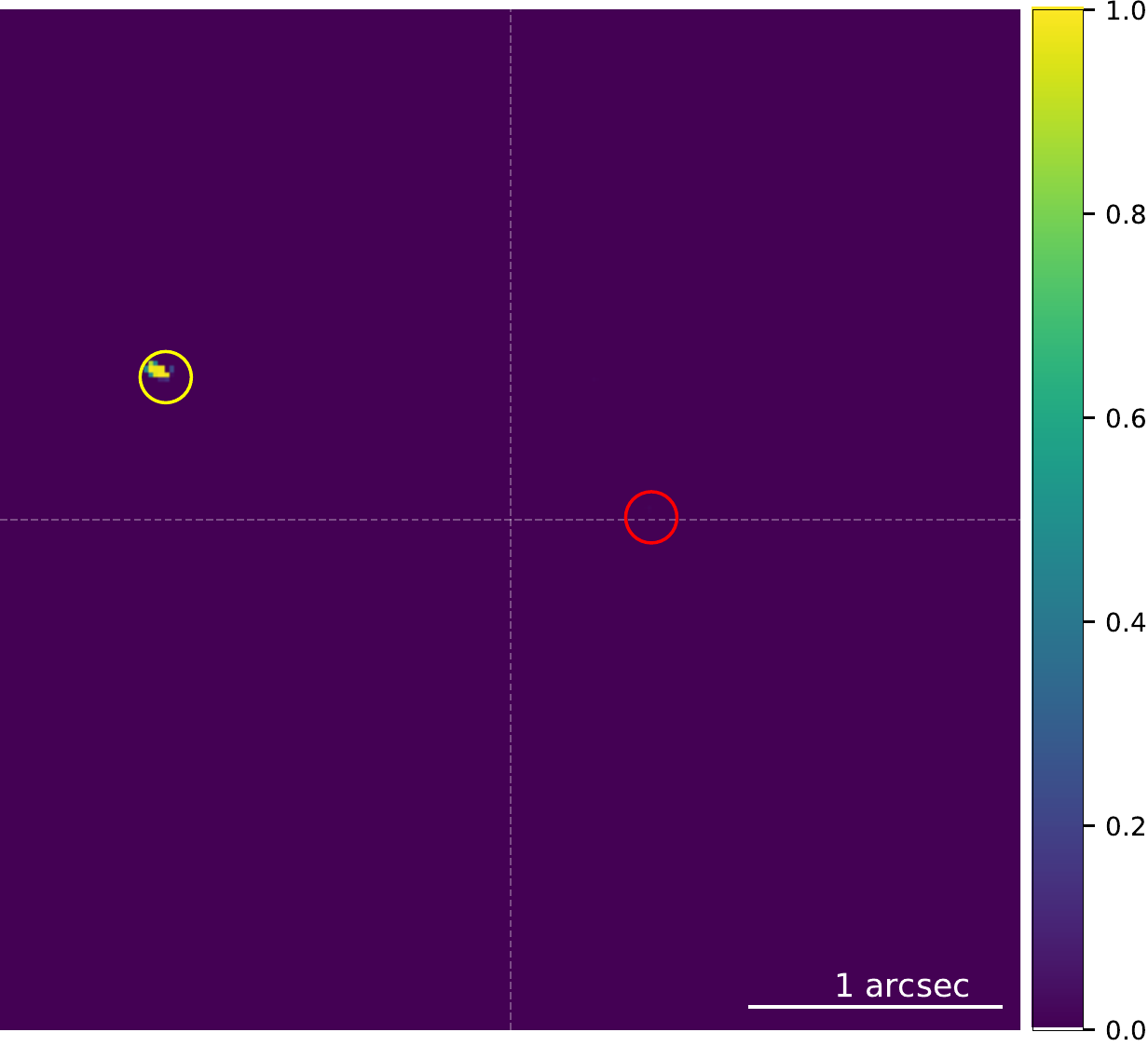}}
    \subfloat[LMIRCam-2 FF BU FB]{\includegraphics[width=115pt]{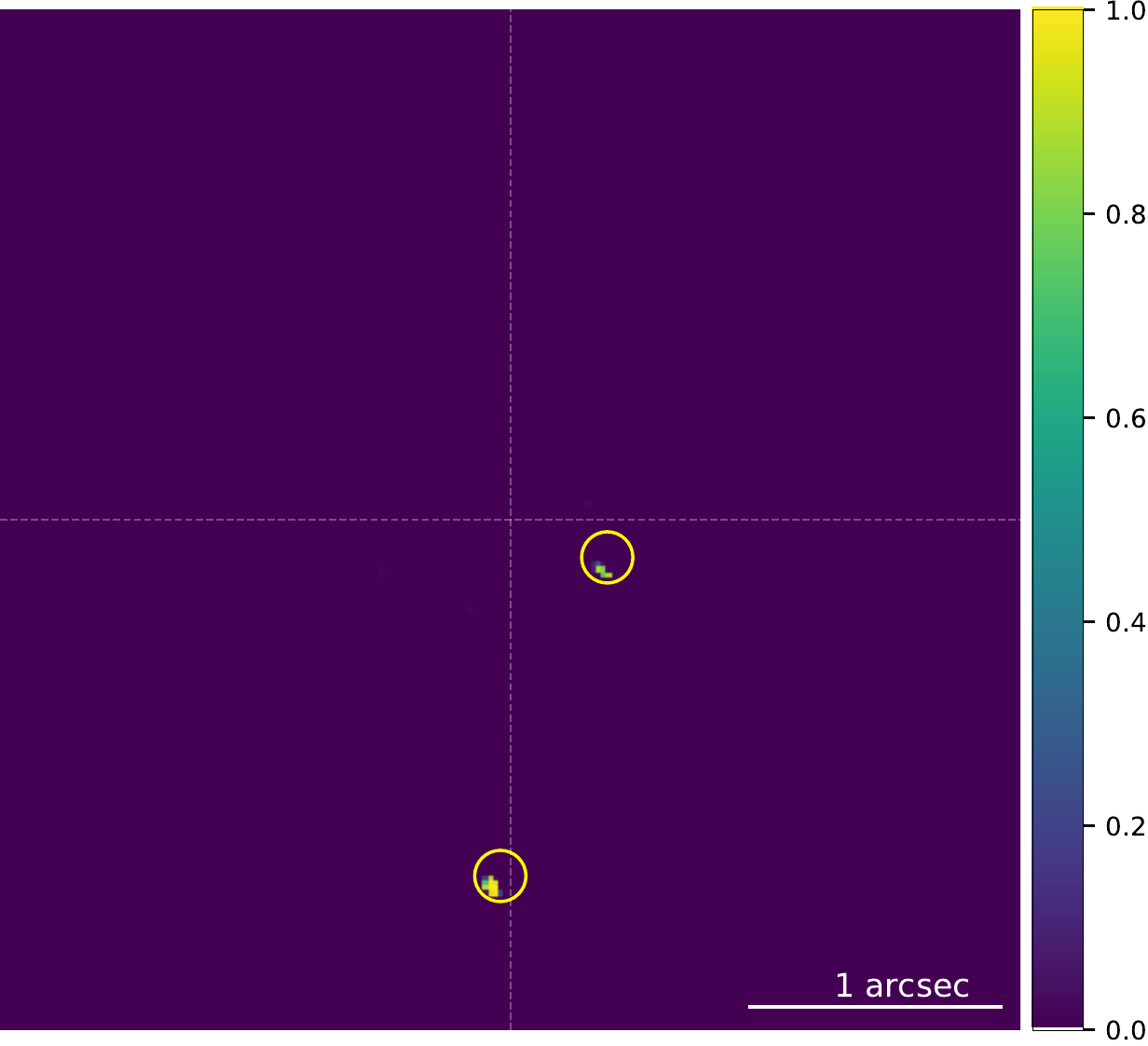}}
        \subfloat[LMIRCam-3 FF BU FB]{\includegraphics[width=115pt]{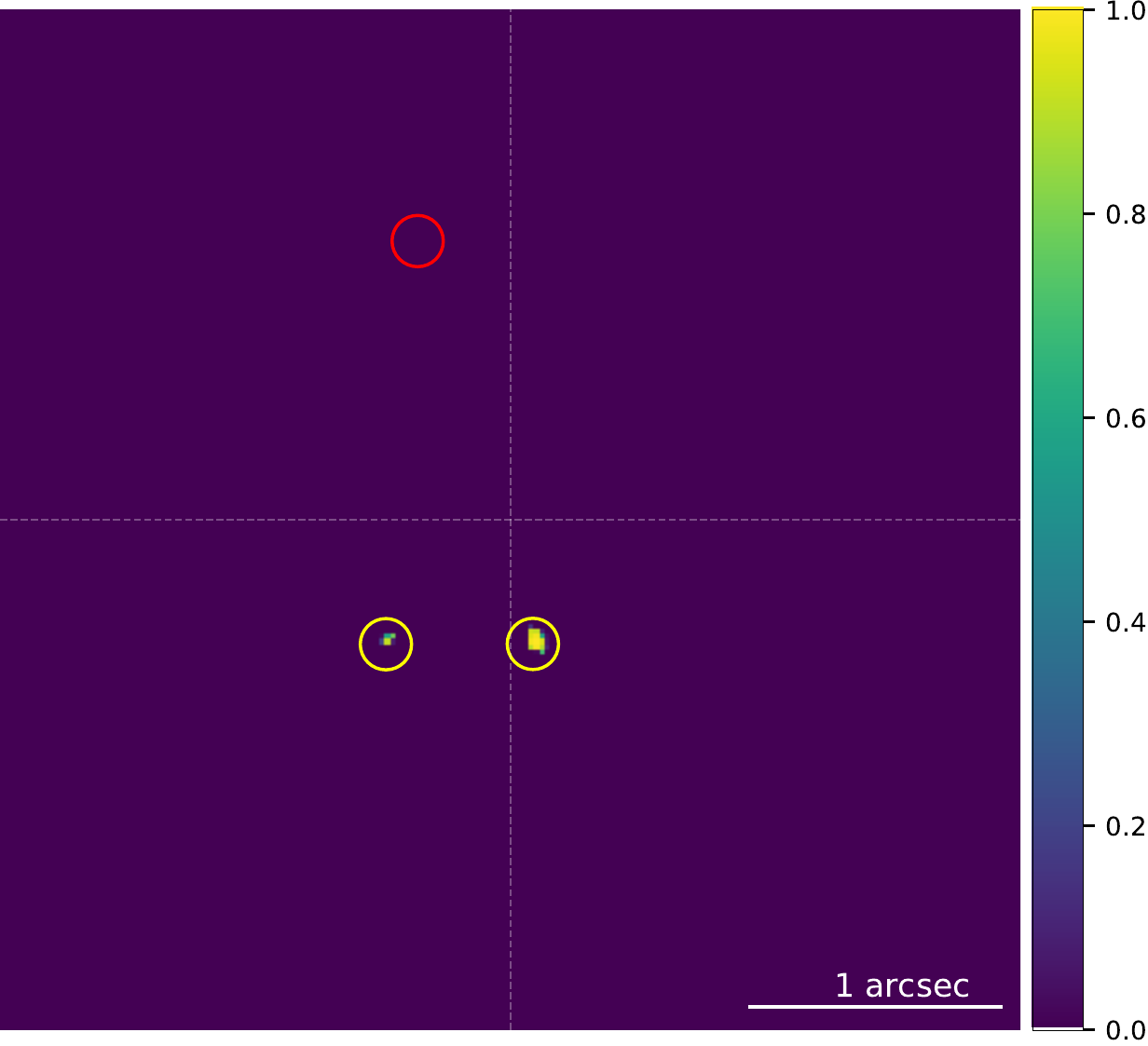}}\\
    \subfloat[LMIRCam-1 A BU F]{\includegraphics[width=115pt]{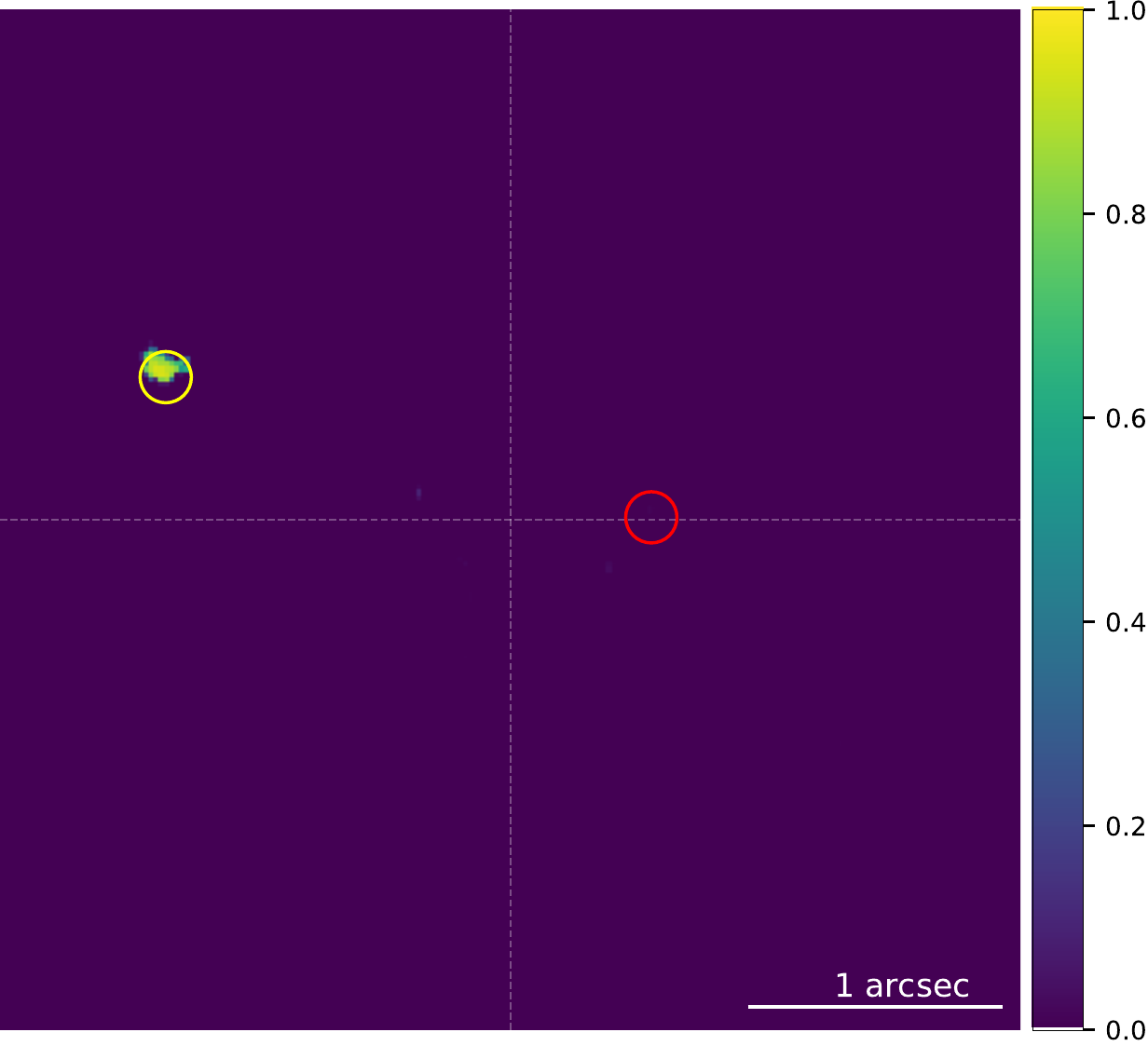}}
        \subfloat[LMIRCam-2 A BU F]{\includegraphics[width=115pt]{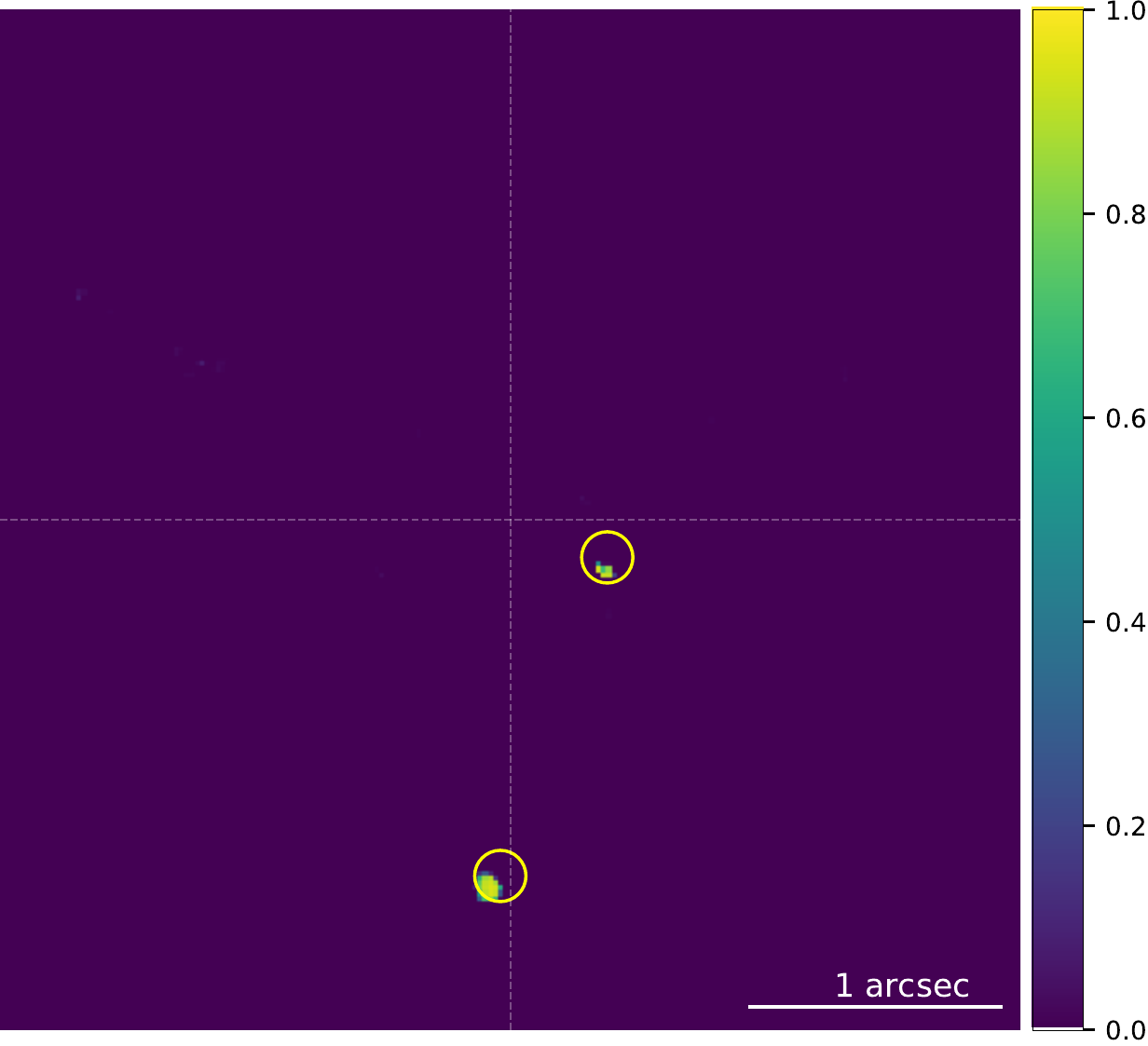}}
                \subfloat[LMIRCam-3 A BU F]{\includegraphics[width=115pt]{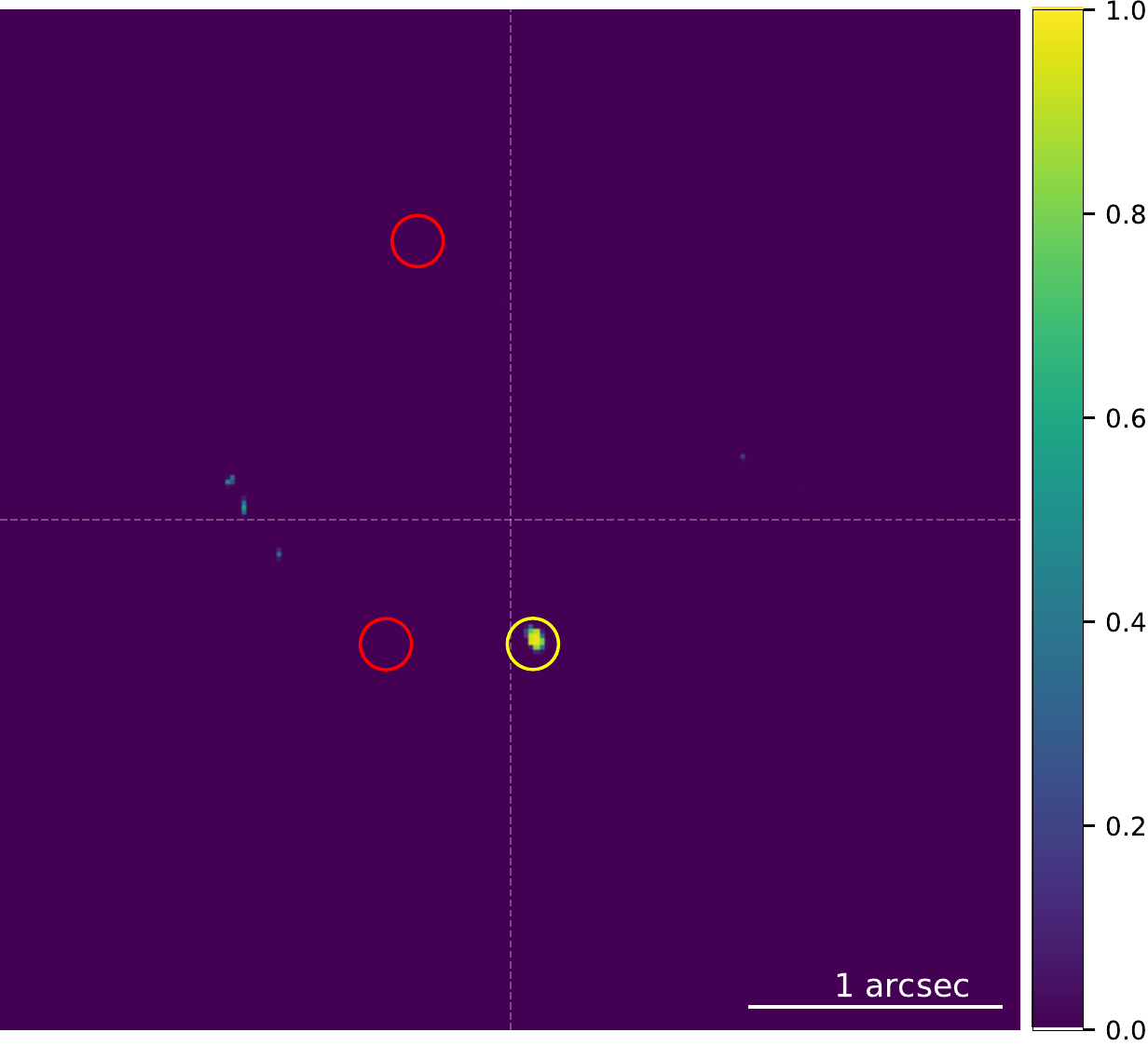}}\\
     \subfloat[LMIRCam-1 AFF BU F]{\includegraphics[width=115pt]{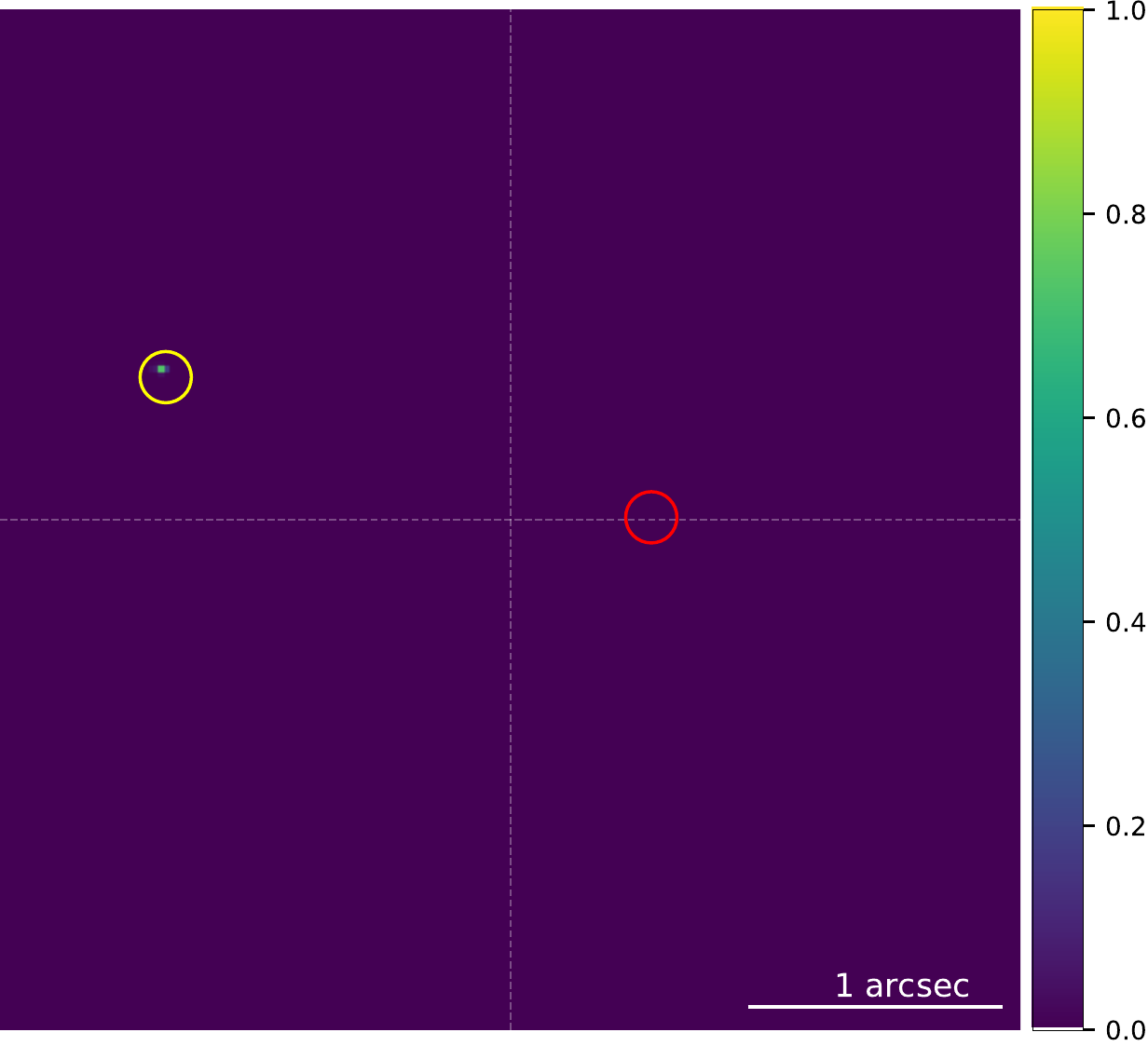}}
      \subfloat[LMIRCam-2 AFF BU F]{\includegraphics[width=115pt]{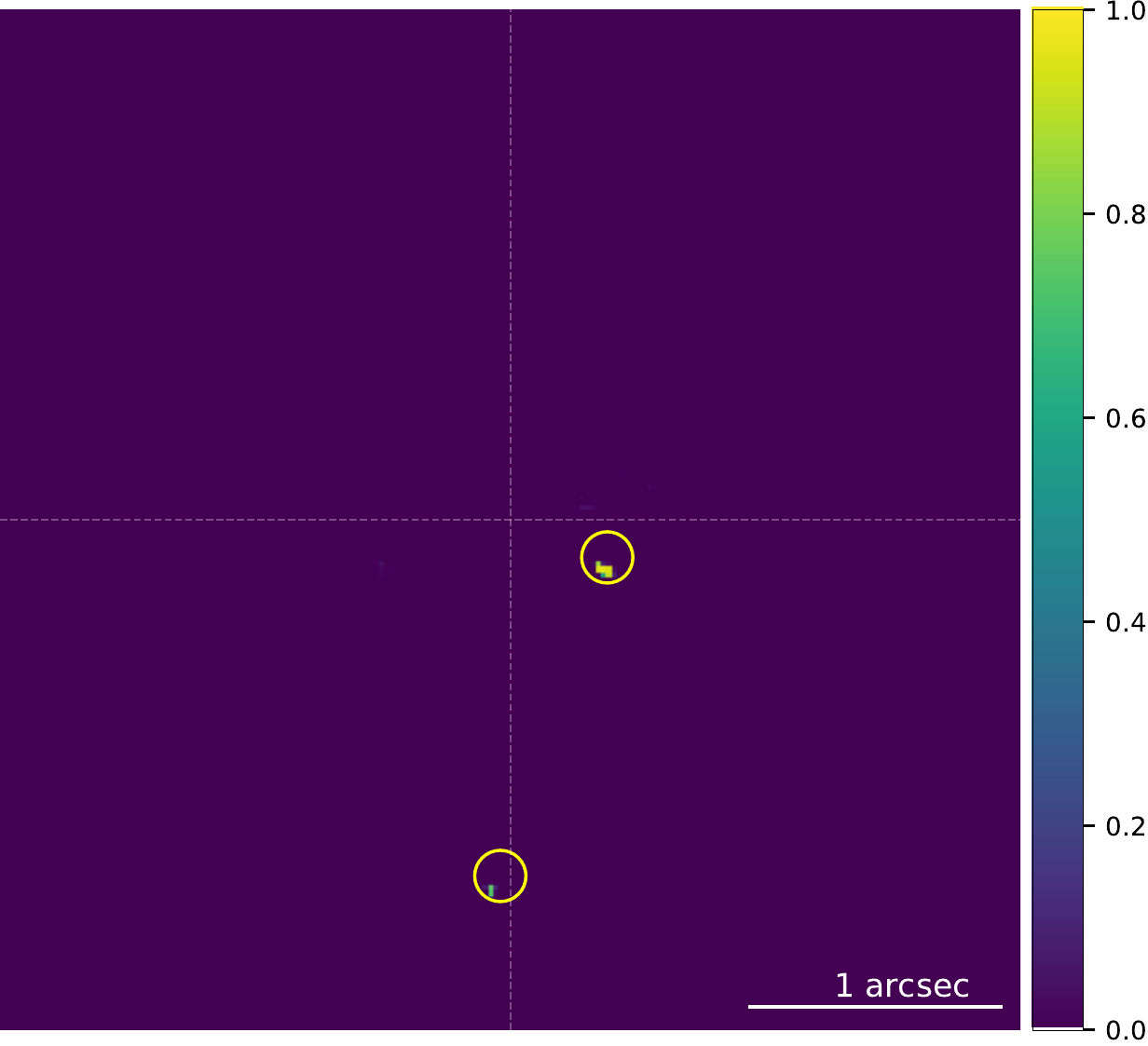}}
          \subfloat[LMIRCam-3 AFF BU F]{\includegraphics[width=115pt]{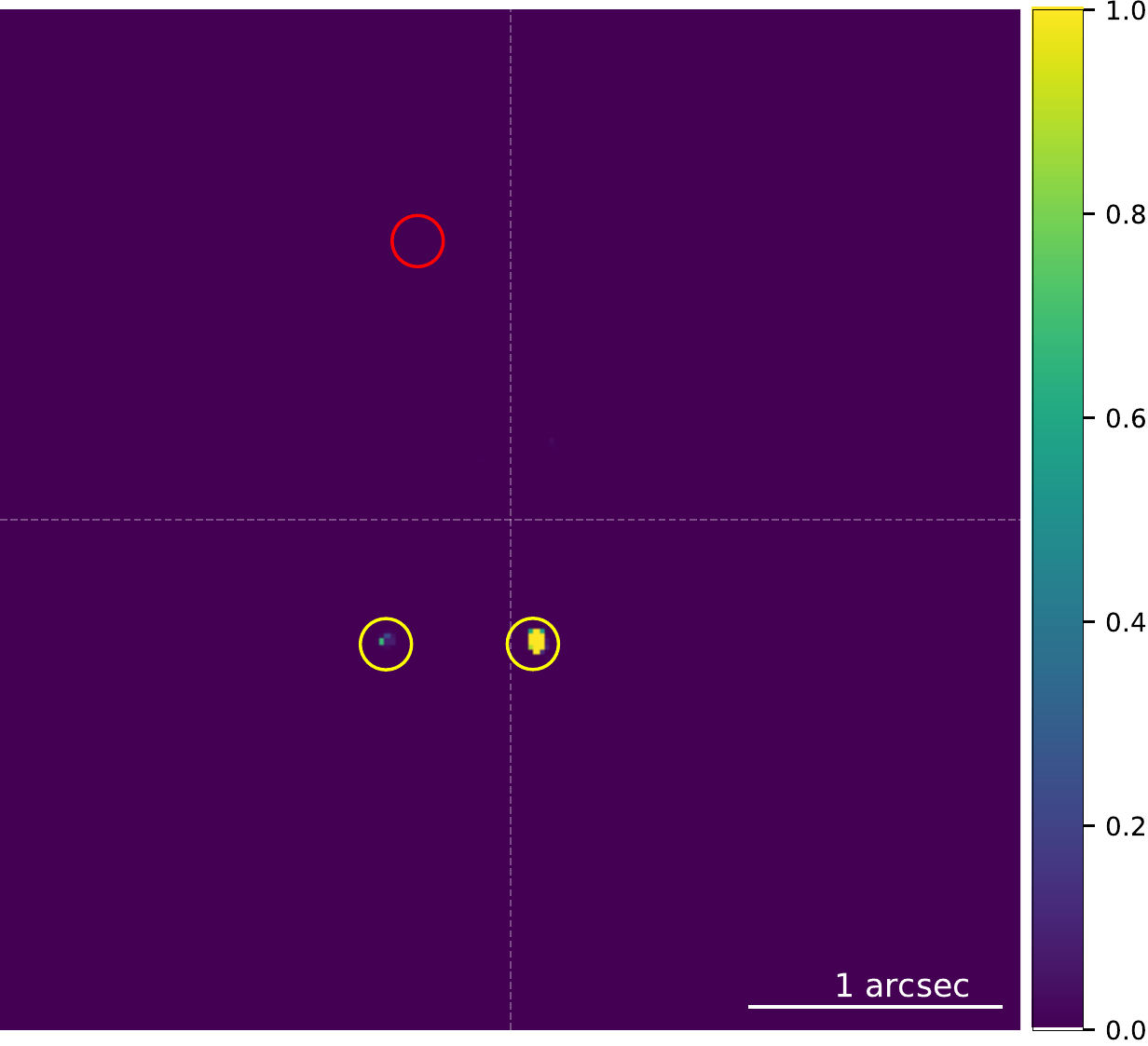}}\\

  \caption{\label{ResRSM2} Detection maps corresponding to the LMIRCam data sets, generated with different parametrisations of the full-frame and annular auto-RSM. See Sect. 4.3.1. for the definition of the acronyms used to characterise the auto-RSM versions. The yellow circles are centred on the true position of the detected targets (TP) and the red circles give the true positions of FNs. }
\end{figure*}

Here, Fig. \ref{ResRSM1} and Fig. \ref{ResRSM2} show the detection maps obtained with four parametrisations of the auto-RSM algorithm for the data sets of the EIDC ADI subchallenge.

\section{Parametrisation for the full-frame auto-RSM}   
\label{paramff} 

Here, Table~\ref{parametersff} regroups the optimal parameters selected with the auto-RSM FF\_BU\_F for the nine ADI sequences of the EIDC ADI subchallenge.
                                
\begin{table*}[!htbp]
                        \caption{Optimal set of parameters for the PSF-subtraction techniques and RSM algorithm for the nine ADI sequences obtained with the auto-RSM FF\_BU\_F.  The `fit' row indicates if the noise properties have been estimated using a best-fit approach while the $\beta$ row indicates if a Gaussian maximum likelihood has been used to compute the intensity parameter. The variance row provides information about the region used for the noise properties computation and translates as follows: ST-Spatio-Temporal estimation, F-Frame based estimation, FM-Frame with mask estimation, SM -Segment with mask estimation, and T-Temporal estimation.}
                        \label{parametersff}
\centering
                        \scriptsize
                        \begin{tabular}{llllllllll}

                        \hline
Parameters/ID &NIRC2-1&NIRC2-2&NIRC2-3&SPHERE-1&SPHERE-2&SPHERE-3&LMIRCam-1&LMIRCam2&LMIRCam-3  \\                             
 \hline         
APCA components& 11     &       17      &       25      &       25      &       18      &       23      &       40      &       42      &       21      \\
APCA segments&4 &       4       &       4       &       2       &       1       &       3       &       4       &       2       &       2       \\
APCA FOV rotation&0.679 &       0.296   &       0.984   &       0.311   &       0.261   &       0.389   &       0.298   &       0.300   &       0.256   \\
NMF components&12       &       2       &       14      &       18      &       18      &       12      &       14      &       20      &       18      \\
LLSG rank& 4    &       5       &       3       &       8       &       8       &       8       &       8       &       6       &       8       \\
LLSG segments & 4       &       4       &       1       &       4       &       2       &       2       &       4       &       2       &       3       \\
LOCI tolerance & 0.00752        &       0.00138 &       0.00425 &       0.00242 &       0.00128 &       0.00887 &       0.00112 &       0.00104 &       0.00218 \\
LOCI FOV rotation & 0.355       &       0.268   &       0.447   &       0.250   &       0.267   &       0.255   &       0.252   &       0.261   &       0.326   \\
APCA $\delta$ & 5       &       5       &       5       &       5       &       5       &       5       &       5       &       5       &       5       \\
NMF $\delta$ &5 &       5       &       5       &       5       &       5       &       5       &       5       &       5       &       5       \\
LLSG $\delta$&5 &       5       &       3       &       4       &       1       &       2       &       4       &       2       &       5       \\
LOCI $\delta$ &5        &       5       &       2       &       5       &       5       &       5       &       4       &       5       &       5       \\
APCA crop size & 3      &       3       &       3       &       3       &       3       &       3       &       3       &       3       &       3       \\
NMF crop size &3        &       3       &       3       &       3       &       3       &       3       &       3       &       3       &       3       \\
LLSG crop size &3       &       3       &       3       &       3       &       3       &       3       &       3       &       3       &       3       \\
LOCI crop size &3       &       3       &       3       &       3       &       3       &       3       &       3       &       3       &       3       \\
APCA Fit & True &       True    &       True    &       True    &       True    &       True    &       True    &       True    &       True    \\
NMF Fit & True  &       True    &       True    &       True    &       True    &       True    &       True    &       True    &       True    \\
LLSG Fit & True &       True    &       True    &       True    &       True    &       True    &       True    &       True    &       True    \\
LOCI Fit & True &       True    &       True    &       True    &       True    &       True    &       True    &       True    &       True    \\
APCA $\beta$ &True      &       True    &       False   &       False   &       False   &       False   &       False   &       False   &       False   \\
NMF $\beta$ &False      &       False   &       False   &       False   &       False   &       False   &       False   &       False   &       False   \\
LLSG $\beta$ &True      &       True    &       False   &       False   &       False   &       False   &       False   &       False   &       False   \\
LOCI $\beta$ &False     &       True    &       False   &       False   &       False   &       False   &       False   &       False   &       False   \\
APCA variance &T        &       FM      &       FM      &       SM      &       T       &       FM      &       ST      &       SM      &       FM      \\
NMF variance &FM        &       FM      &       ST      &       SM      &       FM      &       ST      &       ST      &       SM      &       ST      \\
LLSG variance &FM       &       FM      &       SM      &       SM      &       SM      &       SM      &       ST      &       FM      &       ST      \\
LOCI variance &FM       &       T       &       FM      &       T &     ST      &       SM      &       SM      &       SM      &       SM      \\
\hline
                        \end{tabular}
                                \end{table*}

\end{appendix}

\end{document}